\documentstyle[twoside,12pt]{report}

\newcommand{\be}{\begin{equation}}
\newcommand{\ee}{\end{equation}}
\newcommand{\bea}{\begin{eqnarray}}
\newcommand{\eea}{\end{eqnarray}}
\newcommand{\rref}[1]{(\ref{#1})}
\newcommand{\R}{{\cal R}}

\begin{document}

\begin{titlepage}
\begin{center}
{\large Universit\'e Libre de Bruxelles \\
Facult\'e des Sciences \\
Service de Physique Th\'eorique \\}
\vspace{3cm}

{\Huge \bf
Brane Physics in M-theory }

\vspace{3cm}

Th\`ese pr\'esent\'ee en vue de l'obtention \\
du grade de Docteur en Sciences \\
(grade l\'egal) \\

\vspace{2cm}
{\Large Riccardo Argurio }\\
Aspirant F.N.R.S. \\

\vspace{7cm}
Ann\'ee acad\'emique \\
1997--1998 

\end{center}
\end{titlepage}

\pagestyle{empty}

\mbox{}
\newpage

\setcounter{page}{1}
\vspace*{8cm}
\begin{flushright}
{\em To my parents}
\end{flushright}

\newpage

\mbox{}
\newpage

\mbox{}
\vfill
{\em
Writing the acknowledgements for a thesis which represents four
years of work is a difficult task, and this is why the acknowledgements
are generally written few hours before printing the final version
of the `book'. I apologize in advance for the dullness.

I wish to thank in the first place my advisor, or supervisor, or
directeur de th\`ese, Fran\c{c}ois Englert, for teaching me simply
by his own way of working, what should be the attitude of any theoretical
physicist. To be curious about the new developments, to question 
the foundations of any theory whatever `mainstream' it is, to
appreciate the subtle balance between mathematical structure and
physical insight. It goes without saying that I am also grateful
for his constant advices, for introducing me to quantum gravity
(from black holes to string dualities), for the collaboration
on most of my research, for carefully reading the manuscript of this thesis
and for putting up with my moods and my naivety from time to time.

Special thanks are due to Laurent Houart, who 
has been indispensable for carrying on the research which led to this
thesis, and who has been my main interlocutor on physics (and other)
matters for the last two years. Thanks a lot also for the spectral
reading of the manuscript.

It is also a pleasure to thank Paul Windey for the collaboration
on one article, and all my colleagues at the Service de Physique Th\'eorique,
namely and in random order Marc Henneaux, Christiane Schomblond,
Philippe Spindel, Robert Brout, Jean-Marie Fr\`ere,
Glenn Barnich, Andr\'e Wilch, Bernard
Knaepen, Lubo Musongela, Abdelillah Barkallil, Marianne Rooman,
Pietro Castoldi and Michel Tytgat, not to forget Peter van Driel and
Ruud Siebelink with whom I had many discussion.

Particularly warm thanks go to Carmelo Iannuzzo and Karin Bautier
with whom I shared the office at the beginning and at the end of these
four years respectively. They have proven themselves skillful in enduring my
changing moods.

I would also like to thank my (young) colleagues which I met during schools
and conferences, and especially Edu Eyras and Dimi Matalliotakis for
our virtual physics non-physics correspondence, and Walter Moretti.

Finally I think that an invaluable and possibly the strongest 
support has come from people outside
the {\tt hep-th} community, and since it
would be too long and difficult to specialize this part of the
acknowledgements, I will just say: thank all!!
}

\newpage
\mbox{}
\newpage

\tableofcontents

\pagestyle{headings}

\chapter{Introduction}

M-theory is the name which designates the recent efforts to unify by 
non-perturbative dualities all the known superstring theories.
It is a necessary step if string theories are to be related to a unique theory
of all fundamental forces.

In the attempt to unify all the forces present in Nature, which entails
having a consistent quantum theory of gravity, superstring theories
seem promising candidates. They appear as theories achieving the unification
`from above'. This is certainly appealing from a theoretical point of view,
but it makes much more difficult the task to provide the concrete picture
of how the unification occurs in the `real world'.
The evidence that string theories could be unified theories is provided by
the presence in their massless spectrum of enough particles to account for
those present at low-energies, including the graviton.
At the phenomenological level however, the difficulties associated 
with string theories show up when
one tries to reduce the dimensionality of space-time from ten
(the critical dimension for the superstrings) to four (the number of
dimensions that are accessible to us, and to present-day accelerators),
and when one also attempts to reduce the large amount of supersymmetry
which characterizes the superstring spectrum.

However string theories have also theoretical drawbacks. First of all,
string theories are properly speaking theories in which a single string
is quantized perturbatively. There is no second quantized string theory,
or string field theory. Our knowledge of string interactions, and string
loop expansion, is entirely based on some sort of S-matrix theory.
This means that we are able to compute scattering amplitudes between
string states, but we are not able to formulate a path integral from which
to derive such amplitudes. The inability to formulate a path integral
is merely due to the inexistence of an action (or an Hamiltonian)
describing string field theory. If we restrict our attention
on the massless modes of the string, i.e. we focus on energies well
below the scale fixed by the string tension, we can however use supersymmetry
to argue that the low-energy effective action of these modes
is given precisely by the corresponding supergravity. Such supergravities
are non-renormalizable, but they are relevant essentially for the
classical solutions that they contain, and which are to be identified
with the solitons of the full string theory.

The second main drawback of superstring theories is their number. There are
five string theories which all could pretend to be the unique theory.
Some of them are more realistic than the others, but one would like
to have a more theoretical reason to reduce the number of consistent
string theories. A different way to solve this problem is to show 
that all these five string theories are actually unified in a single theory.

The unification cannot however occur at the perturbative level, because
it is precisely the perturbative analysis which singles out the five
different string theories. The hope is that when one goes beyond this
perturbative limit, and takes into account all non-perturbative effects,
the five string theories turn out to be five different descriptions
of the same physics. In this context, which is now called M-theory,
a duality is a particular relation applying to string theories,
which can map for instance the strong coupling region of a theory to the
weak coupling region of the same theory or of another one, and vice-versa, thus
being an intrinsically non-perturbative relation.
In the recent years, the structure of M-theory has begun to be uncovered,
with the essential tool provided by supersymmetry. Its most striking 
characteristic is that it indicates that space-time should be
eleven dimensional. 

Because of the intrinsic non-perturbative nature of any approach to
M-theory, the study of the $p$-brane solitons, or more simply `branes',
is a natural step to take. The branes are extended objects present
in M-theory or in string theories, generally associated to classical
solutions of the respective supergravities. A 0-brane is simply a
point-like object, a 1-brane is a string, a 2-brane a membrane, and so on.
Leaving aside the fundamental strings which can be called abstrusely 
1-branes, all the
other branes are effectively non-perturbative objects, their mass (or
tension) being proportional to an inverse power of the (string) 
coupling constant.

The study of the branes reveals itself very interesting and full of surprises.
The branes first appear as solutions of the supergravities which are
the low-energy effective theories of the superstring theories, and of
eleven dimensional supergravity which is assumed to be the low-energy
effective theory of M-theory. Most of these solutions were
known before the dualities in string theory became the object of
intensive research. However it is only recently that these objects
have been treated as truly dynamical objects.

Already at the level of the classical supergravities, the analysis
of solutions made up of several different branes gives interesting
results. It is then possible to study some of these solitons directly
in perturbative string theory, i.e. as D-branes. This gives an
enormous amount of information on their dynamics, and some of these
results can even be transposed by dualities to the other branes which
do not benefit from such a `stringy' formulation. Most of the branes turn out
to have non-trivial 
low-energy effective theories defined on their world-volume, and 
these theories can be studied on their own.
This latter property of the branes allows their use in addressing
some very important problems. Indeed, the embedding in M-theory has 
led to serious advances in such problems as the black hole entropy
and the study of non-perturbative aspects of field theories.

The aim of Chapter \ref{MINTROchap} is to give an introduction to
M-theory, and to some of the objects that are relevant to its
definition. Accordingly, the chapter contains a very brief 
introduction to (perturbative) string theory. The notion of duality and
the various dualities in string theory and in M-theory are then reviewed.
The chapter also contains a section on D-brane physics, and another
on the Matrix theory approach to M-theory.

The chapters which follow contain material which is based on original
results.

In Chapter \ref{PBRANEchap} we concentrate on the classical $p$-brane
solutions. The set up is actually more general than the one of a 
particular supergravity, mainly because for most of the time
we only consider bosonic fields. Section \ref{SINGLEsec} is devoted to
reviewing and rederiving the known solutions carrying one antisymmetric
tensor field charge. Both extreme and non-extreme (black) branes are
considered, as well as a self-contained derivation of their
semi-classical thermodynamics.
In Section \ref{INTERsec} we go on presenting the original results
of \cite{rules}. General solutions consisting of several branes intersecting
with definite pairwise rules, and forming extreme marginal bound states, 
are derived from the equations of motion. The intersection rules
thus derived have a compact and very general form, which can be applied
to all the cases of interest in M-theory.
We then present an alternative and original way of deriving the
intersection rules for the branes of eleven dimensional supergravity 
asking that the solution preserves some supersymmetries.
We also derive a solution which is extreme but not supersymmetric,
and thus turns out to be a different case from the previous ones.

In Chapter \ref{OPENchap} we consider in detail the possibility that
some of the intersection rules previously derived can reflect the
property of some branes to open, with the boundaries attached to
some other brane \cite{opening}. We thus check directly that the charge carried 
by the open brane is still conserved in the process. The mechanism
by which the charge is conserved leads to the identification of the 
boundary of the open brane to a charged object on the world-volume 
of the brane which hosts it. Since a charged object couples to
an antisymmetric tensor field, this procedure also fixes the field
content of the world-volume effective theory of the host brane. 
All the relevant cases in M-theory are reviewed.
This chapter and the previous one are also partially based on 
\cite{proceeding}.

Chapter \ref{LITTLEchap} is devoted to the study of the theories
of extended objects that can be defined on the world-volume of
some of the branes of M-theory, when they are decoupled from bulk effects
\cite{little}. These theories, which we call `little theories', have
extended objects corresponding to intersections between the host brane and 
some other brane. We define two different theories of little strings in six 
dimensions, which are obviously non-critical, and a theory in seven 
dimensions which possesses a membrane. Dualities can then be defined which
relate all the above little theories. The low-energy effective theories
of all the little theories are argued not to contain gravity.
The definition of three more little string theories with fewer
supersymmetry and additional group structure allows us to draw
a parallel between these little theories and the ones in ten and
eleven dimensions, based on the fact that the patterns of dualities are
the same.

In Chapter \ref{SCHWARZchap} we study the entropy of the four dimensional
Schwarzschild black hole in the context of M-theory \cite{schwarzbh}. 
By a repeated use of boosts along internal directions and of dualities,
we map a neutral Schwarzschild black brane to a configuration carrying
four charges. In the infinite boost limit, which coincides with 
extremality for the charged configuration, the entropy can be
accounted for by a microscopical counting previously done in the
literature, using D-brane techniques. We argue that the statistical
interpretation of the entropy maps back to the Schwarzschild black hole.

\chapter{What is M-theory?}
\label{MINTROchap}

Rather than answering the question in the title of this chapter,
we will try here to review how this question arose in the very
last few years.
The question above became a relevant one when it was realized
that all superstring theories, and 11 dimensional supergravity,
are actually related by an incredibly extended web of dualities,
and can thus be actually unified in what goes now under
the name of M-theory\footnote{Our ignorance about the complete
structure of M-theory already shows up dramatically in the fact
that we still do not know exactly what M stands for. The extensive
research about this particular problem, which requires notions of history,
philology, and psychology, will not however be reviewed here.}.
The latter, which we will loosely refer to as a `theory' even if it has
not received a firm formulation yet,
seems a very serious candidate for a unique theory
which unifies all known interactions, since it resolves the
problem posed by the existence of several consistent superstring 
theories. 

Up to now, we are however very far from understanding how this unification
concretely takes place. This is because the dualities which relate
the known string theories are mostly intrinsically non-perturbative.
Thus knowing what M-theory really is would be equivalent to 
having a complete mastery of any one of the string theories at all
orders in the coupling expansion, including all non-perturbative
effects. Moreover, the conjectured dualities should become manifest
symmetries of the spectrum. It is worthless
saying that the current research is very far from this point.

In this chapter, we begin by reviewing the history of string theory,
which presents already many interesting aspects. This also betrays
the fact that our current perception of M-theory is still intimately
based on concepts pertaining to superstring theory. Accordingly,
we then proceed to a very sketchy presentation of the five
superstring theories, in their perturbative formulation. In
section \ref{DUALsec} the notion of duality is introduced, and
the interplay with supersymmetry is emphasized. Then the 
history of the so-called `Second Superstring Revolution' is reviewed,
namely all the dualities relating the string theories are presented
in the order of discovery (as far as a conjecture and its evidences
can be discovered). Section \ref{DBRANEsec} considers in more detail
the D-branes, the interest in which stems from the fact that they
are tractable in perturbative string theory. They turn out however
to be crucial in confirming most of the duality conjectures, and
moreover they also led to incredible advances in other interesting
problems in theoretical physics. This is also briefly reviewed.
In section \ref{MATRIXsec}, the Matrix theory attempt at a direct,
non-perturbative formulation of M-theory is discussed.
In order not to deceive the reader, we have to
anticipate that the chapter will end with the same
question with which it began.

All the material presented in this chapter is by no means original.
This is intended to be a very quick introduction to the field
of research, and it aims also at providing an historical
situation to it. The presentation will necessarily be subjective,
and the references are not intended to be exhaustive.
The usual reference for (perturbative) superstring theory is
Green, Schwarz and Witten \cite{gsw}, while a recent review
which also discusses the dualities and M-theory is \cite{kiritsis}.
 
\section{An historical perspective on string theory}
\label{HISTOsec}

The seed from which string theory grew up is commonly assumed
to be the Veneziano amplitude, conjectured in 1968 in the context
of dual models for hadrons.

At that time, there was no quantum theory for the strong interactions,
but there was an experimental output revealing a much more complicated
behaviour with respect to the recently formulated electro-weak interaction. 
Most notably, this intricacy showed up with a proliferation of
particles (actually short-lived resonances) with increasing spin
and mass.
The regularity of the plot of the mass against the spin of these
resonances (i.e. the Regge behaviour), 
and the conjecture of duality between the $s$- and the
$t$-channels for the hadronic amplitudes, led Veneziano to propose
his amplitude \cite{veneziano}, which realizes explicitly both
the duality and the Regge behaviour. 

The most interesting aspect of the Veneziano amplitude is that
it implies an infinity of poles of increasing mass and spin
in both of the (dual) channels. This means that the underlying theory,
whatever it is, must have a spectrum containing an infinite number
of massive excitations (i.e. particles).
This is a clear sign that the theory which produces such an amplitude
is not a conventional quantum field theory.

Soon after Veneziano's proposition, a theory which
reproduced the amplitude, and thus the infinite particle spectrum, was
found in the form of the theory of a quantized string.
This is certainly surprising; one may however consider the infinite
number of particles as a hint to abandon locality, and the string
as the most simple object which gives rise to an infinity of
vibrations of higher and higher energy. In 1970, the Nambu-Goto
action for the (bosonic) string was formulated \cite{nambugoto},
and string theory was officially born. It is amusing to
see already at this stage the incredible series of theoretical
bold conjectures from which string theory sprang up.

A year later, in 1971, the fermionic string was formulated by Ramond and
by Neveu and Schwarz \cite{rns} in order to have a model that describes also
space-time fermions. This implied the introduction of fermions on
the world-sheet of the string that had to be related to the world-sheet
bosons (the embedding coordinates) by a symmetry exchanging bosons
and fermions, supersymmetry. These strings are accordingly called
superstrings. 

An important feature of string theories is that to consistently quantize 
them, the
dimensionality of space-time has to be fixed. This gives $D=26$ for the
bosonic string and $D=10$ for the superstring. Moreover, the same
procedure of quantization fixes completely the spectrum of particles. 
This spectrum has two very particular features. Leaving aside the
infinite tower of massive string modes, we have a massless sector
which always includes a spin 2 particle\footnote{For the open
superstring, this particle only arises at one loop in string amplitudes.}, 
and a tachyon (a particle with imaginary mass) is also necessarily present.
It will turn out that one cannot get rid of the tachyon in the bosonic string,
while one can consistently discard it in the superstring theories, as we will
see shortly. 

However, in the beginning of the 70s, there were two deathly blows for string
theory in general: on the experimental side, the Veneziano amplitude
did not quite fit to some new results in the region of high-energy scattering
at fixed angle; on the theoretical side, quantum chromodynamics (QCD) was
being formulated and revealed itself as a much more promising candidate
for describing strong interactions. 

String theory surprisingly survived to these blows, making a virtue out of
its own problems. One of the most celebrated problems in the string theory
attempt to describe strong interaction was the total lack of experimental
evidence of a massless spin 2 particle. 
However rescaling the string parameter $\alpha'$,
measuring the inverse string tension, from the strong interaction
scale to the Planck scale, i.e. making it much smaller, it may well be that 
the massless spin 2 particle is nothing else than the graviton. This idea
was pushed forward by Scherk and Schwarz in 1974 \cite{scherkschwarz}.
At that moment, string theory ceased to be a candidate for a theory of hadrons,
becoming instead a candidate for a theory describing, and unifying, all known
interactions, including gravity. 

That (super)string theories should describe all interactions is a consequence
of the fact that if the $\alpha' \rightarrow 0$ limit is taken (also called the
zero-slope limit), all the massive string modes decouple and we are left with
the massless sector. This sector includes, besides the graviton, many other
particles with lower spin, which could account (at least in some of the
string theories) for all the known `light' particles, matter and gauge bosons.
Also, it is possible then to write a field theory of these `low-energy' modes,
and the Einstein gravity is correctly recovered.

An important step which led to the removal of all the remaining inconsistencies
in superstring theories was undertaken in 1976 by Gliozzi, Scherk and Olive 
\cite{gso}, who introduced the so-called GSO projection which at the same
time removed the embarassing tachyon and imposed space-time 
supersymmetry. This last property of superstring theories will turn out
to be incredibly important for the present-day developments. It amounts
to saying that, not only the world-sheet fields, but also the string spectrum
in 10 dimensions obeys a symmetry relating bosons to fermions. 
It has to be said that contemporarily to the research in string theory, there
had been in the 70s an extensive research on supersymmetry, and on its
local generalization, supergravity. It suffices to say now that space-time   
(extended) supersymmetries severely constrain the form that an action
can take. Also, a supergravity theory can be formulated in a space-time
with at most 11 dimensions, a dimension more than for superstring theories.

At the beginning of the 80s, there were 3 superstring theories at one's 
disposal:
2 theories of closed superstrings with extended $N=2$ space-time supersymmetry
(one chiral and one non-chiral), and one theory including open strings, with
$N=1$ supersymmetry and with a gauge group which was fixed to be $SO(32)$ 
for consistency.
These theories were shown to be finite at one-loop, and there
was the hope that the finiteness would remain at all loops. This was
because all the massive string modes regulate in such a way the 
effective theories coming from the massless sector, that the ultra-violet
behaviour was much softer than in ordinary field theories. 
These advances greatly benefited from the explicitly space-time 
supersymmetric formulation of the superstrings
given by Green and Schwarz \cite{greenschwarz}.

It is during 1985 that the `First Superstring Revolution' takes place. 
A revival
of interest in string theory takes places after the formulation during 
that year 
of two other kinds of superstrings, called the heterotic strings, by Gross,
Harvey, Martinec and Rohm \cite{heterotic}. These closed string theories
are built asymmetrically combining features of the toroidally 
compactified bosonic
string, and of superstrings. The remaining $N=1$ supersymmetry is enough
to preserve all the nice features of the superstrings, while the bosonic sector
allows the spectrum to fall into representations of a group, respectively
$SO(32)$ and $E_8\times E_8$ (hence the two heterotic string theories). The
massless sector contains all the vectors in the adjoint representation of
the group, which correctly translates into a low-energy effective theory 
consisting of $N=1$ supergravity coupled to a super Yang-Mills (SYM) theory.

The interest raised by the heterotic strings is mainly because they are
promising candidates to incorporate the Standard Model of strong and weak
interactions, thus unifying most simply all the known interactions.
The serious flaw of the previously known closed superstring theories 
was that in the low-energy
sector, there was no chance to find, at the perturbative level, a non-abelian
group. We will see that duality will predict, and evidence has now been found,
that gauge symmetry enhancement nevertheless
occurs in these theories, by non-perturbative
effects. However at that time there was little hope to get, even after
compactification, a viable non-abelian gauge group
from the $N=2$ superstrings. In order to fit
the $SU(3)\times SU(2)\times U(1)$ gauge group of the Standard Model, it
turned out that the $E_8\times E_8$ group of one of the heterotic strings
was the most suitable. This should also explain, besides maybe other
more aesthetical reasons, the focus on heterotic strings more than, say, on
the open-closed $SO(32)$ superstring.

Still, to make contact with phenomenology, one had eventually to compactify
the space-time from 10 to 4 dimensions, and also possibly to reduce the
supersymmetry from the unrealistic $N=4$ (in 4 dimensions) to, say, the more 
workable $N=1$ set up. These two problems could actually be
addressed at the same time, and almost the day after the revolution the
compactifications on Calabi-Yau manifolds \cite{candelasetal} and on orbifolds
\cite{dixonetal} appeared as very promising tools.
Accordingly, the end of the 80s saw a huge literature on string phenomenology
and the hope that string theory was the sought for grand unified theory
(we would think that 
`Theory Of Everything' is too much of a metaphysical notion to be used in
this context).

Nevertheless there were some serious conceptual problems left over, 
leaving aside
the concrete technical complications of the theory. The main problem was
that one had to choose first between the 5 consistent superstring theories,
and secondly between the way one compactifies, and it soon became clear that
there were thousands of ways to do this. This means that the parameters
one had to tune in order to recover the Standard Model physics were much
more than the parameters of the Standard Model itself! This is clearly
not an appealing feature for a theory that pretends to be unique.
Also, even from a mathematical or aestethical point of view, it was
not satisfying that 4 superstring theories, though perfectly consistent,
seemed to be `wasted' by Nature. In this line of thought, 11 dimensional
supergravity had the same uncertain status.
These problems caused a pause in the enthousiasm towards string theory as a
workable grand unified theory at the beginning of the 90s.

The date at which the `Second Superstring Revolution' takes place can 
be situated
at the beginning of 1995, when Witten showed in a very important paper 
\cite{witten}
that most of the strong coupling limits of all the known string theories
could be reformulated through dualities in terms of the weak coupling limits
of some other string theory, or in terms of the 11 dimensional supergravity.
With this paper M-theory was born. However it has to be stressed that the
concept of duality that Witten uses was already introduced and used in 
string theory from the beginning
of the 90s by some authors, in particular Sen \cite{sen} 
who studied and provided evidence for the S-duality of the heterotic 
string theory
in 4 dimensions (the original conjecture was formulated in \cite{fontetal})
and by Hull and Townsend \cite{hulltownsend} who postulated the existence
of U-duality, a key concept in Witten's derivation.

The concept of duality (in the modern sense) already entered high energy
physics when the electric-magnetic duality (which, being a
strong-weak duality, can be realized only at the non-perturbative
level) of 4 dimensional gauge
theories was first conjectured by Montonen and Olive \cite{montonen},
and then shown to be more likely to happen in supersymmetric theories
\cite{osborn}. In string theory, 
it was already known in the late 80s that a duality, called T-duality,
related some of the string theories. T-duality is visible at the 
perturbative level, and acts on the string theories 
when they are compactified on, say, a generic $d$-dimensional torus
(see \cite{tduality} for a review on T-duality).
However the new, non-perturbative, dualities conjectured in 1994-1995
opened up the concrete possibility that all string theories were related
to each other, and to 11 dimensional supergravity as well. It is
this theory that unifies all string theories and is presumably
11 dimensional that we call here `M-theory'. 

It is needless to say that since then, incredible support was gathered
towards this conjecture, and the set up was refined to provide
a very nice picture of the web of dualities, which will be discussed
in Section \ref{DUALMsec}. This same period saw also many other advances
in the understanding of string theory (with the use
of D-branes for instance), and an increasing interrelation
between string theory (or, rather, M-theory) and other theoretical
problems as black hole physics and supersymmetric gauge theories
in various dimensions. 

Another very recent attempt to address directly M-theory is the 
Matrix theory conjecture of Banks, Fischler, Shenker and Susskind
\cite{bfss}, proposed in October 1996. This is however too much of
the present day research to fit into this brief historical review
of string theory, which thus stops here.

Let us end this section with a brief remark on the terminology. 
In the present section and in the following one,
the expression `string theory' refers strictly to its perturbative 
formulation. In the subsequent sections, and in the more general 
considerations, `string theory' will also more widely
indicate both the perturbative and the non-perturbative aspects of the
theory, which eventually lead to the concept of M-theory. Since there
is clearly not a sharp frontier between the two interpretations of
`string theory', we hope it will be clear from the context which one is
referred to.

\section{Perturbative string theory}
\label{PERTUsec}

We review in this section the status of string theory before 
duality came into play. The main point will be to give an account
of the massless particles in the perturbative string spectrum, and
to show the low-energy effective actions that govern their dynamics.
We will focus mainly on type II superstrings, since these are the ones
which will be used in most of the following chapters.
This section is essentially based on \cite{gsw}, where the derivation
of all the results presented here should be found, unless otherwise quoted.

Let us start by writing the world-sheet action of the superstring. The 
easiest action one can write is one in a trivial (flat) background,
in quadratic form, and with a world-sheet metric that has been 
gauged-fixed to a flat 2-dimensional Minkowski metric. The action
for the string reads:
\be
I_{WS}=-{1\over 4\pi \alpha'}\int d^2\sigma\ \eta_{\mu\nu}
\left\{ \eta^{\alpha\beta} \partial_\alpha X^\mu \partial_\beta X^\nu
+i \bar{\psi}^\mu \gamma^\alpha \partial_\alpha \psi^\nu \right\},
\label{wsaction}
\ee
where we have defined the following: $1/ 2\pi \alpha'$ is the string 
tension; $\sigma^\alpha$ (with $\alpha=0,1$) are the world-sheet coordinates;
$X^\mu=X^\mu(\sigma)$ are the embedding coordinates, which are scalars
from the world-sheet point of view; $\psi^\mu=\psi^\mu(\sigma)$ are
world-sheet spinors carrying a vectorial index of the $SO(1,9)$ space-time
Lorentz group. 

In 2 dimensions, one can impose at the same time the
Majorana and the Weyl conditions (see for instance \cite{strathdee}). 
It is thus simplifying to choose a basis
of $\gamma$ matrices in which they are real. Here we take:
\be
\gamma_0= \left[ \begin{array}{cc} 0 & 1 \\ -1 & 0 \end{array}\right], \qquad
\gamma_1= \left[ \begin{array}{cc} 0 & 1 \\ 1 & 0 \end{array}\right], 
\label{gamma2d}
\ee
which verify $\{ \gamma_\alpha, \gamma_\beta\}=2\eta_{\alpha \beta}$.
Since all the $\gamma$ matrices are taken to be real, the matrix $C$
such that $C\gamma_\alpha^T=-\gamma_\alpha C$ is simply $C=\gamma_0$.
If the Majorana and the Dirac conjugate are defined respectively by
$\bar{\psi}_M=\psi^T C$ and $\bar{\psi}_D=\psi^\dagger \gamma_0$, then
the Majorana condition $\bar{\psi}_M=\bar{\psi}_D$ simply translates to
a reality condition on the spinors, $\psi^*=\psi$. The matrix $\gamma_2$
can be defined by:
\be
\gamma_2=\gamma_0 \gamma_1=
 \left[ \begin{array}{cc} 1 & 0 \\ 0 & -1 \end{array}\right],
\label{gamma22d}
\ee
and it correctly verifies $\gamma_2^2=1$, $\gamma_2^\dagger=\gamma_2$.
Accordingly, a spinor $\psi$ can be split into 2 chiral components,
$\psi={1+\gamma_2\over 2}\psi_R+{1-\gamma_2\over 2} \psi_L$; in two component
notation, we have:
\[ \psi=\left(\begin{array}{c} \psi_R \\ \psi_L \end{array} \right). \]
Note that the fermionic term of the action \rref{wsaction} can be
rewritten as:
\be
I_{WS}=\dots-{i\over 2\pi \alpha'}\int d^2\sigma \ \eta_{\mu\nu}\left\{
\psi^\mu_R \partial_+ \psi^\nu_R+ \psi^\mu_L \partial_- \psi^\nu_L \right\},
\label{wschiral}
\ee
where we have defined $\sigma^\pm=\sigma^0\pm \sigma^1$. The equations
of motion thus imply that the fermions with positive and negative chirality
are respectively right and left moving.

The action \rref{wsaction} has a very particular feature.
Since both of the terms
have the same overall coefficient, one can write the following symmetry
relating the bosons and the fermions on the world-sheet:
\be
\delta X^\mu =i\bar{\epsilon} \psi^\mu, \qquad \qquad \delta \psi^\mu=
\gamma^\alpha \partial_\alpha X^\mu \epsilon.
\label{susy2d}
\ee
This is the simplest manifestation of supersymmetry. This symmetry can be
seen to close (on-shell) to the momentum operator.
Supersymmetry is particularly important to carry out properly the quantization
of the superstring, and to fix consistently the dimensionality of
space-time to $D=10$.

When deriving the equations of motion from the action \rref{wsaction},
one sees immediately that, both in the open and in the closed case,
one can impose two different sets of boundary conditions on the world-sheet
fermions $\psi^\mu$. For the open string one has
the choice between imposing at each boundary $\psi_R=\pm \psi_L$. Since
the overall sign does not matter, the condition at one end
is always fixed to have the positive sign. Then if at the other
end there is also a positive sign, the fermionic fields $\psi^\mu$ will
be integer moded. This is called the Ramond (R) sector. If on the contrary
the second boundary condition has a negative sign in it, the $\psi^\mu$
will be half-integer moded. This is the so-called Neveu-Schwarz (NS)
sector.

For a closed string, the discussion is the same, because we have now
a choice in the periodicity condition for the fermions:
$\psi^\mu_R(\sigma=2\pi)=\pm \psi^\mu_R(\sigma=0)$, and similarly
for $\psi^\mu_L$. Again, when choosing the $+$ or the $-$ sign, one is
respectively in the R or in the NS sector. However now we have two
independent choices to make. There will be thus four different sectors
in the end, NS-NS, R-R, R-NS and NS-R.

In the R sector of, say, the open string, the vacuum energy is zero.
The ground state has however to be a representation of 
the algebra of the zero modes
of the fermionic fields $\psi^\mu_0$, which is actually the Clifford
algebra for the space-time Lorentz group. If we classify the states in
terms of the representations of the little group $SO(8)$, we see
that the ground state of the R sector can be chosen to be
either in the ${\bf 8}_+$ or
in the ${\bf 8}_-$ representation (which differ by their chirality).

On the other hand, the ground state of the NS sector has a negative
vacuum energy of $-{1\over 2}$ (when the minimal gap for the
bosonic fields is 1), and it is a singlet of $SO(8)$. This state is hence
a tachyon. When acting on it with a $\psi_{-{1\over 2}}$ mode, a massless
state is created with the indices of the ${\bf 8}_v$ vectorial 
representation of $SO(8)$.
The superstring represents space-time fermions and bosons when both
sectors contribute to the particle spectrum.

Space-time supersymmetry, which requires the same number of on-shell bosonic
and fermionic states at each mass level, is achieved, and the tachyon
is consistently discarded, when the GSO projection is performed. This
projection has as a consequence that the chirality of the R sector ground state
is fixed, and that in the NS sector, all the states
created from the vacuum by an even number of fermionic
operators $\psi_{-r}$ are eliminated
(a more precise definition of the GSO projection 
can be found e.g. in \cite{gsw}). 
This in particular excludes the tachyon from the physical spectrum, and leaves
us with a massless sector consisting of the states ${\bf 8}_v +
{\bf 8}_+$. 

The states in the massless sector realize 10 dimensional $N=1$ chiral
supersymmetry, and actually correspond to a gauge boson and to its
fermionic superpartner (also called gaugino). 

It is however too early to define a theory of open strings, since these
ones necessarily couple to closed strings. Indeed, one-loop
diagrams (e.g. the cylinder) for open strings can be cut in such a way that the internal
propagating string is a closed string. Let us then review before the
theories of closed strings.

The closed string theories are rather easy to build because the right moving
and the left moving sectors decouple, and each one is quantized in
exactly the same way as for the open string.
The massless sector of a closed string theory is then a tensor product
of two copies of the open string massless sector. There are however two
different tensor products that one can make, depending on whether one
takes the Ramond ground states of the opposite chirality on the two sides or
of the same chirality.

If the right and left moving sectors are of opposite chiralities, the
massless particles are classified along the following representations
of $SO(8)$:
\bea
({\bf 8}_v+{\bf 8}_+) \otimes ({\bf 8}_v + {\bf 8}_-)&=&
({\bf 1}+ {\bf 28} +{\bf 35}_v)_{NSNS} + ({\bf 8}_v + {\bf 56}_v)_{RR}
\nonumber \\
& & +({\bf 8}_+ +{\bf 56}_-)_{NSR} + ({\bf 8}_- + {\bf 56}_+)_{RNS}.
\label{iiaspectrum}
\eea 
The bosonic particles of the NSNS sector are respectively a scalar, a two-index
antisymmetric tensor and a two-index symmetric traceless tensor; they are
related to the dilaton field $\phi$, the antisymmetric tensor field $B_{\mu\nu}$
and the metric $g_{\mu\nu}$ associated with the graviton. In the RR sector
we find a vector and a 3-index antisymmetric tensor, which correspond
to the fields $A_\mu$ and $A_{\mu\nu\rho}$. The space-time fermions are found 
in the RNS ans NSR sectors, and the ${\bf 56}_\pm$ representations are
the two gravitini; this is thus a theory with $N=2$
supersymmetry. Moreover, since the gravitini have opposite chiralities,
the theory is non-chiral. This is usually called the type IIA string theory.

To write a low-energy effective action
for the massless fields above is
made easier by the constraint that we should be describing a space-time
supersymmetric theory. Indeed, the action of 10 dimensional
type IIA supergravity is completely fixed. Alternatively, one can 
compute the interactions between the massless fields from string theory
amplitudes, and then extrapolate to the non-linear action that reproduces them.
Still another successfull way to obtain the effective action is to
write the string world-sheet action \rref{wsaction} in curved background
and then to ask that conformal invariance is not broken quantum mechanically,
i.e. the $\beta$-function must vanish.\footnote{ 
Note that a proper definition of the low-energy effective action should
actually only come from a string field theory perspective, which is still
lacking.}
One obtains the following action (we
only write its bosonic sector):
\[
I_{IIA}={1\over 16\pi G_N}\left[\int d^{10}x\ \sqrt{-g} 
\left\{ e^{-2\phi}\left(
R+4(\partial \phi)^2 -{1\over 12} H_3^2\right) -{1\over 4} F_2^2
-{1\over 48} F_4^{'2}\right\} \right. \]
\be
\qquad \qquad \left.+ {1\over 2} \int B_2 \wedge F_4 \wedge F_4 \right].
\label{introiia}
\ee
We have defined the field strengths in the following way: $H_3=dB_2$,
$F_2=dA_1$, $F_4=dA_3$ and $F_4'=F_4+A_1\wedge H_3$. The metric 
in the action above is the string metric, which appears
in the string action in curved background $I_{curved}\sim\int g_{\mu\nu}
\partial_\alpha X^\mu \partial^\alpha X^\nu$. Note that the fields
coming from the NSNS sector have a $e^{-2\phi}$ factor in front of their
kinetic term, while the RR fields not. This is discussed in \cite{witten,tasi}.
The only arbitrary dimensionful parameter of the type IIA string theory
is $\alpha'$, and accordingly the Newton constant appearing in \rref{introiia}
is expressed in terms
of it, $G_N \sim {\alpha'}^4$. The string coupling constant, which
governs the expansion in string loops, is actually the vacuum expectation
value of the dilaton, $g_s=e^{\langle \phi \rangle}$. It is thus
determined dynamically. The effective Newton constant will be affected
by the asymptotic values of such moduli. Note that the action \rref{introiia}
is the leading term in both the $\alpha'$ and the $g_s$ expansion.

An important remark on the closed string spectrum has to be made now.
While there are perturbative string states carrying a charge with respect
to the NSNS fields (i.e. the winding states and the Kaluza-Klein
momentum states when there is a compact direction), 
there are no states carrying
the charge of a RR field. This problem has a technical origin, and has
been a long standing puzzle in perturbative string theory. We now know
that the D-branes, which can be considered as non-perturbative objects,
are the carriers of these charges.

We can now consider the second closed superstring theory, in which 
both left and right moving sectors have the same chirality.
The massless particles are:
\bea
({\bf 8}_v+{\bf 8}_+)\otimes ({\bf 8}_v+{\bf 8}_+)&=&
({\bf 1}+{\bf 28}+{\bf 35}_v)_{NSNS}+({\bf 1}+{\bf 28}+{\bf 35}_+)_{RR}
\nonumber \\
& & +({\bf 8}_- + {\bf 56}_+)_{NSR} + ({\bf 8}_- + {\bf 56}_+)_{RNS}.
\label{iibspectrum}
\eea
The NSNS sector is exactly the same as the one of the type IIA superstring.
The RR sector on the other hand is different; this time we have a
scalar (or 0-form), a 2-form and a 4-form potential. However for the
4-form potential to fit into the ${\bf 35}_+$ representation, its
field strength has to be self-dual. The fields are respectively
$\chi$, $A_{\mu\nu}$ and $A_{\mu\nu\rho\sigma}^+$.
The space-time fermions of the mixed sectors now comprise two gravitini
of the same chirality. We have thus a chiral $N=2$ supersymmetry, and the
theory goes under the name of type IIB string theory.
Note that from the superalgebra point of view, one often refers to this theory
as a $(2,0)$ theory, while the type IIA theory is a $(1,1)$ theory.
This has as a consequence that the fields of the type IIB theory can be
arranged into representations of $SO(2)\equiv U(1)$, i.e. some fields
can be paired in complex fields.

Because of the presence of the self-dual 5-form field strength, the type
IIB supergravity was originally
formulated only in terms of its equations of motion, which
are completely fixed by the (2,0) supersymmetry. We can however write
an action for this theory if we do not ask that the self-duality
condition follows from the variation of the action, but is rather imposed
afterwards. The bosonic part of the low-energy effective action of 
the type IIB superstring thus reads:
\[
I_{IIB}={1\over 16 \pi G_N}\left[\int d^{10} x\ \sqrt{-g} \left\{ e^{-2\phi}
\left( R+4(\partial \phi)^2 -{1\over 12 } H_3^2\right)-{1\over 2}
(\partial \chi)^2 \qquad \qquad \right.\right. \]
\be
\qquad \qquad \qquad \left. \left. -{1\over 12}F_3^{'2}-{1\over 240}
F_5^{'2} \right\} +\int A_4 \wedge F_3 \wedge H_3 \right].
\label{introiib}
\ee
Here the RR field strengths are defined by $F_3'=F_3-\chi H_3$, $F_3=dA_2$
and $F_5'=dA_4+A_2\wedge H_3$, and we impose that $F_5'=*F_5'$.
The same remarks about the type IIA supergravity action \rref{introiia}
apply to this case.

The two actions above \rref{introiia} and \rref{introiib} are the actions
of the only two maximal supergravities in 10 dimensions. Maximal supergravities
exist because if one tries to extend further the supersymmetry, then 
one has more than one graviton in the spectrum, and also massless fields
with spin higher than 2. This is clearly not physical, since we want gravity
to be unique. The total amount of supercharges is thus fixed to be at most
32. This also constrains the dimension of the space-time in which
one wants to build the supergravity, since the number of components of
the smallest spinor representation increases exponentially with $D$ (see
e.g. \cite{strathdee}).
It turns out that the highest dimensional supergravity one can formulate
is in 11 dimensions.
This supergravity is apparently not related to the superstrings, which live
in 10 dimensions. Note however that the Kaluza-Klein reduction of the
11 dimensional supergravity yields exactly the action of the type IIA
supergravity \rref{introiia} (see Appendix \ref{KKapp}). This relation
is now understood to have a very deep meaning. We will come back
to it in Section \ref{DUALMsec}.

We can now go back to discuss (briefly) the open superstring.
We had already stated that the massless modes of the open strings have
$N=1$ supersymmetry, and that they must be coupled to a closed string
sector. However the type II closed strings discussed above cannot
be coupled consistently to the open strings, since they have $N=2$
supersymmetry. One way to divide by two the supersymmetry of the
closed theories is the following: one can identify the left and the 
right moving sectors of the type IIB superstring, thus making a theory
of closed unoriented strings. Only half of the massless
states in \rref{iibspectrum}
survive, and these are the following:
\be
{({\bf 8}_v+{\bf 8}_+)\otimes ({\bf 8}_v + {\bf 8}_+)\over \Omega} =
({\bf 1}+{\bf 35}_v)_{NSNS}+({\bf 28})_{RR}+({\bf 8}_- +{\bf 56}_+)_{RNS-NSR},
\label{ispectrum}
\ee
where we have called $\Omega$ the orientation reversal operator.

The open strings have however a richer structure than the closed ones,
since they can carry charges at their ends. These are called Chan-Paton
factors. If the factors are taken to be in the fundamental representation
of, say, a $U(n)$ group, 
it follows that an open string state generically belongs
to the adjoint representation of that group. If the string is unoriented,
the group becomes either $SO(n)$ or $Sp(n)$ (see \cite{tasi} for
an exhaustive discussion). It turns out that the only anomaly-free
theory is the one with the group $SO(32)$. This is consistent with
the fact that the closed sector is unoriented as well.

The low-energy effective action of the $N=1$ open-closed superstrings,
also called type I superstrings, is then the action of $N=1$ supergravity
in 10 dimensions coupled to a super Yang-Mills theory with gauge group
$SO(32)$. We will not write the action here, however it is worth noting
two features of the type I theory: first of all, as the type II theories,
it has only one free parameter, which is $\alpha'$. Secondly, the bosonic
part of the action has the same structure as the type II ones, 
with a $e^{-2\phi}$ factor
in front of the NSNS fields, and nothing in front of the RR one. Here
however the kinetic term of the 
fields coming from the open sector has a $e^{-\phi}$ factor
in front of it, and this can be traced back to the fact that this part
of the action comes from the amplitude computed on the disk instead of
the sphere.

Let us conclude with a very quick review of the heterotic string theories.
The key idea in constructing these theories is to remark that in closed
string theories the quantization of the right and left moving sectors
can be carried out independently. For both sectors, the quantization
is consistent if the central charge is either 15 if there is supersymmetry
or 26 if there is no such symmetry. Knowing that each boson and each
fermion contributes respectively 1 and $1\over 2$ to the central charge,
we reach the conclusion that $D$ must be 26 for the bosonic string,
and 10 for the superstring. One can however consider that supersymmetry
is present only in, say, the right moving sector. This amounts to take
only the first term in the action \rref{wschiral}. One directly sees
that the world-sheet supersymmetry is realized in this case
with a parameter $\epsilon$
of negative chirality, hence there is one half of the supersymmetry 
of the type II closed superstrings.
This fixes already to 10 the dimensions of the target space-time.

In the left moving sector, we have now the freedom to choose the fields
which will make up the $c=26$ central charge. Ten of these fields
are already fixed to be the (left moving part of the) embedding coordinates.
Then the most natural choice for the other fields is to take
32 (anti)chiral fermions. One can easily guess that these fermions
have an $SO(32)$ global symmetry. Carrying out the quantization
of the left moving sector, one finds that the massless level contains,
besides the vector ${\bf 8}_v$ of the space-time $SO(8)$, also 496
states in the adjoint of $SO(32)$ or $E_8 \times E_8$ depending on
which periodicity conditions have been taken for the fermions. The ground
state of the left moving sector is actually tachyonic, but the level
matching condition with the right moving sector 
excludes it from the physical sector. The supersymmetry of only one of the 
two sectors
is thus enough to ensure a tachyon-free spectrum to the heterotic
strings.

An alternative way to formulate the heterotic string is to take all
the left moving fields to be bosons. However 16 of them must be 
compactified on a 16-dimensional torus $T^{16}$. Since these bosons
are chiral, they satisfy slightly modified commutation relations, and
their quantized momentum is fixed in terms of the winding number.
Moreover, in order for them to contribute to the massless spectrum
(a requirement for one-loop unitarity) the $T^{16}$ has to be chosen
very particularly, i.e. it must be defined by a self-dual lattice.
This results in the same two possible spectra as the ones 
discussed in the previous paragraph. In the latter construction, however,
we have the picture of the heterotic string being a hybrid of the
superstring and of the bosonic string.

We can now recapitulate the massless spectrum of the heterotic strings.
Let us first focus on the tensorial product of the right moving 
sector with the left moving space-time vector. We recover the
same particles as in the closed sector of the type I string, though
they are obtained differently:
\be
({\bf 8}_v+{\bf 8}_+)_R\otimes ({\bf 8}_v)_L =
({\bf 1}+{\bf 28}+{\bf 35}_v)_B+({\bf 8}_- +{\bf 56}_+)_F.
\label{hetspectrum}
\ee
All the particles in the r.h.s. are singlets of the $SO(32)$ or
$E_8\times E_8$ group. The remaining left moving massless states are in the
adjoint of one of the two groups above, and are $SO(8)$ scalars.
When we take the tensorial product of these states with the 
right moving sector, we recover nothing else than the states
which make up the $N=1$ vector multiplet of an $SO(32)$ or $E_8\times E_8$
gauge theory.

The massless particle content of the heterotic string is thus exactly the same
as the one of the type I superstring\footnote{Starting from the first
massive level, the spectra of the heterotic string and of the type I string
strongly differ. This also implies that
the $\alpha'$ corrections to the low-energy effective actions will be
different.}. There are however two versions
of the heterotic string, since the construction is more `flexible' and
allows also for the $E_8\times E_8$ group.

The low-energy effective action for the heterotic strings is also
the one of $N=1$ supergravity coupled to super Yang-Mills, but
it differs from the low-energy effective action of the type I superstring
in that the whole bosonic part of the action has now a factor of $e^{-2\phi}$
in front of it. Duality will eventually relate the $SO(32)$ heterotic 
string to the type I string.

This concludes this short review of all the five consistent supersymmetric
string theories. We will now see how all of these string theories are
related by perturbative and non-perturbative dualities, and are unified
into M-theory, which is effectively an 11 dimensional theory.

\section{The philosophy of duality}
\label{DUALsec}

It is now time to introduce in this section the concept of duality,
which is really the core of the so-called Second Superstring 
Revolution\footnote{Amusingly, string theory originated from a 
conjecture also named duality, as explained in Section \ref{HISTOsec}. 
This was however a duality of another kind with respect to the concept
discussed in this section.}.
Roughly speaking, there is a duality when two seemingly different theories
are actually physically the same. Since showing that two theories produce
exactly the same physics is a tremendous task in most cases, duality
has often the status of a conjecture. 

To be more precise, two theories are related by a duality if there
is a one-to-one map between their physical spectra, and if their dynamics
are equivalent too. From the physical, on-shell, point of view, there
should be no difference between the two theories. The latter are thus
two different descriptions of the same physics. The two descriptions
can be quite different as we will see. Note also that the
two theories can also coincide with the same classical theory
but in two different regimes, e.g. with weak and strong coupling interchanged.

What makes difficult to actually prove duality is that often a duality
conjecture involves a map between a weak and a strongly coupled theory,
and thus only the perfect knowledge of non-perturbative effects could
firmly establish the existence of a duality. Until now, this has been
possible only for the two-dimensional model proposed by Coleman
\cite{coleman}, which is the first and unique case of a proof of 
a non-perturbative duality.

A duality should thus be established properly only by solving
completely the two theories. However the duality conjecture, if supported by
some evidence, can be used as a working hypothesis
to focus alternatively on only one of the two theories, and to consider
them as completely equivalent. Results in one theory can be derived in
the dual one, where they can be more accessible. This is the power of duality.

Having seen how duality acts on two different theories, let us see
the particular, and most interesting case, in which duality
relates two different regimes of the same theory.
The archetype of this duality is the simple observation that the
Maxwell equations are invariant under electric-magnetic duality. This
has been re-inserted in the context of modern quantum field theory
by Montonen and Olive in 1977 \cite{montonen}. 
That the electric-magnetic duality is a non-perturbative duality can
be easily seen considering that it maps particles of the perturbative
spectrum, such as the bosons of the broken gauge symmetry, to magnetic 
monopoles, which are classical, non-perturbative solutions (for this
map to be truly consistent, one has to settle in the framework
of $N=4$ supersymmetric Yang-Mills (SYM) theory \cite{osborn}). Moreover,
under this map the coupling constant $g$ is taken to its inverse,
$g \rightarrow 1/g$. As it will also happen in string theories, 
one can show that the combined possibilities to have dyonic objects
(carrying electric and magnetic charge), axionic $\theta$-terms in the 
action, and the Dirac quantization of the electric charge altogether imply
that the duality group is actually $SL(2,Z)$. We will review in more
detail in the next section how this duality precisely acts.

What kind of symmetry is this $SL(2,Z)$ duality? It states that a
multitude of states in the spectrum are actually the same, 
but it does not require that any physical state should be invariant under it.
It can thus be seen as a broken discrete gauge symmetry \cite{schwarzsdual}.
One has to mod out the spectrum by the $SL(2,Z)$ symmetry before, say,
counting the states. 
Also, the moduli space of all the theories which differ by the
values of some (dynamical) parameters, generically has to be modded
out by the duality group.

The resulting theory is however non-trivial. Let us describe simply how
duality works in the case of $N=4$ SYM. The usual formulation of this theory
is in terms of its coupling $g$ which becomes the electric coupling when
the gauge group is broken down to a $U(1)$. We will call the $W$ bosons
particles with $(1,0)$ charge, where a particle with $(p,q)$ charge
carries $p$ units of electric charge and $q$ units of magnetic charge.
Quantization of this theory is usually performed assuming that $g$ is small,
$g\ll 1$. There are however objects in this theory which cannot be reliably
treated in perturbation theory since they have a mass going like $1/g$.
These are the 't Hooft-Polyakov monopoles, or particles with $(0,1)$ charge.
What the duality conjecture states is that the theory can be reformulated
taking the monopole to be the fundamental object. This new theory is exactly
the same SYM theory, with however a new coupling $\tilde{g}$ given
by $\tilde{g}=1/g$. Furthermore, the configurations with one single
$W$ boson at weak coupling or with one single monopole at strong coupling
are strictly equivalent. Of course, each one of these two configuration is not
duality invariant, rather it breaks the $SL(2,Z)$ duality symmetry. Since
they are mapped one to the other, they belong to the same duality orbit.

Consider now the object with charge $(1,1)$, which is called a dyon.
It belongs also to the same duality orbit as the $W$ boson and the monopole,
and can thus be also considered as the starting point for another perturbation
theory, which will be the same SYM theory. Note however that this dyonic
object is a bound state, and that it differs from a configuration in
which both a $W$ boson and a monopole are present. The latter configuration
would have an energy of the order $E\sim M_W + M_{mon}$, while the dyon
has a mass going like $M_{dy}\sim \sqrt{M_W^2 +M_{mon}^2}$.
This ensures that the theory is still non-trivial, because the 
situation in which a $W$ boson is scattering off a magnetic monopole
is definitely not equivalent to a single $W$ boson elementary state.
The scattering state described here is a configuration in which the
knowledge about the non-perturbative effects introduced by the monopole
is still indispensable. In this case, duality only allows us to
reformulate the problem as, say, the scattering of two dyons with
opposite magnetic charge.

The benefit of duality in this case of $N=4$ SYM theory, is that we no longer
need to develop tools to understand the behaviour at $g>1$, i.e. at 
strong coupling, since we can always reformulate the problem in a
set up in which the coupling constant is smaller than 1. 

The picture above can be roughly transposed to all the other dualities,
which hold in string theory. Sometimes however, as we will see, only
some limits of the dual theories are known and can thus be described.

The non-perturbative nature of the duality that we have just described makes
it very difficult to prove, since this would involve knowing the behaviour
of the strong coupling dynamics.
What we are left to do to establish and support 
a duality conjecture is to provide
evidence for it. It turns out that the most powerful tool to provide such
evidence is supersymmetry.

When one considers a supersymmetry algebra with central charges, it appears
that several kinds of supermultiplets exist (see e.g. \cite{wessbagger}). 
A massless, or lightlike,
supermultiplet is the shortest one, while a generical massive multiplet
has a dimension which is the square of the dimension of the massless 
multiplet. Extended supersymmetry however allows for the presence
of central charges in the superalgebra, and these in turn imply the
existence of massive states which nevertheless fit into short multiplets.
This is so because these massive states still preserve some of the 
supersymmetries. The dimension of the multiplet depends on the amount of
preserved supersymmetry, but it is always smaller than the dimension
of the generic massive multiplet. The shortest massive multiplet has
the same dimension of the massless multiplet, and preserves half
of the supersymmetries.

For a massive state to preserve some supersymmetry, and thus to fit into
a short multiplet, it has to verify a constraint on its mass, which
has to be equal to some of the central charges. Short massive multiplets
are thus associated with Bogomol'nyi-Prasad-Sommerfield (BPS) \cite{bps}
states, which generically saturate a bound on their mass with respect
to their charge(s). These states are thus often called BPS states, even
if the original definition of the latter
did not refer to any supersymmetry property.

The interest of the BPS states resides in the following argument, first
given by Witten and Olive in 1978 \cite{wittenolive}. Since the dimension
of a multiplet cannot change when a continuous parameter, such as the
coupling constant, is varied, a BPS state stays so at any value of the
coupling. Now, the condition for a state to be BPS is that the mass equals
its charge. We have thus the certitude that even at strong coupling,
where we have no handle on the dynamics, a BPS state of a given charge
will have a fixed mass. In other words, this arguments tells us that
the mass-charge relation of a BPS object does not suffer any quantum
correction.
Moreover, in theories with enough supersymmetry also the continuous 
parameters are not renormalized, and this allows us to predict which BPS
states become light at strong coupling.

The utility of BPS states is thus the following. If one restricts the
evidence for a duality to the spectrum of (partially) supersymmetric
states, one does not need any more to compute difficult strong coupling
effects, but instead one can simply analyze the properties of the BPS
states at weak coupling (or, at any rate, in the region where the computations
are possible), and then extrapolate at strong coupling. 
This is in some way a second conjecture to support a first
conjecture, but it has proven very efficient and indeed exact in some
cases. 

We have defined above the set of ideas that characterize what is
meant nowadays by the word `duality'. Let us now emphasize what are
the consequences of a conjectured duality between two theories.

Once one has acquired evidence in support of a duality, one can then
start using the relation between the two theories (or between the strong
and the weak coupling regimes of the same theory) to infer important
and otherwise inaccessible results. In the case of dualities relating
the strong coupling of one theory to the weak coupling of another,
this gives a handle on the non-perturbative effects and on the strong
coupling dynamics of the first theory simply studying the perturbative
effects of the second theory. It is by the exclusive use
of this tool that it has been possible to gain such an enormous
insight into the non-perturbative aspects of string theory in the last
few years.

We review in the next section the web of dualities relating all
the known string theories and 11 dimensional supergravity.

\section{String dualities and the appearance of M-theory}
\label{DUALMsec}

Historically, duality and related ideas already appeared in string
theory before the revolution of 1994-1995. The $SL(2,Z)$ duality of
field theory was transposed to the heterotic string (toroidally) compactified 
to 4 dimensions. This duality, which was basically an electric-magnetic
duality, was conjectured in \cite{fontetal} and
then thoroughly studied by Sen \cite{sen} who provided concrete evidence.
Still in the context of the heterotic string, which attracted much attention
after the First Superstring Revolution of the mid eighties, the
perturbative duality known as target space duality, or T-duality,
was established by Narain \cite{narain} and by Ginsparg \cite{ginsparg}
who showed that the two heterotic string theories are equivalent
when compactified on a circle. Application of T-duality to type II string
theories, which we will review hereafter, was carried out already in 1989 in
\cite{dineetal,daietal}.

Supergravity theories were also intensely studied ever since their
formulation. In particular, the study of the maximal supergravities
uncovered a large symmetry group of the latter, as pioneered
by Cremmer and Julia \cite{cremmerjulia} for $N=8$ supergravity in
4 dimensions (see also \cite{julia} for other cases).
It is the confluence of these ideas which produced the Second 
Superstring Revolution.

\subsubsection*{T-duality}

Let us begin with the review of T-duality for type II theories
(useful references are \cite{tduality,tasi}). T-duality
is the most accessible duality, since it is visible at the perturbative
level. Below we consider its most simple manifestation, namely
the exchange of large and small radii.

In order to see T-duality at work, we have to go back to perturbative
string theory and write the mode expansion of the bosonic fields $X^\mu$.
Their equations of motion imply that they split into left and right
moving parts:
\be
X^\mu=X_L^\mu(\sigma^+)+X_R^\mu(\sigma^-), \qquad\qquad \sigma^\pm=\sigma^0
\pm \sigma^1\equiv \tau\pm\sigma.
\label{bosonsplit}
\ee
Taking into account that $\sigma\sim \sigma +2\pi$ and that $X^\mu$ will
have to obey to some periodicity condition to be defined shortly, the
mode expansions for the closed string are the following:
\bea
X_L^\mu&=&x_L^\mu+\alpha' p^\mu_L \sigma^+ +i\sqrt{\alpha'} \sum_{n\neq 0}
{1\over n} \alpha^\mu_n e^{-in \sigma^+}, \nonumber \\
X_R^\mu&=&x_R^\mu+\alpha' p^\mu_R \sigma^- +i\sqrt{\alpha'} \sum_{n\neq 0}
{1\over n} \tilde{\alpha}^\mu_n e^{-in \sigma^-}, \label{modexp}
\eea
where $x_{L,R}^\mu$ and $p_{L,R}^\mu$ are real and $(\alpha^\mu_n)^\dagger
=\alpha^\mu_{-n}$, and similarly for $\tilde{\alpha}^\mu_n$.
Note that the split of the center of mass $x_0^\mu=x_L^\mu+x_R^\mu$
is arbitrary and thus unphysical, while the relation between
$p^\mu_L$ and $p^\mu_R$ will be fixed by the periodicity condition. The
total momentum carried by the string is $P^\mu=p^\mu_L+p^\mu_R$.

Let us now analyze the periodicity conditions on $X^\mu$. If the
closed string is propagating in flat non-compact spacetime, then
the only choice we have is to impose:
\[ X^\mu(\tau, \sigma+2\pi)= X^\mu(\tau, \sigma). \]
This implies that the sum \rref{bosonsplit} cannot depend linearly on $\sigma$,
and thus fixes $p^\mu_L=p^\mu_R\equiv {1\over 2}P^\mu$. 

Suppose now that the space is still flat, but one of the directions (e.g. 
$X^9$) is
compact, with radius $R$. Space-time has now a topology of $R^9\times S^1$,
and the closed string can wind around this non-trivial 1-cycle.
The periodicity condition for the compact coordinate can then be 
generalized to:
\be
X^9(\tau, \sigma+2\pi)= X^9 (\tau, \sigma)+ 2\pi w R.
\label{winding}
\ee
Here the number $w$ is an integer, and it is clear that it arises as a
classical topological number. Quantum mechanically, the compact direction
has another consequence: the momentum in this direction can no longer
take any value, but is restricted to a set of discrete values in order
for the wave function $e^{i P^9 X^9}$ to be single valued. We thus
have $P^9={n\over R}$, with $n$ an integer (and a truly quantum number).

The center of mass and zero modes of $X^9$ are thus:
\[
X^9=x^9_0 + \alpha' {n\over R} \tau + w R \sigma + oscill.\ terms.
\]
This directly implies that:
\bea
p^9_L &=& {1\over 2}\left( {n\over R} +{w R\over \alpha'}\right), \nonumber \\
p^9_R &=& {1\over 2}\left( {n\over R} -{w R\over \alpha'}\right).
\label{pleftright}
\eea

The classical, bosonic hamiltonian is the following:
\[
H= {\alpha' \over 2} (p_L +p_R)^2 +{\alpha' \over 2} (p_L - p_R)^2
+\sum_{n\neq 0} (\alpha_{-n} \cdot \alpha_n+\tilde{\alpha}_{-n} \cdot
\tilde{\alpha}_n ). 
\]
All the products in the above expression are sums over the 10 space-time
indices.
We neglect the fermionic terms and the (possible) quantum ground state
energy since they are not relevant to the present discussion. 

Inserting the expressions \rref{pleftright} and the
9 dimensional non-compact space-time momentum $P^\mu$, we get:
\be
H={\alpha' \over 2} P^2 +{1\over 2} n^2 {\alpha' \over R^2} +{1\over 2}
w^2 {R^2 \over \alpha'} +oscill.\ terms.
\label{windingenergy}
\ee
Using that $P^2=-M^2$ and that the Hamiltonian must vanish on-shell, 
one can correctly rederive that
the mass of an unexcited wound string is its length divided by its tension.

We have now the full set-up to uncover T-duality. The expression used to
derive the masses of the perturbative string spectrum \rref{windingenergy}
is obviously invariant under the combined exchange:
\bea
R & \leftrightarrow & {\alpha' \over R}, \label{tduality} \\
n & \leftrightarrow & w. \nonumber
\eea
This is a very interesting symmetry in several respects. Although it
is visible perturbatively in string theory, it has some striking
features which have to be attributed to its stringy nature.
The first consequence of \rref{tduality} is that there seems to be
a minimum length in string theory, since whenever a string tries
to probe a distance smaller than $l_s\equiv \sqrt{\alpha'}$, then
we know that we could instead reformulate the problem as a 
(possibly different)
string probing a length scale larger than $l_s$ (note however that
there is now evidence
that non-perturbative objects as D-particles can probe smaller distances).
Furthermore, the second line of \rref{tduality} indicates that
quantum numbers are exchanged with classical, topological ones.
We see already at this stage that string theory is a natural framework
for the duality symmetries to appear. Indeed, T-duality is possible 
because of the existence of winding modes, which are absent in an
ordinary Kaluza-Klein theory.

In terms of the left and right moving momenta \rref{pleftright}, the
T-duality transformation \rref{tduality} becomes:
\[
p^9_L \leftrightarrow p^9_L, \qquad \qquad p^9_R \leftrightarrow -p^9_R.
\]
Transforming also $\tilde{\alpha}_n \leftrightarrow -\tilde{\alpha}_n$ (which
does not change the spectrum),
the T-duality can be rewritten in the following way:
\be
X^9=X_L^9 +X_R^9 \qquad \leftrightarrow \qquad X^{'9}=X_L^9 - X_R^9 .
\label{leftparity}
\ee
We can now see that because of the world-sheet supersymmetry \rref{susy2d},
the fermionic superpartner $\psi^9$ also has to
transform under T-duality, as $\psi^9_L \leftrightarrow \psi^9_L$
and $\psi^9_R \leftrightarrow -\psi^9_R$.

The transformation \rref{leftparity} acts like a space-time parity reversal
restricted to the right moving modes. This is an important remark because it
implies that the chirality of the right-moving Ramond ground state also changes,
and this in turn means that T-duality maps type IIA superstring theory
to type IIB and vice-versa.

We have thus established that type IIA and type IIB theories compactified
on a circle are equivalent at the perturbative level. The equivalence
has of course to be verified also at the non-perturbative level, and
indeed it will be possible to check that the map extends also to all
the non-perturbative objects of the two theories.

At the level of the low-energy effective actions, there is only one
supergravity in 9 dimensions and thus it is obvious that the reduction
of the two 10 dimensional supergravities is equivalent. The concrete
map between the fields of IIA and IIB supergravities (with an isometry)
has been shown in \cite{bergshoeffetal}. 

A last but important remark on T-duality is that the string coupling
constant is actually not inert under it. This can be seen as follows.
For the physics in 9 dimensions to be invariant under T-duality (this
is required by the notion of duality discussed in the previous section),
the Newton constant (which is the only independent coupling constant 
in the low-energy physics) has also to be invariant. Now
the effective 9 dimensional Newton constant derived from, say, \rref{introiia},
is given by:
\[ G_9 \sim {g_A^2 {\alpha'}^4 \over R_A}, \]
where the subscripts indicate that we settle in type IIA theory. The
constants $g_A$ and $R_A$ are simply the asymptotic values of, respectively,
the fields $e^\phi$ and $(g_{99})^{1/2}$, and the way they fit into $G_9$ is
dictated by the form of the action \rref{introiia}.
We now require that the same $G_9$ is given by the IIB theory compactified
on a circle of radius $R_B\equiv {\alpha'\over R_A}$. Since $\alpha'$
cannot change in this problem (the string scale does not change), we
have that:
\be
g_B=g_A {\sqrt{\alpha'}\over R_A}.
\label{tcoupling}
\ee
The same is of course true if IIA and IIB are interchanged.
This formula will turn out to be very useful in determining how the
non-perturbative objects transform under T-duality.

\subsubsection*{S-duality}

We now turn to analyze a symmetry which is particular to type IIB 
supergravity in 10 dimensions, and that can have far-reaching consequences
if extended to the whole type IIB string theory.

As it can be seen in the massless spectrum \rref{iibspectrum}, 
the scalars and the 2-form potentials come into pairs. Due to the particular
r\^ole played by the dilaton in the action \rref{introiib}, it
can be seen that the equations of motion of type IIB supergravity
are invariant under a $SL(2,R)$ symmetry \cite{iibsugra}. The fields
transform as follows. If we arrange the RR scalar $\chi$ and the dilaton
in a complex scalar $\lambda=\chi + ie^{-\phi}$, then the
$SL(2,R)$ symmetry acts like:
\be
\lambda \rightarrow {a\lambda +b \over c\lambda +d},
\label{sl2rlambda}
\ee
where the real parameters are such that $ad-bc=1$. If $B_2$ and $A_2$ are
respectively the NSNS and the RR 2-form potentials, then they also 
transform according to:
\be
\left(\begin{array}{c} B_2 \\ A_2 \end{array} \right) \rightarrow
\left(\begin{array}{cc} d & -c \\ -b & a \end{array} \right)
\left(\begin{array}{c} B_2 \\ A_2 \end{array} \right)=
\left(\begin{array}{c} dB_2 -c A_2 \\ aA_2 -b B_2 \end{array} \right).
\label{sl2rforms}
\ee

Since the NSNS field $B_2$ couples to the fundamental string and 
the corresponding charge is the winding number, which is quantized, the
symmetry group is broken to its discrete subgroup $SL(2,Z)$ when
one asks that the charges also transform in a way similar to \rref{sl2rforms}.
The quantization of the charges can also be seen to arise in 10 non-compact
dimensions because
of the possible presence of magnetically charged 5-branes.

Let us focus on a particular case of the above symmetry. If the RR scalar 
$\chi$ is taken to vanish and recalling that $g=e^{\phi_\infty}$,
one of the above $SL(2,Z)$ transformations is such that:
\be
g  \rightarrow {1\over g}, \qquad B_2  \rightarrow A_2, \qquad
A_2  \rightarrow  -B_2.
\label{sduality}
\ee
This particular transformation is often referred to as S-duality.
It is now trivial to see that this duality is non-perturbative, since
it exchanges weak and strong coupling. However, instead of exchanging
at the same time an electric field with its dual magnetic field
(as in the 4 dimensional dualities), it exchanges a NSNS field with 
a RR one, both electric. 
Now we had already pointed out that there are no RR-charged
states in the perturbative string spectrum. We see that
if the S-duality conjecture is true, then we must find non-perturbative
objects carrying RR-charge.

Note that this $SL(2,Z)$ duality of type IIB strings is a duality relating 
different regimes of the same theory. A general $SL(2,Z)$ transformation
maps the fundamental string, carrying one unit of $B_2$ charge, into
a general $(p,q)$ string carrying $p$ units of NSNS charge and $q$ units
of RR charge (with $p$ and $q$ relatively prime). What the duality 
conjecture tells us is that this $(p,q)$ string can be quantized and
should reproduce, in its own variables, the same type IIB theory
(see \cite{hulltownsend,schwarz}).

Contrary to T-duality, in the present case the fundamental string is mapped
to a `solitonic' string, which must have a tension going like an
inverse power of $g$ (in order not to be seen in perturbation theory).
Since the tension of this dual string gives the string scale of the
dual theory, we see that $\alpha'$ should not be invariant under S-duality.
We can actually determine how it transforms asking that the 10 dimensional
effective Newton constant is invariant. Since we have $G_{10}\sim g^2 
{\alpha'}^4$, we see that under S-duality \rref{sduality} the string
scale must transform as:
\be
\alpha' \rightarrow g\alpha'.
\label{dualscale}
\ee
An immediate consequence of the relation above is that the tension of the
(0,1) string goes like $1/g\alpha'$. This is not what we would expect
from a typically solitonic effect, like the energy of the electro-magnetic
dual of the fundamental string which should go like $1/g^2$ because
of the charge quantization condition. 

\subsubsection*{U-duality}

In fall 1994, the first signal of the Second Superstring Revolution
was sent out with the
conjecture by Hull and Townsend \cite{hulltownsend} that S-duality
and T-duality actually fit into a larger group of duality symmetries,
called U-duality.

Take for instance any of the two type II string theories, compactified
on a $d$-dimensional torus $T^d$. The resulting theory is invariant under
two sets of dualities. Firstly, T-duality extends now to a wider
group, which is $O(d,d;Z)$ \cite{narain,tduality}. 
For $d=1$, we recover the only element described above, namely $Z_2$.
The invariance under $O(d,d;Z)$ can be seen as follows.
If there are $d$ compact directions, then we have $d$ relations
similar to \rref{pleftright}. 
The `compact' momenta can be shown \cite{narain,tduality} to verify:
\[ {\vec{p}_L}^{\ 2}-{\vec{p}_R}^{\ 2}=
{n_i w_i\over \alpha'}, \qquad\qquad i=1\dots d ,\]
which clearly displays $O(d,d;R)$ invariance. This group is broken to
its discrete subgroup asking that the spectrum is mapped to itself.
Note that the larger group
denotes the fact that now one has the freedom to change some of the moduli
of the torus $T^d$, still keeping the $10-d$ dimensional physics invariant.

Secondly, since as soon as there is one compact direction there is no
longer distinction between the two type II theories, any one of these
string theories in less than 10 dimensions inherits the S-duality of
type IIB theory in 10 dimensions.
As we have already seen, S-duality actually extends to a full $SL(2,Z)$
duality group.

The conjecture of Hull and Townsend is that the duality group
of type II string theories compactified on $T^d$ is:
\be
G_d(Z) \supset SL(2,Z) \times O(d,d;Z), 
\label{uduality}
\ee
where $G_d(Z)$ is a discrete subgroup of the Cremmer-Julia group of
the $10-d$ dimensional maximal supergravity. A table with such
groups can be found for instance in \cite{hulltownsend}. If we take
for example type II theories compactified down to 4 dimensions,
then the group is $E_7(Z)$ (a discrete and maximally non-compact version
of $E_7$, also written $E_{7,7}(Z)$).

The group $G_d(Z)$ is in general larger than $SL(2,Z) \times O(d,d;Z)$,
and this implies that the orbits of any object are also much larger,
i.e. U-duality maps fundamental string states to any one
of the non-perturbative objects that can be predicted.

In \cite{hulltownsend}, the authors have been able to find all the
objects belonging to an orbit of a particular class of states, 
the BPS saturated supersymmetric states. What they actually find are
some extreme dilatonic black hole solutions of the dimensionally
reduced supergravity. However some of these black holes are identified
with the winding and momentum modes of the fundamental string.
Furthermore, all the other black holes are reinterpreted as wrapping
modes of some higher dimensional extended objects, called $p$-branes\footnote{
The $p$-branes are extensively studied in the next chapter. We thus
postpone until then a thorough introduction to these objects.}.
Since all of these branes can be mapped by a U-duality to the 
fundamental string, the idea that any one of these objects
can be treated as a fundamental one \cite{townsuperm2,democracy}
appears now natural. The string coupling constant is also mixed 
through U-duality with all the other moduli of the compactified theory,
thus mapping strong coupling physics to some other region in the moduli
space, possibly weakly coupled.

There is however something more which comes out of this conjecture, besided
unifying the two type II theories and putting on the same footing
all the perturbative and non-perturbative objects. It is
well known that all the maximal supergravities (except the 10 dimensional
type IIB) can be derived from the 11 dimensional supergravity
\cite{11dsugra}. The Cremmer-Julia groups arise from this point of
view as the symmetries of 11 dimensional supergravity compactified
on $T^{d+1}$. A remnant of this origin can be seen from the
fact that each of the $G_d$ groups has a subgroup which is
the modular group of the $d+1$ dimensional torus $T^{d+1}$. In the
discretized version, we have:
\[ SL(d+1,Z)\subset G_d(Z). \]
If $G_d(Z)$ is a true symmetry of string theory, and accordingly 
its spectrum fits into representations of this group, then the same spectrum
can be decomposed into representations of $SL(d+1,Z)$. 

The existence of 11 dimensional supergravity has always been intriguing
in a string theory perspective, 
but the fact pointed out above seems a strong hint that
something truly eleven dimensional is actually happening in string
theory itself.

\subsubsection*{11 dimensions and M-theory}

In his famous paper of spring 1995 \cite{witten}, Witten uses the U-dualities
to map all the strong coupling regimes of the known string theories
(with the exception of the $E_8\times E_8$ heterotic string in 10 dimensions)
to some weakly coupled limit of another theory. The most astonishing
surprise is that
for the picture to be self-contained, one has to include in the possible
`dual' theories also the 11 dimensional supergravity. 
It has to be stressed that this is not a mathematical artifact, but 
a clear indication that string dynamics at strong coupling effectively
sees 11 dimensions. This is not in contradiction with the critical dimension
of the superstrings being 10 dimensions. In the weak string coupling, 
perturbative limit, the space-time in which the strings propagate can be
seen to become again 10 dimensional.

Let us now briefly see how 11 dimensional supergravity really comes into
play. It is well known that type IIA supergravity is the dimensional reduction
of 11 dimensional supergravity (see Appendix \ref{KKapp}). It is however
instructing to see how the fields of these two theories are related.
In 10 dimensions, we have only one scalar, the dilaton, which will
control the strength of the string interactions. In 11 dimensions, there
are no scalars, and the only one which is produced by Kaluza-Klein
reduction is the metric component relative to the direction along which
one is reducing. This latter scalar gives rise, by its asymptotic value,
to the compactification modulus
called the radius of the `11th direction', $R_{11}$.
It is thus clear that the string coupling $g$ and the compactification 
radius $R_{11}$ are related.

It is actually easy to derive the exact relation between $g$ and $R_{11}$
using the results of the Appendix \ref{KKapp}. One finds that if
$g_{11,11}=e^{2\sigma}$, the relation between the two scalar fields is
$e^\phi=e^{{3\over 2}\sigma}$, where $\phi$ is the same which appears
in \rref{introiia}. The relation between the string
coupling and the compactification radius (in 11 dimensional Planck units)
is thus:
\be
g\sim \left({R_{11}\over l_p}\right)^{3\over 2}.
\label{grrel1}
\ee
Note that the precise numerical factor in this relation and in the following
ones are not one but can be determined by consistency with the dualities 
(see for instance Appendix \ref{ZOOapp}).
The relation above can be rewritten as:
\be
R_{11}\sim g^{2\over 3} l_p,
\label{grrel2}
\ee
and the 11 dimensional Planck length is simply given in terms
of the 11 dimensional Newton constant by $G_{11}\equiv l_p^9$.

We would like now to relate the 11 dimensional units to the string units,
given by the length $l_s=\sqrt{\alpha'}$. This is straightforward
since the 10 dimensional effective Newton constants have to coincide. We have
thus the following relation:
\[
g^2 l_s^8 \sim G_{10} \sim {G_{11}\over R_{11}} = {l_p^9 \over  R_{11}},
\]
which implies:
\be
l_p \sim g^{1\over 3} l_s, \qquad \qquad R_{11}\sim g l_s .
\label{11iiarel}
\ee
Thus we see that at weak string coupling, $g\ll 1$, both the 11 dimensional
Planck scale and the radius of the 11th direction are small compared
to the string scale\footnote{Note that the two quantities above are also
small compared to the 10 dimensional Planck
scale, which is given by $L_p=G_{10}^{1\over 8}\sim g^{1\over 4}l_s$.}. 
Perturbative
string theory is thus correctly described by a 10 dimensional theory.

However, when we now consider strongly coupled type IIA superstrings, we
see that the radius of the 11th direction $R_{11}$ becomes large
in both string and 11 dimensional units. The 11th direction thus
effectively decompactifies.
This is the core of this duality conjecture and the first glimpse at an
11 dimensional theory underlying all string theories, M-theory. It has
however to be noted that for the present duality, we only have a low-energy
description of the dual theory, since we do not know what is the
quantum theory that reduces to 11 dimensional supergravity at low energies.

To convince oneself that the strong coupling dynamics of type IIA string
theory is truly 11 dimensional, it can be shown that all the branes living
in 10 dimensions have an 11 dimensional ascendant. The most celebrated
example is the 11 dimensional supermembrane which gives the type IIA
superstring upon dimensional reduction along its world-volume 
\cite{duffetal,townsuperm2}.

An even more direct evidence of this duality can be shown following Witten
\cite{witten}. In the limit of a large radius $R_{11}$, the KK modes
propagating in this direction should become light degrees of freedom.
What are the corresponding states which become light on the type IIA string
side? Their mass has to go like:
\be
M= {1\over R_{11}}\sim {1\over g l_s}.
\label{md0}
\ee
They are thus solitonic objects, the mass of which becomes vanishing in the
strong coupling limit. That the mass has the same expression
for all values of the string coupling
is ensured by the fact that these objects can be shown to preserve half
of the 32 supersymmetries, and thus they fit in the shortest massive
supermultiplet.
 
Moreover the KK 2-form field strength under which these 0-branes (since
they are pointlike objects) are charged maps in the IIA language to the
RR 2-form field strength. Thus we find again that a RR-charged object
has a mass going like $1/g$. 

The RR-charged particle discussed here and the RR-charged string discussed
in the previous subsection are actually T-dual to each other. The
quickest way to see it is to suppose that the type IIA theory
has a compact direction and to transform accordingly the mass formula
\rref{md0} under the T-duality \rref{tcoupling}. We obtain:
\[ M'={R_B \over g_B l_s^2}. \]
This is exactly the mass formula for a string-like object with a tension
going like $1/ g\alpha'$, i.e. like the RR-charged (0,1) string of type
IIB theory.

Performing a chain of T-dualities, one can find a series of brane-like
objects, of even dimension in type IIA theory and of odd dimension
in type IIB, which all have a tension going like $1/g$, and which
all have the right dimension to couple to a RR potential. The supergravity
solutions relative to these branes were found in \cite{horostrom}.
It is amusing to see that their tensions perfectly agree with
a Dirac-like quantization condition of their charges (see
\cite{nepomechie,teitelboim}):
\be
T_p T_{6-p}={2\pi \over 16\pi G_{10}},
\label{diracquant}
\ee
where $p$ and $6-p$ are the extensions of the electric and the magnetic
brane, respectively.
The exact value of the Newton constant in 10 dimensions is consistently
fixed to be:
\be
G_{10}= 8\pi^6 g^2 {\alpha'}^4.
\label{g10exact}
\ee
A major breakthrough in string theory was the discovery by Polchinski 
\cite{polcdbranes}
that these RR-charged branes were the Dirichlet branes, or D-branes 
\cite{daietal}. The D-branes, and their many applications,
will be reviewed in the next section. It suffices here to say that the
particular $1/g$ behaviour of the D-brane tensions is the product
of the disk amplitude for the open strings that end on the D-brane.

The way in which all the dualities above and the compactification from
11 to 10 dimensions act on the various branes is summarized in Appendix
\ref{ZOOapp}.

We can now pause a little bit and ask what is M-theory from the
point of view of its duality with type IIA string theory. What
we know about M-theory is that it lives in 11 dimensions, that its
low-energy limit is 11 dimensional supergravity, and that when it is
compactified on a vanishingly small circle it reproduces 
perturbative type IIA string theory. We could also say, as for every
duality, that M-theory is strictly equivalent to type IIA string
theory when all non-perturbative effects are taken into account.
We will now see that more can be said about M-theory, and indeed many
strong coupling regimes of string theories in lower dimensions can
be resolved by M-theory \cite{witten}.

The simplest non-trivial example of the `power of M-theory' is to see
how M-theory and type IIB theory are directly related 
\cite{schwarz,aspinwall}. Consider for instance M-theory compactified
on a 2-torus $T^2$. The membrane of M-theory, or M2-brane, can now
wrap on the $T^2$ to give finite energy modes in the effective
9-dimensional space-time. We also have a set of KK modes for each direction
of the torus. Now, when the volume of the torus uniformly shrinks to zero
size, we have one family of modes which become massless, and two
sets of modes which become very massive. The appearance of massless
modes is the signal of a new direction which opens up and decompactifies,
these modes being identified to the KK modes of this new direction. 
The two sets of infinitely massive modes indicate on the other hand 
that there are two different string-like objects in the dual theory
which cannot wind any more on the new direction. We are clearly
describing type IIB string theory on a circle of growing size.
Moreover, the $SL(2,Z)$ duality of the latter theory is here given
a geometrical interpretation as the modular group of the original torus.
The duality between M-theory on $T^2$ and type IIB theory on $S^1$
can be made more explicit working out the relations between
the couplings and between the other BPS states, but the discussion
above gives already a strong evidence in favour of it.

Dualities in lower dimensions \cite{hulltownsend,witten}
can also involve compactification on manifolds breaking some of the
supersymmetries, like the Calabi-Yau manifolds in 6 and 4 dimensions
(in this latter case they go under the name of $K3$ surfaces).
This allows for dualities between type II and heterotic string theories in
4 and 6 dimensions, and between M-theory and heterotic string theory in
7 dimensions. A crucial fact in establishing these dualities is that the
enhanced gauge symmetry points in the heterotic string moduli space are
reproduced non-perturbatively on the type II and M-theory side by 
branes wrapping on cycles of vanishing area or volume. This is another
example of the branes becoming really elementary particles in some regions
of the moduli space \cite{conifold}.
We will not review further these dualities here.

\subsubsection*{More dualities}

Until now, we have discussed the dualities relating the superstring
theories with maximal supersymmetry ($N=2$) and 11 dimensional supergravity.
We now wish to see how the other three string theories fit into the
same unified picture, and can thus also be covered by the concept
of M-theory.

It was known already in the late 80s that the two heterotic string
theories were related by target space duality 
whenever at least one of the directions
was compact \cite{narain,ginsparg}. In other words, the 10 dimensional
$E_8\times E_8$ and $SO(32)$ heterotic string theories were connected
by a continuous path in the moduli space
of the heterotic string theory compactified on a circle.

The discovery of the r\^ole played by the D-branes allowed for a 
reformulation of type I string theory which displayed
its close relationship with type IIB strings,  and at the same time
provided evidence for a duality with the $SO(32)$ heterotic string.

The description of type I string theory that comes out of the work of
Polchinski \cite{polcdbranes} is the following. Since D-branes
are the only loci in space-time where superstrings can end, the only way
to have open strings propagating in the bulk of space-time is to
fill it with D9-branes. The difference with respect to the traditional
understanding of type I theory is that now the D9-branes are
accompanied by a RR-charge that has to be cancelled in some way.

It turns out that there is an object, called orientifold (see e.g.
\cite{orientifold} and the next section), 
which implements an orientation reversal on the world-sheet.
When the orientifold fills space-time (i.e. it is an orientifold 9-plane),
then the procedure is equivalent to the one performed in 
Section \ref{PERTUsec} to define open superstrings. What one can show
is that this object also carries a RR-charge, which is fixed once for
all. Thus, since we want to have a vanishing RR-charge associated to
these 9-dimensional objects, we have to take an orientifold of $SO$ type
and 16 D9-branes, which together yield a gauge group $SO(32)$ 
(see \cite{tasi} for the details).

We have thus obtained a somewhat more heuristic understanding of type I
string theory: it can really be seen as type IIB theory with some
additional objects in it, and 
the anomaly cancellation condition which fixes the group to $SO(32)$
is reduced to the cancellation of D9-brane RR-charge.

The duality with the $SO(32)$ heterotic string can now be studied in detail
in this framework, as proposed by Polchinski and Witten \cite{polwitt}. 
We know already that the two theories have the same
low-energy field content. What is new in this picture is that type I
theory has a soliton (which survives the projection from IIB theory)
which is the D1-brane, or D-string, or (0,1) string. Contrary to
the fundamental type I strings which, having the possibility to be 
open, do not have a winding number, the D-string is closed and can
wind on compact directions. Moreover, the fundamental open strings
which end on it on one side and on one of the D9-branes at the
other side endow it with a world-sheet structure identical to the
one of the $SO(32)$ heterotic string. For the map between the two
theories to work, one also has to exchange weak and strong coupling
(as already noted in \cite{witten}), the couplings being related by
$g_I={1\over g_{het}}$. This heterotic-type I duality can
actually be seen (with some caution) as a remnant of the S-duality 
of type IIB theory, after the orientifold projection \cite{hull}.

We have thus related type I theory to type IIB, the $SO(32)$ heterotic
string to the type I strings, and the two heterotic strings when they are
compactified on a circle. All the string theories are thus related to each
other, but we are still lacking of an alternative, weakly coupled
description for the strong coupling regime of the
$E_8\times E_8$ heterotic string in 10 dimensions.
Here M-theory comes again into play.

It was shown by Horava and Witten \cite{horavawitten}, still before the end
of 1995, that the strong coupling limit of the $E_8\times E_8$ heterotic 
string can be interpreted as M-theory compactified on a segment, the
length of the segment being related to the heterotic string coupling 
by a relation similar to \rref{grrel2}. This compactification
of M-theory can be considered as an orbifold compactification on
$S^1/Z_2$ (see e.g. \cite{gsw} for a description of the orbifolds in
string theory), and thus it correctly breaks half
of the original 11 dimensional supersymmetry to give 
$N=1$ supersymmetry in 10 dimensions. 10 dimensional physics
is actually associated to the physics on the `ends of the world', i.e.
on the hypersurfaces determined by the two endpoints of the segment.
It can be seen by anomaly cancellation arguments (see \cite{horavawitten}
for all the details on this particular duality)
that each of the two 10-dimensional boundaries must have an $E_8$
gauge theory defined on it. Moreover, the orbifold procedure allows open
membranes of M-theory to stretch from one end-of-the-world to the other,
thus giving effective closed strings in 10 dimensions. These are
to be identified with the fundamental heterotic strings. When the
size of the segment shrinks to zero size, we have a closed string
theory at weak coupling carrying a gauge group which is $E_8\times E_8$.
This is thus the explanation of the duality between M-theory and the
$E_8\times E_8$ heterotic string. Much like the type IIA string,
there is an 11th direction opening up at strong coupling.

Let us conclude this section by some comments on the picture that comes 
out of this web of dualities.

We seem to know 6 limits in the moduli space of what we can now call
M-theory: these are the 5 string theories supplemented by 11 dimensional
supergravity. As far as the string theories are concerned, we
seem to know something more than the low-energy effective action
because we can analyze them perturbatively. On the other hand, we only
know the low-energy effective action of the conjectured 11 dimensional
theory. Note however that this 11 dimensional theory has no natural
dimensionless parameter to define a perturbative expansion, such as
the string coupling $g$ for the 10 dimensional theories.

M-theory represents what stands in the middle, and allows us to relate
all these different limits. It can thus be considered as the region
of intermediate coupling, generically a much wider region than the others.
It is difficult to have an idea of what this unified theory really is,
since it does not have to reproduce all the known string theories at the
same time, but rather its existence implies that all the string theories
describe the same physics.

When we are at intermediate coupling, it is almost an option to say
that we are considering an originally 11 dimensional theory on a circle
of a finite radius, or rather a string theory at finite coupling. 
It is however fair to think about M-theory as an 11 dimensional theory,
since there are effectively regions in the moduli space where
the physics is truly 11 dimensional. On the other hand, nothing
prevents us in principle to choose our favourite string theory and try to
solve it completely, with the goal of knowing what M-theory is.
This approach has already proven to be hopelessly hard.
Nevertheless, there is the hope that M-theory itself can be addressed
directly, short-cutting the traditional string theory path. We will
indeed review in Section \ref{MATRIXsec} a proposal going along this
way.

\section{D-branes and their many uses}
\label{DBRANEsec}

The construction leading to the concept of D-brane is actually very
simple: instead of imposing the usual Neumann boundary conditions
to the open strings, one imposes Dirichlet boundary conditions in some
directions. The D-branes are then the hypersurfaces defined by these
conditions.
It turns out that the open strings constrained to end on 
D-branes, couple consistently to the closed type II strings.
Furthermore, what makes the D-branes interesting from the string duality point
of view is that they precisely couple to the RR bosons of the closed
string, which means that the D-branes are the sought for RR-charged
objects. The D-branes are thus solitonic objects in the sense that
they are not part of the perturbative string spectrum, but their properties
can nevertheless be studied in the framework of perturbative string
theory. Most notably, one finds that the D-branes are dynamical objects.

Let us define the D-branes simply by the boundary conditions one
imposes to the open string sector (the standard reference for
D-branes is \cite{tasi}). 

In order for the action \rref{wsaction} to give the correct equations
of motion for $X^\mu(\tau,\sigma)$ in the case of an open superstring
(we now take $0\leq \sigma \leq \pi$), one
generally imposes Neumann boundary conditions:
\be
\partial_\sigma X^\mu |_{\sigma=0,\pi}=0.
\label{neumann}
\ee
These are the conditions usually imposed when defining type I string theory,
and they imply that the ends of the string are freely moving. In terms
of the left and right moving pieces of $X^\mu$
\rref{bosonsplit}, the conditions rewrite:
\be
\partial_+ X^\mu_L - \partial_- X^\mu_R |_{\sigma=0,\pi}=0.
\label{neumann2}
\ee
The mode expansion of the left and right moving fields $X^\mu_L$ and
$X^\mu_R$ is much similar to \rref{modexp}, except that now there
is only one kind of oscillator, $\alpha^\mu_n=\tilde{\alpha}^\mu_n$, and,
most notably, $p^\mu_L=p^\mu_R$ is imposed regardless of the
fact that a direction is compact or not. The complete expansion of
the bosonic field $X^\mu$ is thus:
\be
X^\mu=x^\mu_0 +2\alpha' p^\mu \tau +2 i \sqrt{\alpha'} \sum_{n\neq 0}{1\over n}
\alpha^\mu e^{-in\tau} \cos n \sigma.
\label{neumexp}
\ee
Note that $x_0^\mu$ and $p^\mu$ are canonical conjugate variables, and 
thus $x_0^\mu$ correctly describes a dynamical variable, i.e. the position
of the string.

Let us now impose Dirichlet boundary conditions instead of the Neumann
ones in the $9-p$ directions $X^i$, with $i=p+1,\dots,9$.
We simply impose:
\be
\partial_\tau X^i |_{\sigma=0,\pi}=0,
\label{dirichlet}
\ee
or, in terms of the left and right moving fields:
\be
\partial_+ X^i_L + \partial_- X^i_R |_{\sigma=0,\pi}=0.
\label{dirichlet2}
\ee
The remaining $p+1$ directions (including the timelike one) still obey
the Neumann boundary conditions \rref{neumann} or \rref{neumann2}.
Now again the mode expansion for $X^i_L$ and $X^i_R$ will be similar
to \rref{modexp}, but in the present case the identifications will
be $\alpha^i_n=-\tilde{\alpha}^i_n$ and $p^i_L=-p^i_R$. Going back
to the expression \rref{pleftright}, this last condition allows
only for a winding number but not for any momentum in the direction $X^i$.
This is fully consistent with the heuristic picture: the Dirichlet
boundary conditions constrain the ends of the open string to live
on a particular $p+1$ dimensional hypersurface (the D-brane). However, since
now the open string can no longer break in two pieces in an arbitrary
locus of space-time, it can indeed wind on a compact direction.
Here one has even an additional possibility, in that the open strings can
stretch between two D-branes separated by some distance (thus giving
some kind of `fractional winding').

A consequence of \rref{dirichlet} is that we can choose the two fixed 
values indicating where the ends of the open string lie:
\[
X^i(\tau,\sigma=0)=x^i, \qquad\qquad X^i(\tau,\sigma=\pi)=y^i.
\]
The mode expansion for $X^i$ is then:
\be
X^i=x^i+ {1\over \pi} (y^i-x^i)\sigma +2\sqrt{\alpha'}\sum_{n\neq0}
{1\over n}\alpha^i_n e^{-in\tau} \sin  n \sigma.
\label{diriexp}
\ee
Since $p^i=p^i_L +p^i_R\equiv 0$,
$x^i$ is now no longer a dynamical variable, and can thus be taken
to vanish for simplicity. If there is only one
D-brane and the direction $X^i$ is compact with radius $R_i$,
then we can replace $y^i$ by $2\pi w R_i$. Note however that
only the mode $w=1$ is topologically conserved, since for higher
$w$ there is always the possibility for the string to break when
it crosses the D-brane.

The fact that the zero modes of the open string in the $X^i$ directions
are not dynamical implies that the low-energy fields corresponding
to the massless modes of the open sector do not depend on these
coordinates. Accordingly, they should be arranged into representations
of $SO(1,p)$ instead of $SO(1,9)$. This means that the low-energy effective
theory of these modes is $p+1$ dimensional. It is however very
easy to determine this effective theory starting from the knowledge
of the effective theory for the open strings with Neumann boundary
conditions, which is $N=1$ super Yang-Mills (SYM) in 10 dimensions.
Indeed, the quantization of the open string with Dirichlet boundary
conditions is exactly the same as for the open sector of the type I 
superstring. For instance, the Dirichlet boundary conditions change
an overall sign for $\psi^i_R$ by world-sheet supersymmetry, and thus
do not change the integer and half-integer moding in the Ramond and
Neveu-Schwarz sectors respectively. The net result is that one
only has to decompose the ${\bf 8}_v +{\bf 8}_+$ states under the
$p+1$ dimensional little group $SO(p-1)$.
The low-energy effective theory one obtains is (to leading order in
$\alpha'$) the SYM theory in $p+1$ dimensions, which is actually
the dimensional reduction of the 10 dimensional one. The gauge field
components with indices in the $X^i$ directions are now scalars.

The D-brane, as we will argue shortly, couples to the closed string
modes of the type II strings. However, by its presence, it breaks 
half of the supersymmetries since this is the maximum space-time
supersymmetry that the open strings (with Dirichlet boundary conditions
in this case) can support. The situation is thus that, far away from
the D-brane, there seem to be two space-time supersymmetries, however
their generators are related to each other on the world-sheet of the
open strings, which are localized near the D-brane (see the discussion
in the original paper of Polchinski \cite{polcdbranes}).

The relation between the two supersymmetry generators can actually be
derived considering how the D-branes behave under T-duality.

Recall that T-duality acted for the closed string as a parity reversal
restricted to the right moving modes \rref{leftparity}. If we do the same
operation on an open string, we see from \rref{neumann2} and 
\rref{dirichlet2} that Neumann and Dirichlet boundary conditions are
interchanged. If we thus start from a theory of open superstrings
and T-dualize on the $9-p$ directions $X^i$, we obtain the same 
D$p$-brane theory as defined above ($p$ is the number of space-like
directions longitudinal to the D-brane). This is actually one of the ways
the D-branes are usually introduced \cite{tasi}. Note however that
in this formulation, the $X^i$ directions have to be compact, while
as we have seen above the D-branes can be defined in 10 non-compact 
directions.

It is now straightforward to see how D-branes transform under T-duality.
If the T-duality is performed along a direction transverse to the
D$p$-brane, then the latter transforms into a D$(p+1)$-brane longitudinal
to the T-dualized direction, while if the T-duality is performed along
a direction longitudinal to the world-volume of the D$p$-brane, then
it becomes a D$(p-1)$-brane transverse to the T-dualized direction.

The supersymmetry projection associated with a D-brane can now
be found. If we start with type I open strings, the Neumann boundary
conditions impose that the two generators of space-time supersymmetry 
$Q_L$ and $Q_R$ must be equal, $Q_L=Q_R$ \cite{gsw,tasi} (up to a
conventional sign).
For this to be consistent, both generators have to have
the same space-time chirality. Type I open superstrings can then be 
(formally) seen as type IIB superstrings coupled to a D9-brane (which
fills space-time).

We can now introduce Dirichlet boundary conditions by the implementation of
space-time parity reversal on the right moving modes. The effect
on the supersymmetry generators
of this operation in, say, the $X^9$ direction is the following:
\be
Q_L'=Q_L, \qquad\qquad Q_R'=P_9 Q_R.
\label{susytdual}
\ee
$P_9$ is an operator which anticommutes with $\Gamma_9$ and commutes
with all the others. It can be taken to be $P_9=\Gamma_9 \Gamma_{11}$,
where $\Gamma_{11}=\Gamma_0\dots \Gamma_9$ is the matrix defining 
chirality.

A D-brane lying in the $X^1\dots X^p$ directions can thus be seen
to impose the following relation between the supersymmetry generators:
\be
Q_L=\pm P_{p+1}\dots P_9 Q_R.
\label{susyproj}
\ee
Using the fact that $Q_R$ is chiral or anti-chiral, the relation
above can be rewritten as:
\be
Q_L=\pm \Gamma_0\dots \Gamma_p Q_R.
\label{susyproj2}
\ee
Note that since T-duality interchanges type IIA and IIB theories,
we only have supersymmetric D-branes of odd dimension in type IIB theory
and of even dimension in type IIA theory.

The computation of Polchinski \cite{polcdbranes} can now be simply
stated. The static force between two parallel
D-branes is given by the amplitude corresponding to a cylinder
stretching between the two D-branes, without any insertion. 
This is a one-loop amplitude from the open string point of view,
but it is a tree-level amplitude in the closed channel which goes from
one D-brane to the other. The result of the computation
is that the amplitude vanishes,
thus implying that there is no static force between parallel D-branes.
The D-branes are truly BPS states, as their supersymmetric properties
might have suggested. 

More can be said \cite{polcdbranes} about this vanishing amplitude. 
In the closed channel
it can be seen that it actually vanishes because two terms cancel:
the exchange of NSNS particles on one side (the graviton and the dilaton),
and the exchange of RR gauge bosons on the other. This is the central result,
i.e. the proof that
the D-branes carry a RR-charge, and that their charge is equal to their
tension. Moreover, the tension can be extracted and it matches
exactly with the one which was guessed by duality arguments in the previous
section:
\be
T_{Dp}={1\over (2\pi)^p g l_s^{p+1}}.
\label{dbranetension}
\ee

Let us now review some generalizations of the theory of D-branes as
described above.

The most straightforward generalization was elaborated on by 
Witten \cite{wittenbound},
and consists in taking a stack of $N$ D-branes of the same kind.
As just noted above, the D-branes are BPS objects and can thus
stay statically at any distance from each other. The presence of $N$
D-branes has the direct consequence of introducing a `quantum number'
associated to each end of the open strings and which distinguishes between
the various D-branes on which the end can be attached. This quantum
number is nothing else than a Chan-Paton factor, as discussed in
Section \ref{PERTUsec}. Here however the situation is slightly 
different. If the $N$ D-branes coincide (i.e. they lie on top of each other),
then the effective theory on the branes is $U(N)$ SYM in $p+1$ dimensions
(in the case of oriented strings). On the other hand, if all the D-branes
are separated by a finite distance, the effective theory at low-energies
becomes the same SYM but now with a group $U(1)^N$.

The mechanism by which the group is broken from $U(N)$ to $U(1)^N$ exactly
maps to spontaneous breaking of local symmetry in the low-energy effective
action. It can be seen that the scalars of the SYM theory actually
represent the location of the branes. In the case of a $U(N)$ gauge theory,
the scalars belong to the vector multiplet of SYM and are thus 
$U(N)$ matrices. The diagonal values can be associated
to the locations in transverse space of the $N$ D-branes.
Pulling the D-branes apart is equivalent to giving an expectation value
to these scalars. Maximal supersymmetry on the effective theory side,
and the BPS property in the D-brane picture, ensure that there
are flat directions in the potential, 
along which such expectation values can be arbitrary.
Since the scalars are in the adjoint representation, the gauge
group can be broken maximally to $U(1)^N$.

The gauge bosons which get masses because of the expectation values of the
scalars are, in the D-brane picture, nothing else than the lowest lying, now
massive,
modes of the open strings connecting two different, separated, D-branes.
The phenomenon of spontaneous breaking of gauge symmetry thus earns
a very beautiful heuristic picture in the D-brane framework.

Let us consider in more detail the discussion above introducing
the reduction to $p$+1 dimensions of the 10
dimensional SYM theory. This is done straightforwardly. The
ten dimensional Yang-Mills field strength is given by:
\[
F_{MN}=\partial_M A_N -\partial_N A_M +[A_M , A_N], \qquad \qquad
M,N=0,\dots,9.
\]
The 10 coordinates are split into two sets, $x^M=\{\xi^\mu, y^i\}$ 
($\mu=0,\dots, p$ and $i=p+1,\dots, 9$), the fields being now
independent of the second set. We thus have:
\begin{eqnarray*}
F_{\mu\nu}&=&\partial_\mu A_\nu-\partial_\nu A_\mu +[A_\mu, A_\nu], \\
F_{\mu i}&=& \partial_\mu A_i +[A_\mu, A_i]\equiv D_\mu A_i, \\
F_{ij}&=& [A_i, A_j].
\end{eqnarray*}
If we rename $A_i\equiv \phi_i$ since they are now scalars, the bosonic part
of the action of SYM in $p+1$ dimensions writes:
\be
I_{p+1}=-{1\over g_{{YM}_{p+1}}^2}\int d^{p+1}\xi\ \left( {1\over 4}
tr F_{\mu\nu}F^{\mu\nu}+{1\over 2} tr D_\mu \phi_i D^\mu \phi^i
+{1\over 4} tr [\phi_i, \phi_j]^2 \right).
\label{symp}
\ee
Because of the presence of a total amount
of 16 supercharges, all the matter fields
present in this SYM theory are actually part of the vector multiplet
containing the gauge field. Therefore all the scalars, as well as the spinors,
belong to the adjoint representation of $U(N)$. The third term in the
action \rref{symp} is the potential, and it has flat directions
for commuting vacuum expectation values of the scalars. Moreover, due
to the large amount of supersymmetry, 
non-renormalization theorems exist (or are conjectured) 
which ensure that these directions remain
flat even at the quantum level. Any vacuum configuration
with diagonal $\phi_i$'s has a vanishing potential, and for generic
values of the diagonal elements the gauge group is broken down to $U(1)^N$.
In the D-brane picture, this corresponds to widely separated D-branes.

The SYM coupling constant $g_{{YM}_{p+1}}$ is fixed in terms of string
variables when one considers that the action \rref{symp} is the leading
term in the expansion of the Born-Infeld action for a D-brane \cite{biaction}.
Since the prefactor of the Born-Infeld action is simply the tension
of the D-brane $T_{Dp}$ as given in \rref{dbranetension}, we have
the following relation (where all numerical factors have been dropped):
\be
g_{{YM}_{p+1}}^2={1\over T_{Dp}{\alpha'}^2}=gl_s^{p-3}.
\label{gym}
\ee
Note that the ${\alpha'}^2$ factor comes in because the fields $A_\mu$ and
$\phi_i$ both have a mass dimension of one. The action \rref{symp} is
thus the leading term for the dynamics of D$p$-branes both in the
$g$ and in the $\alpha'$ expansion.

Another kind of generalization of the D-brane physics is to take, 
for some directions,
Neumann boundary conditions on one side and Dirichlet boundary conditions
on the other side (these directions will be called ND directions). 
This physically corresponds to having two D-branes
of different dimensionality at the same time, and open strings going
from one to the other.

The ND directions now introduce a change of sign in the right moving
(bosonic and fermionic) fields only at one boundary, thus interchanging
integer and half-integer mode expansions. The energy of the NS sector
ground state now depends on the number $\nu$ of these ND directions.
Aside from the previous (trivial) $\nu=0$ case, let us only state here that the
case $\nu=4$ gives a vanishing energy for the NS ground state, and
a reduced number of massless states which correspond to a further
breaking of supersymmetry. Indeed, each of the two D-branes comes
with a different projection of the kind \rref{susyproj2}. The whole
configuration preserves $1/4$ of the original supersymmetry.
The other cases are trickier to treat, it suffices here to say that
in the $\nu=2$ and $\nu=6$ case there is respectively an attractive and
a repulsive static force, while in the $\nu=8$ case there is also a vanishing
static force and the configuration turns out to be supersymmetric
(note that $\nu$ cannot of course be odd because the two D-branes must
belong to the same type IIA or IIB theory). 

There is an object very similar to the D-branes that can also be defined,
that is the orientifold \cite{orientifold,tasi}. Briefly stated, the
orientifold procedure amounts to accompanying an orientation reversal
on the world-sheet by a parity reversal in space-time, and then modding
out by this symmetry. When no parity reversal is taken, then 
we are simply defining unoriented strings. 
When on the other hand we take a parity reversal in $9-p$ directions,
the orientifold procedure $\Omega p$ acts as follows on the closed strings
(it acts similarly on the open ones):
\bea
X^\mu(\tau, \sigma)&=& X^\mu(\tau,2\pi -\sigma), \qquad \mu=0\dots p,
\nonumber \\
X^i(\tau, \sigma)&=& - X^i(\tau,2\pi -\sigma), \qquad i=p+1 \dots 9.
\label{orientifold}
\eea
In the bulk of space-time, i.e. far away from the orientifold $\Omega p$-plane
located at $X^i=0$, the strings are oriented, but \rref{orientifold}
states that they have a `mirror' partner on the other side of
the $\Omega p$-plane. On the $\Omega p$-plane itself, the strings are 
unoriented. The introduction of such an object can be seen
to impose a relation on the supersymmetry generators which is exactly
the one for a D$p$-brane \rref{susyproj}.
Actually, a one-loop computation like the one performed by Polchinski
for the D-branes can show that an orientifold $\Omega p$-plane carries
$2^{p-5}$ units of the same RR-charge carried by a D$p$-brane \cite{tasi}.

An orientifold plane can be combined with D-branes, for instance if one wants
to cancel the RR-charge of the orientifold. If one puts $N$ D-branes
on one side of the orientifold, then one has also to take into account
the existence of the `mirror' D-branes on the other side\footnote{
Even if there is a total amount of $2N$ different Chan-Paton factors,
it is important to note that there are only $N$ dynamical D-branes.}. The gauge
theory one recovers on the world-volume of the D-branes is $SO(2N)$ or
$Sp(2N)$, depending on how the world-sheet orientation reversal
has been defined on the string states.
The main physical difference between an orientifold and a D-brane is that
the former is not dynamical, since it has no low-energy effective
action associated to its world-volume.

Let us now briefly review some of the applications where the D-branes
have proven to be very useful. We single out here three main successes
of the D-branes: proving that the U-duality orbits could be filled;
making the microscopic counting of the entropy of some black holes accessible;
providing a pictorial way to understand some gauge theory phenomena, such
as dualities.

In the preceding section we have reviewed the dualities which led to 
the concept of M-theory. Many of these dualities predicted dual objects
with mass going like $1/g$, and carrying RR-charge. The effective presence
of these objects in the string spectrum, even if non-perturbative, is
crucial in providing evidence for the conjectured unification of string
theories under M-theory. 

The existence of RR-charged solitons was already proved in \cite{horostrom}
at the level of the classical solutions of the supergravities.
This has the same status as 
the existence of the NSNS 5-brane soliton (or NS5-brane),
which is also a matter of having a consistent background for the 
fundamental string \cite{callharvstro}. In this latter case however,
the solitonic 5-brane is the magnetic dual of the fundamental string,
for which a perturbative definition obviously exists. The status of the
RR-charged solitons is somewhat different, in that neither the electric
nor the magnetic object existed in perturbation theory. With the
advent of D-branes, the nature of the RR-charged object has been made more
precise: they appear indeed as solitons, but their quantum mechanical
properties can be analyzed quite in detail from string perturbation 
theory, much more than what can be said about the NS5-brane.
Actually, the particular behaviour of their mass in the string coupling 
constant permits to study them in a flat background, because the strength
of their gravitational interaction is given by $G_N M_{Dp} \sim g$,
contrary to the NS5-brane where the same strength is of order 1 and cannot
be neglected. The D-branes can thus be thought of as the weak coupling
description of the RR-charged solitonic supergravity solutions.

Another feature of D-branes that allows a thorough study of their 
properties is the knowledge of their effective action. For instance,
one can in this context compute very accurately the scattering of two
D-branes (performing a computation either in string perturbation theory,
or in the low-energy effective field theory). An interesting result
\cite{danfersun,kabatpouliot,dkps}
supporting the M-theory conjecture is that the scattering of D0-branes,
or D-particles, displays a characteristic length scale of the order
of the 11 dimensional Planck scale \rref{11iiarel}.

The application of D-brane physics to the problem of the entropy of
black holes has been spectacular \cite{stromingervafa,callanmaldacena}.
The idea is simple and is the following. In the context of the
low-energy effective supergravity theories, one can find all sorts of
black holes, of any space-time dimension. Most notably, any one of
the solitonic solutions discussed above (and treated at length in the next
chapter) can be seen as a black hole in the space transverse to its
world-volume. Now we have just seen that at weak string coupling,
we have a fairly accurate quantum description of some of these solitons,
i.e. those carrying RR-charge. It is thus tempting to study whether
some of the properties of the D-branes can be transposed to properties
of quantum black holes.

If one focuses on the problem of the microscopical origin of the black hole
entropy, there are two more ingredients that one needs. First of all,
if the computation at weak coupling is to teach us something about
the physics of black holes, which are generally defined
when gravitational effects
are important, then we need to be confident that the weak coupling 
computation is not renormalized when extrapolated to strong coupling.
This is done restricting our attention to supersymmetric BPS configurations, 
corresponding to extreme black holes. It is actually the main limitation
of this computation. The second requisite is that the area of the horizon,
which gives the entropy, is still finite in the extreme limit. This
is quite restrictive, so one has to focus only on 5 and 4 dimensional
extreme black holes.

The most accessible setting for the computation of the microscopical entropy is
the one presented in \cite{callanmaldacena}. Here a five dimensional
black hole carrying 3 charges is considered. Two of the charges are of
RR type, and the third one is a KK momentum in an internal direction
(all the directions longitudinal to the D-branes are now taken to be 
compact). The massless modes of the ($\nu=4$) system consisting of $N_1$
D1-branes inside $N_5$ D5-branes are identified, and are taken
to contribute to a total amount of $N_{KK}$ units of internal momentum
(in the direction of the D-string).
In order for the configuration to preserve some supersymmetry and
to be extremal, the massless modes can only be, say, right moving.
The statistical entropy is then given by the logarithm of the degeneracy
of such a configuration, and it exactly matches (including all the
numerical factors) the semiclassical Bekenstein-Hawking entropy
of the corresponding black hole solution.

Note that the entropy, being a physical quantity, must be U-duality invariant.
It can thus be computed once for all in any one of the configurations which
belong to the same U-duality orbit.

Even if away from extremality it is not clear how the strong and weak coupling
pictures can exactly be related, the D-brane picture gives a very nice
heuristic picture of how the Hawking radiation could be explained
in an evidently unitary way. Consider the same 5 dimensional configuration
as before, but now with the massless modes going in both directions
along the D-string. This corresponds to a non-extremal black hole, with
non-vanishing Hawking temperature, and which thus tends to radiate its
energy. The mechanism by which it radiates translates in the D-brane picture
by a process in which two open string modes travelling in opposite
directions collide and produce a closed string which propagates away from
the D-brane compound. The corresponding amplitude is given by the disk
with two insertions at its boundaries and one insertion in the bulk.
The same amplitude gives the reverse process, a particle infalling into
the black hole. These amplitudes have been computed, and matched to the
corresponding semi-classical amplitudes of particle emission and absorption
by black holes \cite{callanmaldacena,emission}.

This microscopic explanation of the black hole entropy with the use
of D-branes is a very important result, but there are still some obscure
points in our understanding of black holes. For instance, it is
still not clear what is the real nature of the event horizon, which
only appears at strong coupling, and is simply not present in the
D-brane picture used to compute the entropy. We will come back to the
problem of black hole entropy in M-theory in Chapter \ref{SCHWARZchap}.

The last topic we will quickly overview is the application of D-branes to
the construction of gauge theories in several dimensions
\cite{hananywitten,elitzur,giveon}. As we have already discussed, the theory
on the world-volume of the branes is at low-energies SYM with gauge
group $U(N)$, where $N$ is the number of parallel branes. The amount
of supersymmetry is equivalent to ${\cal N}=4$ in 4 dimensions
(we now use ${\cal N}$ for the number of supersymmetries in order not to
confuse it with the number $N$ of D-branes). 
One way to reduce the supersymmetry is to suspend the D-branes between
two NS5-branes. It can be seen that the boundary
conditions on the D-brane world-volume break one half of the supersymmetries.
Furthermore, the direction transverse to the NS5-branes is of finite
extent, and it results that D$p$-branes produce a $p$-dimensional theory with
${\cal N}=2$ supersymmetry. If the two NS5-branes are rotated with respect
to each other, one can further break the supersymmetry down to ${\cal N}=1$.
Additional D-branes perpendicular to the original D$p$-branes constituting
the gauge theory provide matter fields arising from the open strings
connecting the two kinds of D-branes.

It can then be shown that moving around the branes of the set-up described
above, and using the dualities of the underlying string theory
(including the possibility to view the configuration as an 11 dimensional
set up \cite{wittenm}), lots of informations on the classical and quantum
moduli space and on the dualities of the gauge theories can be
determined and matched to field theory results, which can thus be
extended in this way. This is actually only one example of how one 
can learn about the non-perturbative aspects of field theories by 
embedding them into M-theory.

To conclude with, we can mention that
one of the applications of D-branes has also been the possibility to 
formulate Matrix theory, which is the subject of the next section.

\section{Matrix theory approach to M-theory}
\label{MATRIXsec}

In this section we discuss the proposal of Banks, Fischler, Shenker
and Susskind \cite{bfss} to treat M-theory directly at the non-perturbative
level by formulating it in the infinite momentum frame.
Some time after the original conjecture of \cite{bfss}, 
which appeared in fall 1996,
the model was given an alternative, but similar, formulation in the
discrete light cone quantization by Susskind \cite{sussdlcq}.
Subsequently,
its rigourous definition emerged from the two contemporary papers
of Sen \cite{senmatrix} and Seiberg \cite{seibergmatrix}.
Some reviews on Matrix theory already exist \cite{matrixrevs}.
It is important to note that unlike the material that has been discussed
until now, Matrix theory as a description of M-theory
is still at a highly conjectural level, and we
will indeed mention some of the problems which might prevent the
establishment of Matrix theory as the `standard' approach to M-theory.

Let us preliminarily discuss what are the infinite momentum frame (IMF)
and the discrete light cone quantization (DLCQ) for a general theory.

The basic idea leading to the formulation of a theory in the IMF is
the following. One takes a particular direction in the (flat) space-time,
say, $\tilde{x}_{11}$. An arbitrary distribution of relativistic particles 
will be described, in particular, by momenta in the 11th direction
$\tilde{p}_{11}$ of both signs and of all magnitudes. Let us now perform a 
boost of parameter $s$ in $\tilde{x}_{11}$:
\be \begin{array}{ccc}
t&=&\cosh s\ \tilde{t} -\sinh s\ \tilde{x}_{11} \\
x_{11}&=&-\sinh s\ \tilde{t} +\cosh s\ \tilde{x}_{11},
\end{array} \label{boost11}
\ee
The momenta transform accordingly:
\be \begin{array}{ccc}
p_0&=&\cosh s\ \tilde{p}_0 +\sinh s\ \tilde{p}_{11} \\
p_{11}&=&\sinh s\ \tilde{p}_0+ \cosh s\ \tilde{p}_{11},
\end{array} \label{boostp}
\ee
Fixing the convention to have $p_0>0$, and choosing the boost parameter $s$
to be also positive, we see that for a sufficiently large boost, $p_{11}$
can always be made positive because for any particle of mass $m$ and
transverse momentum $p_\perp$, the mass-shell relation 
$-\tilde{p}_0^2+\tilde{p}_{11}^2+p_\perp^2=-m^2$ implies 
$\tilde{p}_0\geq |\tilde{p}_{11}|$.
Furthermore, we can also make $p_{11}$ arbitrarily large compared to any
other scale in the problem. When we have done so, the energy of
a particle in the IMF can be extracted from $p_0$:
\be
E=p_0-p_{11}=\sqrt{p_{11}^2+p_\perp^2+m^2}-p_{11}={p_\perp^2+m^2 \over
2p_{11}},
\label{imfenergy}
\ee
in the limit in which $p_{11}$ is very large. We thus see that the
energy acquires a non-relativistic form, with $p_{11}$ playing the
r\^ole of the non-relativistic mass. 

We see that the bulk of the particles of the theory we are considering
have an energy in the IMF which is of ${\cal O}(1/p_{11})$, with
$p_{11}\rightarrow \infty$ denoting now
the generic scale of the momenta in the 11th
direction (as e.g. the total momentum of all the particles). We might be
worried about the particles which after the infinite boost still have a 
vanishing or negative $p_{11}$. Their IMF energy as defined 
in \rref{imfenergy} will be however of ${\cal O}(1)$ or more in $p_{11}$,
which means that they will be very energetic with respect to all the
others. It is in this sense that they can be neglected, even if the
procedure of integrating them out should be by no means trivial.

Further boosts in the 11th direction of the kind \rref{boostp} simply
produce a concomitant rescaling $p_{11}\rightarrow e^s p_{11}$,
$E \rightarrow e^{-s} E$. A boost in the transverse directions $x_\perp$
has to be small with respect to the scale determined by $p_{11}$, and
the consequence is that the system has Galilean invariance instead of
Lorentz invariance in the transverse directions. For a small boost
parameter $v$, the tranverse
momenta can be seen to transform like $p_\perp \rightarrow
p_\perp + p_{11} v$, with $p_{11}$ acting again as the non-relativistic
mass.

In order to regulate the theory, we can further take the direction $x_{11}$
to be compact of radius $R$. The momentum $p_{11}$ is accordingly quantized,
\be
p_{11}={N\over R}.
\label{partons}
\ee
$N$ is the number of partons, i.e. of units of momentum. Since any 
particle in the IMF formulation of the theory has a non-vanishing momentum 
in the 11th direction, it can be said to be made of partons.
The dynamics of partons then describes the whole theory in uncompactified
space-time, when $p_{11}$ is taken to infinity, along with the radius $R$. 
This implies that one has to consider the large $N$ limit of the theory
of partons.

The principle of the IMF formulation of a theory is thus the following:
if one has 
the theory governing the dynamics of partons, then its extrapolation
to large $N$ (in order to have large $p_{11}$) 
and large $R$ should give the complete
non-perturbative description of the original theory.

The basic requirement on the theory of partons
is that it has to have Galilean invariance in the transverse
directions. This invariance is promoted to its supersymmetric version
if the original theory was supersymmetric, however note that generically
the theory of the partons has half of the supersymmetries of the
full theory because the other half is broken by the presence
of the non-vanishing momentum in the 11th direction.

Before going on to see what is the theory of partons for M-theory,
let us briefly discuss the DLCQ formulation for a general theory.
In this case one sets up from the beginning in the light cone
coordinates $x^\pm={1\over 2}(t\pm x_{11})$, and takes $x^+$ to act
as the `new' timelike coordinate. Accordingly, the hamiltonian will be
given by its canonical conjugate variable, $p_+$ (where we have
$p_\pm=p_0\pm p_{11}$). From the mass-shell relation
$-p_+p_-+p_\perp^2 =-m^2$, one extracts
an expression for the light-cone energy:
\be
p_+={p_\perp^2 +m^2 \over p_-}.
\label{lcenergy}
\ee
The momentum $p_-$ now acts as the non-relativistic mass, and we also
see that in this formulation the modes which originally had $p_-=0$
are neglected (the only such modes are massless ones propagating 
along the $x^-$ direction; see \cite{dlcqproblems} for a discussion
of the possible problems in discarding these modes).

We can now periodically identify the lightlike coordinate $x^-$, with
period proportional to $R$ (we do not care about numerical factors
throughout this section). The conjugate momentum is thus quantized,
$p_-={N\over R}$, playing thus the same r\^ole as $p_{11}$ in the IMF.

Note that performing a boost similar to \rref{boost11}--\rref{boostp},
the momenta are rescaled by $p_\pm \rightarrow e^{\pm s}p_\pm$. This
might seem the opposite behaviour with respect to $E$ and $p_{11}$ in
the IMF. It has however the same physical meaning. In the IMF, 
performing a further boost amounted to increase the number
of partons $N$. Only after these rescalings $x_{11}$ could be compactified, 
when the scale of
$p_{11}$ had already been fixed. On the other hand, in the DLCQ the number of
partons is fixed, and one can still perform a boost which has the
effect of rescaling $x^-$ and thus $R$, which goes to $e^s R$. From this
follows
the behaviour of $p_-$ and $p_+$ under the boosts in the light cone
direction.

The DLCQ makes contact with the original theory in non-compact
space-time when we take $R\rightarrow \infty$ and $N\rightarrow \infty$
while keeping $p_-$ fixed.

The same theory of partons can thus describe the original theory 
when the latter is formulated in the DLCQ. The DLCQ is a finite $N$
formulation of the theory in the sense that in the case of the IMF one
has to take $N\rightarrow \infty$ before taking the uncompactified limit
(i.e. even the regulated theory needs the $N\rightarrow \infty$ limit),
while in the present case the large $N$ limit should only be taken
to recover the unregulated theory.

Let us now see why it is convenient to formulate M-theory in the IMF or
in the DLCQ. The main step in the two latter fomulations is having a
good description of the parton dynamics. In the case of M-theory,
we are particularly lucky because even if M-theory itself does not
have a formulation,
an elementary knowledge of string dualities allows us to determine 
completely the theory of partons!

The key remark, which crucially uses the fact that one has to go through
a regulated, compactified theory, is that M-theory with a compact
direction is nothing else than type IIA string theory. Moreover,
momentum in the compact 11th direction maps to the electric charge of
the RR 2-form
field strength of type IIA theory. Now we have already argued in the preceding
sections that there are no perturbative string states carrying such
charge. Thus, the theory of $N$ partons reduces to
the theory of $N$ D0-branes of type IIA theory.
It is important to note that since $p_{11}$ (or $p_-$) is strictly positive,
anti-D0-branes are excluded from the model to begin with.
The type IIA theory in which the D0-branes live is a totally auxiliary theory,
not to be confused with the theory that should be given by M-theory (as 
described by Matrix theory) when compactified on a direction other than 
$x_{11}$ (or $x^-$).

M-theory is then recovered when the number
of D0-branes is taken to infinity, along with the radius $R$. This
should provide a complete and non-perturbative description of M-theory
in 11 non-compact directions (we will see shortly that there are some
non-trivial complications in describing compactifications of M-theory).
In particular, there are other objects in type IIA string theory that
can carry a D0-brane charge, most notably the other D-branes
(this is so because their world-volume action couples to the RR 1-form
potential through Wess-Zumino terms, see e.g. \cite{douglaswithin}).
The D0-brane theory should then reproduce by itself the theory of all
the other D-branes, which are interpreted as M-branes (or branes of 
M-theory) upon decompactification.

The theory of $N$ D0-branes is easily derived from the D-brane physics
described in the preceding section. In the limit of small string coupling
$g$ and large string tension $\alpha'\rightarrow 0$, it is given
by the reduction to 0+1 dimensions of the 10 dimensional super Yang-Mills
(SYM) theory with $U(N)$ gauge group. It is thus some sort of
supersymmetric quantum mechanics of matrices, hence the name Matrix theory,
also spelled M(atrix) theory to indicate that it should describe M-theory.
We will present this SYM theory hereafter. The non-trivial step in the
Matrix theory conjecture is that one can consistently consider the
theory of $N$ D0-brane as decoupled from the rest of the 
type IIA theory, and that this limit correctly reproduces the IMF or DLCQ
limit.

Let us note that Matrix theory is a quite elaborated parton model, since
the set of $N$ parton coordinates is actually promoted to a set
of $N\times N$ non-commuting matrices, thus introducing a lot
of extra modes giving the non-trivial interactions between the partons.

Before going to the precise definition of the model, it is worthwhile
mentioning the concrete facts presented already in \cite{bfss}, and
which supported this bold conjecture.

The first fact is that the D0-brane scattering reproduces the scattering
of gravitons (and their supersymmetric partners) in 11 dimensions. 
That the D0-branes interact at a length scale which is the 11 dimensional
Planck length was already shown in \cite{danfersun,kabatpouliot,dkps},
but in \cite{bfss} the scattering of 11 dimensional massless particles
was reproduced with the exclusive use of the SYM quantum mechanics.
When in the scattering process there is no transfer of momentum in the
11th direction, then the computation is performed in the limit
in which the matrices are commuting, which corresponds to the limit
of widely separated clusters of D0-branes. Note that interactions 
with $p_{11}$ transfer are more difficult to treat because they
involve rearrangements of D0-branes between different clusters.
This first evidence is an important result, even if one could have foreseen it 
simply by duality arguments, for instance noting that after all the D0-brane
and the 11 dimensional massless particles are the same objects
but in two different, dual, pictures. The non-trivial step is really that
the parton model is enough to reproduce the result.

The second, and maybe more astonishing fact, is that the Matrix model
reproduces also the membranes of M-theory, or M2-branes. The fact
that D2-branes could be collective excitations of D0-branes was
already suggested by Townsend \cite{towndfromm}, and both remarks
are actually based on the attempt by de Wit et al. \cite{dewitetal}
to quantize the 11 dimensional supermembrane by formulating
it in a regularized set-up. This regularized supermembrane theory
exactly coincides with the supersymmetric quantum mechanics governing
the dynamics of the D0-branes. The smooth supermembrane is however
recovered when in the $N\rightarrow \infty$ limit the matrices
have a fundamentally non-commuting behaviour. In the context of
Matrix theory, the non-trivial check is that the tension of the matrix
membrane exactly matches the one which is derived for the M2-brane
by duality arguments.
The membrane formalism also provides us with a proof that, at least
at the classical level, Matrix theory has 11 dimensional Lorentz
covariance.

At this stage, one might be puzzled by the following remark: the model
used to describe the dynamics of the D0-branes, i.e. the 0+1 dimensional SYM,
is derived in the string theory context in the limit in which
$g\rightarrow 0$ and $m_s\rightarrow \infty$, where $m_s=1/l_s=
1/\sqrt{\alpha'}$ is the string mass. Now if the D0-branes are identified
with the partons of M-theory in the IMF (or in the DLCQ), the
relations between $g$ and $m_s$ on one side and $R\equiv R_{11}$ 
and $m_p=1/l_p$ (the 11 dimensional Planck mass) on the other are:
\be
g=(Rm_p)^{3\over 2}, \qquad \qquad m_s =(Rm_p^3)^{1\over 2}.
\label{iia11rel}
\ee
These are simply the relations \rref{11iiarel} but expressed the other way
round. In the end, when one is making contact with the original uncompactified
M-theory that the parton model is supposed to describe, one has to take
the limit $R\rightarrow \infty$ while keeping $m_p$ fixed. This seems
to imply that it corresponds to taking the D0-brane theory in
the limit $m_s\rightarrow \infty$ and $g\rightarrow \infty$, in apparent
contradiction with the limit in type IIA theory discussed above. Moreover,
when one is actually doing computations with the parton model, one generally
wants to take $m_p$ and $R$ fixed at a finite value. This also implies
finite values for $g$ and $m_s$, and not the limiting values which
give the SYM theory. How can the Matrix theory conjecture be correct?

The argument which solves this apparent problem comes actually in two
steps, each one of which has been emphasized respectively by Sen 
\cite{senmatrix} and by Seiberg \cite{seibergmatrix} in two contemporary
papers in fall 1997, one year after the original Matrix theory proposal.

The first step has been essentially analyzed by Sen, and amounts to 
finding the exact limit in which the SYM quantum mechanics of the 
D0-branes is a good approximation, but nevertheless yields non-trivial
dynamics. In other words, one is searching for the limit in which the
D0-branes decouple from the rest of the type IIA string theory, but
still produce an interacting theory. As in the two papers
\cite{senmatrix,seibergmatrix}, we also take the occasion to discuss
the case in which some of the transverse directions are compact.

Let us specialize the general action for a D$p$-brane \rref{symp}--\rref{gym}
to our case of interest, that is the action for the
D0-branes. In this case, actually one can fix the gauge in such a way
that only the scalars survive. The residual gauge symmetry then enters
through a constraint on the dynamics of the model. The bosonic action
we are left with is:
\be
I_{D0}={1\over gm_s^3}\int dt\ \left({1\over 2} tr \dot{\phi}_i^2
-{1\over4} tr [\phi_i, \phi_j]^2 \right).
\label{symqm}
\ee

We are now at the core of the problem of how the dynamics of the D0-branes
decouples from the rest of the physics of type IIA string theory. For the
action above to be a good description of the D0-brane dynamics, we have
to take the limits
\be
g\rightarrow 0,\qquad \qquad m_s \rightarrow\infty 
\label{d0limit}
\ee
in the string theory. However for the same action to yield non-trivial
informations, one has to keep fixed the value of the coupling constant
$g_{{YM}_{0+1}}$, and the scale given by the expectation values
of the scalars. This second requirement implies that if there are
some compact directions, we want the scale of the interval over
which the $\phi_i$ are defined to be fixed. If the compact directions
have radii $\tilde{R}_i$, then the scale to be fixed is $\tilde{R}_i m_s^2$.

Since $g_{{YM}_{0+1}}^2=gm_s^3$, fixing the SYM coupling constant is
consistent with the limit \rref{d0limit}. In order to keep $\tilde{R}_i m_s^2$
fixed in the same limit, we see that we must take $\tilde{R}_i\rightarrow 0$,
and actually $\tilde{R}_i$ has to take substringy values, 
$\tilde{R}_i\ll l_s$. 

Let us now view the string theory parameters as coming from 11 dimensional
ones, which we call for the moment $R_s$ and $\tilde{m}_p$ to
distinguish them with the ones which define the Matrix model, $R$ and $m_p$,
and which must be finite. From \rref{11iiarel} we see that the
limit \rref{d0limit} implies:
\be
R_s\rightarrow 0, \qquad\qquad  \tilde{m}_p \rightarrow\infty. 
\label{d0limit2}
\ee
The SYM coupling however is the only combination
of these two quantities which enters in the action \rref{symqm},
and is given by:
\[
g_{{YM}_{0+1}}^2=gm_s^3=(R_s \tilde{m}_p^2)^3.
\]
The quantity to be fixed is thus $R_s \tilde{m}_p^2$. As for the compact
directions, one can also see that:
\[
\tilde{R}_i m_s^2=\tilde{R}_i R_s \tilde{m}_p^3 =(R_s \tilde{m}_p^2)
\tilde{R}_i\tilde{m}_p,
\]
so that one fixes $\tilde{R}_i\tilde{m}_p$, 
the transverse radii in 11 dimensional Planck units, 
when taking the limit \rref{d0limit2}.

Sen argues in his paper \cite{senmatrix} that the Matrix theory
conjecture amounts to postulating that the same action \rref{symqm} describes
the physics of the DLCQ partons of M-theory with finite
parameters $R$, $m_p$ and
$R_i$, provided one makes the following identifications:
\bea
R_s \tilde{m}_p^2&=& R m_p^2, \label{rmp2} \\
\tilde{R}_i\tilde{m}_p&=& R_i m_p. \label{rimp}
\eea

Seiberg has actually shown in his paper \cite{seibergmatrix} how the
two pictures are physically related. The basic result is that
the theory in the DLCQ is equivalent, by a large boost, to the same
theory in the IMF and compactified on a vanishingly small circle.

Let us see, following Seiberg, how we can reach the DLCQ formulation
by boosting a theory compactified on a space-like circle.
The `space-like' theory is defined by the following identifications:
\be
\tilde{t}\sim \tilde{t}, \qquad\qquad \tilde{x}_{11}\sim \tilde{x}_{11}+R_s.
\label{spacelikerad}
\ee
When we perform a boost like \rref{boost11}, the identifications on the 
boosted coordinates become:
\[
t\sim t -\sinh s\ R_s, \qquad \qquad x_{11}\sim x_{11} +\cosh s\ R_s.
\]
This gives for the light like coordinate $x^-\sim x^-+e^s R_s$. 
Choosing the boost parameter to be:
\be
e^s ={R\over R_s},
\label{boostfixed}
\ee
we recover the DLCQ identifications in the limit $R_s\rightarrow 0$:
\be
t\sim t-{1\over 2} R, \qquad \qquad x_{11}\sim x_{11} +{1\over 2} R, 
\label{lightlikerad}
\ee
with $R$ finite. Note that in this limit the velocity of the boost
is:
\[
v= \tanh s =1-2{R_s^2\over R^2}+\dots
\]
We have thus shown that the theory we started with has to be compactified on a
vanishingly small space-like circle.

Remember now that we had already argued that it was still possible in the
DLCQ to perform boosts in the lightcone direction, which amounted to rescale
the radius by $R\rightarrow \lambda R$, and accordingly the DLCQ momentum
and energy by $p_-\rightarrow {1\over \lambda} p_-$ and
$p_+ \rightarrow \lambda p_+$ respectively. This means that $p_-$ correctly
scales as $1/R$, and that the typical DLCQ energy must scale linearly in $R$, 
i.e. it is of the order:
\be
p_+ \sim R m_p^2,
\label{dlcqenergy}
\ee
as suggested also by \rref{lcenergy}.

When we `undo' the boost from the light cone set up \rref{lightlikerad}
to the space-like compactification \rref{spacelikerad}, the momentum and
the energy are accordingly rescaled by the factor \rref{boostfixed}.
This gives at leading order in $R_s$ the following behaviour:
\[
|\tilde{p}_{11}|\simeq {R\over R_s}p_- ={N\over R_s},
\]
\be
\tilde{p}_+=\tilde{p}_0-|\tilde{p}_{11}|\equiv E ={R_s \over R} p_+
\sim R_s m_p^2. \label{unboostedenergy}
\ee
We thus observe that all the relevant energies of the DLCQ formulation
are mapped to vanishingly small energies in the space-like 
set up \rref{spacelikerad}.

The benefit of considering the original DLCQ M-theory in the equivalent
formulation \rref{spacelikerad} is that now the compactification
gives perturbative string theory. Since however we are keeping $m_p$
fixed, the string theory parameters given by \rref{iia11rel}
are such that $g\rightarrow 0$ but also $m_s\rightarrow 0$. This is
not what we really want in order to make contact with the D0-brane action
\rref{symqm}. Note however that from \rref{unboostedenergy} we
learn that we are actually focusing on a very particular range of energies.

One way to simplify the problem is the following. Instead of fixing
the Planck scale and looking at very small energies, we can take the
energies to be finite and rescale the Planck mass to be very large.
In doing this, we are actually changing the theory in which we are
working, in some ways we are transposing the problem to an auxiliary
M-theory. There is however a very controlled way to do this transposition.
We simply choose the rescaled Planck mass $\tilde{m}_p$
in such a way that the energies
in the auxiliary theory are matched to the energies in the DLCQ of the
original theory:
\[ \tilde{E}\sim R_s \tilde{m}_p^2 =R m_p^2 \sim p_+. \]
Also we want the transverse geometry to be invariant under the rescaling,
so that if there are compact directions with radii $R_i$ in the original
theory, these are rescaled to the values $\tilde{R}_i$ which
satisfy $\tilde{R}_i\tilde{m}_p=R_im_p$.

We are obviously repeating the procedure which led to the identifications
\rref{rmp2}--\rref{rimp}, but we have now a clear picture of how the
two theories are related and why should they be equivalent.

Let us recall the steps which lead to Seiberg's formulation of Matrix
theory and which we have discussed above:
\begin{itemize}
\item We start with M-theory in a DLCQ formulation, with finite
parameters $R$, $m_p$ and $R_i$, and with energies which scale as $p_+\sim
Rm_p^2$. 
\item We can see the preceding formulation as the infinite boost limit
of the same M-theory (thus characterized by the same $m_p$ and $R_i$), but
this time compactified on a space-like circle with radius $R_s$ which
is vanishingly small, $R_s \rightarrow 0$. The energies are also very small,
since they now scale like $E\sim R_s m_p^2$.
\item The relevant states of the M-theory of the previous step are 
identified to the states of a new, auxiliary $\tilde{\mbox{M}}$-theory, with
parameters $\tilde{m}_p$ and $\tilde{R}_i$ satisfying 
\rref{rmp2}--\rref{rimp}. The energies are now finite.
\item Since now in $\tilde{\mbox{M}}$-theory we have $R_s \rightarrow 0$
and $\tilde{m}_p \rightarrow \infty$, we can turn to the string theory
picture and describe the dynamics of
the partons by the D0-brane mechanics in the limit $g\rightarrow 0$,
$m_s \rightarrow \infty$, which is correctly given by the action
\rref{symqm}.
\end{itemize}

Using the relations \rref{iia11rel} together with \rref{rmp2}--\rref{rimp},
we can now precisely derive the behaviour of the parameters in the
string theory in which the D0-branes live when, say, $R_s \rightarrow 0$.
For the string coupling and the string mass, we find:
\bea
g&=&(R_s \tilde{m}_p)^{3\over 2}=R_s^{3\over 4} (R m_p^2)^{3\over 4}, 
\label{gbehav} \\
m_s&=& (R_s \tilde{m}_p^3)^{1\over 2}=R_s^{-{1\over 4}}
(R m_p^2)^{3\over 4}. \label{msbehav}
\eea 
The radii of the transverse directions behave like:
\be
\tilde{R}_i={1\over \tilde{m}_p} R_i m_p = R_s^{1\over 2} \left(
{R_i^2 \over R}\right)^{1\over 2},
\label{ribehav}
\ee
and thus are smaller than the string length $l_s=1/m_s$, as already noted.

What the dualities have taught us to do in this case, is to perform
a T-duality on all the compact directions (we now assume that M-theory
was compactified on a $p$-torus $T^p$). The string scale \rref{msbehav}
is not changed under T-duality. On the other hand, the radii of the
dual torus are given by \rref{tduality}, which translates here into:
\be
\tilde{R}_i\quad \rightarrow \quad\Sigma_i={1\over \tilde{R}_i m_s^2}=
{1\over R_i R m_p^3}.
\label{dualtorus}
\ee
The dual torus ${T'}^p$ has thus a finite volume. Note that the finite
values of the $\Sigma_i$ are the inverse of the scale of the vacuum 
expectation values of the scalars in \rref{symqm}, which we wanted
to keep fixed.

Under T-duality, the string coupling also changes according to 
\rref{tcoupling}. Performing the transformation in the $p$ directions,
one obtains:
\be
g\quad \rightarrow\quad g'=g\prod_i{1\over \tilde{R}_i m_s}=
R_s^{3-p\over 4}(R m_p^2)^{3-p\over 4} \prod_i(R_i m_p)^{-1}.
\label{gdualbehav}
\ee
The behaviour of the dual string coupling thus depends on the number
of compact directions. Note also that the $p$ T-dualities change
the D0-branes into D$p$-branes wrapped on the dual torus ${T'}^p$.
We thus see that for $p\leq 3$, we have a theory of D-branes at
weak or finite string coupling, but as soon as we consider
compactifications with $p>3$, we have a theory of D-branes at strong
coupling. We postpone until Chapter \ref{LITTLEchap} the detailed discussion
on how to make sense of these D-brane theories embedded in a 
strongly coupled string theory. Indeed, string dualities and
elevation to (yet another auxiliary) M-theory will allow us to
give a proposal for the cases $p=4,5$ and 6.

We can still compute the coupling of the $p+1$ dimensional SYM theory,
using \rref{gym}. We obtain:
\be
g^2_{{YM}_{p+1}}=g' m_s^{3-p}=(Rm_p^2)^{3-p}\prod_i(R_i m_p)^{-1}.
\label{gymbehav}
\ee
This finite result, as the one for the dual radii \rref{dualtorus}, is
a direct consequence of, and actually the motivation for,  the
prescription \rref{rmp2}--\rref{rimp}.
Note also that the two finite values \rref{gymbehav} and \rref{dualtorus}
are the same as the ones that would have been obtained if one 
had naively computed these values directly from the DLCQ set up,
without caring about the limit \rref{d0limit}.

As a last remark on the Matrix description of M-theory compactifications,
we should point out that new light BPS states arise from
the wrapping modes of the branes of M-theory. This is a non-trivial test
for the Matrix model, because if it is to decribe M-theory in the
full non-perturbative regime, then all these modes have to be contained
in Matrix theory compactifications as given by the prescription above.
We will see in Chapter \ref{LITTLEchap} that indeed the theories on the
D-branes reproduce all these BPS states, at the price of being rather
unusual theories for $p\geq 4$.

Let us conclude this section on Matrix theory by mentioning the problems 
that are still unresolved, and which might prevent it from being
the favourite approach to M-theory.

A problem that has been addressed ever since the original proposal, but
that nevertheless has still not been settled is the proof of 11 dimensional
Lorentz invariance at the quantum level. More recently, there have been
two papers stating that some of the predictions of Matrix theory did
not reproduce some non-trivial but standard supergravity results,
namely three graviton scattering \cite{dineraja} and two graviton scattering
on a (weakly) curved background \cite{dougooguri}.

Another problematic issue is the background dependence of the model.
For instance, if one wants to adapt the Matrix model to the presence
of a M5-brane wrapped on the IMF or DLCQ direction (or longitudinal
5-brane), then in \cite{berkoozdouglas} it was shown that the 
Hamiltonian has to be supplemented by new pieces. In the string
picture, these new terms are related to the fields corresponding to the 
open strings connecting the D0-branes to the D4-brane (which is
the longitudinal 5-brane in the IIA picture). A somewhat orthogonal
approach was undertaken in \cite{bss}, where the central charges
of the supersymmetry algebra of the Matrix model were computed and
shown to reproduce the charges of the membrane and of the longitudinal
5-brane. This result would thus imply that the description of
the branes is contained in the Matrix model from the beginning.
Yet another problem about the branes in Matrix theory is that contrary
to the longitudinal one, the transverse 5-brane (i.e. an M5-brane
not wrapping on the 11th direction) still remains very elusive.

\section{No conclusion}

This section is not a conclusion since there is obviously no definite answer
to the question at the head of the chapter, `What is M-theory?'. We will
instead try to collect and summarize the facts presented in this chapter
which suggest that string theories are unified at the non-perturbative
level under an 11 dimensional theory.

The 5 consistent superstring theories presented in Section \ref{PERTUsec}
are all related to each other in the following way: type IIA and type IIB
theories are related by T-duality, which acts at the perturbative level
as soon as one of the 9 space-like directions is compact. These two
theories can thus be seen as two different limits in the moduli space 
of compactifications of a unique, type II theory. Then we have seen that
type I theory can be seen as type IIB theory with the addition of a
set of D9-branes and an orientifold 9-plane. In this respect type I theory
is `contained' in type IIB theory, even if this requires the knowledge
of the non-perturbative aspects of the latter, such as the physics of
D-branes. Type I theory is further related to the $SO(32)$ heterotic
string theory by a duality interchanging strong and weak coupling.
To end with, again T-duality relates the two heterotic string theories
as it did for the two type II theories. Further direct relations between,
say, type II and heterotic string theories exist when these theories
are compactified to lower dimensions, possibly on manifolds breaking
some of the supersymmetries.

The above dualities already tell us that there is a unique theory
with many perturbative string theory limits, 
with the fundamental string of one theory
often being a complicated soliton in the dual theory. 
There are however two more dualities which relate 10 dimensional string
theories to an 11 dimensional theory, the low energy effective action of
which is 11 dimensional supergravity. For both type IIA string  theory and
the $E_8\times E_8$ heterotic string theory, the strong coupling
limit can be reinterpreted as the low-energy limit of a theory with
one more dimension. The unique theory
behind all string theories, i.e. M-theory, 
appears thus to be an 11 dimensional theory.

There is little we can say directly about M-theory. The main outcome of
the string dualities is that its low-energy limit should be 11 dimensional
supergravity. This is appealing since this latter theory is totally
constrained and furthermore it is the endpoint of the supergravity program,
i.e. no supergravity in more than 11 dimensions can be constructed.
In order to reproduce also the non-perturbative spectrum of the string
theories, M-theory should contain some BPS extended objects, or branes. 
There is however little hope that M-theory could come out of the quantum
theory of one of these objects. Leaving aside the technical reasons, it appears
that in M-theory the choice  between the M2-brane and the
M5-brane as the fundamental object of the theory is somewhat arbitrary, 
simply because there is actually no
perturbative coupling constant in M-theory, such as the string coupling
constant $g$ in string theories.

It should also be mentioned that M-theory has recently been shown
to possess more exotic branes, such as the KK6 monopole (which is properly
defined only when one of the directions is compactified on a circle)
and the M9-brane, which is still at a very conjectural level 
(see \cite{hull} for a discussion), and which should be related to 
the Horava-Witten construction of M-theory compactified on the
orbifold $S^1/Z_2$. The M-branes have been a powerful tool to provide
evidence for dualities involving directly (the 11 dimensional manifestation
of) M-theory.

Matrix theory is a direct approach to M-theory, even if string theories
are used as auxiliary theories in which to find the appropriate tools
for this approach. It has however some serious flaws, such as the difficulty
to recover all the BPS branes, and the indirect way in which the 
supergravity low-energy limit is recovered. For instance, it seems
unlikely or very difficult to prove that Matrix theory could reproduce
the equations of motion of 11 dimensional supergravity, as string
theories do in 10 dimensions.

After all, string theories as we know them now, are very nice constructions,
despite the lack of a second quantized string theory.
In their almost 30 years of living, they have been cast in a well-defined
set of operating rules which gives beautiful results. 
If there was only one consistent
realistic string theory instead of five, the construction
described in Section \ref{PERTUsec} is really what a theoretical physicist
would call a unified theory based on first principles. What M-theory
teaches us is that if non-perturbative effects are taken into account,
this has a chance to be true, even if we have five (and maybe more)
starting points.

The Second Superstring Revolution, besides promoting M-theory to the
leading r\^ole, has also made the string theories much more powerful.
The exploration of their non-perturbative aspects has been enormously
facilitated by the accessibility of the physics 
of some of their solitons through D-brane
physics, and by the dualities which allow to study the strong coupling
limit of a string theory in terms of the perturbative regime of a dual
theory. It is maybe this enhanced understanding of the non-perturbative
aspects of string theories that can be generically called M-theory.

What we should not forget when discussing a theory unifying all interactions,
is that one should indeed be able to make contact with
`all the interactions'.

The aim of a unique theory of all the interactions should thus be to meet
the basic facts that are accessible in everyday's life. In other
words, M-theory should produce a picture consistent with the world as
we know it. A minimal requirement is that M-theory should predict
that the world in which we live is 4 dimensional, and that the
low-energy physics should be given by the Standard Model. Moreover,
the constants entering in the Standard Model should also be predicted
by M-theory. Now, string phenomenology, and its M-theory extensions,
even if they can reproduce some of the features of the Standard Model,
suffer from an initial problem: in order to reproduce the phenomenology,
one has to introduce many things by hand, first of all compactification to
4 dimensions, which includes supersymmetry breaking.

There are thousands of ways in which one can compactify string or M-theory
to 4 dimensions, and one is generally reduced to study the more realistic
model. The problem which is however much more difficult to address is
how M-theory chooses the particular compactification which is closest to
the Standard Model. String dualities and M-theory have led to the picture
in which all the different compactifications are related. This would
mean that all the vacua of M-theory are connected, and that it would
thus be possible, in principle, to find the global  minimum, or at least
some (long-lived) local minima. Then showing that the Standard Model
indeed corresponds to one of these minima should be the ultimate
evidence that M-theory predicts the world as we see it. 
Present-day research is clearly still very far from this ultimate goal.

M-theory could well not be the Theory Of Everything, or at least we are still
not convinced of it. 
The study of M-theory is however highly interesting in two respects.
First of all, M-theory has a deep structure that still waits to be uncovered,
and this can be approached from the several sides and aspects of the theory.
Secondly, M-theory reveals itself as an incredibly powerful framework
to address many other relevant problems in high energy physics.

\chapter{Classical $p$-brane solutions and their intersections}
\label{PBRANEchap}
\markboth{CHAPTER \protect \ref{PBRANEchap}. CLASSICAL P-BRANE SOLUTIONS}{}

The aim of this chapter is to give a classical description of the 
solitons that are relevant to the study of the non-perturbative
aspects of string theories and of M-theory. These solitons are identified
with the classical $p$-brane solutions of the  supergravities
which are the low energy effective actions of the above-mentioned theories.
A large part of the derivation of the $p$-brane solutions and of their
intersections will be actually done in the more general setting of
a purely bosonic theory with arbitrary couplings of the antisymmetric
tensor fields to the dilaton. Then for some particular values of these
couplings we will reduce to the cases of 10 and 11 dimensional
supergravities. In these latter cases, one can perform an analysis
of the supersymmetric properties of the solutions previously found, and
sometimes find an alternative derivation of the solutions, the
main drawback of this second approach being its lack of generality.
Supersymmetry is however often helpful in motivating the form of the
sought for solutions, and above all it is the main ingredient in
allowing us to consider these solitons of the supergravities as 
solitons of the full string or M-theories.

The chapter is organized as follows. In Section \ref{WHYsec} the problem
of finding classical solutions of some supergravity theories is put
in the perspective of addressing issues pertaining to the 
non-perturbative aspects of superstring theories. Section \ref{SINGLEsec}
is devoted to finding singly charged, spherically symmetric black
$p$-brane solutions. This derivation is interesting with two
respects: it will allow us to classify all the most simple
$p$-brane solutions in 10 and 11 dimensional supergravities, and
thus all the basic solitons of M-theory; we will use techniques
and find some characteristics of these solutions that will guide
us in finding more involved solutions carrying several charges.
In Section \ref{INTERsec} extreme solutions carrying several charges are found.
The supersymmetry of the 11 dimensional solutions is also carefully
examined. In addition, a different kind of solution carrying two
charges is also derived, and it is shown to break all the supersymmetries.
In the Section \ref{CONCLUsec} we end with a discussion and with the review of
some more general solutions that can be found.

The results of the paper \cite{rules} (see also \cite{proceeding}) are
included in Section \ref{INTERsec}.

\section{Why $p$-branes are interesting}
\label{WHYsec}

The name `$p$-brane' actually covers a good deal of different objects.
Most generally, it should be used to indicate, in the context
of a theory containing gravity,  a classical solution which is extended
in $p$ directions, i.e. which has
$p$ spacelike translational Killing vectors. In general $p$-branes
will carry the charge of an antisymmetric tensor field, and those
for which the mass saturates a lower bound given by the charge are 
called extremal.
Black holes and, extrapolating to theories without gravity, magnetic
monopoles are prototypes of $p$-branes for the case $p=0$.
As they are classical solutions, $p$-branes are often called `solitons'.
This is essentially in view of the r\^ole they play in the quantum
theory which is behind the classical theory of which they are solutions.

$p$-branes gain a lot of interesting properties when supersymmetry
is a symmetry of the theory. When enough supersymmetry is present,
we are allowed to consider that some relevant properties of the (extremal)
$p$-brane solitons do not change when the coupling constant of the theory is
varied. At strong coupling the mass, or the tension, of these objects
becomes very small and the solitons become the fundamental quantum
objects of a new, dual theory at strong coupling. The properties
of the classical $p$-brane solutions can be used to identify in this
way some of the properties of this new quantum objects and of their
quantum theory. This is most interesting when one finds hints that
the dual strong coupling theory is actually an already known and well-defined
theory, thus giving a handle on the strong coupling dynamics of the
original theory.

The first example of a theory for which a strong coupling formulation
was found owing to the properties of its solitons was $N=4$ Super
Yang-Mills (SYM) theory in 4 dimensions. This theory has magnetic
monopoles with mass $M_{mon}\sim M_W /g^2$, where $g$ is
the coupling constant of the SYM theory and $M_W$ is the mass of the
massive gauge bosons in the broken gauge symmetry phase. These monopoles
preserve half of the supersymmetries of the $N=4$ SYM and can be shown 
to fall into supermultiplets exactly similar to the ones to which belong
the massive $W$ gauge bosons \cite{osborn}. The conjecture, following
the ideas of \cite{montonen}, is that the strong coupling dynamics
of $N=4$ SYM is exactly reproduced by the same theory where however
the coupling constant is the `magnetic' charge $g_{dual}=1/g$.
The main ingredient of this conjecture is that $N=4$ supersymmetry  
prevents the renormalization of the relation between the mass of the 
monopole and its magnetic charge, thus allowing its interpretation
as a `magnetic' massive gauge boson.

Although the interplay between supersymmetry and duality has produced
incredible advances in the understanding of 4 dimensional gauge theories
(since the work of Seiberg and Witten \cite{seibergwitten}), it is really
in the higher dimensional theories that the pattern of solitonic extended
objects becomes crucial in determining the properties of the underlying
quantum theory. 

The superstring theories, which have a critical dimension $D=10$, have 
as low energy effective actions of their massless modes the $N=1$ or
$N=2$ ten dimensional supergravities\footnote{ 
Though non-renormalizable, these latter theories should give a fairly
accurate description of the dynamics at energies well below the string
scale, which is fixed by ${\alpha'}^{-1/2}$.}. The supergravity
theories have a well-defined set of (abelian) antisymmetric tensor
fields with rank which can be taken to be $\le 5$. To each of these
$n$-form fields correspond two charged objects, one electric and the
other magnetic.

Suppose we have in $D$ dimensions an antisymmetric tensor field strength
$F_n$ of rank $n$, deriving from a potential such that $F_n=d A_{n-1}$.
Then this potential can couple minimally to the world-volume of an
object which thus has to be a $p$-brane with $p=n-2$ spatial extended
directions:
\be \mu_{n-2} \int_{W^{n-1}} A_{n-1}. \label{elcharge} \ee
Here $\mu_{p}$ denotes the charge of the $p$-brane under the $(p+2)$-form
field strength, and $W^{p+1}$ is the world-volume swept by the $p$-brane. 
The term displayed in \rref{elcharge} is an electric
coupling. However one can always define the Hodge dual of the form $F_n$
which will be the $(D-n)$-form field strength $\tilde{F}_{D-n}=*F_n$.
Outside electric sources one can always define a dual potential by
$\tilde{F}_{D-n}=d \tilde{A}_{D-n-1}$. This magnetic potential can now
couple to a $p'$-brane, with $p'=D-n-2$, by the following term:
\be \mu_{D-n-2} \int_{W^{D-n-1}} \tilde{A}_{D-n-1}. \label{magcharge} \ee
To summarize, every $n$-form field strength should imply the 
existence of an electric $p$-brane and a magnetic $p'$-brane, 
with $p=n-2$ and $p'=D-n-2$. Note that $p+p'=D-4$, and that the existence
of both of these objects imposes a Dirac-like quantization condition
on their charges \cite{nepomechie,teitelboim}.

According to the remarkable conjecture of Hull and Townsend 
\cite{hulltownsend} and of Witten \cite{witten}, 
all string theories, together with 11 dimensional supergravity, are related by
dualities, and therefore all strong coupling limits are reformulated
in terms of the perturbative dynamics of a dual theory \cite{witten}.
The essential assumption leading to this conjecture is that the full
non-perturbative duality group of a string theory in 10 dimensions
or lower, coincides with the discretized version of the symmetry group
of the corresponding supergravity. This enlarged duality, generalizing
and unifying the strong-weak S-duality (much similar to the 
electric-margnetic duality in 4 dimensions) and the perturbative
T-duality exchanging momentum and winding modes in the string spectrum,
has been called U-duality \cite{hulltownsend} (see Section \ref{DUALMsec}).

One of the main consequences of U-duality is to put on an equal footing
all the gauge fields appearing in an effective (maximal) supergravity
obtained by trivial (i.e. toroidal) compactification. These gauge fields
arise from the off-diagonal parts of the metric and from partial
compactification of the antisymmetric tensors of the higher dimensional
supergravities. The relation between the gauge fields extends to
their charged objects, and thus to their 10 (or 11) dimensional
$p$-brane ascendants. All $p$-branes should then be considered on
an equal footing, i.e. equally `potentially' fundamental \cite{democracy}.

There is, in string theories, a subtlety in the solitonic `spectrum',
which is most easily seen in type II string theories. In the latter 
theories, the fields coming from the Ramond-Ramond (RR) sector have
the peculiarity that there are no states carrying their charge in the
perturbative string spectrum. However U-duality relates this sought for
states to elementary string states, thus requiring the presence of
the former. This means
that solitons of the effective supergravities must exist, which
correspond to both electric and magnetic RR-charged states.
Now classical U-duality in turn implies the existence of solitons
carrying the same charge as some perturbative string states. The
consistency of the U-duality picture imposes the identification of
these classical solutions to the corresponding perturbative string states.
In conclusion, all possible charged objects must exist as classical solutions
in the effective supergravities.

One of the interests in finding all the $p$-brane solutions is thus 
to corroborate the duality conjectures and to properly find all the
objects belonging to the U-duality multiplets. It is worth saying
that there is by now overwhelming evidence for U-duality. 

Even more interesting, searching for the way the classical solutions
interact, for instance by building solutions with intersecting 
$p$-branes, is relevant for checking the consistency with, 
and sometimes deriving, the interactions of the quantum objects. It is also
important that all these intersecting 
solutions realize the duality conjectures.

Another important issue involving intersections of $p$-branes is
the search for supersymmetric extremal black holes with a non-vanishing
horizon area. It will turn out that this is possible only in 4 and 5
non-compact spacetime (including time). Then going on to the identification
of these $p$-brane configurations with systems of objects in
perturbative string theory, i.e. mainly systems involving D-branes \cite{tasi},
it has been possible to give a microscopic counting of states
strictly reproducing the semiclassical black hole entropy (see
\cite{stromingervafa,callanmaldacena} and the following, considerable,
literature).

\subsection{The starting point: the supergravity actions}

We now go on to the formulation of the problem of finding $p$-brane
solutions in classical supergravities.

We will be primarily interested in the $p$-branes of the maximal 
supergravities in 10 and 11 dimensions, because in principle all
lower dimensional configurations can be constructed by trivial
compactification. Note that compactifications which break 
some of the supersymmetries have a richer structure, and it is
more involved to analyze which brane configurations they are
compatible with. We will not consider such compactifications here.

Despite the differences between the supergravity theories discussed
above, namely 11 dimensional supergravity, type IIA non-chiral 
and type IIB chiral 10 dimensional 
supergravities, searching for solutions  of the equations of motion
will be essentially the same. It is actually a matter of 
selecting particular values of some parameters. It is thus
worthwhile attempting to formulate these equations of motion
for a general theory, and then specialize to the particular cases
of interest after the solutions have been found.
At the end of this subsection we will present a general action
in $D$ dimensions, with general couplings to the dilaton and with
an arbitrary number of antisymmetric tensor fields of arbitrary
rank $n$, which can be straightforwardly reduced to any one of
the cases above.

This generalization to arbitrary dimension, for instance, is of course 
possible because we will consider only a bosonic action. In the
case of the supergravities, we thus consider the truncation to their
bosonic sector. This is what is  ordinarily done when one is
searching for classical solutions, since in the latter the expectation values
of the fermionic fields are generically taken to vanish. 
This fact actually allows for an alternative way to find a particular
class of solutions of the supergravity theories, namely solutions
which preserve some of the supersymmetries: since all the fermionic
fields must vanish, the condition of preserved supersymmetry reduces
to the condition that the supersymmetric variation on the fermionic
fields must vanish also, $\delta_{susy} \psi=0$. However the
equations one obtains in this way depend crucially on all the parameters
of the theory, and thus cannot be straightforwardly generalized.
In Section \ref{INTERsec} we will perform this calculation
in the context of 11 dimensional supergravity.

Let us now review the actions in $D$=10 and 11 and see how they can
be captured in a single general action.

The bosonic part of the action of $D$=11 supergravity is the simplest
one, because there are only two fields present: the metric and a 4-form
field strength $F_4=dA_3$. We use conventions for the
$n$-forms such that $F_n={1\over n!}F_{\mu_1\dots \mu_n} dx^{\mu_1}\wedge
\dots \wedge dx^{\mu_n}$. The action is the following \cite{11dsugra}:
\be
I_{11}={1\over 16\pi G_{11}}\left(\int d^{11}x \sqrt{-g}\left\{ R
-{1\over 48} F_4^2\right\} +{1\over 6}\int A_3\wedge F_4 \wedge F_4 \right).
\label{11action}
\ee
The Newton constant $G_{11}$ defines the 11 dimensional Planck length
by $G_{11}=l_p^9$. The second term in \rref{11action} is the Chern-Simons-like 
term necessary for supersymmetry to hold \cite{11dsugra}. We will
disregard it in most of the rest of this chapter, but it can be checked
`a posteriori' that the solutions we will find are not affected by the
reintroduction of this term.

The (bosonic part of the) action of $D$=10 type IIA supergravity 
\cite{iiasugra} is, written in the so-called string frame (see for instance
\cite{kiritsis}):
\[
I_{IIA}={1\over 16\pi G_{10}}\int d^{10}x \sqrt{-g_S}\left\{e^{-2\phi}
\left(R_S +4 \partial_\mu \phi \partial^\mu \phi - {1\over 12}
H_3^2\right) -{1\over 4}F_2^2 -{1\over 48}F_4^{'2}\right\}\]
\be
+{1\over 16\pi G_{10}} {1\over 2}\int B_2 \wedge F_4 \wedge F_4.
\label{iiastring}
\ee
Here we have used the following definitions: $H_3=dB_2$, $F_2=dA_1$,
$F'_4=F_4+A_1\wedge H_3$ and $F_4=dA_3$. 
The fields which have a factor of $e^{-2\phi}$ in front of their
kinetic term are those arising from the Neveu-Schwarz-Neveu-Schwarz
(NSNS) sector of the closed string, while the others arise from the
Ramond-Ramond (RR) sector. The 10 dimensional Newton constant is
defined in terms of the string tension $G_{10}\sim \alpha'^4$ and
we recall that $e^{\phi_\infty}=g$, the string coupling constant.
It is in general convenient to subtract from the dilaton field
its constant part at infinity $\phi_\infty$ and to insert it
into the value of $G_{10}$, which becomes $G'_{10}\sim g^2 \alpha'^4$.
If we want to keep a unique coupling constant in front of the whole
action, we also have to rescale the RR fields. The new dilaton
$\phi'=\phi-\phi_\infty$ is thus vanishing at infinity.
It is always this dilaton and Newton constant that we will use
hereafter, and we will accordingly drop the primes.

We now rewrite the action \rref{iiastring} in the Einstein frame,
where the gravitational term is canonical. In order to do this
we have to perform a Weyl rescaling of the metric:
\be
g^S_{\mu\nu}=e^{\phi\over 2}g^E_{\mu\nu}. 
\label{stringeinstein}
\ee
Note that, following the discussion of the previous paragraph,
this Einstein metric actually coincides with the string metric at infinity.
Using the formula \rref{weylcurv} of Appendix \ref{KKapp}, we can
rewrite the action in the following way:
\[
I_{IIA}={1\over 16\pi G_{10}}\int d^{10}x \sqrt{-g_E}\left\{R_E 
-{1\over2}\partial_\mu\phi \partial^\mu \phi - {1\over 12} e^{-\phi}
H_3^2 -{1\over 4} e^{{3\over2}\phi}F_2^2 -{1\over 48}e^{{1\over2}\phi}
{F'}_4^2\right\}\]
\be
+{1\over 16\pi G_{10}} {1\over 2}\int B_2 \wedge F_4 \wedge F_4.
\label{iiaeinstein}
\ee
It can now be seen, using again Appendix \ref{KKapp}, that 
this action is the dimensional reduction of the 11 dimensional
action \rref{11action}. This is actually the simplest way to
determine the action of type IIA supergravity.

Let us now turn to the action of the chiral type IIB theory in $D$=10. 
In the string frame it reads:
\[  
I_{IIB}={1\over 16\pi G_{10}}\int d^{10}x \sqrt{-g_S}\left\{e^{-2\phi}
\left(R_S +4 \partial_\mu \phi \partial^\mu \phi - {1\over 12}
H_3^2\right) -{1\over 2}\partial_\mu \chi \partial^\mu \chi\right.\]
\be
\left.-{1\over 12} F_3^{'2}-{1\over 240}F_5^{'2}\right\}
+{1\over 16\pi G_{10}} \int A_4\wedge F_3 \wedge H_3.
\label{iibstring}
\ee
Here the RR fields have the following definitions: $F'_3=F_3-\chi H_3$,
$F_3=dA_2$ and $F'_5=dA_4+A_2\wedge H_3$. To recover the true
equations  of motion of type IIB supergravity \cite{iibsugra}, 
one has to impose
a self-duality condition on the 5-form field strength: $*F'_5=F'_5$.
Strictly speaking, the equations of motion are thus not derived from
the action \rref{iibstring} alone. It will be however convenient
in the following to work with such an action, and to impose the
self-duality condition only at the end (actually on the solutions).

If we go to the Einstein frame with the same transformation above
\rref{stringeinstein}, we obtain:
\[I_{IIB}={1\over 16\pi G_{10}} \int d^{10}x \sqrt{-g_E}\left\{R_E
-{1\over2}\partial_\mu\phi \partial^\mu \phi - {1\over 12} e^{-\phi}
H_3^2 -{1\over 2} e^{2\phi}\partial_\mu \chi \partial^\mu \chi\right.\]
\be
\left.-{1\over 12} e^{\phi}F_3^{'2}-{1\over 240}F_5^{'2}\right\}
+{1\over 16\pi G_{10}} \int A_4\wedge F_3 \wedge H_3.
\label{iibeinstein}
\ee

We can now easily write an action in $D$ dimensions which generalizes
the ones above \rref{11action}, \rref{iiaeinstein} and \rref{iibeinstein}.
However, in order to do this we have to neglect all the Chern-Simons-like
terms in the actions above and in the definitions of the field strengths. 
Luckily, they will turn out to be
completely irrelevant in the case of the solutions we will be searching 
for.

Consider a theory including gravity, a dilaton field and ${\cal M}$
antisymmetric tensor fields of arbitrary rank $n_I$, $I=1\dots {\cal M}$.
Then the most general action, written in the Einstein frame, is the
following \cite{rules,proceeding}:
\be
I={1\over 16\pi G_D}\int d^D x \sqrt{-g}\left\{ R -{1\over2}\partial_\mu \phi
\partial^\mu \phi-{1\over2} \sum_I {1\over n_I!} e^{a_I\phi}F_{n_I}^2
\right\}. \label{genaction}
\ee
Note that some of the forms can be of the same rank, with nevertheless
a different coupling to the dilaton $a_I$ which distinguishes them.

The reduction to the cases discussed above is straightforward:
for $D$=11 supergravity, we only have a 4-form and $a_4=0$, which means
that the dilaton is irrelevant in this theory. For the type II
supergravities in $D$=10, we have $a_3=-1$ for the NSNS 3-form,
and $a_n={5-n\over 2}$ for any RR $n$-form.

We can also specialize to the other string theories: the low energy
effective action of type I string theory can be obtained with
the same value of the coupling for the RR 3-form, and with a coupling
$a_2={1\over2}$ for the 2-form field strengths coming from the open sector
(we only consider the 16 fields relative to the abelian subgroup). 
This coupling derives from the fact that the latter fields are
multiplied, in the string frame action, by a factor of $e^{-\phi}$,
characteristic of the disk amplitude.
For the heterotic string theories, all the fields arise from a closed
NSNS-like sector and the couplings are thus $a_3=-1$ for the 3-form
and $a_2=-{1\over2}$ for the 16 (abelian) 2-forms.

Note that if we wanted to reproduce also lower dimensional supergravities,
we should have added more than one scalar, i.e. instead of only one
dilaton we should have had several moduli. However all the solutions
in higher dimesions can be trivially compactified to reproduce
solutions in lower dimensions. Though straightforwardly generalizable,
the action \rref{genaction} is a compromise in favour of the treatability
of the equations and of the readability of their solutions.

The equations of motion one derives from \rref{genaction} are:
\bea \lefteqn{
R_{\mu\nu}-{1\over2}g_{\mu\nu}R-{1\over2}\left(\partial_\mu \phi
\partial_\nu \phi -{1\over2}g_{\mu\nu} \partial_\lambda \phi
\partial^\lambda \phi \right)
} \nonumber \\
& &-{1\over2}\sum_I {1\over n_I!}
e^{a_I \phi}\left(n_I F_{\mu \lambda_2 \dots \lambda_{n_I}}
{F_\nu}^{\lambda_2 \dots \lambda_{n_I}}-{1\over2}g_{\mu\nu} F_{n_I}^2
\right)=0, \label{eomrfirst}
\eea
\be
{1\over \sqrt{-g}} \partial_\mu\left(\sqrt{-g}g^{\mu\nu}\partial_\nu \phi
\right)-{1\over2}\sum_I {a_I \over n_I!}e^{a_I \phi}F_{n_I}^2=0,
\label{eomphifirst}
\ee
\be
{1\over (n-1)!}{1\over \sqrt{-g}} \partial_\mu \left(\sqrt{-g}
e^{a_I \phi}F^{\mu \nu_2 \dots \nu_{n_I}}\right)=0.
\label{eomformfirst}
\ee
They are better recast in the following form:
\be
{R^\mu}_\nu={1\over2}\partial^\mu \phi\partial_\nu \phi+{1\over2}\sum_I 
{1\over n_I!} e^{a_I \phi}\left(n_I F^{\mu \lambda_2 \dots \lambda_{n_I}}
F_{\nu\lambda_2 \dots \lambda_{n_I}}-{n_I-1 \over D-2} \delta^\mu_\nu
F_{n_I}^2 \right),
\label{eomeinstein}
\ee
\be
\Box\phi={1\over2}\sum_I {a_I \over n_I!}e^{a_I \phi}F_{n_I}^2,
\label{eomphi}
\ee
\be
\partial_\mu \left(\sqrt{-g}e^{a_I \phi}F^{\mu \nu_2 \dots \nu_{n_I}}\right)=0.
\label{eommaxwell}
\ee
The statement that the $n$-form derives from a potential can be replaced
by the imposition, at the level of the equations of motion, 
of the Bianchi identities:
\be
\partial_{[\mu_1}F_{\mu_2 \dots \mu_{n_I+1}]}=0.
\label{bianchi}
\ee

The four sets of equations above, \rref{eomeinstein}--\rref{bianchi},
are the ones which we will use extensively in this chapter to 
find the $p$-brane solutions.

\section{Singly charged extremal and black $p$-branes}
\label{SINGLEsec}

In this section we focus on a particular case of the action \rref{genaction},
where there is a single $n$-form field strength. 
In this framework, we can implement several symmetries in the problem,
which will allow us to constrain the fields in such a way that it will
be possible to find a solution to the equations derived from the action.
The solutions found in this way will then turn out to be useful guides
when we will search for more general solutions in problems with less
symmetries.

The material presented in this section is well-known, and one can find
most of these results in \cite{stelle,gibbmaeda,horostrom,dufflupope}.
We have included it because it is very useful to be familiar
with the single brane solutions before considering the configurations
with several branes, i.e. the multiply charged solutions. The presentation
followed here is however original, and takes into account the structure of
the whole chapter.

We can rewrite 
the equations of motion and Bianchi identities for this simpler case,
containing only gravity, a $n$-form field strength and the dilaton:
\be
{R^\mu}_\nu={1\over2}\partial^\mu \phi\partial_\nu \phi+{1\over 2\ n!}
e^{a\phi} \left(n F^{\mu \lambda_2 \dots \lambda_n}
F_{\nu\lambda_2 \dots \lambda_n}-{n-1 \over D-2} \delta^\mu_\nu
F_n^2 \right),
\label{eomeinstsingle} 
\ee
\be
\Box\phi={a\over2\ n!}e^{a \phi}F_n^2,
\label{eomphisingle}
\ee
\be
\partial_\mu \left(\sqrt{-g}e^{a \phi}F^{\mu \nu_2 \dots \nu_n}\right)=0,
\label{eommaxsingle} 
\ee
\be
\partial_{[\mu_1}F_{\mu_2 \dots \mu_{n+1}]}=0.
\label{bianchisingle}
\ee
Let us note that the equations above have an interesting property. One
can reformulate them in terms of a Hodge dual $(D-n)$-form field 
strength. However to correctly have a generalized electric-magnetic
duality, the definition of the Hodge dual must contain a dilaton
dependent factor. A good definition of the dual field strength is thus
the following:
\be
\sqrt{-g} e^{a\phi}F^{\mu_1 \dots \mu_n}={1\over (D-n)!}
\epsilon^{\mu_1 \dots \mu_n \nu_1 \dots \nu_{D-n}} \tilde{F}_{
\nu_1 \dots \nu_{D-n}}, \qquad \epsilon^{0\dots D-1}=1.
\label{genhodge}
\ee
It is easy to see that using this definition of $\tilde{F}_{D-n}$, 
the equations \rref{eomeinstsingle}--\rref{bianchisingle} rewritten
in terms of it have exactly the same form with however $F_n$ replaced
by $\tilde{F}_{D-n}$, $n$ replaced by $D-n$ and $a$ replaced by $-a$
(or equivalently $\phi$ replaced by $-\phi$).
In this process, the Bianchi identity for $F_n$ becomes
the `Maxwell-like' equation for $\tilde{F}_{D-n}$ and vice-versa. Note
also that the replacement $a \rightarrow -a$ is valid everywhere since
$F_n^2$ is proportional to $-\tilde{F}_{D-n}^2$.

The electro-magnetic duality described above is a symmetry of the 
equations of motion supplemented by the Bianchi identity, but not
of the action. We shall exploit this duality to give a unified treatment
of electrically and magnetically charged $p$-branes.

\subsection{The symmetries of the $p$-brane ansatz and the general 
equations of motion}
\label{KILLssec}

We now consider the symmetries of the problem, in order to simplify
the field equations above by restricting us to some particular
field configurations.

A $p$-brane solution living in a $D$ dimensional space-time
is characterized by the fact that $p$ space-like directions can be
considered `longitudinal' to the brane, while the remaining
$d\equiv D-p-1$ space-like direction are taken to be `transverse'
to the $p$-brane. Since we are considering solutions carrying a single
charge, and thus by definition composed of a single $p$-brane, it is
natural to suppose that the $p$ longitudinal directions are all 
equivalent, i.e. there is nothing in the solution that distinguishes
between them. The timelike direction, which can be considered
as being also longitudinal to the world-volume of the brane, will
not be considered as equivalent to the other longitudinal directions,
at least in the general case.

Since the $p$-brane is taken to be a uniform object, i.e. nothing should 
single out particular points or regions of its volume, the $p$
space-like longitudinal directions define as many translational invariant
directions of the solution. When one considers static objects, as we will
do here, another translational invariant direction is the timelike one.

In the transverse space, the $p$-brane is taken to be localized at a
particular point. Invariance under translations is thus broken. If however
the brane is truly static and not endowed with any angular momentum,
it is natural to postulate spherical symmetry in the $d$ dimensional
transverse space. These are the maximal symmetries we can 
postulate for a single $p$-brane ansatz.

Let us start with a general metric in $D$ dimensions, namely:
\be
ds^2= g_{\mu\nu}dz^\mu dz^\nu, \qquad \mu,\nu=0 \dots D-1. \label{generalissim}
\ee
Having in mind the discussion above, we split the coordinates
accordingly in three sets, $z^\mu=\{t, y^i, x^a\}$, with $i=1\dots p$ and
$a=1\dots d$ (thus $1+p+d=D$). The $y$'s span the directions longitudinal
to the brane, and the $x$'s span the transverse space. We have singled
out the time-like direction $t$.

It is now easy to translate the discussion above on the expected symmetries
of the solutions in terms of Killing vectors of the $p$-brane
geometry. 

We can define the following sets of Killing vectors:
\begin{itemize}
\item The geometry should be invariant under time translations, i.e. it should
describe a static configuration. The corresponding Killing vector
is $\xi_t=\partial_t$ (which in components means $(\xi_t)^\mu=\delta^\mu_0$).
\item There should be invariance under translations in the $y^i$ directions,
along the extension of the brane. The Killing vectors are $\xi_i=\partial_i$.
\item There should be invariance under $SO(d)$ rotations in the 
transverse space of the $x$'s. The Killing vectors are 
$\xi_{ab}=x^a\partial_b -x^b\partial_a$, or in components
$(\xi_{ab})^\mu=x^a\delta^\mu_b-x^b\delta^\mu_a$.
\item Invariance under $SO(p)$ rotations in the longitudinal space
leads to the Killing vectors $\xi_{ij}=y^i\partial_j -y^j\partial_i$.
\end{itemize}
A word of caution has to be said about the last set of Killing vectors.
Generally, we would like eventually to be able to compactify the longitudinal
space, and take it to have the topology of a $p$-torus $T^p$. In this
case these last Killing vectors can only be defined locally but not
globally. It is however useful to assume that they exist, at the price
of considering at a first stage truly infinite branes.

The statement that a vector field is a Killing vector means that the
Lie derivative of any tensor field along this vector field must vanish.
The Lie derivative along $\xi^\mu$ is defined by:
\bea
{\cal L}_\xi {T_{\mu_1\dots \mu_r}}^{\nu_1\dots \nu_s}&=&
{{T_{\mu_1\dots \mu_r}}^{\nu_1\dots \nu_s}}_{,\rho}\xi^\rho+
{T_{\rho \mu_2 \dots\mu_r}}^{\nu_1\dots \nu_s} {\xi^\rho}_{,\mu_1}+\dots
+{T_{\mu_1 \dots\mu_{r-1}\rho}}^{\nu_1\dots \nu_s} {\xi^\rho}_{,\mu_r}
\nonumber \\ & &
-{T_{\mu_1\dots \mu_r}}^{\rho\nu_2\dots \nu_s}{\xi^{\nu_1}}_{,\rho}-
\dots -{T_{\mu_1\dots \mu_r}}^{\nu_1\dots \nu_{s-1}\rho}{\xi^{\nu_s}}_{,\rho}.
\label{liederiv}
\eea

The translational Killing vectors $\xi_t$ and $\xi_i$, being constant
vector fields, imply as expected that an invariant
tensor (and particularly the metric tensor $g_{\mu\nu}$) should not
depend on the $t$ and $y^i$ coordinates respectively.

The Killing vectors of $SO(p)$ symmetry $\xi_{ij}$, because of the
simultaneous presence of the $\xi_i$ symmetries, are very restrictive.
In particular, they imply that an invariant symmetric 2-index tensor
must be proportional to the identity $\delta_{ij}$, and that the only
invariant antisymmetric tensor is the one proportional to the
Levi-Civita $p$-form $\epsilon_{i_1\dots i_p}$.

The $\xi_{ab}$ Killing vectors of the $SO(d)$ symmetry in the
transverse space give rise to more complicated equations since
the tensor components still depend on the $x^a$ coordinates.

For a 2-index symmetric tensor these equations can be solved without much
effort and when we apply the result to the metric we obtain:
\bea
ds^2&=&-B^2(r)dt^2+C^2(r)\delta_{ij}dy^i dy^j+h(r)\delta_{ab}x^a dx^b dt
+k(r)\delta_{ab}dx^a dx^b
\nonumber \\ & &
+j(r) \delta_{ac}\delta_{bd}x^a x^b dx^c dx^d,
\qquad r^2=\delta_{ab}x^a x^b.
\label{firstmetric}
\eea
We can now go to spherical coordinates, for which:
\[
\delta_{ab}dx^a dx^b=dr^2+r^2 d\Omega^2_{d-1}, \qquad
d\Omega^2_{d-1}=d\theta_1^2+\sin^2\theta_1 d\theta_2^2+\dots+
\sin^2\theta_1 \dots \sin^2\theta_{d-2} d\theta_{d-1}^2.
\]
The precise change of coordinates and the range of the angular variables
is given in the Appendix \ref{RICCIapp}, eqs. \rref{sphtox}--\rref{xtotheta}.
Note also that $rdr=\delta_{ab}x^a dx^b$.
As far as the metric is concerned, we can still use diffeomorphisms to
put the metric in a simpler form. We can indeed redefine the $t$ 
coordinate adding to it an $r$-dependent piece in order to eliminate
the $g_{tr}$ element of the metric. 

We eventually obtain the following metric, which is the most general one
that incorporates all the symmetries discussed above:
\be
ds^2=-B^2 dt^2 + C^2 \delta_{ij}dy^i dy^j + F^2 dr^2 +G^2 r^2 d\Omega^2_{d-1},
\label{gensphmetric}
\ee
with all the functions depending only on $r$.

We can still perform reparametrizations of $r$, which means that the
two functions $F(r)$ and $G(r)$ introduce a redundance. However it depends
from case to case whether one gauge choice (e.g. $F=G$ or $G=1$) is better
than the other, and 
actually we will discuss a case for which it is better not to fix
the `$r$-gauge' at all.

It is worth making a remark now that will only be justified in Section 
\ref{INTERsec},
when going to the discussion of solutions charged under several tensor
fields. Even in cases where the $SO(p)$ symmetry of the `internal' space can no
longer be invoked, we will still postulate that the metric is diagonal
(this being now truly an ansatz on the field configuration), taking a
form which is a slight generalization of \rref{gensphmetric}:
\be
ds^2=-B^2 dt^2 + \sum_{i=1}^p C_i^2 (dy^i)^2 +
F^2 dr^2 +G^2 r^2 d\Omega^2_{d-1}. 
\label{thegensphmetric}
\ee
We will actually use this slightly more general metric to compute
the Ricci tensor in Appendix \ref{RICCIapp} and the thermodynamics of general
black $p$-branes in this section.

We now go on characterizing which components of the $n$-form field
strength are invariant under the Killing vectors and are relevant 
to the $p$-brane solutions. 

One is guided intuitively by the simplest case, which is the 
Reissner-Nordstr\"om charged black hole in 4 dimensions. In this problem
it is quite easy to constrain the form of the 2-form Maxwell field strength.
Requiring time translation invariance and $SO(3)$ rotational symmetry, the
only surviving components are, in spherical coordinates,
$F_{tr}={\cal E}(r)$ and $F_{\theta\varphi}={\cal B}(r)r^2 \sin \theta$. 
This gives in
cartesian coordinates, $F_{ta}=\partial_a E(r)$ and 
$F_{ab}=\epsilon_{abc}\partial_c \tilde{E}(r)$ respectively, where
$E'={\cal E}$ and $\tilde{E}'={\cal B}$.

What we learn in this case is that as far as the physically interesting cases 
are concerned, either only one index relative to the $x^a$ coordinates
appears in the electric case, either all but one of the same indices
appear in the magnetic case, the tensor now being proportional
to the Levi-Civita $d$-form ($d=3$ in the Reissner-Nordstr\"om case above).

If we recall that by $SO(p)$ and translation invariance in the internal
space either all the $y^i$ indices appear or none,
we are thus led to formulate the following two ans\"atze for a
$n$-form field strength. A $(p+2)$-form will be said to satisfy 
an electric ansatz if it is of the form:
\be
F_{ti_1 \dots i_p a}=\epsilon_{i_1 \dots i_p}\partial_a E(r).
\label{electricansatz}
\ee
It is straightforward to check that with such an ansatz the Bianchi
identities \rref{bianchisingle} are trivially satisfied. The
expression \rref{electricansatz} can be rewritten in spherical
coordinates as $F_{ti_1 \dots i_p r}=\epsilon_{i_1 \dots i_p} E'(r)$.

A magnetic ansatz is satisfied by a $(d-1)$-form of the form:
\be
F^{a_1 \dots a_{d-1}}=\epsilon^{a_1 \dots a_d}
{1\over \sqrt{-g}} e^{-a\phi}\partial_{a_d} \tilde{E}(r).
\label{magneticansatz}
\ee
Such a $(d-1)$-form satisfies trivially its equations of motion
\rref{eommaxsingle},
as it should be to correctly represent a magnetic charge. In spherical
coordinates, the only non-vanishing component is
$F_{\theta_1 \dots \theta_{d-1}}\sim f(r)
\sin^{d-2}\theta_1 \dots \sin \theta_{d-2}$.

It can indeed be checked that the $n$-forms with only non-vanishing 
components satisfying either the electric \rref{electricansatz} or
the magnetic \rref{magneticansatz} ansatz are invariant under all
the Killing vectors discussed above, and in particular under the
ones related to the $SO(d)$ spherical symmetry.

If a $n$-form is such that $n=D/2$, it can have at the same
time non-vanishing electric and magnetic components. Moreover, if
the form is (anti-)self-dual, the two ans\"atze above single out
the same component.

The requirement that an electric component of a $n$-form field 
strength must be proportional to $\epsilon_{i_1 \dots i_p}$ ceases 
to hold if there is no longer $SO(p)$ symmetry in the internal space.
The absence of the $\xi_{ij}$ Killing vectors will allow us in
Section \ref{INTERsec} to consider both electric and magnetic components
depending on an arbitrary number of $y^i$ indices.

It is important to note that once the $n$-form field
is fixed to correspond to the expression \rref{electricansatz} or
\rref{magneticansatz}, then the other equation
linear in the $n$-form (the equation of motion in the electric case
and the Bianchi identity in the magnetic one) can always be solved
in the first place. 

Let us end this subsection rewriting the field equations, taking into
account the symmetries of the problem. We will do it for an electrically
charged $p$-brane, and thus for a $(p+2)$-form field strength, but
the equations we will obtain will also be valid for a $p$-brane
carrying the magnetic charge of a $(d-1)$-form, by virtue of the
relation \rref{genhodge} and the discussion that follows it.

In spherical coordinates, the only non-vanishing components of the
$(p+2)$-form are $F_{t i_1 \dots i_p r}=\epsilon_{i_1 \dots i_p}E'$.
Taking into account the metric \rref{gensphmetric}, the remaining equation
\rref{eommaxsingle} for the antisymmetric tensor becomes:
\[
\left(e^{a\phi}{(Gr)^{d-1}\over BC^p F}E'\right)'=0,
\]
which gives the solution:
\be
F_{t i_1 \dots i_p r}=\epsilon_{i_1 \dots i_p}BC^pFe^{-a\phi}
{Q\over (Gr)^{d-1}},
\label{nformsolution}
\ee
where $Q$ is an integration constant, and it is proportional
to the electric charge (or alternatively to the magnetic one in what follows).

Using \rref{nformsolution}, we are now able to compute the various terms
appearing on the r.h.s. of the equations \rref{eomeinstsingle} and
\rref{eomphisingle}. For instance, we have:
\[
F_{p+2}^2=(p+2)!\ F_{t y_1 \dots y_p r}F^{t y_1 \dots y_p r} =
-(p+2)!\ e^{-2a\phi}{Q^2\over (Gr)^{2(d-1)}}.
\]
By virtue of the fact that the components in \rref{nformsolution} are
the only non-vanishing ones, we have the following relation:
\[
F^{\mu\lambda_2\dots \lambda_{p+2}}F_{\nu\lambda_2\dots\lambda_{p+2}}=
\bar{\delta}^\mu_\nu (p+1)!\ F_{t y_1 \dots y_p r}F^{t y_1 \dots y_p r}= 
-\bar{\delta}^\mu_\nu(p+1)!\ e^{-2a\phi}{Q^2\over (Gr)^{2(d-1)}},
\]
where $\bar{\delta}^\mu_\nu=\delta^\mu_t \delta^t_\nu+\delta^\mu_i
\delta^i_\nu +\delta^\mu_r \delta^r_\nu$.

The last field we have to consider before writing the remaining equations
of motion is the dilaton, which depends only on $r$ by spherical symmetry.
Note first of all that in the r.h.s. of \rref{eomeinstsingle}
the $\partial^\mu \phi \partial_\nu \phi$ term contributes only to
the $({}^r_r)$ equation by the term ${1\over F^2}{\phi'}^2$.

The Dalembertian operator can be written in our set up in the following
form:
\[
\Box \phi={1\over \sqrt{-g}}\partial_r \left( \sqrt{-g}g^{rr}\partial_r
\phi\right)={1\over BC^pF(Gr)^{d-1}}\left[BC^pF(Gr)^{d-1}{1\over F^2}
\phi'\right]'
\]
\[
={1\over F^2}\left\{\phi''+{d-1\over r}\phi'+\phi'\left[
(\ln B)'+p(\ln C)' -(\ln F)' +(d-1)(\ln G)'\right]\right\}.
\]

We are now ready to collect all the above results and to write 
the Einstein and the dilaton equations for the metric 
\rref{gensphmetric} and the $(p+2)$-form \rref{nformsolution}.
The Einstein equations are totally diagonal in these coordinates, and we
list hereafter the equations in the following order:
$({}^t_t)$, $({}^{y_i}_{y_i})$, $({}^r_r)$, 
$({}^{\theta_\alpha}_{\theta_\alpha})$ and $\Box \phi$.
\[
{1\over F^2}\left\{-(\ln B)''-(\ln B)'\left[(\ln B)'+p(\ln C)' -(\ln F)'
+(d-1)(\ln G)'+{d-1\over r}\right]\right\}
\]
\be =
-{d-2\over 2(D-2)} e^{-a\phi}{Q^2\over (Gr)^{2(d-1)}}, 
\label{einstsphtt} 
\ee
\[
{1\over F^2}\left\{-(\ln C)''-(\ln C)'\left[(\ln B)'+p(\ln C)' -(\ln F)'
+(d-1)(\ln G)'+{d-1\over r}\right]\right\}
\] \be =
-{d-2\over 2(D-2)} e^{-a\phi}{Q^2\over (Gr)^{2(d-1)}}, 
\label{einstsphii} \ee
\[
{1\over F^2}\left\{-(\ln B)''-p(\ln C)''-{(\ln B)'}^2-p{(\ln C)'}^2
+(\ln B)'(\ln F)'+p(\ln C)'(\ln F)' {{}\over{}}
\right. \]
\[ \left.
-(d-1)\left[(\ln G)''+{(\ln G)'}^2
+{2\over r}(\ln G)'-(\ln G)'(\ln F)'-{1\over r}(\ln F)'\right]\right\}\]
\be =
{1\over 2}{1\over F^2}{\phi'}^2-{d-2\over 2(D-2)}
e^{-a\phi}{Q^2\over (Gr)^{2(d-1)}},
\label{einstsphrr} \ee
\[
{1\over F^2}\left\{-\left[(\ln G)'+{1\over r}\right]\left[(\ln B)'+p(\ln C)'
-(\ln F)'+(d-1)(\ln G)'+{d-1\over r}\right]
\right. \]
\be \left.
-(\ln G)''+{1\over r^2}
+(d-2) {F^2\over r^2 G^2}\right\}=
{p+1\over 2(D-2)}e^{-a\phi}{Q^2\over (Gr)^{2(d-1)}},
\label{einstsphtheta} \ee
\[
{1\over F^2}\left\{\phi''+\phi'\left[
(\ln B)'+p(\ln C)' -(\ln F)' +(d-1)(\ln G)'+{d-1\over r}\right]\right\}\]
\be =
-{a\over 2}e^{-a\phi}{Q^2\over (Gr)^{2(d-1)}}.
\label{dilatonsph}
\ee
We have used the equations \rref{sphricciunik} of Appendix \ref{RICCIapp},
where the computation of the Ricci tensor (i.e. the l.h.s. of the Einstein
equations) is performed. 
We are here in the particular case where all $C_i=C$.

\subsection{The extremal dilatonic $p$-brane}
\label{EXTREMssec}

We now set up for finding our first solution. It will become clear
very shortly that the further restrictions that we will implement
on the fields will actually confine the solution to be in a particular
class, i.e. to be extremal. This case is however interesting because
first of all the equations are easily solved, and also because
these extreme $p$-branes turn out to be physically most interesting,
being identified in some cases to fundamental objects.

The first simplification to the equations of motion
\rref{einstsphtt}--\rref{dilatonsph} comes actually by choosing a
definite gauge for the radial coordinate $r$. We fix what we call
the `isotropic gauge', i.e. we require $F=G$. With this choice,
the metric can be written with the function $G^2(r)$ multiplying a flat
metric for the $d$ dimensional transverse space.
Let us stress that this is not a restriction but a choice of 
coordinates which allows the equations of motion to take a simpler
form for this working case.

The following restrictions, which are now physically sensible,
rely on the fact that we focus on extremal solutions. Extremality
roughly corresponds  to the statement that the mass of the object
is equal (or related) to its charge. This means that the object
is fully characterized (in the absence of angular momenta) by only
one parameter, which we take to be its charge. Moreover, in the
context of a supersymmetric theory, extremality is most of the
time\footnote{There are some extremal solution which are
not supersymmetric. However this happens either in theories which
do not have maximal supersymmetry, either for more complicated solutions,
like the one discussed at the end of Section \ref{INTERsec}. Here we suppose
that there are no such complications.} related to the fact that the solutions
preserves a portion of supersymmetry. 

We call
this kind of configuration a Bogomol'nyi--Prasad--Sommerfield (BPS)
state \cite{bps}. It will become clear that this class of BPS states have
a complete balance of forces between each other, i.e. two such states
can stay statically one next to the other. This in turn implies
that all the forces between them either vanish either are proportional
and eventually cancel. Since the dilaton is also responsible of one of 
these forces, the above no-force condition implies that the dilaton
cannot introduce a new parameter.
The conclusion of this discussion is that as far as an extremal
solution is concerned, all the fields should depend on only one 
parameter. Note that the asymptotical values at infinity are not parameters of
this sort because they are fixed from the beginning.
Here we will always take $e^\phi$ and the metric components to be 1 at
infinity, and the $n$-form field strength to vanish there.

The main restriction, or ansatz, that we take for the moment 
is partly motivated 
by the structure of the equations of motion \rref{einstsphtt} and
\rref{einstsphii}: since their r.h.s. are identical, they are fully
compatible with taking $B=C$.
This could also be motivated requiring $SO(1,p)$ Lorentz invariance
along the whole world-volume of the $p$-brane.

That extremality implies Lorentz invariance on the world-volume can
be heuristically understood as follows: a configuration for which
the mass equals the charge, and thus saturates a lower bound, 
is seen from the point of view
of the world-volume of the $p$-brane as a configuration carrying no
energy at all. It seems thus logic to describe this configuration
as flat space from the $(p+1)$ dimensional point of view. Any mass
excitation, or in $D$ dimensional language any departure from
extremality, breaks the Lorentz invariance in $p+1$ dimensions and thus
should lead to $B\neq C$.

We are left with a set of 4 equations, named respectively
$({}^t_t)$, $({}^r_r)$, $({}^\alpha_\alpha)$ and $\Box \phi$:
\[
-(\ln B)'' -{d-1 \over r}(\ln B)'-(\ln B)'\left[ (p+1)(\ln B)'
+(d-2)(\ln G)'\right] 
\]
\be
\qquad\qquad\qquad\qquad=-{d-2\over 2(D-2)} e^{-a\phi}G^{-2(d-2)}
{Q^2\over r^{2(d-1)}},
\label{extremett} 
\ee
\[
- (p+1)(\ln B)''-(d-1)(\ln G)''-(p+1){(\ln B)'}^2 {{}\over {}}
+(p+1)(\ln B)'(\ln G)'
\]
\be
\qquad\qquad\qquad-{d-1 \over r}(\ln G)'-{1\over 2}{\phi'}^2=
-{d-2\over 2(D-2)} e^{-a\phi}G^{-2(d-2)}{Q^2\over r^{2(d-1)}},
\label{extremerr} 
\ee
\[
-\left[(\ln G)'+{1\over r}\right]\left[(p+1)(\ln B)'+(d-2)(\ln G)'\right]
-(\ln G)''-{d-1 \over r}(\ln G)'
\]
\be
\qquad\qquad\qquad\qquad={p+1\over 2(D-2)} e^{-a\phi}G^{-2(d-2)}
{Q^2\over r^{2(d-1)}},
\label{extremetheta} 
\ee
\be
\phi''+{d-1 \over r}\phi'+\phi'\left[(p+1)(\ln B)'+(d-2)(\ln G)'\right]=
-{a\over 2}e^{-a\phi}G^{-2(d-2)}{Q^2\over r^{2(d-1)}}.
\label{extremephi}
\ee

As one can be convinced by a first glance at the equations above, 
and as already noted in Appendix \ref{RICCIapp}, the
quantity $\varphi=(p+1)\ln B + (d-2)\ln G$ plays a crucial r\^ole.
Noting the structure of the r.h.s. of the equations \rref{extremett}
and \rref{extremetheta}, one can actually obtain an equation for $\varphi$.
Taking the combination of the equations 
$(p+1)({}^t_t)+ (d-2)({}^\alpha_\alpha)$ one obtains:
\be
-\varphi''-{\varphi'}^2-{2d-3\over r}\varphi'=0.
\label{eqforvarphi}
\ee
The simplest way to satisfy this equation is to take $\varphi=0$ or,
restoring its definition:
\be
(p+1)\ln B + (d-2)\ln G=0, \qquad \qquad B^{p+1}G^{d-2}=1.
\label{extremecondition}
\ee
This can be justified as follows. The asymptotic value of $\varphi$ is 
fixed to zero by the asymptotic values of $B$ and $G$. The second integration
constant coming from the equation \rref{eqforvarphi} must not be arbitrary
since an extreme solution should
depend only on one parameter, and by the equation \rref{nformsolution}
we have already introduced the parameter $Q$. We thus fix this second
constant to be zero.

We have now reduced the problem to the three following equations for
$G$ and $\phi$:
\bea
(\ln G)''+{d-1 \over r}(\ln G)'&=&-{p+1\over 2(D-2)}
e^{-a\phi}G^{-2(d-2)}{Q^2\over r^{2(d-1)}}, \label{eqforg}\\
\phi''+{d-1 \over r}\phi'&=&-{a\over 2}e^{-a\phi}G^{-2(d-2)}{
Q^2\over r^{2(d-1)}}, \label{eqforphi} \\
{(D-2)(d-2)\over p+1}{(\ln G)'}^2+{1\over 2}{\phi'}^2&=&{1\over 2}
e^{-a\phi}G^{-2(d-2)}{Q^2\over r^{2(d-1)}}. \label{eqforboth}
\eea
The last equation \rref{eqforboth} is obtained from \rref{extremerr}
taking into account \rref{extremecondition} and \rref{eqforg}.

Again, we fix the following quantity to vanish in order
not to introduce a new parameter in the solution: 
\[
a\ln G -{p+1\over D-2}\phi=0.
\]
The two equations \rref{eqforg} and \rref{eqforphi} become thus redundant, and
the equation \rref{eqforboth} becomes a first order differential
equation for the remaining function, which we take to be $\phi$:
\be
{\Delta\over a^2(D-2)}{\phi'}^2={1\over 2}e^{-{2\Delta\over a (D-2)}\phi}
{Q^2\over r^{2(d-1)}}, 
\label{almostfinaleq}
\ee
where:
\be
\Delta=(p+1)(d-2)+{1\over 2}a^2 (D-2).
\label{deltissim}
\ee
Taking the square root of the equation \rref{almostfinaleq}, we obtain:
\be
\left(e^{{\Delta\over a (D-2)}\phi}\right)'=\pm\sqrt{
\Delta\over 2(D-2)}{|Q|\over r^{d-1}}.
\label{finaleq}
\ee
This is straightforwardly solved to give:
\be
e^{{\Delta\over a (D-2)}\phi}=1+{1\over d-2}\sqrt{\Delta\over 2(D-2)}
{|Q|\over r^{d-2}}.
\label{solutionforphi}
\ee
We have chosen one of the two signs appearing in \rref{finaleq}
requiring that the solution is not singular for $r>0$. The asymptotic
value of the dilaton is fixed to zero according to the
discussion in Section \ref{WHYsec}. Note that in going from \rref{finaleq}
to \rref{solutionforphi}, we have implicitly assumed that $d>2$, which
is the condition required for having a constant dilaton 
(and an asymptotically flat geometry) at infinity.

One can show that the equation \rref{eqforphi}, when $G$ is
replaced for $\phi$, is a direct consequence
of \rref{finaleq}. We have thus solved completely the problem of
finding a non-trivial solution depending on only one parameter, $Q$.

We can characterize completely the solution in terms of
the function (which is harmonic in $d$-space):
\be
H=1+{1\over d-2}\sqrt{\Delta\over 2(D-2)}{|Q|\over r^{d-2}}\equiv
1+{h^{d-2}\over r^{d-2}}.
\label{harmonicfn}
\ee
The metric components, dilaton and $(p+2)$-form are thus:
\be
B=H^{-{d-2\over \Delta}}, \quad G=H^{p+1\over \Delta}, \quad
e^\phi=H^{a{D-2\over \Delta}}, \quad F_{ty_1\dots y_p r}=(\pm)\sqrt{2(D-2)\over
\Delta}\left(H^{-1}\right)',
\label{thesphextremesolution}
\ee
where $(\pm)=Q/|Q|$.

The metric of a spherically symmetric, extremal $p$-brane is thus:
\be
ds^2=H^{-2{d-2\over \Delta}}\left(-dt^2+dy_1^2+\dots+dy_p^2\right)+
H^{2{p+1\over \Delta}}\left(dr^2+r^2d\Omega_{d-1}^2\right).
\label{metricextreme}
\ee

There are now some intereresting remarks to make about the solution we have
just found. 

The first one arises when we note that the function $H$ which characterizes
the solution diverges at $r=0$. For the region of space-time determined
by $r=0$ to coincide actually with the origin of the spherical coordinates,
and thus with the location of the $p$-brane, a necessary condition
is that the radius of the $(d-1)$-sphere shrinks to zero size there.
Now the radius of the $S^{d-1}$ in the geometry \rref{metricextreme}
is given by $Gr$. If we consider qualitatively that near $r=0$ the
function $H$ behaves like $H\sim r^{-(d-2)}$, then we have
that $Gr\sim r^{-{(p+1)(d-2)\over \Delta}+1}=r^{a^2{D-2\over 2\Delta}}$.
This means that the radius is vanishing at $r=0$ whenever $a\neq 0$, i.e.
for any truly dilatonic $p$-brane. For $a=0$, the fact that
$Gr\rightarrow cst$ is a hint that 
something different is happening at $r=0$.

However the radius of the $S^{d-1}$ is not all, one would like to know 
whether the locus $r=0$ is a singularity of the geometry or not. This 
actually can be easily checked using the components of the Riemann
tensor in the orthonormal frame, which are given in 
Appendix \ref{RICCIapp}, eqs.
\rref{isoriemtitj}--\rref{isoriemabcd}. One can consider the 
following component:
\[
{{\cal R}^{\hat{t}\hat{\imath}}}_{\hat{t}\hat{\imath}}=
-{1\over G^2}{(\ln B)'}^2=
-{(d-2)^2Q^2\over 2\Delta (D-2)}\left[r^{d-2}+{1\over d-2}
\sqrt{\Delta\over 2 (D-2)}|Q|\right]^{-2\left(1+{p+1\over \Delta}\right)}
r^{-a^2{D-2\over \Delta}}.
\]
It clearly diverges at $r=0$ when $a\neq 0$. This in turn implies
the divergence of the square of the Riemann tensor, i.e. the scalar
$K={\cal R}_{\mu\nu\rho\sigma}{\cal R}^{\mu\nu\rho\sigma}$. 

We can thus conclude that for the dilatonic extreme $p$-branes, the
curvature diverges at the origin, i.e. at the location of the brane
itself. This could actually have been guessed from the fact that, for
instance, the dilaton field is singular at $r=0$, and provides a singular
source term to the Einstein equations.

Before drawing conclusions on the physical consequences of this singular
behaviour, we should consider the non-dilatonic case. Here the locus
$r=0$ yields a finite curvature, and the radius of the sphere does
not vanish. It can be shown in general \cite{gibbhorotown}, 
and thus for some particular cases of interest
in string/M-theory \cite{stelle}, that the geometry in this case is
regular at $r=0$, which becomes thus an horizon much similar to the
one of the extreme Reissner-Nordstr\"om black hole.

One can rewrite the $a=0$ metric in a Schwarzschild-like radial
coordinate defined by $R^{d-2}=r^{d-2}+h^{d-2}$:
\be
ds^2=\left(1-{h^{d-2}\over R^{d-2}}\right)^{2\over p+1}
\left(-dt^2+dy_1^2+\dots+dy_p^2\right)+\left(1-{h^{d-2}\over R^{d-2}}\right)^{
-2}dR^2+R^2d\Omega_{d-1}^2.
\label{extremeschw}
\ee
This metric is in the $G=1$ gauge, and now the horizon is at $R=h$. Note
that for $p=0$ we find the $d+1$ dimensional Reissner-Nordstr\"om
black hole. 
We will not discuss here the global structure of the solution
above, and we refer to \cite{gibbhorotown,stelle}. 

We will see that when considering black $p$-branes, i.e. non-extremal
ones, also the dilatonic ones possess a regular external horizon. This
external horizon coincides with the $r=0$ singularity in the extremal
limit. It is thus tempting to call in what follows the locus $r=0$ 
the horizon even for the extremal dilatonic $p$-branes.

One should not discard these
solutions because of their singular behaviour; rather, they should be
considered as reproducing the long-distance, low-energy fields of
a quantum object. One could think of them in the same way as of the 
classical electric field produced by, say, an electron, which is singular
at the (classical) location of the electron. Note that for
the $p$-branes of 10 dimensional supergravities, one can alternatively
consider their 11 dimensional origin \cite{gibbhorotown} and find
that they all descend from regular `solitonic' solutions. The
complemetarity of these two points of view must be ascribed to duality.

We postpone until the derivation of the formula giving the ADM mass the
discussion on the relation between the mass and the charge of the 
above solution \rref{thesphextremesolution}, and we will prove
in Section \ref{INTERsec} that, in the $D=11$ case, the extreme solutions
of the class described above preserve half of the supersymmetries.

\subsection{The general non-extremal black $p$-brane}

In this subsection we generalize the extremal $p$-brane solutions to
the non-extremal ones, depending now on two parameters. The reason
why we took this path and not the opposite one, consisting in deriving
the general solution and then specializing to the extremal one, is
that the derivation in the general case in more involved. 
Moreover, our approach to the present problem
will be strongly guided by the results of the
previous subsection.

We start from the general spherically symmetric metric \rref{gensphmetric}
and from the usual electric ansatz for the $(p+2)$-form
\rref{nformsolution}. The equations are thus simply 
\rref{einstsphtt}--\rref{dilatonsph}.

The form of the above mentioned equations suggests that we introduce
a new function:
\be
BC^pF^{-1}G^{d-1}=f.
\label{fdef}
\ee
Note that in the extremal case of the previous subsection, we have
that $f=B^{p+1}G^{d-2}=1$ by virtue of \rref{extremecondition}.

If we also define the following shorthand for the r.h.s.:
\[ 
{\cal S}^2={1\over 2(D-2)}e^{-a\phi}F^2G^{-2(d-1)}{Q^2\over r^{2(d-1)}},
\]
the equations can be rewritten as:
\be
(\ln B)''+{d-1\over r}(\ln B)'+(\ln B)'(\ln f)'=(d-2){\cal S}^2,
\label{blacktt} \ee
\be
(\ln C)''+{d-1\over r}(\ln C)'+(\ln C)'(\ln f)'=(d-2){\cal S}^2,
\label{blackii} \ee
\[
(\ln f)''+(\ln F)''-(\ln F)'(\ln f)'-{(\ln F)'}^2+{(\ln B)'}^2
+p{(\ln C)'}^2+(d-1){(\ln G)'}^2
\]
\be
\qquad\qquad\qquad+2{d-1\over r}(\ln G)'-{d-1\over r}(\ln F)'
+{1\over 2}{\phi'}^2=(d-2){\cal S}^2,
\label{blackrr} \ee
\[
(\ln G)''+{d-1\over r}(\ln G)'+(\ln G)'(\ln f)'+{1\over r}(\ln f)'
+{d-2\over r^2}\left(1-{F^2\over G^2}\right)\]
\be \qquad\qquad\qquad\qquad\qquad\qquad
=-(p+1){\cal S}^2,
\label{blacktheta} \ee
\be
\phi''+{d-1\over r}\phi'+\phi'(\ln f)'=-a (D-2) {\cal S}^2.
\label{blackphi}
\ee

We now redefine once more the variables, taking:
\[
\ln B=\ln C+\ln \bar{B}, \qquad \qquad \ln F=\ln G+\ln \bar{F}.
\]
Again, the extremal case corresponds to $\bar{B}=\bar{F}=1$.
Reexpressing \rref{blacktt} in these variables, and taking into
account \rref{blackii}, we obtain the homogeneous equation:
\be
(\ln\bar{B})''+{d-1\over r}(\ln\bar{B})'+(\ln\bar{B})'(\ln f)'=0.
\label{blackbbareq}
\ee
Another equation with vanishing r.h.s. can be obtained taking
the combination $(p+1)$\rref{blackii}$+(d-2)$\rref{blacktheta}:
\be
\varphi''+{d-1\over r}\varphi'+\varphi'(\ln f)'+{(d-2)\over r}\left[
(\ln f)'+{(d-2)\over r}\left(1-\bar{F}^2\right)\right]=0,
\label{blackfbareq}
\ee
where as usual $\varphi=(p+1)\ln C+(d-2)\ln G$. Note that:
\be
\ln f =\varphi +\ln \bar{B}-\ln \bar{F}.
\label{fvarphieq}
\ee

We are now ready to reduce the variables by taking some combinations
of the functions above to vanish. We wish to keep in the end two independent
functions, because we expect the non-extremal solution to depend on
two parameters, the charge and the mass (or the internal and the external
horizons).

An interesting reduction is the following: guided by the previous, extremal,
case, we take $\varphi=0$. This relates $C$ to $G$. We will in the end
also relate $\phi$ to $G$. The second independent function is provided
by $f$. Indeed, since $\ln\bar{B}$, $\ln\bar{F}$ and $\ln f$ all vanish
in the extremal limit, it is consistent to take $\ln\bar{B}$ and 
$\ln\bar{F}$ proportional to $\ln f$:
\be
\ln\bar{B}=c_B \ln f, \qquad\qquad \ln\bar{F}=c_F  \ln f.
\label{bfred}
\ee
The relation \rref{fvarphieq} constrains $c_B -c_F=1$.
With these simplifications, one can now solve for $f$, $\bar{B}$ and 
$\bar{F}$. The equation \rref{blackbbareq} becomes a second order 
differential equation for $f$, and the equation \rref{blackfbareq},
which becomes a first order one, will be used to fix $c_F$.
Note that the solution will contain an arbitrary integration constant,
not related to the charge appearing in ${\cal S}^2$.

The equation \rref{blackbbareq} rewrites:
\[
(\ln f)''+{(\ln f)'}^2+{d-1\over r}(\ln f)'=0,
\]
which gives:
\be
f=1-{2\mu\over r^{d-2}}.
\label{fsolution}
\ee
The second integration constant has been set to one by the asymptotic value
of the metric. $\mu$ is thus the non-extremality parameter.

To determine $c_F$, the equation \rref{blackfbareq} gives for $\varphi=0$:
\[
(\ln f)'+{d-2\over r}\left(1-f^{2c_F}\right)=0.
\]
Using the fact that $f'={d-2\over r}(1-f)$, the equation above holds
if and only if $c_F=-{1\over 2}$. We thus have the
solution:
\be 
\bar{B}=f^{1\over 2},\qquad \qquad \bar{F}=f^{-{1\over 2}}.
\label{barssolution}
\ee

The two equations \rref{blacktheta} (where we take into account
\rref{blackfbareq} for $\varphi=0$) and \rref{blackphi} again suggest,
as in the extremal case, that $\ln G$ and $\phi$ are proportional,
actually through the same relation $a\ln G ={p+1\over D-2}\phi$.

This leaves us with the equation for $\phi$:
\be
\phi''+{d-1\over r}\phi'+(\ln f)'\phi'=-{a\over 2f }e^{-{2\Delta\over
a(D-2)}\phi} {Q^2\over r^{2(d-1)}}.
\label{blackeqforphi}
\ee
$\Delta$ has been defined in \rref{deltissim}, and we have used 
all the above relations to rewrite ${\cal S}^2$ in terms of $\phi$
and $f$ only.

Performing the same game of substitutions in the equation \rref{blackrr}, and
using \rref{blacktheta} and the equations satisfied by $f$, we finally
obtain:
\be
{\Delta\over a (D-2)}{\phi'}^2-\phi'(\ln f)'={a\over 2f }e^{-{2\Delta\over
a(D-2)}\phi} {Q^2\over r^{2(d-1)}}.
\label{blackeqforphi2}
\ee

The sum of the two above equations is homogeneous and dictates that the
function $e^{{\Delta\over a(D-2)}\phi}$ is harmonic. The solution is
thus of the same form as in the extreme case:
\be
e^{{\Delta\over a(D-2)}\phi}=1+{h^{d-2}\over r^{d-2}}\equiv H,
\label{blacksolphi}
\ee
where $h$ is an arbitrary (positive) parameter.  
However the parameter $h$ has to be determined from 
\rref{blackeqforphi} or \rref{blackeqforphi2}, which must reduce to 
algebraic equations for $h$. 
Substituting in \rref{blackeqforphi} or in \rref{blackeqforphi2}
the expressions \rref{blacksolphi}
and \rref{fsolution}, we indeed obtain the same algebaic equation:
\be
\left(h^{d-2}\right)^2+2\mu h^{d-2}={\Delta\over 2(D-2)(d-2)^2}Q^2.
\label{eqforh}
\ee
The solution is thus:
\be
h^{d-2}=\sqrt{{\Delta\over 2(D-2)(d-2)^2}Q^2+\mu^2}-\mu,
\label{solforh}
\ee
which correctly reduces to the extremal value of $h$ which can be
extracted from the equation \rref{harmonicfn} when the non-extremality
parameter $\mu$ vanishes.

The solution is thus fully characterized in terms of the two functions
$f$ and $H$, and thus in terms of the two parameters $Q$ (or $h$) and $\mu$:
\[
B=f^{1\over 2}H^{-{d-2\over \Delta}},\qquad C=H^{-{d-2\over \Delta}},\qquad
F=f^{-{1\over 2}}H^{p+1\over \Delta}, \qquad G=H^{p+1\over \Delta},
\]
\be
e^\phi=H^{a{D-2\over \Delta}},\qquad F_{ty_1\dots y_p r}=(\pm)\sqrt{2(D-2)\over
\Delta}\sqrt{1+{2\mu\over h^{d-2}}} \left(H^{-1}\right)'.
\label{blacksolution}
\ee

It is straightforward to check that for $\mu=0$, $f=1$, the above solution
becomes the extremal one \rref{thesphextremesolution}.
The metric writes \cite{dufflupope}:
\be
ds^2=H^{-2{d-2\over \Delta}}\left(-fdt^2+dy_1^2+\dots+dy_p^2\right)
+H^{2{p+1\over \Delta}}\left(f^{-1}dr^2+r^2d\Omega_{d-1}^2\right).
\label{blackmetric}
\ee
Note that for vanishing charge, $Q=0$, we have also $h=0$ and thus
$H=1$. The metric above thus reduces to the metric of a Schwarzschild
black hole in $d+1$ dimensions, multiplied by a flat $R^p$ space.

The space-time structure of the above geometry can also be analyzed
(see for instance \cite{stelle}), and one finds that the locus $r^{d-2}=2\mu$
is a regular horizon, while $r=0$ is a singularity whenever $a\neq 0$.
As already stated in the previous subsection, the horizon coincides with
the singularity in the extremal limit for a dilatonic brane.

We can perform a change of coordinates defining $R^{d-2}=r^{d-2}+h^{d-2}$.
If we define:
\[
H_\pm=1-{h_\pm^{d-2}\over R^{d-2}}, \qquad h_+^{d-2}=h^{d-2}+2\mu, \qquad
h_-=h,
\]
we have that $H=H_-^{-1}$, $f=H_+ H_-^{-1}$, $r=RH_-^{1\over d-2}$ and
${dr\over dR}=H_-^{-{d-3\over d-2}}$. This readily gives the following 
metric:
\[
ds^2=-H_+ H_-^{2{d-2\over \Delta}-1}dt^2+H_-^{2{d-2\over \Delta}}\left(
dy^2_1+\dots+dy^2_p\right) \qquad\qquad\qquad\qquad 
\]
\be \qquad\qquad\qquad\qquad
+H_+^{-1}H_-^{-2{p+1\over \Delta}-{d-4\over d-2}}
dR^2+H_-^{a^2{D-2\over \Delta (d-2)}}R^2d\Omega_{d-1}^2.
\label{blacksolution2}
\ee
This metric reduces to the one found in $D=10$ and in the string frame
by Horowitz and Strominger \cite{horostrom}. Note that in this picture
$\mu$ measures the difference
between the external and the internal horizons, namely
$2\mu=h_+^{d-2}-h_-^{d-2}$. 

For the non-dilatonic
branes $a=0$, we can reproduce the metric discussed in \cite{gibbhorotown}:
\be
ds^2=-H_+ H_-^{-{p-1\over p+1}}dt^2+H_-^{2\over p+1}\left(
dy^2_1+\dots+dy^2_p\right)+H_+^{-1}H_-^{-1}dR^2+R^2d\Omega_{d-1}^2.
\label{ghtmetric}
\ee
This metric is in the $G=1$ gauge. We could have actually found this
metric solving the equations in this specific gauge. This is indeed
possible, however this approach only works in the non-dilatonic
case, and this is why we did not consider it here.

\subsection{Semi-classical thermodynamics and the mass of the black
$p$-branes}

The aim of this subsection is to provide us with a formula for the
mass of a $p$-brane or, more generally, of a configuration of
$p$-branes giving rise to a metric like \rref{thegensphmetric}. 

Such a formula for the mass could be simply obtained using the
ADM (Arnowitt--Deser--Misner) formalism (see e.g. \cite{admformalism}), 
and this has already been
derived for $p$-branes in \cite{admforbranes}.
Here we prefer to rederive the ADM formula in our particular case,
and we base our derivation on considerations pertaining to 
semi-classical thermodynamics of black objects.
Almost as a by-product, we will obtain also useful formulas for the
Hawking temperature and for the entropy of the same configurations.

Once the formula for the mass will have been found, we will show that
the extreme $p$-branes \rref{thesphextremesolution} saturate a lower
bound for their mass, which is indeed proportional to their charge.

Let us now introduce some aspects of the semi-classical thermodynamics
of gravitating systems that we will need to derive the ADM mass formula.
The general idea is the following: one wants to compute the partition
function of a system including gravitation. If one is only interested
in the contribution to the partition function of the (classical) 
background, then only the leading term of the saddle point approximation
is relevant, which means that all fluctuation determinants are neglected.
We call this approximation the semi-classical one.

As shown by Gibbons and Hawking \cite{gibbhawk},
the semi-classical contribution to the partition function is non-trivial
when the background gravitational field has a horizon, i.e.~when it defines
a black object. However to correctly compute this contribution, one
has to pay much attention to the boundary terms which have to be added to
the action \rref{genaction} in order to implement the boundary
conditions on the fields. The analysis of which boundary terms are needed
in general,
and how to regularize them, can be carried out (see for instance
\cite{hawkhoroross,hawkross}) but it involves the introduction
of quite a lot of formalism.

Here we will use only some useful aspects of the formalism, and we will
short-cut the rest. Inspired by \cite{btz94}, the short-cut exploits
some simplifications introduced by the Hamiltonian formalism.

In order to compute a partition function in field theory, one has to
define a path integral over field configurations which are periodic
in the imaginary time $\tau= i t$, with period $\beta$. In theories
including gravitation, this step means that we are really 
considering a Euclidean section of a complexified metric
\cite{gibbhawk}. Here comes the first surprise: when the background
geometry is the one of a non-extreme black object, the Euclidean section
stops at the locus where the Lorentzian section defines a horizon, and
it is smooth provided the period $\beta$ has a definite value. Moreover,
the value of $\beta$, which is the inverse temperature, is in perfect
agreement with the Hawking temperature $T_H$. 
The conclusion is that the Euclidean
section of a black object is a smooth manifold with only one boundary
at infinity, and that this procedure singles out
a definite temperature $T_H$.

If the integrand in the path integral is $e^{-I_E}$, the Euclidean action
$I_E$ is simply minus the Lorentzian one \rref{genaction}, where
however the metric is replaced by one of Euclidean signature.
The gravitational part thus reads:
\be
I_E=-{1\over 16 \pi G_D}\int d^D x \sqrt{g}R+ boundary\  terms\ 
at\ \infty.
\label{euclideanaction}
\ee
We will no longer include the matter terms as in \rref{genaction}
since they are not relevant to the following discussion, centered
on the derivation of the ADM mass formula. These terms are however
important for a general treatment of the thermodynamics of black objects,
and in particular for the distinction between the canonical and the
grand canonical ensembles, distinction which can be subtle
when dealing with electrically and/or magnetically charged objects
\cite{hawkross}. 

The action $I_E$ can be used to define the partition function in the
canonical ensemble in the following way \cite{gibbhawk}:
\be
-\ln Z(\beta)=\beta F =\beta M -S \approx I_E,
\label{canonik}
\ee
where the symbol $\approx$ denotes that the action is evaluated on the
equations of motion (and also that the path integral is reduced to
its saddle point value). 
$M$ is the ADM mass, $S$ is the entropy and $\beta$ is
the inverse temperature, $\beta=T^{-1}$.
Using the first law of (black hole) thermodynamics, $\delta M=T \delta S$,
it is straightforward to show that we must have:
\be
\delta I_E \approx M\delta \beta,
\label{canvary}
\ee
which is consistent with the fact that the canonical partition function 
is a function of the temperature, and that the equations of motion are
derived from $I_E$ taking the variation at fixed $\beta$.

Consider now the Hamiltonian action.
For a static space-time like \rref{thegensphmetric} 
and in the Euclidean section it can be written as:
\be
I_h=\int d^D x N {\cal H},
\label{hamiltaction}
\ee
where $N\equiv B$ is the so-called lapse function, which does not 
appear in the Hamiltonian density
${\cal H}$ and thus acts as a Lagrange multiplier. Accordingly, the
action $I_h$ vanishes on-shell.

Now, following \cite{reggeteit}, this action cannot give the true Hamiltonian
precisely because it always vanishes on-shell. Moreover, its variation
yields a non-vanishing boundary term. One thus has
to add a boundary term to $I_h$ which all at once cancels, when varied,
the non-trivial variation of $I_h$, and when evaluated yields the energy
of the configuration. 
The variation at infinity of $I_h$ is thus \cite{reggeteit}:
\be
\delta I_h |_\infty \approx -\beta \delta M.
\label{hamvar}
\ee

We have thus now the procedure to compute the ADM mass. First we compute
$I_h$ simply taking \rref{euclideanaction} and expurging from it all
terms containing derivatives of $B$. This will be done pushing away these
terms into boundary integrals that will all be discarded. Then we 
compute the variation of $I_h$, knowing that we will find, besides
the equations of motion, the boundary term at infinity \rref{hamvar}.
The boundary term of the variation at the horizon on the other hand will
give us an interesting relation from which we will extract the expressions
for both $T$ and $S$.

Let us thus start by computing the action $I_E$. We use the results
\rref{sphricciunik} of the Appendix \ref{RICCIapp}. Then if we take the by now
familiar function $\varphi=\ln B +\sum_i \ln C_i -\ln F +(d-1)\ln (Gr)$,
and consider that $R=R^{\hat{t}}_{\hat{t}}+\sum_i 
R^{\hat{\imath}}_{\hat{\imath}}+R^{\hat{r}}_{\hat{r}}+(d-1)
R^{\hat{\alpha}}_{\hat{\alpha}}$, we obtain:
\[
I_E={1\over 16 \pi G_D}\int d^D x\ \omega_{d-1} e^\varphi \left\{
2\varphi''+{\varphi'}^2+2(\ln F)'' +{(\ln B)'}^2 +\sum_i {(\ln C_i)'}^2
-{(\ln F)'}^2 \right.
\]
\be \qquad \qquad \qquad \left.
+(d-1) {(\ln Gr)'}^2 -(d-1)(d-2) {F^2\over G^2 r^2}\right\},
\label{ieexpression}
\ee
where $\omega_{d-1}=\sin^{d-2}\theta_1 \dots \sin \theta_{d-2}$. We now 
wish to single out all dependence on $B$. Accordingly, we introduce
the new auxiliary function $\gamma$ such that $\varphi=\ln B +\gamma$.
The action $I_E$ can thus be rewritten in two pieces:
\[ I_E=I_B+I_h. \]
The first piece contains all the terms with derivatives of $B$, and is
actually a total derivative:
\begin{eqnarray*}
I_B&=&{1\over 8 \pi G_D}\int d^D x\ \omega_{d-1} B\ e^\gamma \left\{
(\ln B)'' +{(\ln B)'}^2 +(\ln B)' \gamma'\right\} \\ &=&
{1\over 8 \pi G_D}\int d^D x\ \omega_{d-1} \left( e^\gamma B'\right)'.
\end{eqnarray*}
The second piece is thus the sought for Hamiltonian action:
\[
I_h={1\over 16 \pi G_D}\int d^D x\ \omega_{d-1} B\ e^\gamma \left\{
2 \gamma'' +{\gamma'}^2+2(\ln F)'' +\sum_i {(\ln C_i)'}^2
-{(\ln F)'}^2 \right.
\]
\be \qquad \qquad \qquad \qquad \left.
+(d-1) {(\ln Gr)'}^2 -(d-1)(d-2) {F^2\over G^2 r^2}\right\}.
\label{ihexpression}
\ee
Note that the action above should be completed with other bulk terms arising
from the matter terms in the action \rref{genaction}, which would correctly
yield $I_h\approx 0$. However these terms are not relevant for the present
discussion, since we are not interested in recovering the equations of
motion.

We now take the variation of the action \rref{ihexpression} and we care
only about the boundary terms which are produced integrating by parts. We
obtain:
\begin{eqnarray*}
\delta I_h &\approx& {1\over 8 \pi G_D}\int d^D x\ \omega_{d-1}\left\{
B\ e^\gamma \left[ \delta \gamma'- (\ln B)'\delta \gamma +\delta(\ln F)'
-(\ln B)'\delta (\ln F) {{}\over {}} \right.\right. \\ & &
\left. \left.
-\gamma' \delta (\ln F) +\sum_i(\ln C_i)' \delta
(\ln C_i) -(\ln F)'\delta(\ln F) +(d-1)(\ln Gr)'\delta(\ln Gr)\right]\right\}'.
\end{eqnarray*}
We can now perform the integral and we obtain two contributions, one from
infinity and one from the horizon $r=r_h$. If we also substitute $\gamma$
for its definition, we have:
\bea
\delta I_h &\approx& {\beta L^p \Omega_{d-1} \over 8\pi G_D} 
BC_1\dots C_p {1\over F}(Gr)^{d-1} \left[ \sum_i {1\over C_i}\delta C_i'
+(d-1) {1\over G}\delta G'  \right.\nonumber \\
& & +(d-1){1\over Gr} \delta G -\sum_i {B'\over B C_i}\delta C_i
-(d-1){B'\over B G}\delta G \nonumber \\
& & \left.\left. 
-\sum_i {C_i'\over C_i F}\delta F -(d-1){G'\over FG}\delta F
-(d-1){1\over F r}\delta F\right] \right|^{r\rightarrow \infty}_{r=r_h}.
\label{hamvarexpre}
\eea
Here $\beta$ is the period of the Euclidean time, $L^p$ is the volume 
of the `longitudinal' space spanned by the $y$'s and $\Omega_{d-1}$
is the volume of the $(d-1)$-sphere $S^{d-1}$.

Let us focus first on the boundary term at infinity. We have to consider
the behaviour of the functions appearing in \rref{hamvarexpre}, and to
implement the requirement of asymptotic flatness (recall that now we
have crucially $d\geq 3$). By inspection of the metrics
\rref{blackmetric} and \rref{blacksolution2}, 
we see that all the black $p$-brane
solutions are such that:
\be
B, C_i, F, G \equiv {\cal F}=1+{\cal O}\left( {1\over r^{d-2}}\right).
\label{asymptcond}
\ee
The leading term of the functions appearing in \rref{hamvarexpre} is
thus respectively the following:
\[
{\cal F}=1, \quad \delta {\cal F}={\cal O}\left( {1\over r^{d-2}}\right),
\quad {\cal F}',\delta {\cal F}'={\cal O}\left( {1\over r^{d-1}}\right).
\]
When taking into account this behaviour, the only terms in \rref{hamvarexpre}
which do not die off at infinity are the following:
\[
\delta I_h|_\infty\approx{\beta L^p \Omega_{d-1} \over 8\pi G_D} \left.
r^{d-1}\left[ \sum_i \delta C_i'+(d-1) \delta G' +(d-1){1\over r} \delta G
-(d-1){1\over r}\delta F\right]\right|_{r\rightarrow \infty}.
\]
Now, recalling the equation \rref{hamvar} relating the variation 
of the action $I_h$ at infinity to the variation of the mass, we
derive the expression for the ADM mass:
\be
M={L^p \Omega_{d-1} \over 8\pi G_D}\left. r^{d-1}\left[{d-1\over r}
(F-G) -\sum_i C_i'-(d-1) G'\right]\right|_{r\rightarrow \infty}.
\label{admmassformula}
\ee
Note that we do not have to subtract 
any constant part from \rref{admmassformula}
since the expression vanishes for flat space, i.e.~when all functions
are equal to 1.

The formula \rref{admmassformula} can also be shown to be invariant
under reparametrizations of $r$ which respect the behaviour at infinity
\rref{asymptcond}, i.e.~ reparametrizations such that:
\[
\bar{r}=r\left[ 1+{\cal O}\left( {1\over r^{d-2}}\right)\right].
\]

Before evaluating the expression of the ADM mass for the black $p$-branes
of the previous subsection, we take a quick look at the second term
in \rref{hamvarexpre}, the variation at the horizon $\delta I_h|_{r_h}$.

There is however a subtlety when considering the variation of the action
at the horizon. One in fact needs a definition of the location of the
horizon which is independent from the parameters of the metric that
one is varying. Without delving into a truly diffeomorphism-invariant
formulation of the boundary term at the horizon, we simply
use the possibility to operate a reparametrization of the radial coordinate
to always bring back the location of the horizon to, say, $\xi =0$.
The new radial  coordinate should also be such that the functions
defining the metric have a controllable behaviour at the horizon.

Since the Euclidean topology of a black hole is $R^2 \times S^{d-1}$
(see for instance \cite{btz94}), and thus that of a black $p$-brane
is $R^2\times T^p \times S^{d-1}$, where the Euclidean time $\tau$
and the coordinate $\xi$ are respectively
the angular variable and the radial coordinate on the $R^2$ factor, 
all the functions in the metric can be taken of order 1,
$C_i, F, (Gr)={\cal O}(1)$, except $B$ which has to be 
$B={\cal O}(\xi)$. We have implicitly used the fact that now $(Gr)\equiv
R(\xi)$ denotes a function of $\xi$, $r$ being the original, non-gauge
fixed radial variable.

In order to avoid conical singularity at the origin of the $R^2$ plane,
which coincides with the location of the horizon, the 
angular variable $\tau$ has to have a definite periodicity.
If the metric near the horizon is:
\be
ds^2 \simeq {B'_h}^2 \xi^2 d\tau^2 +F_h^2 d\xi^2 +\dots,
\label{xigauge}
\ee
where the subscript $h$ denotes the value of the function at $\xi=0$,
we can define $\bar{\xi}=F_h \xi$ such that:
\[ ds^2 \simeq \left({B'_h\over F_h}\right)^2 \bar{\xi}^2 d\tau^2 
+ d\bar{\xi}^2 +\dots .
\]

We define the surface gravity at the horizon by:
\be
\kappa = \left. \left( {B'\over F} \right) \right|_{r=r_h}.
\label{kappa}
\ee
Then there is no conical singularity at the horizon if $\tau\sim \tau+\beta_H$,
with $\beta_H={2\pi \over \kappa}$. The Hawking temperature
is thus the inverse of $\beta_H$:
\be
T_H={\kappa \over 2\pi}={1\over 2\pi}\left. \left( {B'\over F} \right) 
\right|_{r=r_h}.
\label{hawkingtemp}
\ee
Note that although we have derived this expression in the particular
`gauge' \rref{xigauge}, the above expression for $T_H$ is invariant
under reparametrizations of the radial coordinate, and it is thus
a good expression for a general $r$.

We can now reconsider the horizon term of the variation \rref{hamvarexpre}.
The only terms that survive are the ones containing $B'$, all the
others vanish due to $B|_{r_h}=0$:
\bea
\delta I_h|_{r_h} &\approx& {\beta L^p \Omega_{d-1}\over 8\pi G_D}
C_1\dots C_p {1\over F} (Gr)^{d-1} B'\left.\left[\sum_i {1\over C_i}
\delta C_i +(d-1){1\over Gr} \delta (Gr)\right] \right|_{r=r_h}
\nonumber \\
& \approx& {\beta L^p \Omega_{d-1}\over 8\pi G_D} \left.{B'\over F}
\delta \left( C_1\dots C_p (Gr)^{d-1}\right) \right|_{r=r_h}.
\label{horvar}
\eea

As noted in \cite{btz94}, this term is of the form:
\be
\delta I_h|_{r_h} \approx {\beta\over \beta_H} {\delta A_h\over 4 G_D},
\label{hamhorvar}
\ee
where $A_h$ is the volume of the horizon (the $(D-2)$-dimensional
manifold defined by $r=r_h$).
Such a term in the variation would break covariance, since it singles
out a definite location on the (Euclidean) manifold, i.e.~ the horizon.
However, we still have to implement that the periodicity has to be
$\beta=\beta_H$ in order to have a truly smooth manifold. That can be
done simply adding a term to $I_h$ such that:
\be
I_h'=I_h -{A_h\over 4 G_D}.
\label{micro2}
\ee
The variation at the horizon of the new action is thus:
\[ \delta I_h'|_{r_h} \approx \left({\beta\over \beta_H}-1\right)
{\delta A_h\over 4 G_D},
\]
which enforces $\beta=\beta_H$ if $A_h$, which is a function of the metric,
is to vary freely.
If we also add the boundary term at infinity $\beta M$ to $I_h'$, we obtain
once again the action $I_E$. Indeed, one can check that the variation of
$I_h'+\beta M$ yields the equations of motion when $\beta$ is kept fixed.
Taking into account \rref{canonik}, \rref{micro2} and the fact that 
$I_h\approx 0$, we can equate the on-shell values of $I_E$ and of 
$I_h'+\beta M$, thus obtaining the familiar formula for the entropy:
\be
S={A_h\over 4 G_D}.
\label{preentropy}
\ee
More explicitly, we have the following expression for it:
\be
S={L^p\Omega_{d-1}\over 4 G_D} \left. C_1\dots C_p (Gr)^{d-1}\right|_{r=r_h}.
\label{entropy}
\ee
It is also an $r$-reparametrization invariant expression.

Let us now finally apply the formulas \rref{admmassformula}, 
\rref{hawkingtemp} and \rref{entropy} to the black $p$-brane solution
\rref{blacksolution}. 

For the ADM mass, we have:
\be
M_p={L^p\Omega_{d-1}\over 8\pi G_D} \left[ (d-1)\mu +
(d-2){D-2\over \Delta}h^{d-2} \right].
\label{blackadmmass}
\ee
Using the expression \rref{solforh} for $h^{d-2}$, \rref{blackadmmass}
can be rearranged as:
\[
M_p={L^p\Omega_{d-1}\over 8\pi G_D} \left\{{1\over \Delta}\left[
{1\over 2}a^2(D-2)(d-1)+p(d-2)^2\right] \mu\right.  \qquad\qquad\qquad
\]
\[ \qquad\qquad\qquad\qquad\qquad \left.
+(d-2){D-2\over \Delta}
\sqrt{{\Delta\over 2(D-2)(d-2)^2}Q^2+\mu^2}\right\}.
\]
For fixed $Q$, the bound $\mu \geq 0$ yields for the ADM mass:
\be
M_p\geq {L^p\Omega_{d-1}\over 8\pi G_D} \sqrt{D-2\over 2\Delta}|Q|,
\ee
which is naturally saturated by the $\mu=0$, extreme $p$-brane solutions 
\rref{thesphextremesolution}.

In order to compare the ADM mass with the physical charge, we compute
the charge density in the canonical way:
\be
{\cal Q}={1\over 16 \pi G_D} \int_{S^{d-1}} *F_{p+2},
\label{chargedensity}
\ee
where $*F_{p+2}\equiv \tilde{F}_{d-1}$ as defined in \rref{genhodge}, and
the charge density has the right dimensions of energy per unit $p$-volume.

If we compute the total amount of charge carried by the solution
\rref{blacksolution}, we find:
\be
|q| = L^p |{\cal Q}|={L^p \Omega_{d-1}\over 16\pi G_D}|Q|.
\label{totalcharge}
\ee
The final form of the bound on the ADM mass, or Bogomol'nyi bound, is thus:
\be
M_p\geq \sqrt{2(D-2)\over \Delta} |q|.
\label{bogomolnyi}
\ee

We now compute the Hawking temperature $T_H$ for the black $p$-branes of
the solution \rref{blacksolution}. The behaviour of $T_H$ when the 
extremal limit $\mu \rightarrow 0$ is taken will deserve some comments.

In terms of the functions $f$ and $H$, see \rref{fsolution} and 
\rref{blacksolphi}, we have the following surface gravity \rref{kappa}:
\[
\kappa =\left. {1\over 2}f' H^{-{D-2\over \Delta}}\right|_{r=r_h}.
\]
Note that the horizon is at $r_h=(2\mu)^{1\over d-2}$, i.e.~at $f=0$.
When we substitute for $f$ and $H$, we find for the Hawking temperature:
\bea
T_H&=&{d-2\over 4\pi}{1\over (2\mu)^{1\over d-2}}\left(1+{h^{d-2}\over 2\mu}
\right)^{-{D-2\over \Delta}} \nonumber \\
&=&{d-2\over 4\pi}(2\mu)^{-{1\over d-2}+{D-2\over \Delta}}
\left(2\mu +h^{d-2}\right)^{-{D-2\over \Delta}}.
\label{blackhawking}
\eea
In the limit $\mu \rightarrow 0$, $h$ is finite as seen from its definition
\rref{solforh}, thus the second term in the above expression is finite. 
The first factor on the other hand gives a vanishing, finite or 
infinite Hawking temperature at extremality depending whether the combination
$-{1\over d-2}+{D-2\over \Delta}$ is respectively positive, zero or
negative.

Let us now see which branes of type II/M-theory give rise to which behaviour
at extremality. In 10 or 11 dimensional maximal supergravities, we will see
in the next subsection
that $\Delta$ as given in \rref{deltissim} is always
such that $\Delta=2(D-2)$. Then the dependence of the temperature $T_H$
on $\mu$ is governed by the combination $-{1\over d-2}+{1\over 2}=
{d-4\over 2(d-2)}$. This gives the following behaviours (see also
\cite{branetemp}):
all the branes of type II/M-theory have a vanishing extremal 
Hawking temperature except the 10 dimensional 5-branes, which
have a finite temperature in the extremal limit, and the D6-brane
of type IIA theory which has an infinite extremal temperature.
We are of course only considering branes which allow for an asymptotically
flat geometry, $d\geq 3$.

The fact that the D6-brane is singled out, and that the 5-branes
of type II theories are the limiting cases, is a pattern that will
again show up when we will consider in Chapter \ref{LITTLEchap} the 
decoupling of the physics on the world-volume of the brane from
the bulk supergravity dynamics. It is tempting to think that these
two features (the decoupling from the bulk and the vanishing of the
Hawking temperature at extremality) are closely related. Indeed 
the presence of the Hawking temperature signals, from the brane world-volume
point of view, that quanta are being exchanged with the surrounding
space, i.e.~the bulk. However a more detailed analysis is required,
and we will get back to this problem when we will consider the
world-volume effective actions of the branes.

The last quantity we wish to compute for the solutions \rref{blackmetric}
is the entropy \rref{entropy}. We have that:
\[
S=\left.{L^p \Omega_{d-1}\over 4 G_D}H^{D-2\over \Delta}r^{d-1}\right|_{r=r_h}.
\]
The expression is independent of $f$. Plugging in \rref{blacksolphi} and
$r_h=(2\mu)^{1\over d-2}$, we obtain:
\bea
S&=& {L^p \Omega_{d-1}\over 4 G_D}\left(1+{h^{d-2}\over 2\mu}
\right)^{D-2\over \Delta} (2\mu)^{d-1\over d-2} \nonumber \\
&=& {L^p \Omega_{d-1}\over 4 G_D}(2\mu)^{{d-1\over d-2}-{D-2\over \Delta}}
\left(2\mu+h^{d-2}\right)^{D-2\over \Delta}.
\label{blackentropy}
\eea
Since we have the relation ${d-1\over d-2}-{D-2\over \Delta}=
{a^2(D-2)(d-1)\over 2\Delta(d-2)}+{p(d-2)\over \Delta}$, the entropy always
vanishes in the extremal limit, except for the non-dilatonic black hole, i.e.
the 0-brane with $a=0$. It is amusing to see that the Reissner-Nordstr\"om 
black hole falls into this last, peculiar, class.

\subsection{The branes of M-theory}
\label{MBRANESssec}

As a physical conclusion to this rather mathematical section, we list
and discuss hereafter the extremal branes that are found in M-theory, or
more precisely in 11 dimensional supergravity and in type II string
theories. However, this is not merely a zoological 
review of some weird specimens.
Indeed, as we will show in the next section, these extremal branes
will turn out to be the building blocks of a lot of new, more complicated
configurations, which carry more than one charge. Moreover, some of these new
configurations will show new features like a non-vanishing entropy in the
extremal limit.
The focus on extremal branes is also motivated by the conjectured 
identification of some, and possibly all, of these solitons with 
(BPS saturated) fundamental objects.

Let us rewrite the solution for clarity (we everywhere replace $d=D-p-1$):
\be
ds^2=H^{-2{D-p-3\over \Delta}}\left(-dt^2+\sum_i dy_i^2\right)
+H^{2{p+1\over \Delta}} \delta_{ab}dx^a dx^b,
\label{conclmetric}
\ee
\be
e^\phi=H^{a{D-2\over \Delta}}, \qquad F_{ty_1\dots y_p r}=E', \quad
E\equiv A_{ty_1\dots y_p}=\sqrt{2(D-2)\over \Delta}(H^{-1}-1),
\label{conclfieldselec}
\ee
with:
\be
H=1+{1\over D-p-3}\sqrt{\Delta\over 2(D-2)}{Q\over r^{D-p-3}},
\qquad \Delta=(p+1)(D-p-3)+{1\over 2}a^2 (D-2).
\label{concldefini}
\ee
All the discussion in the previous subsections was based on the above
electric solution. However we now exploit the symmetry of the
equations of motion diplayed in \rref{genhodge} to obtain a magnetic
solution. This is straightforward: we simply take the above solution
and we replace $a$ by $-a$ and $F$ by $\tilde{F}$. Using \rref{genhodge}
we can then compute the magnetic $(D-p-2)$-form field strength:
\[
F^{\theta_1\dots \theta_{D-p-2}}={1\over \sqrt{-g}}e^{-(-a)\phi}
\tilde{F}_{ty_1\dots y_p r}={1\over \sqrt{-g}}e^{a\phi}E',
\]
which thus gives instead of the line \rref{conclfieldselec}:
\be
e^\phi=H^{-a{D-2\over \Delta}}, \qquad 
F_{\theta_1\dots \theta_{D-p-2}}=Q \omega_{D-p-2} \qquad (\omega_{D-p-2}
=\sin^{D-p-3}\theta_1\dots \sin\theta_{D-p-3}).
\label{conclfieldsmag}
\ee

The first quantity one has to compute to write down a specific solution
is $\Delta$. Now, since for an electric-magnetic pair of branes deriving
from the same field strength we have $p_{el}+p_{mag}=D-4$, we see that
$\Delta$ is the same for both and thus characterizes the $n$-form.
We will now see that for all the cases of interest below, $\Delta$ will
always satisfy $\Delta=2(D-2)$.

Let us start with 11 dimensional supergravity. Here we have only a 4-form
field strength and since there is no dilaton, we put $a=0$.
The 4-form gives rise to 2- and 5-branes, called respectively M2 and M5.
As anticipated, we have $\Delta = 3\cdot 6=18$.

The two solutions are thus:
\be
ds_{M2}^2=H^{-{2\over 3}}\left(-dt^2+dy_1^2+dy_2^2\right)+H^{1\over 3}
\left(dx_1^2+\dots+ dx_8^2\right), \qquad F_{ty_1 y_2 r}=(H^{-1})',
\label{m2metric}
\ee
and:
\be
ds_{M5}^2=H^{-{1\over 3}}\left(-dt^2+dy_1^2+\dots+dy_5^2\right)+H^{2\over 3}
\left(dx_1^2+\dots+ dx_5^2\right), \qquad F_{\theta_1\dots \theta_4}=
Q\omega_4.
\label{m5metric}
\ee
Since there is no dilaton, the Einstein metric coincides with the original
metric as in \rref{11action}. These solutions were found respectively
in \cite{duffstelle} and \cite{guven}.

Going to type II theories in 10 dimensions, we can separate the fields
in two classes, those coming from the NSNS sector and those coming from
the RR one. 

Beginning with the NSNS 3-form field strength, we have $a=-1$
and it yields an electric 1-brane which we call F1 and a magnetic 
5-brane referred to as NS5.  The F1 is identified to the fundamental
string since it couples to the same 2-form potential which couples to
the world-sheet of the type II string. The NS5 has been called alternatively
the solitonic 5-brane, and also the neutral 5-brane (when it is embedded
in heterotic string theory).

The value of $\Delta$ is straightforwardly $\Delta=2\cdot 6+{1\over 2} 
1\cdot 8=16$. We then have the following metric and dilaton for
the F1 and NS5:
\be
ds^2_{F1}=H^{-{3\over 4}}\left(-dt^2+dy_1^2\right)+H^{1\over 4}
\left(dx^2_1+\dots+dx_8^2\right), \qquad e^{\phi}=H^{-{1\over 2}},
\label{f1metric}
\ee
\be
ds^2_{NS5}=H^{-{1\over 4}}\left(-dt^2+dy_1^2+\dots+dy_5^2\right)+H^{3\over 4}
\left(dx^2_1+\dots+dx_4^2\right), \qquad e^{\phi}=H^{1\over 2}.
\label{ns5metric}
\ee
The 3-form can be easily deduced from \rref{conclfieldselec} and
\rref{conclfieldsmag}.

The two metrics above are again written in the Einstein frame. However in
string theory, we can define another natural metric, which is the string 
metric and is obtained by a rescaling of the Einstein one. 
We rewrite the equation
\rref{stringeinstein}:
\[\hat{g}_{\mu\nu}=e^{\phi\over 2}g_{\mu\nu}, \]
the metric in the string frame carrying (temporarily) a hat.

In the string frame, the metrics above become:
\bea
d\hat{s}^2_{F1}&=&H^{-1}\left(-dt^2+dy_1^2\right)+
dx^2_1+\dots+dx_8^2, 
\label{f1smetric}
\\
d\hat{s}^2_{NS5}&=&-dt^2+dy_1^2+\dots+dy_5^2+H 
\left(dx^2_1+\dots+dx_4^2\right).
\label{ns5smetric}
\eea
These metrics are solutions of the actions \rref{iiastring} and 
\rref{iibstring}. The above solutions were discussed respectively
in e.g. \cite{dabol} and \cite{callharvstro}.

The RR $n$-forms all have a coupling to the dilaton such that
$a_n={5-n\over 2}$. This gives for an electric $(p+2)$-form
$a_{p+2}={3-p\over 2}$ and for a magnetic $(8-p)$-form
$a_{8-p}=-{3-p\over 2}$. We can thus write in a unified way:
\[
\varepsilon a={3-p\over 2},
\]
where $\varepsilon$ is $+(-)$ if the $p$-brane is electric (magnetic).
Again it can be shown that $\Delta=(p+1)(7-p)+(3-p)^2=16$.

The metric of the $p$-branes charged under the RR forms, which we call
D$p$ in view of their relation with the D-branes, is:
\be
ds^2_{Dp}=H^{-{7-p\over 8}}\left(-dt^2+dy_1^2+\dots+dy_p^2\right)+
H^{p+1\over 8}\left(dx^2_1+\dots+dx_{9-p}^2\right), \qquad
e^\phi=H^{3-p\over 4}. 
\label{dpmetric}
\ee
 
In the string frame, it gives \cite{horostrom}:
\be
d\hat{s}^2_{Dp}=H^{-{1\over 2}}\left(-dt^2+dy_1^2+\dots+dy_p^2\right)+
H^{1\over 2}\left(dx^2_1+\dots+dx_{9-p}^2\right).
\label{dpsmetric}
\ee

For all the solutions discussed above, the mass saturates the bound
\rref{bogomolnyi} and moreover
the mass is strictly equal to the total charge:
\[
M=|q|,  \qquad\qquad q={L^p \Omega_{D-p-2}\over 16\pi G_D}Q.
\]

With the branes described above we have exhausted all the objects
carrying a charge in 11 or 10 dimensions. When one considers
a toroidal compactification of M-theory, all these charges are related
by U-duality. However, in order to have a complete U-duality multiplet
of charges, one has to include also the Kaluza-Klein (KK) charges
generated by the dimensional reduction procedure.

Properly speaking, the KK charged objects are branes living in the
compactified space-time and carrying a charge (electric or magnetic)
with respect to the 2-form field strength generated by dimensional
reduction, see Appendix \ref{KKapp}. 
If after compactification on one direction the space-time dimension is $D$, 
we then have an electric KK 0-brane and a magnetic KK $(D-4)$-brane.
These two objects correspond, in the $D+1$ dimensional uncompactified
space-time to configurations where the only non-trivial field is the 
metric, and which are identified respectively to a KK wave and a KK
monopole. Note that the metric in the uncompactified space necessarily
has non-trivial off-diagonal terms, and thus is not of the class
\rref{gensphmetric}
considered in this section. 

Nevertheless, we derive hereafter the metric of the KK wave and KK monopole
simply by KK `oxidation', i.e. by reversing the KK procedure to
derive a $D+1$ dimensional metric from a $D$ dimensional field
configuration.
Since the only relevant field in $D+1$ dimensions is the metric, we can work
for general $D$.

Let us start with the KK wave. In $D$ dimensions, the KK 2-form has
a coupling to the (KK) dilaton $a=\sqrt{2(D-1)\over D-2}$, 
see Appendix \ref{KKapp}
and more precisely the discussion leading to the equation \rref{kksinglecan}.
For the electric KK 0-brane we thus have $\Delta=(D-3)+{1\over 2}
{2(D-1)\over D-2}(D-2)=2(D-2)$.

The solution in $D$ dimensions is thus:
\be
ds^2_D=-H^{-{D-3\over D-2}}dt^2 +H^{1\over D-2}\left(dx^2_1+\dots +
dx_{D-1}^2\right),
\label{kkelmetric}
\ee
\be
e^\phi=H^{a\over 2}, \qquad A_t=H^{-1}-1, \qquad H=1+{1\over D-3}{Q\over
r^{D-3}}.
\label{kkelfields}
\ee
The first step before going to $D+1$ dimensions, is to rescale the metric
\rref{kkelmetric} to put it in the $D+1$ dimensional Einstein frame. Using
the formula \rref{kkweyl}, we have that:
\[ \hat{g}_{\mu\nu}=H^{-{1\over D-2}}g_{\mu\nu}, \]
where we have already used that $h_{yy}\equiv e^{2\sigma}=e^{{2\over a}\phi}
=H$. $y$ is the compact direction.

Collecting the above results, the $D+1$ dimesional metric can be written as:
\[
ds^2_{D+1}=-H^{-1}dt^2+dx^2_1+\dots +dx_{D-1}^2+H\left[ dy +(H^{-1}-1)dt
\right]^2. 
\]
Note that all dependence on $D$ has disappeared from the exponents.
If we define the new harmonic function $K$ such that $H=1+K$, we arrive
at the final form for the metric of a KK wave:
\be
ds^2_{W}=-dt^2+dy^2 -K(dt-dy)^2+dx^2_1+\dots +dx_{D-1}^2.
\label{wavemetric}
\ee

Consider now our last object, the KK monopole. Since in the reduced 
$D$ dimensional picture
it has to be magnetically charged under a 2-form field strength, it
is a $(D-4)$-brane. We can thus simply consider it in its most
simple manifestation, when it is a 0-brane in 4 dimensions. Going to
higher dimensions is then simply a matter of adding to the $D+1$ dimensional
metric $p$ flat compact
directions with trivial metric (e.g. a trivial $T^p$ factor).

In 4 dimensions we have a dilaton coupling of $a=\sqrt{3}$ and
$\Delta=4$. The magnetic solution is the following:
\be
ds^2_4=-H^{-{1\over 2}}dt^2+H^{1\over 2}(dr^2+r^2d\Omega_2^2),
\label{kkmagmetric}
\ee
\be
e^\phi=H^{-{\sqrt{3}\over 2}},\qquad F_{\theta \varphi} =Q\sin \theta, \qquad
H=1+{Q\over r}.
\label{kkmagfields}
\ee
From \rref{kkmagfields} we can write a magnetic potential:
\be
A_\varphi=Q(1-\cos \theta).
\label{kkmagpot}
\ee
Again, we find that $h_{yy}=H^{-1}$ and that the rescaling to the 5
dimensional Einstein metric leads to 
$\hat{g}_{\mu\nu}=H^{1\over 2}g_{\mu\nu}$.
The 5 dimensional metric is thus:
\[
ds^2_5=-dt^2+H^{-1}\left[ dy+Q(1-cos\theta) d\varphi\right]^2
+H(dr^2+r^2d\Omega_2^2).
\]

In $D$ dimensions, we finally have for the metric of a KK monopole:
\[
ds^2_{KKM}=-dt^2 +dy^2_1+\dots +dy^2_{D-5}+H^{-1}(dz+A_i dx^i)^2+
Hdx_idx^i, \qquad i=1\dots 3, 
\]
\be
\partial_i A_j-\partial_j A_i=-\epsilon_{ijk}\partial_k H.
\label{kkmmetric}
\ee
We recognize that the above metric describes a geometry which is 
$R_{(t)}\times T^{D-5}\times TN$, where $TN$ is the Euclidean Taub-NUT
space \cite{refontaubnut1}. The direction $z$ (which was formerly $y$)
is called the NUT direction, and has to be periodically identified 
with period $4\pi Q$ in order to avoid a (physical) conical singularity
at the origin $r=0$. Note that the NUT direction is generally not
considered as a direction longitudinal to the KK monopole (even
if its compactness is necessary for the existence of the monopole),
and thus the KK monopole is formally a $(D-5)$-brane (see e.g. \cite{hull}).
This means that in 11 dimensional supergravity we have a KK6, and
in type II theories we have a KK5.

This completes the review of all the branes which populate the menagerie
of string/M-theory.

\section{Intersections of branes}
\label{INTERsec}

In this section we search for new, more complex, solutions to the
equations of motion derived from the general action \rref{genaction},
which includes several antisymmetric tensor fields. The new solutions
we are now focusing on will be different with respect to the ones
of Section \ref{SINGLEsec} in that they carry several charges. 
Another new feature will be that the harmonic function characterizing
the extremal solutions will be allowed to be a sum of an arbitrary number
of contributions, centered at different points. The price to pay will
be the loss of spherical symmetry. 

The aim of finding new solutions, which are more general, is not merely
mathematical. The solutions carrying several charges physically represent
elementary $p$-branes, the building blocks discussed in subsection
\ref{MBRANESssec}, which intersect each other with some definite rules.
From the way they intersect we will find hints for what
are their quantum interactions. Moreover, we will be able to give a
prescription for building supersymmetric black holes with 
non-vanishing semi-classical entropy.

As it is already clear from above, we will focus essentially on 
extremal configurations. This is motivated by supersymmetry. Indeed,
results obtained at the low-energy, classical level can be better
extrapolated to the quantum theory if they are somehow protected
by some supersymmetry properties, which highly constrain the quantum
corrections that could spoil the classical result.
Nevertheless, we will present at the end of this section a solution 
which, even in its extremal limit, breaks all supersymmetries. Its
relevance in M-theory is still unclear.

Historically, the prototypes of intersecting brane solutions were
the solutions found by G\"uven \cite{guven} in 11 dimensional supergravity.
More recently, Papadopoulos and Townsend \cite{papatown} 
correctly reinterpreted these solutions as intersecting membranes, and 
furthermore they used the same principle to build new solutions
with intersecting fivebranes. At the same time, general stringy concepts such
as the various dualities \cite{hulltownsend,witten,schwarz,bergshoeffetal}
and the recently discovered D-branes \cite{polcdbranes} led to 
several conjectures about the intersections of the branes of M-theory,
mainly based on supersymmetry arguments \cite{tasi} and on the
extrapolation, using dualities, of D-brane physics 
\cite{stromopen,towndfromm,beckerbound}. This work culminated in
the proposal by Tseytlin of the `harmonic superposition rule' 
\cite{tseytlinharm}, which generalized further the solutions of
\cite{papatown} to include an independent harmonic function for
each set of non-parallel branes. Tseytlin also applied all sorts
of dualities to the solutions he presented, thus obtaining a classification
of intersecting brane solutions. Other papers that appeared soon after
Tseytlin's proposal are \cite{gaunkasttras}, where the harmonic 
superposition rule was found independently, and \cite{balalars} where
it was used in order to discuss four dimensional black holes.
A thorough classification of all intersections down to 2 effective space-time
dimensions was given in \cite{bdejv1}. The common approach of all these
papers was to deduce the expressions for the fields 
from the harmonic superposition rule and then prove that they are consistent
solutions, this task being facilitated either by dualities or by supersymmetry.
A somewhat different approach was proposed again by Tseytlin 
\cite{tseytlinnoforce}, using brane probes in brane backgrounds and
asking that the probe feels no attraction nor repulsion. In this way,
the intersection rules for the marginal bound states of the branes of M-theory
were derived.

The approach followed here, and based on \cite{rules}, is different
from the ones described above in that it aims at deriving the harmonic 
superposition rule from the equations of motion of a general low-energy
effective theory
and only in the end it is specialized to the cases of interest in M-theory.
In this purely classical (and bosonic) context, we find solutions and
general intersection rules which are fully consistent
with the ones previously found (or guessed) by duality arguments or by
other means. 
The main advantage of our approach is that all the branes are treated
manifestly on the same footing.
Note that contemporarily a similar
approach was undertaken in \cite{arefevaint}.

We now set up for solving the equations of motion 
\rref{eomeinstein}--\rref{bianchi} (including the Bianchi identities for
the $\cal M$ $n_I$-form field strengths). The first step, as at the beginning
of Section \ref{SINGLEsec}, is to write the metric in a simplified form.
However we now want to find solutions not necessarily constrained by the
large amount of symmetries discussed in subsection \ref{KILLssec}.
In particular, we still want our solutions to be static and 
translational invariant in the longitudinal space, but we want to drop
the $SO(p)$ `internal' symmetry and the $SO(d)$ `external' symmetry.
This is to allow respectively for anisotropic combinations of 
different branes, and for several branes not necessarily located at the same
point in the transverse space.

An important remark is now in order, for the notation not to be confusing.
We will call $p$ the number of directions which are translational 
invariant, also called the longitudinal directions. This however does
no longer mean that we necessarily have a $p$-brane in the solution,
i.e. that the solution is charged under an electric $(p+2)$-form field
strength or its magnetic dual. It suffices that we have a collection
of $\cal N$ $q_A$-branes, all with $q_A\leq p$, which fill out a total
internal space of dimension $p$, when all their intersections are taken
into account. All these branes will have a common $d$ dimensional
transverse space (also called the overall transverse space).
The internal directions which are longitudinal to some branes and 
transverse to some others, are in general called relative transverse
directions. 
Note that the branes are often generally referred to as `$p$-branes', and
in this case the `$p$' only reflects that we are talking of objects
with arbitrary spatial extension.

The general metric we will take is the following:
\be
ds^2=-B^2(x^a)dt^2 +\sum_{i=1}^p C_i^2(x^a)dy_i^2 +G^2(x^a)\delta_{ab}
dx^adx^b, \qquad a,b=1\dots d.
\label{generalisotropic}
\ee
This is a generalization with respect to \rref{thegensphmetric} in that
the spherical symmetry is no longer required, and with respect to
\rref{gensphmetric} also the internal $SO(p)$ symmetry is broken.
Let us point out that unlike the metric \rref{gensphmetric} which
was found in this form imposing its symmetries and using some of the
remaining diffeomorphisms, the metric \rref{generalisotropic} is
really an ansatz inspired by the solutions found in the previous
section. In particular, imposing isotropicity in the overall transverse
space is a very strong ansatz when there are no symmetries there
(at least when $d\geq 3$). Also, we impose the metric to be diagonal,
while we could have allowed for off-diagonal elements. We thus
exclude some of the solutions described in the subsection \ref{MBRANESssec},
namely the waves and the KK monopoles.
This is mainly a concession to treatability of the equations of motion.
Also, the procedure of KK oxidation, i.e. of elevating to higher 
dimensions a solution, will not be possible because we included only
one scalar field, the dilaton. Solutions with several scalars which
can thus correctly take into account all the moduli arising by dimensional
reduction can be found in the literature, see e.g. \cite{stelle,kklp}.
It should be noted however that all these solutions with off-diagonal
terms in the metric can generically be related to diagonal configurations
by U-dualities, and more precisely T-dualities. Now it is well known
\cite{bergshoeffetal} how T-duality acts on the fields of the 10 dimensional
supergravities, and this provides a way of indirectly deducing 
field configurations which include waves and/or KK monopoles. 
The dual configurations with and without KK waves and monopoles
can be guessed using the dualization properties of the branes listed
in Appendix \ref{ZOOapp}. 

A last remark on the metric \rref{generalisotropic} is that all the functions
are taken to depend only on the overall transverse space. This means
that for a generic brane involved in the configuration, the fields it
produces will not depend on the relative transverse directions. We will see,
once the solutions will have been found, how it is possible to link 
these solutions with reduced dependence on the transverse space with
those discussed in the preceding section.

\subsection{Finding the classical solutions and the intersection rules}

As in Section \ref{SINGLEsec}, the first step towards solving the 
equations of motion will be to focus on the antisymmetric tensor 
fields. However now we will not solve completely the set of equations
of motion and Bianchi identities \rref{eommaxwell} and \rref{bianchi},
rather we make a suitable ansatz to render one of the two equations
for each $n$-form completely trivial. The other equation is then
a non-trivial equation for the $n$-form.

In the following, we call an electric ansatz for a $n$-form an ansatz
which trivially solves its Bianchi identities \rref{bianchi}, while
we call a magnetic ansatz an ansatz which  solves trivially its
equations of motion \rref{eommaxwell}.
The ans\"atze will be more general than the ones for a singly charged
brane, \rref{electricansatz} and \rref{magneticansatz}. This is because
there is no more $SO(p)$ internal symmetry, and thus the constraints
on the presence of all or none of the internal indices are relaxed.

Let us start with the electric ansatz. A straightforward generalization
of \rref{electricansatz} is the following:
\be
F_{ti_1\dots i_{q_A}a}=\epsilon_{i_1\dots i_{q_A}} \partial_a E_A(x).
\label{genelectricansatz}
\ee
This is thus an ansatz for a $(q_A+2)$-form which gives an electric charge
to a $q_A$-brane, and the index $A$ runs over all the $\cal N$ branes.
Note that the number of branes can be (and generally is) larger than the
number $\cal M$ of antisymmetric tensor fields. Several branes charged
under the same field strength will be distinguished by the directions
in which they lie, i.e. the directions labelled by $\{i_1\dots i_{q_A}\}$
in the case above. The ansatz above is inspired by the fact that the
$(q_A+1)$-form potential must couple to the world-volume of the brane,
and should thus contain all the indices of the directions tangential
to it. 

A magnetic ansatz generalizing \rref{magneticansatz} is readily found
considering the Hodge duality \rref{genhodge}. The $(D-q_A-2)$-form
field strength must contain $d-1$ indices of the overall transverse space
and all the indices of the relative transverse (internal) space:
\be
F^{i_{q_A+1}\dots i_p a_1\dots a_{d-1}}=\epsilon^{i_{q_A+1}\dots i_p}
\epsilon^{a_1\dots a_d}{1\over \sqrt{-g}}e^{-a_A\phi}\partial_{a_d} E_A(x).
\label{genmagneticansatz}
\ee
With respect to \rref{magneticansatz} we have dropped the tilde on the
function $E_A$, because it will soon become clear that it will play
exactly the same r\^ole as the function appearing in \rref{genelectricansatz}.
The coupling $a_A$ refers to the field strength, but for simplicity
it carries the label of the brane. The above ans\"atze can be trivially
seen to solve respectively the Bianchi identities \rref{bianchi} and
the equations of motion \rref{eommaxwell}.

In order to rewrite the equations of motion derived from the action
\rref{genaction} taking into account the ans\"atze \rref{generalisotropic}
and \rref{genelectricansatz} and/or \rref{genmagneticansatz}, we first
introduce some notation. Let us define the two quantities:
\be
V_A=B\prod_{i\parallel A}C_i, \qquad \qquad V_A^{\perp}=\prod_{i\perp A}C_i,
\label{defvolparallperp}
\ee
where the first product runs over all the internal directions parallel
to the brane labelled by $A$ and the second one over the internal directions
transverse to it. Let us also define the following quantities:
\be
\varepsilon_A=\left\{
\begin{array}{l} + \quad \mbox{if $q_A$-brane is electric} \\
- \quad \mbox{if $q_A$-brane is magnetic}
\end{array} \right.
\label{varepsilon}
\ee
and:
\be
\delta_A^i=\left\{
\begin{array}{l} D-q_A-3 \quad \mbox{for}\ i\parallel A \\
-(q_A+1) \quad \mbox{for}\ i\perp A
\end{array} \right.
\label{deltaia}
\ee

The set of equations \rref{eomeinstein}--\rref{bianchi} then becomes:
\be
R^t_t=-{1\over 2}\sum_A {D-q_A-3\over D-2} (V_A G)^{-2}e^{\varepsilon_A a_A
\phi} (\partial E_A)^2, \label{einsteintt} \ee
\be
R^i_i=-{1\over 2}\sum_A {\delta_A^i\over D-2}(V_A G)^{-2}e^{\varepsilon_A
a_A\phi} (\partial E_A)^2, \label{einsteinii} \ee
\be
R^a_b= {1\over 2}G^{-2}\partial_a \phi \partial_b \phi+{1\over 2}\sum_A 
(V_A G)^{-2}e^{\varepsilon_Aa_A\phi} \left[-\partial_a E_A \partial_b E_A +
{q_A+1\over D-2} \delta_{ab} (\partial E_A)^2 \right], \label{einsteinab} \ee
\be
\Box \phi = -{1\over 2}\sum_A \varepsilon_A a_A (V_A G)^{-2}
e^{\varepsilon_Aa_A\phi} (\partial E_A)^2, \label{boxphi} \ee
\be
\partial_a \left( {V_A^{\perp}\over V_A}G^{d-2} e^{\varepsilon_A a_A \phi}
\partial_a E_A \right)=0. \label{maxbianchi}
\ee
We use the components of the Ricci tensor as given in the equations
\rref{isoriccitt}--\rref{isoricciab} of Appendix \ref{RICCIapp},
we have defined $(\partial E_A)^2=\partial_a E_A \partial_a E_A$ and 
$\Box \phi$ is given by:
\[
\Box \phi={1\over G^2}\left\{\partial_a \partial_a \phi +\partial_a \phi\
\partial_a\left[\ln B+\sum_i \ln C_i +(d-2) \ln G\right]\right\}.
\]
The last equation \rref{maxbianchi} derives either from the equations
of motion \rref{eommaxwell} or from the Bianchi identities \rref{bianchi},
depending on the electric or magnetic nature of the brane labelled by $A$.

One remark is in order before we continue. In deriving \rref{einsteinii},
we have implicitly assumed that there are no two branes in the configuration
which are both charged under the same (electric or magnetic) field strength
and which intersect over all but one of their world-volume directions. Such
a configuration would give off-diagonal elements to the $\left({}^i_j\right)$
Einstein equations. However our metric ansatz \rref{generalisotropic}
excludes non-trivial solutions of this kind, since the Ricci tensor components
$R_j^i$ are diagonal.

We are now ready to perform some further simplifications, or further
ans\"atze, in order to actually find a class of solutions to the
equations above. The first simplification is motivated by the structure
of the Ricci tensor as given in \rref{isoriccitt}--\rref{isoricciab}, and
is indeed a generalization of the relation \rref{extremecondition} in
the case of a singly charged brane. We impose on the metric components
the following condition:
\be
BC_1\dots C_pG^{d-2}=1.
\label{genextremecondition}
\ee
This condition greatly simplifies the expression for the Ricci tensor,
and moreover it is consistent with the equations of motion simply
by the fact that we will indeed find some solutions. It is sligthly
more complicated to see that this condition singles out (at least from
the known solutions) the brane configurations which are extremal. To
see this, however, it is better to restore spherical symmetry and to
rewrite the condition in a $r$-reparametrization covariant way,
suitable for a general metric like \rref{thegensphmetric}.
One finds that the condition \rref{genextremecondition} is equivalent
to imposing that:
\[
{B'\over B}+\sum_i{C_i'\over C_i}+(d-2)\left({G' \over G}+{1\over r}
-{F\over G r}\right) =0.
\]
It is easy to check that the above condition holds for the metric
\rref{blackmetric} or \rref{blacksolution2} if and only if the non-extremality
parameter $\mu$ vanishes. We will thus often refer to 
\rref{genextremecondition} as the `extremality condition'.

The equations \rref{einsteintt}--\rref{maxbianchi} now become:
\be
\partial_a\partial_a \ln B={1\over 2}\sum_A {D-q_A-3\over D-2} S_A^2
(\partial E_A)^2, 
\label{einsteintt2}
\ee
\be
\partial_a\partial_a \ln C_i = {1\over 2}\sum_A {\delta_A^i\over D-2}
S_A^2 (\partial E_A)^2, 
\label{einsteinii2}
\ee
\[
\partial_a \ln B\ \partial_b \ln B +\sum_i \partial_a \ln C_i\ \partial_b 
\ln C_i +(d-2)\partial_a \ln G\ \partial_b \ln G+{1\over 2}
\partial_a\phi\ \partial_b \phi
\]
\be
+\delta_{ab} \partial_c\partial_c \ln G
={1\over 2}\sum_A S_A^2 \left[\partial_a E_A \partial_b E_A -
{q_A+1\over D-2} \delta_{ab} (\partial E_A)^2 \right], 
\label{einsteinab2}
\ee
\be
\partial_a \partial_a \phi=
-{1\over 2}\sum_A \varepsilon_A a_A S_A^2 (\partial E_A)^2, 
\label{boxphi2}
\ee
\be
\partial_a (S_A^2 \partial_a E_A)= 0,
\label{maxbianchi2}
\ee
where we have defined:
\be
S_A^2= {1\over V_A^2}e^{\varepsilon_A a_A \phi}.
\label{defsa}
\ee
It is worth noting that the condition \rref{genextremecondition} can be
written for each brane $V_A V_A^\perp G^{d-2}=1$, and it is indeed by virtue
of this relation that the same combination $S_A$ appears both in the
Einstein equations and in the Maxwell-like equation \rref{maxbianchi2}.

The second, crucial, simplification is to reduce the number of 
independent functions to $\cal N$. This is motivated by the requirement
that all the branes constituting the configuration are extremal, and
that moreover there is no binding energy between them \cite{rules,proceeding}. 
They thus define a BPS marginal bound state obeying the no-force condition 
\cite{tseytlinnoforce}. 
This means that each brane can be pulled apart from the others at zero
cost in energy, until the fields around it are a good approximation
of the fields describing a singly charged brane solution. Now such a solution
is fully characterized by a single (harmonic) function. It is thus
expected that the solution with $\cal N$ intersecting branes is
characterized by $\cal N$ independent (harmonic) functions.

Since we have already $\cal N$ functions $E_A$, it is natural to take them
for the moment as the only independent functions in the problem. For the
Maxwell-like equations \rref{maxbianchi2} not to couple these functions
in a non-trivial way, we have to assume that the combinations $S_A$ only
depend on the $E_A$ carrying the same label:
\[ S_A \sim E_A^\gamma. \]
Then \rref{maxbianchi2} directly implies that $E_A^{2\gamma+1}$ is a 
harmonic function (for $\gamma\neq -{1\over 2}$).
Now, according to the discussion above, if all the other functions have
to be expressed in terms of the $E_A$'s, and should reproduce the single
brane solutions in the appropriate limit, then we have to take for
$B$:
\[
\ln B=\sum_A \tilde{\alpha}_A^{(B)} \ln E_A,
\]
and similarly for the other functions $C_i$, $G$ and $\phi$.
The equation \rref{einsteintt2} reduces to an algebraic equation for
the coefficients $\tilde{\alpha}_A^{(B)}$ if and only if $\gamma=-1$, i.e.
if $S_A^2 (\partial E_A)^2 \sim (\partial \ln E_A)^2$. 
If we had $\gamma\neq -1$,
the $E_A$'s would be either constant or not truly independent.

We are thus forced to take the following ansatz, which introduces the
new functions $H_A$:
\be
E_A=l_A H_A^{-1}, \qquad \qquad S_A=H_A.
\label{secondansatz}
\ee
The constant $l_A$ is to be determined shortly, by the equations of motion.

The equations \rref{maxbianchi2} simply state now that the $H_A$'s are
harmonic:
\be
\partial_a \partial_a H_A=0.
\label{harmeq}
\ee
The most general solution to the above equation is the following:
\be
H_A=1+\sum_k c_A {Q_{A,k}\over |x^a-x^a_k|^{d-2}},
\label{harmsol}
\ee
where, requiring asymptotic flatness, we fix the overall factor to
1; the $c_A$'s are constants to be determined when one computes the charge
of the branes, and $\{x_k\}$ are the locations in transverse space of
several branes of the same type, i.e. parallel. The harmonic functions above
thus represent a multi-center solution.

Note that the solution \rref{harmsol} assumes that $d\geq 3$. We will comment
afterwards on the cases $d=2$ and $d=1$.

We now express the functions $B$, $C_i$ and $\phi$ in terms of the $H_A$'s:
\bea
\ln B&=& \sum_A \alpha_A^{(B)} \ln H_A, \nonumber \\
\ln C_i &=& \sum_A \alpha_A^{(i)} \ln H_A, \label{bcphidef} \\
\phi &=& \sum_A \alpha_A^{(\phi)} \ln H_A. \nonumber 
\eea
The function $G$ will be determined later, using \rref{genextremecondition}.
The equations \rref{einsteintt2}, \rref{einsteinii2} and \rref{boxphi2}
reduce to equations for the coefficients appearing in \rref{bcphidef},
when \rref{harmeq} are taken into account. We find that:
\bea 
\alpha_A^{(B)}&=& -{D-q_A-3\over D-2} \alpha_A, \nonumber \\
\alpha_A^{(i)} &=& -{\delta^i_A \over D-2} \alpha_A, \label{coeffsol} \\
\alpha_A^{(\phi)}&=& \varepsilon_A a_A \alpha_A, \nonumber
\eea
where:
\be
\alpha_A={1\over 2} l_A^2.
\label{alphaandl}
\ee
For $d\neq 2$, \rref{genextremecondition} and \rref{bcphidef} imply that:
\be
\ln G=\sum_A {q_A+1 \over D-2} \alpha_A \ln H_A.
\label{gdef}
\ee
This trivially solves the part in the equations \rref{einsteinab2} which is 
proportional to $\delta_{ab}$. The remaining piece of the same equations
\rref{einsteinab2} gives us the value of $\alpha_A$, and 
surprisingly gives as a set of consistency conditions the pairwise
intersection rules for the branes involved in the configuration.

The set of equations \rref{einsteinab2} becomes:
\[
\sum_{A,B}\alpha_A \alpha_B \partial_a \ln H_A \ \partial_b \ln H_B \left[
{(D-q_A-3)(D-q_B-3)\over (D-2)^2}+\sum_{i=1}^p {\delta^i_A \delta^i_B \over
(D-2)^2} \right. \qquad\qquad \]
\be \left. \qquad\qquad
+(d-2) {(q_A+1)(q_B+1)\over (D-2)^2}+{1\over 2} \varepsilon_A a_A
\varepsilon_B a_B\right]=\sum_A \alpha_A \partial_a \ln H_A \ \partial_b
\ln H_A.
\label{fatherofrules}
\ee
It can be rewritten as:
\[
\sum_{A,B}(M_{AB}\alpha_A-\delta_{AB})\alpha_B \partial_a \ln H_A \
\partial_b \ln H_B=0. \]

For the $H_A$'s to be truly independent, we must have:
\[M_{AB}\alpha_A=\delta_{AB}, \]
which implies two sets of conditions, $M_{AA}\alpha_A=1$ and $M_{AB}=0$
for $A\neq B$. The first set of equations assigns a value to $\alpha_A$:
\be
\alpha_A=(M_{AA})^{-1}={D-2\over \Delta_A},
\label{alphasol}
\ee
where we have defined as before:
\be
\Delta_A=(q_A+1)(D-q_A-3)+{1\over 2}a_A^2 (D-2).
\label{deltissima}
\ee
We have now an expression for all the parameters in the solution, and
we can thus use \rref{alphasol}, \rref{alphaandl} and \rref{coeffsol}
together with \rref{secondansatz}, \rref{bcphidef} and \rref{gdef}
to characterize completely the fields of the solution:
\be
B=\prod_A H_A^{-{D-q_A-3\over \Delta_A}}, \qquad 
C_i=\prod_A H_A^{-{\delta^i_A\over \Delta_A}}, \qquad
G=\prod_A H_A^{q_A+1\over \Delta_A}, \qquad
\label{intsolution}
\ee
\be
e^\phi=\prod_A H_A^{\varepsilon_A a_A {D-2\over \Delta_A}},\qquad \qquad
E_A=\sqrt{2(D-2)\over \Delta_A} H_A^{-1}.
\label{intsolution2}
\ee
The $E_A$'s coincide with the potentials, up to a constant if the latter are
to vanish at infinity. We have fixed the overall sign to a positive value.
Note that there is only one sign to fix for each set of branes. Accordingly,
all the branes of each type have to have only positive (or only 
negative) charges.

The above derivation thus proves the `harmonic superposition rule' as
formulated by Tseytlin in \cite{tseytlinharm}. Note that the $\Delta_A$
are strictly the same as in the singly charged solution
\rref{thesphextremesolution}, thus the solution above is really
the superposition of single brane solutions.

We are however left now  with an additional set of consistency conditions,
i.e. the conditions $M_{AB}=0$. Since there are no more parameters
to adjust, these conditions will really tell us whether the above solution
is a solution of the equations of motion or not.
Happily, it is easy to see that we can give a simple expression for $M_{AB}$
when $A\neq B$:
\bea
M_{AB}&=&{1\over (D-2)^2}\left[ (D-q_A-3)(D-q_B-3)+\bar{q} {{}\over {}}
(D-q_A-3)(D-q_B-3)
\right. \nonumber \\ & & \left.
-(q_A-\bar{q})(D-q_A-3)(q_B+1)-(q_B-\bar{q})(q_A+1)
(D-q_B-3) {{}\over {}} \right.\nonumber \\ & & \left. 
+(p-q_A-q_B+\bar{q})(q_A+1)(q_B+1)+(D-p-3)(q_A+1)(q_B+1){{}\over {}}\right] 
\nonumber \\ & &
+{1\over 2} \varepsilon_A a_A \varepsilon_B a_B \nonumber \\
&=& \bar{q}+1 -{(q_A+1)(q_B+1)\over D-2 } 
+{1\over 2} \varepsilon_A a_A \varepsilon_B a_B, 
\label{mabsol}
\eea
where we have defined $\bar{q}$ the dimension of the intersection
between the brane labelled by $A$ and the one labelled by $B$ 
($\bar{q}\leq q_A, q_B$), and we have rewritten $d=D-p-1$.

The consistency condition is thus a restriction on the dimension of
the intersection for each pair of branes in the compound:
\be
\bar{q}+1={(q_A+1)(q_B+1)\over D-2 }-
{1\over 2} \varepsilon_A a_A \varepsilon_B a_B.
\label{intersectionrules}
\ee
This relation can be interpreted as an equation for $\bar{q}$. It thus
gives the dimension of the intersection of two branes, given only the 
dimension of the branes, their coupling to the dilaton and the requirement
that they should form a bound state with vanishing binding energy. Note
that the intersection rules above are perfectly pairwise, in the sense 
that they are not influenced by the presence of other branes in the
configuration. For instance, we see that in \rref{mabsol}, even
the value of $p$, the total number of `internal' directions, disappears
from the final expression.

Note that asking that $S_A$ as given in
\rref{defsa} verifies $S_A=H_A$, and making use of the relations
\rref{bcphidef} and \rref{coeffsol}, we hopefully obtain
the same equations \rref{alphasol} and \rref{intersectionrules}.

\paragraph{The $d=2$ and $d=1$ cases}

Let us now comment the cases that we have temporarily excluded from the
derivation above, i.e. the cases for which $d=2$ or $d=1$. These cases
are relevant for the physics in an effective three and two dimensional
space-time respectively. However, despite the somewhat pathological physics
in these lower dimensions,
the results will be very similar to the $d\geq 3$ case. Most of all,
the intersection rules \rref{intersectionrules} are insensitive to
this issue.

The derivation of the intersection rules in the case $d=2$ is a little 
bit more involved, mainly because of the several $(d-2)$ factors that appear
in the equations. In particular, the function $G$ does not appear any
more in the `extremality' ansatz \rref{genextremecondition}. This has as
a consequence that the relation \rref{gdef} is no longer implied by the
other relations \rref{bcphidef} and \rref{coeffsol}, but has to be derived
directly from \rref{einsteinab2}, noting that if the off-diagonal part
of this set of equations is to vanish generically, then its part
proportional to $\delta_{ab}$ also has to vanish independently.
Once this trick is understood, the solution is then easily shown to be 
exactly the same as before, with the exception that now the harmonic
solutions of \rref{harmeq} have a logarithmic behaviour:
\[
H_A^{(d=2)}=h_A+\sum_k c_A Q_{A,k} \ln |x^a-x^a_k|.
\]
The solutions can be expressed formally as in \rref{intsolution}, however
one has to keep in mind that the harmonic functions $H_A$ diverge at
infinity. This points towards a more careful treatment of this problem,
possibly along the lines of \cite{dinstantons,sevenbrane}.
For the extrapolation of the intersection rules from supergravity to
M-theory, it will be nevertheless useful to know that these rules still hold 
in this case.

The $d=1$ case is technically simpler than the preceding one. 
Here the derivation
is exactly the same as in the $d\geq3$ case, except that the partial 
derivatives are now simple derivatives. 
The intersection rules and the solution \rref{intsolution} are (formally)
unchanged, provided we now write for $H_A$:
\[ H_A^{(d=1)}=h_A+\sum_k c_A Q_{A,k} (x-x_k). \]
Note that branes with only one transverse direction are
actually domain walls, and these in general require the supergravity
of which they are solutions to be massive. If however there is no
such $(D-2)$-brane in the configuration, no cosmological constant
seems to be required.

\paragraph{Euclidean branes}

As a last remark on the derivation of the solution 
\rref{intsolution}--\rref{intsolution2} and
of the intersection rules, we note that it extends also to Euclidean
configurations of branes. The time is no longer singled out, and
it can be considered as one of the $y_i$ coordinates.
There is thus no longer a timelike direction that has to be common to
all the branes.
The equations one obtains are exactly the same as 
\rref{einsteintt2}--\rref{maxbianchi2}, except that there is always an
additional factor of $-\varepsilon_A$ appearing inside the sum in 
each r.h.s. This
is because, since now $g_{tt}>0$, the stress-energy tensor of an electric
field has the opposite sign with respect to the Lorentzian case.

The derivation goes along the same way as for the Lorentzian case, and 
the only result which differs is \rref{alphaandl}. It becomes:
\[ \alpha_A=-\varepsilon_A {1\over 2} l_A^2. \]
As a consequence, the fields $E_A$ are given by:
\[ E_A=\sqrt{-\varepsilon_A {2(D-2)\over \Delta_A}}H_A^{-1}. \]
This means that the electric fields are purely imaginary. This is however
a well-known trick to make sense of electrically charged black holes
when one goes to Euclidean gravity (see e.g. \cite{hawkross}
or \cite{dinstantons}).

Note that the condition on the intersection \rref{intersectionrules} has
now potentially more solutions, since $\bar{q}$ can now take also the
value $\bar{q}=-1$ (thus defining an intersection on a point in Euclidean
space). Also, a single brane solution which only exists
in the Euclidean formulation is the $(-1)$-brane, or instanton.

\subsection{Further general considerations on intersecting brane solutions}

In this subsection we try to collect some of the direct
physical consequences of the intersecting brane solutions found
previously. In the next subsection we discuss the intersection rules in the
context of string/M-theory.

We have first to elaborate a little bit on the remark that the solution
\rref{intsolution}--\rref{intsolution2} is a superposition of single
extremal brane solution as the one treated in subsection \ref{EXTREMssec}.
There is a difference between the single extremal brane
solution \rref{thesphextremesolution} and the solution 
\rref{intsolution} in which all but one of the branes have been put
to zero charge: in the second solution the dimension of the transverse
space $d$ can be smaller than $D-q-1$, where $q$ is the dimension of
the brane. Are both solutions physically meaningful and how are they
related to each other?

The answer to this question is well known and the explanation involves
taking the transverse space of the $q$-brane to have some compact
directions. We present hereafter the argument for one additional compact
direction, the generalization is straightforward by induction.

Suppose we have a $q$-brane and we have defined $\bar{d}=D-q-1$.
However only $d=\bar{d}-1$ of the space-like directions
are non-compact, while the extra direction is periodically identified
$\bar{x}\sim \bar{x}+L$. The topology of the transverse space is now
$S^1\times R^d$. A $q$-brane sitting at one point of this transverse
space is seen from the point of view of the covering space
of the $S^1$ factor as an equally spaced array of $q$-branes in the $\bar{x}$
direction, at the same point in the $R^d$ space.

This is correctly described by a multi-center solution obtained from
\rref{intsolution}--\rref{intsolution2} keeping only one non-trivial charge.
The harmonic function characterizing the solution reads:
\[ H=1+\sum_k{Q\over |x-x_k|^{\bar{d}-2}}=
1+\sum_k{Q\over \left[r^2+(\bar{x}-kL)^2\right]^{\bar{d}-2\over 2}}, \qquad
r^2=x_1^2+\dots+x_d^2.
\]
One can approximate the sum assuming that the distance in non-compact space
is much larger than the size of the compact direction, ${L\over r}\ll 1$.
Indeed we have:
\[
\sum_k{1\over \left[r^2+(\bar{x}-kL)^2\right]^{n\over 2}}=
{1\over r^n} \sum_k {1\over \left[1+\left({\bar{x}\over r}
-k{L\over r}\right)^2\right]^{n\over 2}} 
\] \[\simeq
{1\over r^n} {r\over L}\int_{-\infty}^\infty du {1\over (1+u^2)^{n\over 2}}
={1\over L r^{n-1}} \int_0^\pi d\theta \sin^{n-2}\theta,
\]
where we have made the change of variables $u=k{L\over r}-{\bar{x}\over r}$ 
and then
$u=\cot \theta$. The first change of variables, due to the smallness
of ${L\over r}$, changes the sum in a continuous integral.
The integral:
\[
I_m=\int_0^\pi d\theta \sin^m\theta
\]
verifies the following identity:
\[ I_m={m-1\over m} I_{m-2}, \qquad \qquad m\geq 2. \]
The two basic values are $I_1=2$ and $I_0=\pi$.
We can also express the volume of a $(d-1)$-sphere in terms of it:
\[
\Omega_{d-1}=2I_{d-2}I_{d-3}\dots I_1 I_0={2 \pi^{d\over 2}\over 
\Gamma\left({d\over 2}\right)}.
\]
The harmonic function then becomes:
\[ H=1+{I_{d-3}\over L}{Q\over r^{d-2}}. \]
We thus see that `compactifying' one of the transverse directions,
and assuming that we are not too close to the location of the brane,
we recover a harmonic function in the non-compact space. 

We can also show that the charge density, computed as in \rref{chargedensity},
is the same for the uncompactified and the compactified spaces if the
harmonic functions are related as above. Indeed,
\[ {\cal Q}={1\over 16\pi G_D}\int_{S^{\bar{d}-1}}*F_{q+2}=
(d-1)\Omega_d \tilde{Q}, \qquad 
\tilde{Q} = {1\over 16\pi G_D}\sqrt{2(D-2)\over \Delta}Q,
\]
for the brane in non compact $R^{\bar{d}}$ space. In the compact
space on the other hand:
\[
{\cal Q}'={1\over 16\pi G_D}\int_{S^1\times S^{d-1}} *F_{q+2}=
(d-2) L \Omega_{d-1} {I_{d-3}\over L}\tilde{Q}.
\] 
The two charge densities ${\cal Q}$ and ${\cal Q}'$ are indeed equal:
\[ {\cal Q}'={\cal Q} {(d-1)\Omega_d \over (d-2) \Omega_{d-1} I_{d-3}}=
{\cal Q} {(d-1)I_{d-1} \over (d-2)I_{d-3}}={\cal Q}. \]

The procedure that we have shown above can be straightforwardly
continued to other compact transverse directions. 
Note that the solutions found in the previous subsection are exact
solutions of the classical equations of motion. 
In this case, 
it is often said that the solutions are `smeared' over the transverse
compact directions, in order to eliminate the dependence of the
fields in these directions.
The procedure discussed above shows that
the `smeared' solutions are classical approximations to the (quantum)
solutions, in the sense that an average is taken on the location in the
compact transverse directions.
Properly speaking, the above solutions are good `long distance' 
approximations only
up to distances of the order of the compactification length.
Alternatively, one can distribute the branes along the compact directions,
keeping the total charge fixed.
In this way, the charge will look like a uniform distribution at a closer
distance.
See \cite{gibbhorotown} where the procedure above, considered however
in the reverse order, leads to the higher dimensional resolution of
otherwise singular geometries.

After this rather lengthy but necessary parenthesis, we are now 
totally able to discuss the physics of the intersecting brane configurations.

Let us recall the metric:
\be
ds^2=\prod_A H_A^{-2{D-q_A-3\over \Delta_A}}dt^2+\sum_i\prod_A H_A^{-2
{\delta^i_A\over \Delta_A}}dy_i^2+\prod_A H_A^{2{q_A+1\over \Delta_A}}
\delta_{ab}dx^a dx^b,
\label{intsolution3}
\ee
and the field strengths:
\bea
F_{ti_1\dots i_{q_A}a}&=&\epsilon_{i_1\dots i_{q_A}}\sqrt{2(D-2)\over \Delta_A}
\partial_a H_A^{-1}, \qquad \mbox{if}\ \varepsilon_A=(+), 
\label{intsolutionel} \\
F_{i_{q_A+1}\dots i_p a_1\dots a_{d-1}}&=&-
\epsilon_{i_{q_A+1}\dots i_p}\epsilon_{a_1\dots a_d}\sqrt{2(D-2)\over \Delta_A}
\partial_{a_d}H_A, \qquad \mbox{if}\ \varepsilon_A=(-),
\label{intsolutionmag}
\eea
where the harmonic functions are as in \rref{harmsol}. The charge densities are
defined as in the preceding section by:
\bea
{\cal Q}_A&=&{1\over 16 \pi G_D}\int_{T^{p-q_A}\times S^{d-1}} *F_{q_A+2}
, \qquad \mbox{if}\ \varepsilon_A=(+),
\label{elchargedensity} \\
{\cal Q}_A&=&{1\over 16 \pi G_D}\int_{T^{p-q_A}\times S^{d-1}} 
F_{D-q_A-2}, \qquad \mbox{if}\ \varepsilon_A=(-).
\label{magchargedensity}
\eea
Both of the above definitions give, when taking into account respectively
\rref{intsolutionel} and \rref{intsolutionmag}, the following value
for the charge density:
\be
{\cal Q}_A={L^{p-q_A}\Omega_{d-1} \over 16 \pi G_D}(d-2)
\sqrt{2(D-2)\over \Delta_A}c_A \sum_k Q_{A,k},
\label{multichargedensity}
\ee
where $L$ is the (common) size of all the internal directions. As noted
before, all the charges $Q_{A,k}$ have to be of the same sign\footnote{
This can also be motivated by supersymmetry (see subsection \ref{SUSYssec}),
and from D-brane physics \cite{tasi,bankssuss}.}, which here
is taken to be positive.

It is then straightforward to compute the ADM mass as in \rref{admmassformula}
(the multi-center solutions are approximately spherically symmetric when
the branes are confined to a certain region of transverse space) and
to express it in terms of the charge densities \rref{multichargedensity}:
\be
M=\sum_A \sqrt{2(D-2)\over \Delta_A} L^{q_A}{\cal Q}_A.
\label{multiadmmass}
\ee
This formula actually proves the assumption that our solutions represent
bound states of branes with vanishing binding energy. Indeed, each term
in the sum above is nothing else than the mass of each constituent 
extremal $q_A$-brane, as in \rref{bogomolnyi}.

This is an important result: the two mathematical conditions of imposing
first the relation \rref{genextremecondition} and then the formulation
in terms of $\cal N$ independent function were correctly interpreted
as enforcing extremality and the no-force condition.

If we consider a configuration in which all the branes are located at
the same point in the overall transverse space (this is truly a bound
state), we have a spherically symmetric configuration and we can
moreover compute the entropy using the formula \rref{entropy}, with
the horizon coinciding with $r=0$.           
We find:
\be
S=\left.{L^p\Omega_{d-1}\over 4G_D}\prod_A H_A^{D-2\over \Delta_A}
r^{d-1}\right|_{r\rightarrow 0}=
\left.{L^p\Omega_{d-1}\over 4G_D}\prod_A (c_AQ_A)^{D-2\over \Delta_A}
r^{d-1-\sum_A (d-2){D-2\over \Delta_A}}\right|_{r\rightarrow 0}.
\label{multientropy}
\ee
The actual value  of the power of $r$ depends on the number and 
(presumably) on the nature of the branes involved in the bound state,
and it determines whether the solution has a finite entropy or not.
We will see shortly how to achieve a finite entropy in configurations
of extreme branes in M-theory.

\subsection{Applications to M-theory}

It is now time to specialize to the intersecting brane solutions
that one finds in 11 dimensional supergravity and in 10 dimensional
string theories. We will first consider extensively the theories
with maximal supersymmetry ($D=11$ and type II theories), and in the 
end comment on the theories with less supersymmetry. 

Let us start with the intersections of the branes of 11 dimensional
supergravity, also called M-branes. 
Taking into account that we only have one 4-form field strength 
and no dilaton, the formula for the intersection 
\rref{intersectionrules} rewrites simply:
\be
\bar{q}+1={(q_A+1)(q_B+1)\over 9}.
\label{mintrules}
\ee
Adopting the obvious notation $q_A\cap q_B=\bar{q}$,
we have the following intersections
involving the two M-branes, M2 and M5:
\be
M2\cap M2=0,\qquad M5\cap M5=3,\qquad M2\cap M5=1.
\label{mintersections}
\ee
The supergravity solutions representing special cases of the first
two intersections above were derived in \cite{papatown}, using the
solutions previously found in \cite{guven}. The third intersection
was postulated in \cite{stromopen,towndfromm} from D-brane and duality
arguments, and the classical
solution relative to it was shown in \cite{tseytlinharm,gaunkasttras}.

The most intriguing intersection is the third one. 
According to the discussion in the previous subsection,
the supergravity solution is `smeared' over all the longitudinal 6-dimensional
space. The M2-branes are not localized on the M5-brane and vice-versa.
If however we extrapolate the solution to a configuration in which both
branes are infinite and well-localized, 
then we arrive at the picture of an M5-brane
cutting an M2-brane in two pieces, since the intersection has the
dimension of the boundary of the M2-brane. A second, more speculative 
step is to consider that the two parts of the M2-brane can break up,
leaving two open M2-branes with their boundaries attached to the M5-brane.
This is a familiar picture in D-brane physics, the novelty being that
also non-perturbative objects like the M-branes behave like the fundamental
strings. In the next chapter this picture will be made more precise,
but additional input will be needed.

Note that in \cite{gaunkasttras} a configuration giving M5$\cap$M5=1 is
discussed. However the harmonic functions have a different dependence than
in the solutions discussed in this section. We will comment on
this different case in Section \ref{CONCLUsec}.

Getting one dimension lower, we find ourselves in the low-energy effective
theories of type II strings. Consider first of all the intersections
of the RR-charged branes. Since the coupling to the dilaton is
$\varepsilon a={3-q\over 2}$, the intersection rules can be rewritten
as:
\be
\bar{q}={1\over 2}(q_A+q_B-4).
\label{dintrules}
\ee
This rule gives the following intersections (where the D is conventional):
\bea
Dq\cap Dq&=&q-2,  \label{dintequal}\\
D(q-2)\cap Dq&=&q-3, \label{dintopen} \\
D(q-4)\cap Dq&=& q-4. \label{dintwithin}
\eea
The intersection \rref{dintopen} was also conjectured in \cite{stromopen}, 
and the same discussion as for the M2$\cap$M5=1 case applies.
The last intersection \rref{dintwithin} actually represents a D$(q-4)$-brane
within a D$q$-brane, as suggested in \cite{douglaswithin}.

There are some interesting remarks that can be made about the above
intersections of branes. Note first that the intersection
D0$\cap$D6 makes no sense at all, since it would give $\bar{q}=1$!
This means that the D0-brane and the D6-brane cannot form a marginal
bound state, and this applies also to all the configurations which
are dual to this one. The second remark is that there are several
intersections giving $\bar{q}=-1$, e.g.  D1$\cap$D1=$-$1. Though 
meaningless in ordinary space-time, these configurations are 
nevertheless relevant in Euclidean space, and can for instance represent
instantons.

The last set of branes we consider are the NSNS ones, for which $a=-1$.
We have the following intersections involving the fundamental string F1
and the solitonic fivebrane NS5:
\be
F1\cap F1=-1,\qquad F1\cap NS5=1,\qquad NS5\cap NS5=3,
\label{nsnsint}
\ee
\bea
F1\cap Dq&=&0,
\label{f1dint} \\
Dq \cap NS5&=&q-1, \qquad q\leq 6.
\label{ns5dint}
\eea
The three intersections \rref{nsnsint} are perfectly consistent through
type IIB S-duality with the intersections involving the D1 and the D5. 

The relation \rref{f1dint} seems really a classical `hint' that fundamental
strings can end on RR-charged branes, which should then consequently
be called D(irichlet)-branes. Of course historically the discovery was
done the other way round, defining the D-branes in (perturbative) 
string theory and then showing that they are the carriers of RR charge
\cite{polcdbranes}. However here the last relation \rref{ns5dint}
seems to imply further that all the D-branes themselves can have their
boundaries on the NS5. Again, this will be elaborated in the next chapter.

The concluding remark about all the intersections that we have discussed
above is that they were all derived in a strikingly similar way,
simply applying the formula \rref{intersectionrules} to the
pair of branes at hand. All the intersections are on an equal footing
within this classical approach, while they have a clearly different
origin in any one of the theories in which they can be embedded. 
Note the other approach discussed in \cite{tseytlinnoforce},
which however only applies to the string/M-theory framework.

We only have presented above pairwise intersections. As pointed out in
the derivation of \rref{intersectionrules}, one can build configurations
with more than two branes simply imposing the intersection rules on each
pair of branes in the compound.

It is straightforward but lengthy to prove that the Chern-Simons terms
present in the original actions \rref{11action}, \rref{iiaeinstein}
and \rref{iibeinstein} modify the equations of motion with terms
which vanish identically for the solutions corresponding to the
intersections above. We can thus confidently consider the intersecting
brane solutions as configurations in M-theory. The Chern-Simons terms
will however turn out to play a non-trivial r\^ole when we will
discuss more accurately the possibility of having open branes.

We now turn to the problem of finding bound states of branes with a
non-vanishing entropy \rref{multientropy}.

As far as branes in M-theory are concerned, we had already noted in
subsection \ref{MBRANESssec} that they all verify $\Delta_A=2(D-2)$.
Note that this is insensitive to the possible existence of additional
compact directions transverse to the particular brane labelled by $A$.
This fact simplifies the expression for the entropy which becomes,
when $\cal N$ branes participate to the bound state:
\[
S\sim \left. \prod_A Q_A^{1\over 2}r^{d-1-{1\over 2}{\cal N}(d-2)}
\right|_{r\rightarrow 0}.
\]
If we define $\bar{D}=D-p=d+1$, the condition for a finite entropy is:
\be
{\cal N}=2{\bar{D}-2\over \bar{D}-3}.
\label{entropcond}
\ee
For $4\leq\bar{D}\leq 11$ there are only two cases which give an integer
$\cal N$, and these are:
\be
\bar{D}=5, \quad{\cal N}=3\qquad \mbox{and} \qquad \bar{D}=4, \quad{\cal N}=4.
\label{onlytwocases}
\ee
Restoring the right coefficients, the entropy in these two
cases is:
\be
S={L^p\Omega_{d-1}\over 4G_D}\prod_A (c_AQ_A)^{1\over 2}. 
\label{finiteentropy}
\ee

As a straightforward exercise, we can write the metrics of a $\bar{D}=5$
and a $\bar{D}=4$ black hole. For the $\bar{D}=5$ configuration,
we take the one in 11 dimensions consisting of three different types
of M2-branes, all intersecting each other in one point. The metric
reads:
\bea
ds^2_5&=&-(H_1H_2H_3)^{-{2\over 3}}dt^2+H_1^{-{2\over 3}}(H_2 H_3)^{1\over 3}
(dy_1^2+dy_2^2)+H_2^{-{2\over 3}}(H_1 H_3)^{1\over 3}(dy_3^2+dy_4^2)
\nonumber \\& &+H_3^{-{2\over 3}}(H_1 H_2)^{1\over 3}(dy_5^2+dy_6^2) 
+(H_1H_2H_3)^{1\over 3}(dr^2+r^2d\Omega_3^2).
\label{222blackhole}
\eea
For the $\bar{D}=4$ black hole, we take its realization in type IIA theory
as a bound state of three D4-branes, intersecting each other over
2 dimensions, and one D0-brane:
\bea
ds^2_4&=&-H_{D0}^{-{7\over 8}}(H_{{D4}_1}H_{{D4}_2}H_{{D4}_3})^{-{3\over 8}}
dt^2+H_{D0}^{1\over 8}(H_{{D4}_1}H_{{D4}_2})^{-{3\over 8}}
H_{{D4}_3}^{5\over 8}(dy_1^2+dy_2^2)\nonumber \\& &+
H_{D0}^{1\over 8}(H_{{D4}_1}H_{{D4}_3})^{-{3\over 8}}
H_{{D4}_2}^{5\over 8}(dy_3^2+dy_4^2)+
H_{D0}^{1\over 8}(H_{{D4}_2}H_{{D4}_3})^{-{3\over 8}}
H_{{D4}_1}^{5\over 8}(dy_5^2+dy_6^2)\nonumber \\& &+
H_{D0}^{1\over 8}(H_{{D4}_1}H_{{D4}_2}H_{{D4}_3})^{5\over 8}
(dr^2+r^2d\Omega_2^2).
\label{4440blackhole}
\eea
Starting from these two solutions, and using the U-dualities, one can
obtain all the configurations with respectively 3 charges in 5
dimensions and 4 charges in 4 dimensions. Note two interesting facts:
first, the equation \rref{entropcond} implies that there are no
`stringy' extreme black holes with non-vanishing entropy in $\bar{D}\geq 6$;
secondly, the intersection rules are such that there are no allowed
configurations for which the entropy diverges, i.e. it
is impossible to achieve ${\cal N}>2{\bar{D}-2\over \bar{D}-3}$.

Let us now digress on a non-orthodox case.
We did not consider up to now the branes of the non-maximally
supersymmetric string theories, namely type I string theory and the
two heterotic string theories. All of these theories include 
a Super Yang-Mills sector with a gauge group of rank 16. 
We consider here only the fields relative to the
abelian part of the gauge group. 

Consider one of the heterotic theories. The coupling of the 2-form field
strengths to the dilaton is $a_2=-{1\over 2}$. There are electrically
charged 0-branes and magnetically charged 6-branes, which we call
respectively H0 and H6. We now encounter the first difference with
respect to the maximally supersymmetric theories: if we compute
the $\Delta$ for these branes, according to \rref{deltissim}, we
find $\Delta=1\cdot 7+{1\over 2}{1\over 4}8=8$. Remember that previously
we always had $\Delta=16$ in $D=10$, and that moreover this last value
holds for the F1 and NS5 which are also present in the heterotic theories.

The intersections that one finds are the following:
\be
H0\cap H0=-1,\qquad H0\cap H6=0,\qquad H6\cap H6=5,
\label{hhint}
\ee
\be
H0\cap F1=-1, \qquad\qquad H0\cap NS5=0,
\label{h0nsint}
\ee
\be
H6\cap F1=1, \qquad\qquad  H6\cap NS5=4.
\label{h6nsint}
\ee

Going to type I theory, we have that the coupling of the 2-forms is
given by $a_2={1\over 2}$. We have thus exactly the same pattern
of intersections, where however the NSNS branes are replaced by
the D1 and the D5. This is consistent with Heterotic/type I duality.

These 0- and 6-branes of the 10 dimensional $N=1$ theories are seemingly
different from the ones encountered in the maximal supergravities. 
Indeed, a quick look at the $N=1$ superalgebra in
$D=10$ \cite{democracy} directly tells us that there is no central charge
corresponding to their (electric or magnetic) charge. This implies that
the above configurations cannot be supersymmetric.
Their treatability and their potential r\^ole in the wider context
of the underlying string theories is thus uncertain, and probably limited.
Note that the H0 and H6 branes, and their intersections, provide examples
of extremal branes which are nevertheless non-supersymmetric.

Our last comment is about what we did not do. The derivation of
the intersecting brane configurations is straightforwardly generalizable
to the case were there are several scalars instead of only one, the dilaton.
The general solution and the intersection rules would have come out
in quite the same form, with now the coupling to the scalars being a
vector instead of a single value. The interest of doing this, besides
finding solutions in a more general theory, would have been that now
some solutions could have been extrapolated to 10 or 11 dimensional
solutions which included waves and/or KK monopoles. Also some
intersection rules for these two objects would have been derived.
However, the relation between the 10 dimensional couplings and the ones
in the necessarily reduced space-time grow more and more complicated
as one reduces further dimensions. At a certain moment, it becomes
more convenient to apply the T-duality rules as decribed in 
\cite{bergshoeffetal}. The approach using several scalars has already
been followed in \cite{kklp,stelle,arefeva}. Here,
we have preferred to give the
derivation in a simpler, even if incomplete, case.

\subsection{Supersymmetry and the branes}
\label{SUSYssec}

We now consider in more detail one of the most interesting properties
of the branes and of their intersections, that is supersymmetry. 
We have already elaborated on the power of supersymmetry in extrapolating
low-energy,
classical results to the quantum underlying theory. We had also anticipated
that all the extremal branes presented in subsection \ref{MBRANESssec}
were supersymmetric, i.e. that they preserve one half of the space-time
supersymmetries of the theory in which they live.

An interesting problem is the following. The intersecting brane
solutions presented in this section combine several of these 
half-supersymmetric branes. It is thus natural to ask wheter the
intersecting configuration is supersymmetric and how many
supersymmetries it preserves. Here also, we had anticipated that
the solutions are indeed supersymmetric. 

In this subsection, we will consider the supersymmetric properties
of the branes of 11 dimensional supergravity. We will first prove
that the basic branes, the M2 and the M5-brane, preserve half of
the supersymmetries, and then we will consider the most general 
configuration of intersecting branes. The intersection rules
\rref{intersectionrules} will, as expected, play a crucial r\^ole
in establishing the existence and the amount of preserved supersymmetries.

We focus on 11 dimensional supergravity for two reasons. One is that
it is a simpler theory with respect to 10 dimensional type II theories
(it has only one antisymmetric tensor field and no dilaton). The
second reason is that all the branes, and the intersecting 
configurations, are related by dualities. Since the dualities preserve
the supersymmetry, proving the supersymmetry for
a configuration of M-branes proves it also for the other 10 dimensional
configurations.
Of course this is the drawback of supersymmetry, one can no longer
work in a general setting as with \rref{genaction}.

What we will actually do is to start with a fairly general configuration
of fields, and then impose that some
supersymmetries are preserved. We will get in the end the same
solution as we found solving the equations of motion. However in
the present approach the ans\"atze will come out of the equations
rather than being imposed. If we were only interested in solutions
in 11 dimensional supergravity, we could have indeed chosen this
approach from the beginning. The advantage of the `bosonic' approach
is nevertheless that we have a much wider range of application of our
results, and a very concise formula for the intersection rules.

Let us begin writing the supersymmetric variation of the
gravitino $\psi_\mu$ in 11 dimensional supergravity \cite{11dsugra}
(in a vanishing gravitino background):
\be
\delta \psi_\mu=D_\mu \eta+{1\over 288}\left(\Gamma_{\mu\nu\rho\sigma\lambda}
-8g_{\mu\nu}\Gamma_{\rho\sigma\lambda}\right)F^{\nu\rho\sigma\lambda}\eta,
\label{deltapsi}
\ee
\be
D_\mu \eta=\partial_\mu\eta +{1\over 2}\omega^{\hat{\rho}\hat{\sigma}}_\mu
\Sigma_{\hat{\rho}\hat{\sigma}} \eta.
\label{spinderiv}
\ee
The 32-component Majorana spinor $\eta$ is the supersymmetry parameter. 
The number of
its independent arbitrary components measures the portion of preserved
supersymmetry.
As in Appendix \ref{RICCIapp}, the hatted indices are (flat) Lorentz indices
while the unhatted ones are covariant indices. A $\Gamma$ matrix with
a covariant index is defined by:
\be
\Gamma_\mu =e^{\hat{\sigma}}_\mu \Gamma_{\hat{\sigma}}, \qquad \mbox{where}
\qquad \Gamma_{\hat{\mu}}\Gamma_{\hat{\nu}}+\Gamma_{\hat{\nu}}
\Gamma_{\hat{\mu}}=2 \eta_{\hat{\mu}\hat{\nu}}.
\label{gammastuff}
\ee
$\Gamma$ matrices with several indices are antisymmetric products of
simple $\Gamma$ matrices:
\[ \Gamma_{\mu_1\dots \mu_n}=\Gamma_{[\mu_1}\dots \Gamma_{\mu_n]}.\]
The spin connection $\omega^{\hat{\rho}\hat{\sigma}}_\mu$ is
discussed also in Appendix \ref{RICCIapp}, and the matrices:
\[
\Sigma_{\hat{\mu}\hat{\nu}}={1\over 4}(\Gamma_{\hat{\mu}}\Gamma_{\hat{\nu}}
-\Gamma_{\hat{\nu}}\Gamma_{\hat{\mu}})={1\over 2}\Gamma_{\hat{\mu}\hat{\nu}}
\]
are the $SO(1,10)$ Lorentz generators acting on the spinor representation. 

We do not need to write the supersymmetric variations of the bosonic
fields because, since the fermionic fields (the gravitino) have 
vanishing expectation value in the classical background we are considering,
they are trivially zero. Asking that $\delta\psi_\mu=0$ will be
sufficient to prove that the solution is unchanged with respect to that
particular supersymmetry transformation. 

Consider first the M2-brane. The metric and 4-form field strength we
start with are:
\[
ds^2=-B^2dt^2+\sum_{i=1}^2 C^2 dy_i^2 +G^2 \delta_{ab}dx^a dx^b, \qquad
a,b=1\dots 8, 
\]
\[ F_{tij a}=\epsilon_{ij} \partial_a E.
\]
The spin connection relative to this geometry is given in Appendix
\ref{RICCIapp}, eqs. \rref{isospinconta}--\rref{isospinconab}.

Supposing that the supersymmetry parameter also depends only on the
$x$'s, we obtain after some algebra (the index 0 stands for $t$):
\bea
\delta\psi_t&=&{1\over 6C^2 G}\Gamma_{\hat{0}}\left(3C^2 \partial_a B \
\Gamma_{\hat{a}}-\partial_aE\ \Gamma_{\hat{a}} \Gamma_{\hat{0}\hat{1}\hat{2}}
\right)\eta, \label{deltam2t} \\
\delta\psi_i&=&{1\over 6BCG}\Gamma_{\hat{\imath}}\left( 3BC\partial_a C\
\Gamma_{\hat{a}}-\partial_aE\ \Gamma_{\hat{a}} \Gamma_{\hat{0}\hat{1}\hat{2}}
\right)\eta, \label{deltam2i} \\
\delta\psi_a&=&\partial_a \eta+{1\over 2G}\partial_b G \ 
\Gamma_{\hat{a}\hat{b}}\eta+{1\over 12 BC^2}\partial_b E\ 
\Gamma_{\hat{a}\hat{b}} \Gamma_{\hat{0}\hat{1}\hat{2}}\eta
-{1\over 6BC^2}\partial_b E\ \Gamma_{\hat{0}\hat{1}\hat{2}}\eta.
\label{deltam2a}
\eea

Begin with imposing $\delta\psi_t=0$. It implies:
\be
3C^2 \partial_a B=\pm \partial_aE, \qquad \qquad (1\mp
\Gamma_{\hat{0}\hat{1}\hat{2}})\eta=0.
\label{solpsit}
\ee
This can be seen as follows. If $M=\not\! u +\not\! v \tilde{\Gamma}$, where
$\not\! u =u_a \Gamma_a$ (we drop hats) and $\tilde{\Gamma}$ is such that 
$\tilde{\Gamma}^2=1$, $tr \tilde{\Gamma}=0$, $\tilde{\Gamma}\Gamma_a=
- \Gamma_a\tilde{\Gamma}$ and $\tilde{\Gamma}^T=
\tilde{\Gamma}$, as
$\Gamma_{\hat{0}\hat{1}\hat{2}}$ (for definiteness, we are in a basis
in which all the $\Gamma$'s are real), then we have:
\[ \det M=(\det M^T\det M)^{1\over 2}=(\det M^TM)^{1\over 2}=
\left[\det (u^2+v^2+ 2u\cdot v \tilde{\Gamma})\right]^{1\over 2}\] \[
=(u^2+v^2+2u\cdot v)^8(u^2+v^2-2u\cdot v)^8=
|u+v|^{16}|u-v|^{16}. \]
The condition $M\eta=0$ has non-trivial solutions if and only if $\det M=0$,
and this in turn implies $u_a=\pm v_a$. The spinor $\eta$ is then solution
of $(1\pm \tilde{\Gamma})\eta=0$, 
which projects it onto a subspace of 16 (real) dimensions over 32.

The equation $\delta\psi_i=0$ then becomes, making use of \rref{solpsit}:
\[ {C\over 2 G}\Gamma_{\hat{\imath}}\left( \partial_a \ln C-
\partial_a \ln B\right)\Gamma_{\hat{a}} \eta=0,
\]
which thus gives:
\be
B=C \qquad \qquad \Rightarrow \qquad E=\pm B^3.
\label{solpsii}
\ee
We thus see that $SO(1,2)$ Lorentz invariance on the world-volume of
the M2-brane is imposed on us by supersymmetry. Note that if we wanted
an asymptotically vanishing potential $E$, we should write, instead
of the second relation above, $E=\pm(B^3-1)$;  the
conclusions are obviously unchanged.

Asking that $\delta\psi_a=0$, and using \rref{solpsit} and \rref{solpsii},
we arrive at the expression:
\[
\partial_a \eta-{1\over 2} \partial_a \ln B\ \eta+{1\over 2}\left(\partial_b
\ln G+{1\over 2}\partial_b\ln B\right)\Gamma_{\hat{a}\hat{b}}\eta=0.
\]
This equation is solved\footnote{That the solution \rref{solpsia} is
the only one can be proved \cite{duffstelle} using arguments relative
to the integrability of the supersymmetric condition $\delta\psi_\mu=0$,
see for instance \cite{duffnilssonpope}.} for:
\be
BG^2=1, \qquad \qquad \eta=B^{1\over 2}\eta_{(0)},
\label{solpsia}
\ee
where $\eta_{(0)}$ is a constant spinor. Note that the first relation
above can be rewritten as $B^3G^6=1$, i.e. exactly in the same form
as for the `extremality' condition \rref{extremecondition}.

Using the equations of motion for the 4-form field strength one
can now with a minimum effort solve completely the problem,
finding that $\partial_a\partial_a E^{-1}=0$. If one
writes the solution in terms of the harmonic function $H=E^{-1}$, then
the solution \rref{m2metric} is found once again.

We have thus proved that the solution \rref{m2metric} is preserved
by the supersymmetries with parameter $\eta$ such that:
\be
\eta = H^{-{1\over 6}}\eta_{(0)}, \qquad \qquad 
\eta_{(0)}={1+\Gamma_{\hat{0}\hat{1}\hat{2}}\over 2}\eta_{(0)}.
\label{susym2}
\ee
We thus say that the M2-brane preserves half of the supersymmetries.

Let us now consider the M5-brane. Its metric and field strength are taken
to be generally:
\[
ds^2=-B^2dt^2+\sum_{i=1}^5 C^2 dy_i^2 +G^2 \delta_{ab}dx^a dx^b, \qquad
a,b=1\dots 5,
\]
\[ F^{abcd}={1\over BC^5 G^5}\epsilon^{abcde} \partial_e \tilde{E}.
\]
The variation of the gravitino is then the following:
\bea
\delta\psi_t&=&{1\over 12 C^5 G}\Gamma_{\hat{0}}( 6C^5\partial_a B\
\Gamma_{\hat{a}}+\partial_a\tilde{E}\ \Gamma_{\hat{a}} \Gamma_{(5)})\eta,
\label{deltam5t} \\
\delta\psi_i&=&{1\over 12 B C^4 G}\Gamma_{\hat{\imath}}(
6BC^4\partial_a C\ \Gamma_{\hat{a}}+
\partial_a\tilde{E}\ \Gamma_{\hat{a}} \Gamma_{(5)})\eta,
\label{deltam5i} \\
\delta\psi_a&=&\partial_a \eta+{1\over 2G}\partial_b G\ \Gamma_{\hat{a}\hat{b}}
\eta +{1\over 12BC^5}\partial_a\tilde{E}\ \Gamma_{(5)}\eta
-{1\over 6BC^5}\partial_b\tilde{E}\ \Gamma_{\hat{a}\hat{b}}\Gamma_{(5)}\eta.
\label{deltam5a}
\eea
We have used the following properties of the $\Gamma$
matrices (all indices hereafter are flat):
suppose $\Gamma_{(d)}={1\over d!}\epsilon^{a_1\dots a_d}\Gamma_{a_1
\dots a_d} \equiv \Gamma_1\dots \Gamma_d$; we then
have:
\be
\epsilon^{a b_1\dots b_{d-1}}\Gamma_{b_1\dots b_{d-1}}=(d-1)!\ \Gamma^a 
\Gamma_{(d)}, \qquad \epsilon^{a b c_1\dots c_{d-2}}\Gamma_{c_1\dots c_{d-2}}
=-(d-2)!\ \Gamma^{ab}\Gamma_{(d)}.
\label{gamma5rel}
\ee

The condition $\delta\psi_t=0$ implies in a way exactly similar as 
for the M2-brane (except that we now have $\Gamma_{(5)}\Gamma_a=\Gamma_a
\Gamma_{(5)}$):
\be
6C^5 \partial_a B=\pm \partial_a \tilde{E}, \qquad \qquad (1\pm\Gamma_{(5)})
\eta =0.
\label{solm5t}
\ee
Then $\delta\psi_i=0$ in turn gives:
\be
B=C \qquad \qquad \Rightarrow \qquad \tilde{E}=\pm B^6.
\label{solm5i}
\ee
The last condition is $\delta\psi_a=0$, and it is solved by:
\be
B^2 G=1, \qquad \qquad \eta=B^{1\over 2}\eta_{(0)}.
\label{solm5a}
\ee
Again, the first relation above can be seen to be equivalent to the
extremality condition \rref{extremecondition} for a 5-brane, $B^6G^3=1$.

The Bianchi identities for the 4-form impose straightforwardly
that $\tilde{E}^{-1}\equiv H$ is harmonic. We have thus rediscovered
the M5-brane solution as in \rref{m5metric}. The M5-brane is therefore
invariant under a supersymmetry variation of parameter:
\be
\eta=H^{-{1\over 12}}\eta_{(0)}, \qquad \qquad \eta_{(0)}=
{1-\Gamma_{(5)}\over 2}\eta_{(0)}.
\label{susym5}
\ee
Note that the sign in front of $\Gamma_{(5)}$ is a consequence of the
conventions.
Using the properties of the $\Gamma$ matrices in 11 dimensions, most notably
the fact that $\Gamma_0\dots \Gamma_{10}=1$, the condition on the constant
spinor can be rewritten in the more familiar form:
\[
\Gamma_{\hat{0}\hat{1}\hat{2}\hat{3}\hat{4}\hat{5}}\eta_{(0)}=\pm \eta_{(0)},
\]
where all the indices now lie along the world-volume of the M5-brane.

We are now ready to go to the general case, where we have an arbitrary
number of (intersecting) M2- and M5-branes. Accordingly, we start with
a general metric as \rref{generalisotropic}, and with a set of non-trivial
electric and magnetic components of the 4-form field strength:
\be
F_{ti_1 i_2 a}=\epsilon_{i_1 i_2} \partial_a E_N, \qquad  \qquad
N=1\dots {\cal N}_{el}, \label{multim2fs} \ee \be
F^{i_6\dots i_p a_1 \dots a_{d-1}}={1\over BC_1\dots C_pG^d}
\epsilon^{i_6\dots i_p} \epsilon^{a_1 \dots a_d} \partial_{a_d}\tilde{E}_M,
\qquad  \qquad M=1\dots {\cal N}_{mag}. \label{multim5fs}
\ee
Note that using \rref{defvolparallperp}, we can rewrite $BC_1\dots C_p=
V_M V_M^{\perp}$ and $F^{ti_1 i_2 a}=-{1\over V_N^2 G^2}
\epsilon_{i_1 i_2} \partial_a E_N$.

It is useful to keep in mind the previous computations for a single M2-brane
and a single M5-brane to compute the gravitino variation in this general
case. We obtain:
\bea
\delta \psi_t&=& {1\over 2G} \partial_a B\ \Gamma_{\hat{0}}\Gamma_{\hat{a}}\eta
-{1\over 6}\sum_N{B\over V_N G}\partial_a E_N \Gamma_{\hat{0}}
\Gamma_{\hat{a}} \Gamma_N \eta +{1\over 12}\sum_M {B\over V_M G}
\partial_a \tilde{E}_M \Gamma_{\hat{0}}\Gamma_{\hat{a}} \Gamma_M^\perp \eta,
\nonumber  \\
\delta \psi_i&=& {1\over 2G} \partial_a C_i \Gamma_{\hat{\imath}}
\Gamma_{\hat{a}}\eta -{1\over 6}\sum_{N\parallel i} {C_i\over V_N G}
\partial_a E_N\Gamma_{\hat{\imath}} \Gamma_{\hat{a}} \Gamma_N \eta 
+{1\over 12} \sum_{N\perp i}{C_i\over V_N G} \partial_a E_N
\Gamma_{\hat{\imath}} \Gamma_{\hat{a}} \Gamma_N \eta
\nonumber \\ & &
+{1\over 12} \sum_{M\parallel i} {C_i\over V_MG} \partial_a \tilde{E}_M
\Gamma_{\hat{\imath}} \Gamma_{\hat{a}}\Gamma_M^\perp \eta
-{1\over 6}\sum_{M\perp i} {C_i\over V_MG} \partial_a \tilde{E}_M
\Gamma_{\hat{\imath}} \Gamma_{\hat{a}}\Gamma_M^\perp \eta, 
\nonumber \\
\delta \psi_a &=& \partial_a \eta+{1\over 2G} \partial_b G\ 
\Gamma_{\hat{a}\hat{b}}\eta +{1\over 12} \sum_N {1\over V_N} \partial_b E_N
\Gamma_{\hat{a}\hat{b}}\Gamma_N \eta -{1\over 6}\sum_N {1\over V_N} 
\partial_a E_N\Gamma_N \eta 
\nonumber \\ & &
+{1\over 12} \sum_M {1\over V_M} \partial_a
\tilde{E}_M \Gamma_M^\perp \eta -{1\over 6}\sum_M {1\over V_M} \partial_b
\tilde{E}_M\Gamma_{\hat{a}\hat{b}}\Gamma_M^\perp \eta.
\label{deltamulti}
\eea
We have written $\Gamma_N$ for the product of $\Gamma$ matrices longitudinal
to the M2-brane labelled by $N$, and $\Gamma_M^\perp$ for the 
product of $\Gamma$ matrices transverse to the M5-brane labelled by $M$.
The notation $A\parallel i$ ($A\perp i$) denotes that the brane is 
longitudinal (transverse) to the direction $y_i$.

The set of conditions $\delta\psi_t=0$ and $\delta\psi_i=0$ can be combined
in the ${\cal N}_{el}+{\cal N}_{mag}$ following equations:
\[
\partial_a \ln V_N \Gamma_{\hat{a}}\eta-{1\over V_N}\partial_a E_N
\Gamma_{\hat{a}} \Gamma_N \eta-{1\over 2}\sum_{N'\neq N} \bar{q}_{NN'}
{1\over V_{N'}}\partial_a E_{N'} \Gamma_{\hat{a}} \Gamma_{N'}\eta \qquad\qquad 
\]
\[ \qquad \qquad \qquad \qquad  
+{1\over 2}\sum_M(\bar{q}_{NM}-1){1\over V_M} \partial_a\tilde{E}_M
\Gamma_{\hat{a}}\Gamma_M^\perp \eta=0,
\]
\[
\partial_a \ln V_M \Gamma_{\hat{a}}\eta-{1\over V_M} \partial_a\tilde{E}_M
\Gamma_{\hat{a}}\Gamma_M^\perp \eta -{1\over 2}\sum_N (\bar{q}_{NM}-1)
{1\over V_N}\partial_a E_N\Gamma_{\hat{a}} \Gamma_N \eta \qquad\qquad  \]
\[ \qquad \qquad \qquad \qquad +
{1\over 2}
\sum_{M'\neq M} (\bar{q}_{MM'}-3){1\over V_{M'}}\partial_a\tilde{E}_{M'}
\Gamma_{\hat{a}} \Gamma_{M'}^\perp \eta=0.
\]
As before, $\bar{q}_{AB}$ is the dimension of the intersection of the two
branes in its label. The equations above are easily solved as follows.
We have to take $\bar{q}_{NN'}=0$, $\bar{q}_{NM}=1$ and $\bar{q}_{MM'}=3$,
i.e. exactly the intersection rules \rref{mintersections}. Note however
that we do not recover the general formula \rref{mintrules}. The equations
are then completely solved by:
\bea
V_N=\pm E_N, \qquad & & \qquad (1\mp \Gamma_N)\eta=0, \nonumber \\
V_M=\pm \tilde{E}_M, \qquad & & \qquad (1\pm \Gamma_M^\perp )\eta =0.
\label{solmultiv}
\eea
We will comment shortly on the compatibility of the several projections
on the supersymmetry parameter $\eta$.

The relations above however do not completely solve the conditions
$\delta\psi_t=0$ and $\delta\psi_i=0$, since typically we have
$p+1\geq {\cal N}_{el}+{\cal N}_{mag}$ (taking into account the intersection
rules). Using \rref{solmultiv}, they further impose:
\[
{B\over 2G}\Gamma_{\hat{0}}\left[\partial_a \ln B -{1\over 3}\sum_N \partial_a
\ln E_N  -{1\over 6} \sum_M \partial_a \ln \tilde{E}_M\right] 
\Gamma_{\hat{a}}\eta=0 
\]
\[
{C_i\over 2G}\Gamma_{\hat{\imath}}\left[\partial_a \ln C_i -{1\over 3}
\sum_{N\parallel i}\partial_a\ln E_N +{1\over 6} \sum_{N\perp i}
\partial_a\ln E_N \right. \qquad \qquad \]
\[ \qquad\qquad\qquad\left. 
-{1\over 6} \sum_{M\parallel i} \partial_a \ln \tilde{E}_M
+{1\over 3} \sum_{M\perp i} \partial_a \ln \tilde{E}_M \right]
\Gamma_{\hat{a}}\eta=0
\]
The solution to this equations gives the expression of $B$ and $C_i$ in
terms of the $E_N$ and $\tilde{E}_M$:
\be
B=\prod_N E_N^{1\over 3} \prod_M \tilde{E}_M^{1\over 6}, \qquad
C_i=\prod_{N\parallel i}E_N^{1\over 3}\prod_{N\perp i} E_N^{-{1\over 6}}
\prod_{M\parallel i}\tilde{E}_M^{1\over 6}\prod_{M\perp i}\tilde{E}_M^{-
{1\over 3}}. \label{solmultiti}
\ee

In order to determine the expression for $G$ and for the preserved
spinor $\eta$, we solve the last set of equations $\delta \psi_a=0$, which
gives:
\[ \partial_a \eta-{1\over 6}\sum_N \partial_a \ln E_N \eta -{1\over 12} \sum_M
\partial_a \ln \tilde{E}_M \eta \qquad\qquad\qquad\qquad\qquad \]
\[ \qquad\qquad\qquad +{1\over 2}\left(
\partial_b \ln G+{1\over 6} \sum_N \partial_b \ln E_N +{1\over 3}
\sum_M \partial_b \ln \tilde{E}_M \right) \Gamma_{\hat{a}\hat{b}} \eta =0.
\]
The solution is thus:
\be
G=\prod_N E_N^{-{1\over 6}}\prod_M \tilde{E}_M^{-{1\over 3}}, \qquad \qquad
\eta= \prod_N E_N^{1\over 6}\prod_M \tilde{E}_M^{1\over 12}\eta_{(0)},
\label{solmultia}
\ee
where $\eta_{(0)}$ is the usual constant spinor. It can be
trivially checked that the functions above verify
the extremality ansatz $BC_1\dots C_p G^{d-2}=1$.
The equations of motion and Bianchi identities for the 4-form field strength
are now simple to solve and they yield that $H_N=E_N^{-1}$ and 
$H_M=\tilde{E}_M^{-1}$ are harmonic functions.

In terms of these harmonic functions, we have the same solution
as in \rref{intsolution}--\rref{intsolution2} applied to the 11 dimensional
supergravity. We have thus shown above that this solution preserves
the supersymmetries with parameter constrained by:
\be
\eta=\prod_N H_N^{-{1\over 6}} \prod_M H_M^{-{1\over 12}}\eta_{(0)},
\qquad \Gamma_N \eta_{(0)}=\eta_{(0)}, \quad \Gamma_M^\perp \eta_{(0)}
=-\eta_{(0)}. 
\label{susymulti}
\ee

We might be worried that the projections on the constant spinor $\eta_{(0)}$
eventually project it onto a zero-dimensional subspace. This would be
the case if there were pairs of $\Gamma_A$
matrices ($A=N$ or $M$) which anticommuted.
It would have the consequence that:
\[ \eta_{(0)}=\Gamma_A \eta_{(0)}=\Gamma_A \Gamma_B \eta_{(0)}
=-\Gamma_B \Gamma_A \eta_{(0)}=- \eta_{(0)} \equiv 0. \]
Happily, and owing to the intersection rules \rref{mintersections}, all
such pairs of matrices commute. 
If there are $\cal N$ branes in the configuration, then $\eta_{(0)}$ is
projected by:
\be
{\cal P}_{\cal N}\eta_{(0)}=
{1 +\Gamma_{[1]}\over 2}{1 +\Gamma_{[2]}\over 2}\dots
{1 +\Gamma_{[{\cal N}]}\over 2} \eta_{(0)}=\eta_{(0)}.
\label{projector}
\ee
The dimension of the subspace onto which $\eta_{(0)}$ is projected can
be computed performing the trace of the projector, which is always:
\be tr {\cal P}_{\cal N} \geq {1\over 2^{\cal N}}(32). \label{projbound} \ee
The equality is obtained whenever there are no products of the $\Gamma_A$
matrices appearing in \rref{projector} that give the unity.
If on the other hand we have, for instance, $\Gamma_{[1]}\Gamma_{[2]}\dots
\Gamma_{[{\cal N}-1]}=\Gamma_{[{\cal N}]}$, then it is easy to
show that we have ${\cal P}_{\cal N}={\cal P}_{{\cal N}-1}$.

We end this subsection giving two examples, one which saturates the bound
\rref{projbound} and the other which does not.
The first one is the configuration already displayed in \rref{222blackhole},
consisting of three sets of M2-branes lying respectively in the directions
$\hat{1}\hat{2}$, $\hat{3}\hat{4}$ and $\hat{5}\hat{6}$. This 
configuration can be trivially seen to preserve ${1\over 8}$ of the 
supersymmetries.

The second configuration is the following: take two sets of M5-branes
lying in the directions $\hat{1}\hat{2}\hat{3}\hat{4}\hat{5}$ and
$\hat{1}\hat{2}\hat{3}\hat{6}\hat{7}$, and two sets of M2-branes
along $\hat{4}\hat{6}$ and $\hat{5}\hat{7}$. Note that the intersection
rules are respected. We can write the following table:
\[
\begin{array}{cccccccccccc}
 & 0 & 1& 2& 3& 4& 5&6&7&8&9&10 \\
M5 & \times &\times &\times &\times &\times &\times & -&-&-&-&- \\
M5 & \times &\times &\times &\times & -&-&\times &\times &-&-&- \\
M2 & \times & -&-&-&\times &-&\times &-&-&-&- \\
M2 & \times & -&-&-&-&\times &-&\times &-&-&- 
\end{array}
\]
It is now easy to see that, if we define $\bar{\Gamma}=\Gamma_8 \Gamma_9
\Gamma_{10}$, we have:
\[ (\Gamma_6 \Gamma_7 \bar{\Gamma})(\Gamma_4 \Gamma_5\bar{\Gamma})
(\Gamma_0 \Gamma_4 \Gamma_6)=-\Gamma_0 \Gamma_5 \Gamma_7.
\]
Note that the sign of the fourth projector is however determined
by the signs of the three other ones. This means that if we had only
the first three sets of branes, we could add the fourth one without
breaking additional supersymmetry, but one has to be careful not to
add anti-branes in the last step otherwise the supersymmetry is
completely broken (note that chosing which are the branes and which
the anti-branes is a matter of convention).
The conclusion is thus that the above configuration, which is dual to
the IIA configuration shown in \rref{4440blackhole}, preserves also
${1\over 8}$ supersymmetries.

This, and the dualities, imply that the 5 dimensional extremal black holes
with 3 charges, and the 4 dimensional ones with 4 charges, the same
that have a non-vanishing entropy, all are supersymmetric and preserve
${1\over 8}$ of the 32 supersymmetries. This is crucial in establishing
the relation between the semi-classical entropy and the microscopic
counting of states in the D-brane picture.

\subsection{A non-supersymmetric, extremal solution}
\label{D0D6ssec}

We conclude this section on intersecting branes with an odd case.
The solution that we present in this subsection shows that the 
penalty for violating the intersection rules is the loss of 
supersymmetry. 

Keeping the things simple we only consider a solution carrying two
charges, thus a bound state of two (sets of) branes. We also
still focus on extremal solutions, though this case could be
easily extended to a non-extremal version.

A naive way to look for a solution is to try to formulate the problem
in the most simple way. Most notably, what complicates the solutions
from the point of view of the reduced $(D-p)$-dimensional space-time,
is the presence of non-trivial moduli fields, such as the dilaton and
the metric in the compact space longitudinal to the branes. We can thus
search for solutions with trivial moduli, i.e. for which $\phi$ and
$C_i$ are constant and thus keep the value that is imposed at infinity
throughout all the space.

In order to do this, we can
take advantage of the structure of the r.h.s. of the
equations \rref{einsteintt}--\rref{maxbianchi}, in particular the
following ones:
\bea
R^i_i&=&-{1\over 2} {\delta^i_1\over D-2} 
(V_1 G)^{-2}e^{\varepsilon_1 a_1 \phi}
(\partial E_1)^2 -{1\over 2} {\delta^i_2\over D-2} (V_2G)^{-2}e^{\varepsilon_2
a_2 \phi} (\partial E_2)^2, \nonumber  \\
\Box \phi&=& -{1\over 2} \varepsilon_1 a_1(V_1 G)^{-2}
e^{\varepsilon_1 a_1 \phi} (\partial E_1)^2 -{1\over 2}\varepsilon_2a_2
(V_2G)^{-2}e^{\varepsilon_2a_2 \phi} (\partial E_2)^2 . \nonumber
\eea
The labels 1 and 2 refer to the two branes present in the problem.
Both of the r.h.s. above are sums of terms with coefficients which can
have an arbitrary sign.
If we now implement in the equations that the moduli are constant, i.e. we take
$C_i=1$ and $\phi=0$, the l.h.s. become trivially vanishing. 
The equations reduce to:
\bea
{1\over 2(D-2) (BG)^2}
[ -\delta^i_1(\partial E_1)^2-\delta^i_2(\partial E_2)^2]&=&0,
\label{2chargesii} \\
{1\over 2 (BG)^2}[ -\varepsilon_1 a_1(\partial E_1)^2-
\varepsilon_2a_2(\partial E_2)^2]&=&0. \label{2chargesbox}
\eea

At a first glance, we see that for the equations \rref{2chargesii} to be
satisfied, $\delta^i_1$ and $\delta^i_2$ must be of opposite sign
for each direction $y_i$. This means, if we recall the definition
\rref{deltaia}, that each direction has to be longitudinal to one
brane and transverse to the other. In other words, the two branes
intersect in a completely orthogonal way, $q_1\cap q_2=0$. 
If for the moment we assume that both branes are extended objects ($q>0$),
the equations \rref{2chargesii} imply two relations:
\bea
(D-q_1-3)(\partial E_1)^2 -(q_2 +1) (\partial E_2)^2&=&0, \label{zorrro1}\\
(q_1+1) (\partial E_1)^2 -(D-q_2-3) (\partial E_2)^2&=&0. \label{zorrro2}
\eea
The sum of these two equations gives the following highly constraining 
equation:
\[
(D-2)[(\partial E_1)^2-(\partial E_2)^2]=0. 
\]
We thus take both functions to be equal:
\be E_1=E_2\equiv E. \label{equalfns} \ee
This physically means that both branes will have the same total 
amount of charge. Note that since the functions coincide, in particular
the centers will coincide also.

Plugging \rref{equalfns} in \rref{zorrro1} or \rref{zorrro2}, we have the
following constraint on the extension of the branes:
\be
q_1+q_2=D-4.
\label{constrem}
\ee
They have thus the right dimensions to be an electric-magnetic dual pair
of branes. This is confirmed reinspecting the equation \rref{2chargesbox},
which forces to have $\varepsilon_1 a_1=-\varepsilon_2a_2$. Since
conventionally \rref{constrem} implies that $\varepsilon_1$ and
$\varepsilon_2$ will be of opposite sign, we must have $a_1=a_2$, i.e.
the two branes are really an electric-magnetic dual pair.

This conclusion is unchanged if one of the branes is a 0-brane (say $q_1=0$), 
since then
its dual is a $(D-4)$-brane, and the equation \rref{zorrro2}, which
is the only one, is still satisfied.

Note that in the case $q_1=0$, we could have more general cases, not 
necessarily leading to $a_1=a_2$. However none of these cases applies
directly to string/M-theory.

We can now solve completely the problem, writing the remaining equations:
\begin{eqnarray*}
R^t_t&=&-{1\over 2 }(BG)^{-2} (\partial E)^2, \\ 
R^a_b&=&(BG)^{-2}\left\{-\partial_a E\ \partial_b E+{1\over 2} \delta^a_b
(\partial E)^2\right\},\\
\partial_a \left( {G\over B} \partial_a E \right)&=& 0.
\end{eqnarray*}
Note that necessarily, \rref{constrem} implies that we have $d=3$.
Accordingly, the extremality condition \rref{genextremecondition} becomes
$BG=1$. The equations above become:
\bea
\partial_a \partial_a \ln B&=& {1\over 2B^2 } (\partial E)^2, \nonumber \\
2\partial_a \ln B\ \partial_b \ln B& =&{1\over B^2} \partial_a E\ \partial_b E,
\label{2chargeseom} \\
\partial_a \left( {1\over B^2} \partial_a E \right)&=& 0. \nonumber 
\eea
The solution is easily found as for \rref{intsolution}--\rref{intsolution2},
and it yields:
\be
E=\sqrt{2} H^{-1}, \qquad B=H^{-1}, \qquad G=H, \label{2chsol} 
\ee
where $H$ is the harmonic function:
\be
H=1+\sum_k {Q_k \over |x-x_k|}. \label{2chharm}
\ee

As already noted, the fact that we obtain a multi-centered solution should
not be misunderstood. Our construction 
forbids us to separate the $q_1$-branes from the $q_2$-branes in each 
bound state. However there is a vanishing force between two (or more) such
bound states.

The metric thus writes simply:
\be
ds^2=-H^{-2}dt^2 +\sum_{i=1}^{q_1+q_2} dy_i^2 +H^2 dx_a dx_a. \label{2chmetric}
\ee
In the simplest case in which we have only one center, $H=1+{Q\over r}$,
we can perform the change of coordinates $R=r+Q$, moving the horizon
to the value $R=Q$. The metric becomes:
\be
ds^2=-\left(1-{Q\over R}\right)^2dt^2 +\sum_{i=1}^{q_1+q_2} dy_i^2 +
\left(1-{Q\over R}\right)^{-2} dR^2+R^2 d\Omega_2^2. \label{2chmetric2}
\ee
This is exactly the metric of a 4 dimensional extreme Reissner-Nordstr\"om
black hole, multiplied by a flat $p$-dimensional internal space.
The generalization to the non-extreme Reissner-Nordstr\"om black hole
is straightforward, but the extreme version is enough for the present
discussion.

Let us now compute the mass-charge relation of the above solution. To fix
the notations, we set ourselves in 10 dimensional type IIA theory, and
we consider a bound state of D0- and D6-branes. Remember that these
two branes were unable to form a bound state with vanishing binding
energy. 

The RR 2-form has the following two non-trivial components:
\[ F_{ta}=\sqrt{2} \partial_a H^{-1}, \qquad \qquad F_{ab}=-\sqrt{2}
\epsilon_{abc} \partial_c H. 
\] 
Note that with respect to the solution \rref{intsolution2} applied to
the D0- and D6-brane, we have an extra factor of $\sqrt{2}$.

The charges carried by the solution \rref{2chsol} are thus:
\be
{\cal Q}_0={1\over 16 \pi G_{10}} \int_{T^6\times S^2} *F_2= \sqrt{2}
{L^6 \over 4 G_{10}} Q, \qquad q_0={\cal Q}_0, \label{2chq0}
\ee
\be
{\cal Q}_6={1\over 16 \pi G_{10}} \int_{S^2}F_2= \sqrt{2}
{1\over 4 G_{10}} Q, \qquad q_6=L^6 {\cal Q}_6. \label{2chq6}
\ee
As expected, the solution carries the same amount of charges, $q_0=q_6$.
This does not necessarily mean that, if we consider the elementary branes
with quantized charge, we should have an equal number of both kinds.
Rather, the ratio will be fixed by the volume of the $T^6$ in the
relevant units.

The ADM mass is easily obtained computing the formula \rref{admmassformula}
for the metric \rref{2chmetric}:
\[
M={L^6 \over G_{10}}|Q|.
\]
We have now to recall that in section \ref{MBRANESssec} we established
that for all the branes of M-theory, we had that $M_p=|q_p|$.
The expression for the ADM mass can thus be rewritten as:
\be
M=2\sqrt{2} |q_0| =\sqrt{2}(|q_0| +|q_6|) \equiv \sqrt{2}(M_0+M_6).
\label{2chmass}
\ee

This result is interesting. 
The energy of the bound state is bigger than the energy
of the two constituent objects. 
This should be ascribed to the fact that from the perturbative string theory
point of view, D0- and D6-branes exert a repulsive force on each other
\cite{lifshytz}. One has then to supply extra energy to bind them.
We thus do not seem to be in presence of a stable bound state. At best,
it could be a metastable one. Some of the properties
of this bound state were analyzed in \cite{sheinblatt}
(see also \cite{khuriortin}, and \cite{adhering} where the D-brane
description of the bound state is given, along with evidence for its
metastability).

Another feature of this solution is that, as one can easily guess from
\rref{2chmetric2}, it has a non-vanishing entropy. This also violates the
rule for the supersymmetric black holes, which in four dimensions need
4 charges to have a non-vanishing entropy.

The two remarks above are strong hints against the preservation
of supersymmetry by the solution \rref{2chsol}. It is actually very easy
to prove it directly. 

In order to be able to use the equations of the preceding subsection,
we have first to perform some dualities to reformulate the problem in
11 dimensional supergravity. The D0+D6 bound state can be T-dualized to
a D1$\cap$D5=0 intersection in IIB theory, then S-dualized to
a F1$\cap$NS5=0 configuration, and eventually elevated 
(after another T-duality) to a M2$\cap$M5=0
bound state in 11 dimensions. Note that the latter configuration is
also a solution of the kind described above. The chain of dualities
ensures that the two solutions have the same supersymmetry properties.

Let us suppose that the M2-brane is stretched in the $\hat{1}\hat{2}$
directions, while the M5 lies in $\hat{3}\hat{4}\hat{5}\hat{6}\hat{7}$.
Then supersymmetry imposes,  for instance, $\delta \psi_{y_1}=0$. 
Plugging the solution \rref{2chsol} in the second of the expressions
\rref{deltamulti}, we get the following expression:
\begin{eqnarray*}
\delta\psi_{y_1}&=&-{1\over 6BG} \partial_a E\ \Gamma_{\hat{1}}\Gamma_{\hat{a}}
\Gamma_{M2}\eta -{1\over 6BG} \partial_a E\ \Gamma_{\hat{1}}\Gamma_{\hat{a}}
\Gamma_{M5}^\perp \eta \\
&=& -{\sqrt{2}\over 6} \partial_a H^{-1} \Gamma_{\hat{1}}\Gamma_{\hat{a}}
(\Gamma_{M2}+\Gamma_{M5}^\perp)\eta.
\end{eqnarray*}
Here we have defined $\Gamma_{M2}=\Gamma_0\Gamma_1
\Gamma_2$ and $\Gamma_{M5}^\perp=\Gamma_1\Gamma_2
\bar{\Gamma}$, with $\bar{\Gamma}=\Gamma_8\Gamma_9 \Gamma_{10}$ (we have
dropped hats).

The condition for preserved supersymmetry would thus read:
\be
\tilde{\Gamma}\eta =\eta, \qquad \qquad \tilde{\Gamma}=-\Gamma_{M2}
\Gamma_{M5}^\perp.  \label{2chsusy}
\ee
However it is easy to see that $\tilde{\Gamma}^2=-1$. This implies
that the condition \rref{2chsusy} has no non-trivial solutions.
Therefore, the bound state \rref{2chsol} breaks all supersymmetries.

We have thus presented in this subsection an example of a solution
of a different kind of the ones presented in the previous subsections.

\section{Conclusions and related work}
\label{CONCLUsec}

In this last section we recollect the results obtained in this chapter, 
and we (rapidly) review the recent work on classical $p$-brane
solutions other than the ones considered here.

The main objective of this chapter was to study the basic $p$-branes
that one encounters in M-theory, and to treat them in a unified way.
The need to unify the treatment 
is inspired by U-duality \cite{hulltownsend,witten,democracy},
which states that from the effective lower dimensional space-time 
point of view, all
the charges carried by the different branes are on the same footing.
While string theory `breaks' this U-duality symmetry, choosing the 
NSNS string to be the fundamental object of the perturbative theory,
the supergravity low-energy effective theories realize the U-duality
at the classical level. 

Here we have pushed the `unification' further choosing a general action
\rref{genaction} which can be reduced to the bosonic sector of any one
of the supergravities of interest in 10 or 11 dimensions.
The evidence that it was indeed worthwhile considering this unified picture,
is that we could very easily recover all the branes of M-theory in
section \ref{MBRANESssec}, and moreover we were able to present the
very general intersection rule \rref{intersectionrules}. This is an
interesting fact since it relates the properties of the various branes
directly in 11 and 10 dimensions, while U-duality stricly speaking
needs that we carry out compactification.

Supersymmetry is an odd ingredient in the present chapter. Its main
disadvantage is that it is impossible to formulate a supersymmetric
version of the general theory that we have considered. On the other
hand, the supersymmetric properties of the branes are crucial in
relating the classical solutions to the quantum objects.
Here we have proved, in the context of 11 dimensional supergravity,
that for the `elementary' branes, extremality coincides with
supersymmetry. Also, the proof that the extremal intersecting brane
bound states leading 
to a non-vanishing semi-classical entropy are indeed supersymmetric
is very important in view of their relevance in black hole physics.

One could wonder whether the ans\"atze made in order to solve
the (bosonic) equations of motion (as \rref{genextremecondition} and
the reduction to $\cal N$ independent functions) were actually motivated by 
supersymmetry. It should be more precise to say that they were
motivated by extremality. Indeed, the no-force condition between the
branes participating to the configuration is a consequence of
extremality, as the same is true for BPS monopoles \cite{bpsmonopoles}.
As they are, the BPS conditions for monopoles and, here, for branes
do not need supersymmetry, but coincide with the preservation of
some supersymmetries in a (sufficiently) supersymmetric theory.
Supersymmetry thus generically prevents the BPS conditions to be
corrected by quantum effects when one goes beyond the classical solutions.
A good example for this issue is to consider brane configurations
in a theory with less than maximal supersymmetry. There are for instance
configurations in type I string theory which are closely related to
configurations in type IIB theory. Namely, the D1-brane, the D5-brane,
the wave and the KK-monopole are exactly the same in both theories.
It has been shown in \cite{dabhononsusy} that a configuration which
preserves $1/8$ of type II supersymmetry can be transposed to a 
non-supersymmetric, but still extremal configuration in type I, satisfying the
no-force condition. Leaving aside the considerations on the relevance of this
bound state
in the quantum theory, we see that imposing the no-force condition
and extremality does not single out only supersymmetric solutions.

It should also not be underestimated that the derivation of the intersecting
solutions presented in this chapter is a thorough consistency check
of all the dualities acting on, and between, the supergravity theories.
It is straightforward to check that, starting from one definite configuration,
all its dual configurations are also found between the solutions presented
here (with the exception of the solutions involving waves and KK monopoles).
In this line of thoughts, we presented a recipe for building 
five and four dimensional extreme supersymmetric black holes. Some of these
black holes were used in the literature
to perform a microscopic counting of their entropy,
as in \cite{stromingervafa,callanmaldacena} for the 5-dimensional ones. 
Actually, the only (5 dimensional)
black holes in the U-duality `orbit' that were counted were the ones
containing only D-branes and KK momentum.
It is still an open problem to directly count the microscopic states
of the same black hole but in a different M-theoretic formulation.

Some of the intersection rules \rref{intersectionrules} point towards
an M-theory interpretation in terms of open branes ending on other
branes. This idea will be elaborated and made firmer in the next chapter.
It suffices to say here that this interpretation is consistent with
dualities if we postulate that the `open character' of a fundamental
string ending on a D-brane is invariant under dualities. S-duality
directly implies, for instance, that D-strings can end on NS5-branes 
\cite{stromopen}.
Then T-dualities imply that all the D-branes can end on the NS5-brane.
In particular, the fact that the D2-brane can end on the NS5-brane
should imply that the M5-brane is a D-brane for the M2-branes 
\cite{stromopen,towndfromm,beckerbound} (this could also be extrapolated
from the fact that a F1-string ends on a D4-brane).
In the next chapter we will see how these ideas are further supported
by the presence of the Chern-Simons terms in the supergravities, and
by the structure of the world-volume effective actions of the branes.

\vspace{1cm}

We end this chapter mentioning some of the generalizations that have been
done concerning intersecting branes, and some related solutions that
were also studied. 

The most logical generalization of the present work is to include internal
waves (or KK momenta) and KK monopoles to the brane configurations. This
problem was already mentioned before in this chapter. There are several
ways to do this. The first one, which does not really rely on solving
the equations of motion, consists in performing dualities on the
`diagonal' configurations like the ones presented here. One can then
classify all the BPS configurations including all the branes of
subsection \ref{MBRANESssec}, and this has indeed been done in \cite{bdejv2}.
Another way to include the branes with KK origin, is to generalize
the metric ansatz to include at least some off-diagonal elements.
However this seems to be a major complication to the equations discussed here.
A short-cut to this complication is to generalize instead to a model
comprising several scalars \cite{kklp,stelle,arefeva}. 
These extra scalars can now be taken to be
some of the moduli fields arising from the internal metric, while the
KK 2-form field strengths are easily inserted between the other
antisymmetric tensor fields, their coupling to the scalars being a trace
of their origin. The other tensor fields have also to be `multiplied'
according to the KK reduction procedure, as shown in the Appendix \ref{KKapp}.
This leads to a proliferation of `matter' fields, and thus of branes,
which complicates the
intersection rules that one can indeed find. Reading off the particular
intersection rules of a single, 10- or 11-dimensional object becomes
thus rather cumbersome. This would be however the price to pay to really
unify all the branes of M-theory.

In this chapter we presented only extremal configurations of intersecting
branes. The natural further step to take would be to consider also
non-extremal configurations of intersecting branes. There is however
a subtlety: there could be a difference between intersections of
non-extremal branes, and non-extremal intersections of otherwise extremal
branes. 
If we focus on bound states (and thus not on configurations of well 
separated branes), it appears that a non-extremal configuration
would be characterized for instance by $\cal N$ charges
and by its mass. There is only one additional parameter with
respect to the extremal configurations. Physically, we could have hardly
expected to have, say, as many non-extremality parameters as the number of
branes in the bound state. Indeed, non-extremality can be roughly associated
to the branes being in an excited state, and it would have thus been very 
unlikely that the excitations did not mix between the various branes
in the bound state. Non-extremal intersecting brane solutions were
found first in \cite{cvetictseytlin}, and were derived from the
equations of motion following a similar approach as here in \cite{ivanov,ohta}.

It is at first puzzling that the 
solutions found in the papers cited above follow
exactly the same intersection rules as the ones for extremal bound states,
\rref{intersectionrules}. One could have hoped to recover now a much
larger class of intersection, unconstrained by supersymmetry for instance.
However what these `black intersections' describe are compounds of
branes which exert no force on each other, which have been gathered to the
same location in transverse space and have now an excess
of mass (with respect to the sum of their extremal energies). 
For intersections violating \rref{intersectionrules} there is
generically a non-vanishing attractive or repulsive force. This prevents
the participating branes to form simple bound states as the ones
described in \cite{cvetictseytlin,ivanov,ohta}. 
The solution presented in subsection \ref{D0D6ssec}, and its generalizations
\cite{bisy,dharmandal}, show that the bound states of branes that actually
repel each other are indeed very different. Bound states
of branes which attract each other also have different features, and will be
discussed below.
Note that the attractive, repulsive or vanishing nature of the force between
two branes can be quite accurately accounted for in the case where the two 
branes are D-branes \cite{tasi,lifshytz}.

The intersections presented in this chapter are actually orthogonal in the
sense that each brane lies along some definite directions. It is however
possible to consider branes lying at an angle with respect to some of 
the directions. Take for instance a string, or 1-brane, lying in the
$\hat{1}$-$\hat{2}$ plane, at an angle $\theta$ from the $\hat{1}$ direction.
If the two directions are both compact (and of the same size), then the
string will have a non-trivial winding number in both directions.
Taking the mass of the configuration to be proportional to the total length
of the string (which is quantized), 
the two charges will be respectively $q_1\sim M \cos \theta$
and $q_2 \sim M \sin \theta$. We thus see that an `angled' brane should
not be confused with a marginal bound state of two strings, since
the mass will go like $M\sim \sqrt{q_1^2+q_2^2}$ instead of $\sim |q_1|+|q_2|$.

From D-brane considerations, it is possible to show \cite{berkdougleig}
that some configurations of several D-branes intersecting at angles
are still supersymmetric. This requires that the rotations applied to,
say, the second brane with respect to the first one, are not general but
of a restricted type. For example, if we consider two D2-branes, and
we start with a configuration in which they are parallel to each other, the
most general rotation belongs to $SO(4)$. It turns out that the only
rotations consistent with supersymmetry are the ones belonging to a particular
$SU(2)$ subgroup. For instance, one can go continuously from the
D2$\parallel$D2 configuration to the D2$\cap$D2=0 one, but the D2$\cap$D2=1
is never allowed.

Supergravity solutions corresponding to D-branes at angles were found
in \cite{breckmimy,hambli,balalarsleig}. The resulting solutions
contain as expected off-diagonal elements in the internal metric, and
the derivation from the equations of motion as in \cite{hambli} is
accordingly rather intricated.

We still have to discuss the bound states of the configurations of branes
which exert an attractive force between each other. These are truly 
bound states, in the sense that there is a non-vanishing binding energy
(i.e. the mass of the bound state is lower than the sum of the masses
of the constituents). They are also called non-marginal bound states,
in opposition with the marginal ones treated in this chapter.
The archetype of these bound states is the dyon of a Yang-Mills theory,
which has a mass going like $M\sim \sqrt{q_{el}^2+q_{mag}^2}$. If this
dyon is embedded in a supersymmetric theory (SYM), it turns out
that it preserves $1/2$ of the supersymmetries. This is a higher 
fraction than for the marginal bound states, which preserve $1/4$
in a generic configuration with two objects.

Such dyonic branes were presented in \cite{izquierdo}, where it is
further shown that the configuration M2$\cap$M5=2, or in another notation 
M2$\subset$M5, is half-supersymmetric, and its mass is given by
$M\sim \sqrt{q_2^2+q_5^2}$. A particular feature of this solution
is that, in order to prove that it solves the equations of motion,
the Chern-Simons term in \rref{11action} plays a crucial r\^ole. 
Moreover, if the M5 lies
in the directions from $y_1$ to $y_5$, and the M2 in the $y_1$ and $y_2$
directions, then the 4-form field strength has also 
a non-trivial $F_{y_3y_4y_5 a}$
component. Intersections of branes where the constituents
are these kind of M2$\subset$M5 dyonic branes were considered in
\cite{costa1} for the extremal case, and in \cite{costa2} in the 
non-extremal generalization. The intersection rules have to be satisfied
for both the components of each dyonic brane.

Compactifying and applying dualities to the M2$\subset$M5 bound state one
can classify all such non-marginal bound states. These comprise the
configurations D$p\subset$D$(p+2)$, F1$\subset$D$p$ ($p\geq 1$),
D$p$$\subset$NS5 ($p\leq 5$) and the KK momentum orthogonal to any
one of the branes (except the KK monopole). These configurations
all preserve the same amount of supersymmetry as, for instance, the
bigger brane alone. From the point of view of the world-volume of the latter,
the smaller brane can be interpreted as a flux of a field strength of the
world-volume effective action (see \cite{gura} for the D-branes). 
Note that other configurations with intersections which seem to violate
\rref{intersectionrules} are actually non-marginal bound states of this
kind. Take the D2$\subset$D4 bound state and T-dualize in one direction of
the compact space
transverse to the D2 and longitudinal to the D4. We obtain a configuration
such that D3$\cap$D3=2, with an off-diagonal element in the internal metric
directly related by dualities to the $F_{y_3y_4y_5 a}$ component in
the M2$\subset$M5 configuration. In fact, this clearly represents a D3-brane
wrapped at an angle on the $T^4$ torus. The configurations of `angled'
branes and of dyonic branes thus belong to the same duality orbit, as the
formulas for their mass might have suggested.

Other half-supersymmetric bound states of this class are the $SL(2,Z)$
multiplets of 1- and 5-branes in type IIB theory \cite{schwarz,wittenbound}, or
more precisely the configurations F1$\parallel$D1 and NS5$\parallel$D5,
also called $(p,q)$ 1- and 5-branes, where $p$ is the NSNS charge
and $q$ the RR charge of the compound. The classical solutions
corresponding to this latter case were actually found more simply performing an
$SL(2,Z)$ transformation on the F1 or NS5 solutions.

The main unsatisfactory point in the intersecting brane solutions
presented in this chapter is that they are not localized in the relative
transverse space. This is disappointing if one wanted, for instance,
to have a supergravity description of the brane configurations used to
derive field theory results, as in \cite{hananywitten,elitzur}.

In \cite{gaunkasttras} (inspired by \cite{khuri}) a solution is presented
which corresponds to a M5$\cap$M5=1 configuration, which follows
the harmonic superposition rule, provided however that the harmonic
functions depend on the respective relative transverse space (i.e. they are 
functions of two different spaces). 
The problem now is that the harmonic functions 
do not depend on the overall transverse space (which is 1-dimensional in
the case above), the configuration thus not being localized there.
A method actually inspired by the one presented here to derive the
intersecting brane solutions, has been applied in \cite{edelstein} to
the intersections of this second kind. Imposing that the functions
depend on the relative transverse space(s) (with factorized dependence)
and not on the overall one, the authors of \cite{edelstein} arrive
at a formula for the intersections very similar to \rref{intersectionrules},
with $\bar{q}+3$ on the l.h.s. This rule correctly reproduces the
M5$\cap$M5=1 configuration, and moreover also all the configurations
of two D-branes with 8 Neumann-Dirichlet directions, which preserve
$1/4$ supersymmetries but were excluded from the intersecting
solutions derived in this chapter (only the configurations with 4 ND
directions were found as solutions). One such configuration is e.g.
D0$\subset$D8.

Very recently, some approximate solutions were found in \cite{lasttseytlin}
which represented marginal bound states like,
for instance, D2$\subset$D6 and NS5$\subset$D6,
not only localized in transverse space, but where the smaller brane
is additionally localized within the D6-brane.
An analogous configuration representing F1$\subset$NS5 was further
shown to satisfy the `modified' Laplace equation which was introduced
in \cite{tseytlincompo} to implicitly characterize the solutions with
localized intersections. Unfortunately, the solutions discussed 
in \cite{lasttseytlin} are valid only near
the location of the bound state, and for branes within branes.

The main goal when considering intersecting brane
solutions is certainly to find a general solution in which the
intersection is localized in both relative transverse directions, and in
the overall transverse space. 
Finding solutions representing branes with fully localized
intersections is most interesting with respect to several issues
as the world-volume dynamics of the branes involved, and also 
to analyze the possible corrections and the decoupling limit when 
field theory phenomena are derived from brane configurations as in e.g.
\cite{hananywitten,elitzur}.

\chapter{The dynamics of open branes}
\label{OPENchap}

In this chapter we will take a step beyond the classical description
of $p$-branes. We aim at considering the dynamics of the quantum objects
which are described at low energies by the classical $p$-brane solutions
discussed in the previous chapter. It is now crucial that we specialize
to the branes of M-theory, since it is the structure of the latter theory
that we eventually want to uncover, and also because its particular
features will turn out to be necessary ingredients in the discussion
presented in this chapter.

The particular aspect of brane dynamics that we will focus on is the
possibility that the branes of M-theory have to open, provided their
boundary is restricted to lie along the world-volume of another brane.
The original example, and by far the most thoroughly understood, is
the type II fundamental string that can end on the D-branes, which
are the (non-perturbative) objects carrying Ramond-Ramond (RR)
charge. The D-branes were introduced in Section \ref{DBRANEsec}.
We will show that the D-branes themselves can in some cases be open and
have their boundaries attached to some other brane. It thus follows
that there are no two distinct classes of branes, one with branes that
can open and the other with branes which collect the boundaries of the
open branes. Rather, most of the branes display both aspects.

Here we will adopt a different strategy than the one which led to the
definition of D-branes, and which required a microscopic description
of the open brane (the fundamental string in that case). We will
start from the classical solutions which could lead to an interpretation
in terms of branes ending on other branes, and check that when the
equations of motions are completed with the correct source and Chern-Simons
like terms, then the opening of the branes is always consistent with charge
conservation. In the course of this derivation, we will see that many
ingredients like supersymmetry, gauge invariance and world-volume effective
actions are necessary for the subtle consistency of the whole picture.

The chapter is organized as follows. In Section \ref{LIGHTsec} we suggest
the transition between configurations with intersecting branes and
configurations with open branes. In Section \ref{CHCONSsec} we
provide evidence that charge conservation allows for such an interpretation.
The proof is a combination of arguments relying on the Chern-Simons terms
of supergravity origin and on restoration of gauge invariance. Both
arguments lead to the determination of the world-volume fields for the
`host' brane on which the open brane ends, as reviewed in Section 
\ref{OPENREVsec} for type II and M theories. Section \ref{TWOMOREsec} 
contains the discussion of two limiting cases. The last section
contains brief speculations.

This chapter is essentially based on \cite{opening} (and also on part of
\cite{proceeding}).

\section{A different light on some intersections}
\label{LIGHTsec}

General intersection rules between extremal $p$-branes can be derived
for type II string theories and for M-theory, the latter being
the (hypothetical) 11 dimensional theory the low energy effective action 
of which is 11 dimensional supergravity. The complete set of rules was
presented in \cite{papatown,tseytlinharm,gaunkasttras,tseytlinnoforce,rules},
and we gave a thorough account in the previous chapter on how these
rules can be altogether derived from the low-energy supergravity equations
of motion. The intersections we are concerned with are orthogonal
(in the sense that each brane of the configuration is either parallel or
perpendicular to each direction) and the bound states they describe have
vanishing binding energy, i.e. they are marginal bound states. This latter
property makes them supersymmetric, and thus leaves us the hope that
some of the features of the quantum bound states are related to the classical
supergravity solutions discussed in the previous chapter.

Let us recall the condition that guarantees that a multiple brane
configuration has zero binding energy (see \cite{rules} and Section
\ref{INTERsec}). 
It is sufficient that each intersecting pair satisfies the following
rule: the intersection of a $p_a$-brane with a $p_b$-brane must
have dimension
\begin{equation}
\bar{q}=\frac{(p_a+1)(p_b+1)}{D-2}-1-\frac{1}{2}\varepsilon_a a_a
\varepsilon_b a_b , \label{intformula}
\end{equation}
where $D$ is the space-time dimension, $a_a$ and $a_b$ are the couplings of
the field strengths to the dilaton in the Einstein frame and
$\varepsilon_a$ and $\varepsilon_b$ are $+1$ or $-1$ if the corresponding
brane is electrically or magnetically charged.  In eleven dimensional
supergravity, the $a$'s are equal to zero while in ten dimensions $\varepsilon
a=\frac{1}{2}(3-p)$ for Ramond-Ramond (RR) fields and $a=-1$ for the
Neveu-Schwarz-Neveu-Schwarz (NSNS) three form field strength. Note that
we introduced a slight change of notations with respect to 
\rref{intersectionrules}.

The cases in which we will be primarily concerned here are the ones in
which $\bar{q}$ has the same dimension as the boundary of one of the
two constituent branes, i.e. $p_a-1$ or $p_b-1$. We can single out such
cases from the complete list discussed in Section \ref{INTERsec}.
In M-theory, we have only one case, namely:
\be
M2\cap M5=1.
\label{mopen}
\ee
The prefix to M2 and M5 stands for their 11 dimensional (M-theoretic)
origin.
In ten dimensional type II theories, we have three different series of
cases:
\bea
F1\cap Dp&=&0, \qquad \qquad \qquad p=0,\dots, 8, \label{fundopen} \\
Dp\cap D(p+2)&=& p-1, \qquad \qquad p=1,\dots, 6, \label{ddopen} \\
Dp\cap NS5&=& p-1, \qquad \qquad p=1, \dots, 6. \label{dnsopen}
\eea
Above, F1 stands for the fundamental string, D$p$ stands for a $p$-brane
carrying RR charge and NS5 stands for the solitonic NSNS 5-brane.
Of all the cases above, the $p=0$ case of \rref{fundopen} and the
$p=6$ case of \rref{dnsopen} are particular in that the `host'
brane (the brane on which the open brane should end) has the dimension of the
boundary itself. We will have to treat these two cases separately.
Also, the cases in which a D8-brane participates to the configuration
should be treated in the framework of massive type IIA supergravity
\cite{romans}.

All the cases above, \rref{mopen}--\rref{dnsopen}, will turn out to
effectively correspond to a well-defined possibility for the first
brane to be open and to end on the second one. It is certainly surprising
that such a good guess is essentially based on the knowledge of the classical
solutions, which are actually `smeared' over all the relative transverse
directions, and thus represent at best delocalized intersections.
In the following, we will represent a configuration in which a 
$p_a$-brane (the open brane) ends on a $p_b$-brane (the host brane) 
by the symbol $p_a \mapsto p_b$.
Accordingly, the intersections \rref{mopen}--\rref{dnsopen}
correspond to the following open brane configurations: M2$\mapsto$M5,
F1$\mapsto$D$p$, D$p$$\mapsto$D$(p+2)$ and D$p$$\mapsto$NS5.
Note that the $p=1$ case of \rref{fundopen} and the $p=5$ case of
\rref{dnsopen} both seem to give rise to a second possibility, respectively
D1$\mapsto$F1 and NS5$\mapsto$D5.

That the fundamental type II strings can be open, with their ends tied
to the world-volume of a D-brane, is the first appearance of `open branes'
of the kind we are considering here. Indeed, the fundamental type II strings
are generically closed, and they carry a charge (the electric charge
of the NSNS 3-form field strength) that would not be conserved if they
were to open freely at any point of space-time. Charge conservation 
however allows the strings to open provided their ends lie along the
world-volume of a D-brane \cite{polcdbranes}. 
The presence of the RR-charged D-branes is required by the conjectured 
U-duality non-perturbative symmetry of the theory.
The first case which was considered, and the only one to have a microscopic
description, is thus F1$\mapsto$D$p$.

The repeated use of dualities allows then to predict all the other 
configurations with open branes, along the lines of 
\cite{stromopen,towndfromm}. The configurations with D$p$$\mapsto$D$(p+2)$ and 
D$p$$\mapsto$NS5 are obtained by S- and T-dualities, while the case
M2$\mapsto$M5 is obtained considering, for instance, the strong
coupling, 11 dimensional description of the type IIA configuration
F1$\mapsto$D4.

From the classical point of view, the main obstacle towards the opening
of branes is charge conservation. Generically and in $D$-dimensional
flat space-time, the charge of a $p$-brane is measured performing an
integral of the relevant field strength on a $(D-p-2)$-dimensional
sphere $S^{D-p-2}$ surrounding the brane in its transverse space:
\[
Q_p \sim \int_{S^{D-p-2}} *F_{p+2}.
\]
If the brane is open, then we can slide the $S^{D-p-2}$
off the loose end and shrink it to zero size. This implies that the
charge must vanish. This conclusion is avoided if in the process
above, the $S^{D-p-2}$ 
necessarily goes through a region in which the equation
\be
d*F_{p+2}=0
\label{chacons}
\ee
no longer holds. 
Only when the r.h.s. of this equation is strictly vanishing we are allowed
to deform
the $S^{D-p-2}$ while keeping the charge unchanged (and thus conserved).
Thus, any higher dimensional source in the r.h.s. of \rref{chacons}, 
other than the source of the $p$-brane itself, is a potential
topological obstruction to sliding off the $S^{D-p-2}$. 
This is the core of the description 
of how extended objects carrying a conserved
charge can be open, provided they end on some other higher dimensional 
object.

The matter of concretely providing the source term in \rref{chacons}
which ensures charge conservation for the open branes 
can be addressed in different ways.
The approach of Strominger \cite{stromopen} was based on the knowledge
of the world-volume effective actions, and more precisely of the
couplings between world-volume fields and (pull-backs to the world-volume of)
space-time fields. The boundaries of the open branes are then identified
with sources for the world-volume fields.
An alternative approach was given by Townsend \cite{townsurgery}
who showed that the
presence of Chern-Simons type terms in supergravity allows for charge
conservation for well defined pairings of open and `host' branes.

In the rest of this chapter, we will show that in every case where
the zero binding energy
condition is satisfied and where the intersection has dimension $p_a-1$,
it is
possible to open the $p_a$-brane along its intersection with the $p_b$-brane.
This means that charge conservation is always preserved.
The general case will be $p_a\leq p_b$, while the case $p_a=p_b+1$ will
necessitate a special treatment.
The boundary of each open brane configuration
carries a charge living in the closed brane on which it terminates. 
To each such
configuration corresponds a second one. Their respective boundary charges  are
electric-magnetic dual of  each other in the host brane and are  coupled to
field strengths in that brane. The fields pertaining to all branes are
all related by their origin, and can be seen as Goldstone bosons of 
broken supersymmetry.

More precisely, we shall first review and apply the
analysis of charge conservation in terms of the  Chern-Simons type terms
\cite{townsurgery} present in supergravity to the zero binding energy
configurations. We shall then complete this analysis with gauge invariance and
supersymmetry considerations and obtain all world-volume
field strengths and their coupling to boundary charges.

Note that configurations with branes ending on other branes were crucially
used in \cite{hananywitten} and in all the subsequent literature, where
some issues related to charge conservation in the opening mechanism were
also considered.

\section{Charge conservation for the open branes}
\label{CHCONSsec}

In order for a $p_a$-brane to open along its
intersection with a $p_b$-brane, its boundary must behave, by charge
conservation, as an induced charge living in the $p_b$-brane.  
Considering the general case for which $p_a\leq p_b$, this charge
will act as a source for a field strength on the $p_b$-brane
world-volume. As shown below this field strength is related to
(and thus determined
by) the Chern-Simons type term present in the supergravity action.

Since the interplay of the different ingredients is rather subtle,
we present in the first place a specific example, and afterwards we proceed
to the general case.

\subsection{A D2-brane ending on a D4-brane}

We work out in this subsection the example of an open D2-brane ending
on a D4-brane in the framework of type IIA string theory

The equation of motion for the 4-form field strength has to be supplemented
by the Chern-Simons term present in the full IIA supergravity action
\rref{iiaeinstein},
and by the source due to the presence of the D2-brane. The equation thus
reads, neglecting the dilaton and all numerical factors:
\be
d*F_4=F_4 \wedge H_3 +\mu_2 \delta_7. \label{eqcssource}
\ee
The tension of the (elementary) D2-brane is thus $\mu_2$, and the $\delta_7$
localizes the D2-brane in the transverse space.
Since there is also a D4-brane in the configuration, the Bianchi
identity for $F_4$ is also modified by a source term:
\be dF_4=\mu_4 \delta_5. \label{bisource} \ee
On the other hand, due to the absence of NS5-branes, $H_3$ can be
globally defined as $H_3=dB_2$. The equation \rref{eqcssource} can be
rewritten as:
\be
d(*F_4 -  F_4 \wedge B_2)= \mu_2 \delta_7 - \mu_4 \delta_5 \wedge B_2.
\label{eqcssource2} \ee
We can now integrate both sides of this equation over a 7-sphere $S^7$
which intersects the D2-brane only once (this is possible only if the
D2-brane is open). The result is:
\be 0=\mu_2-\mu_4 \int_{S^2} \hat{B}_2, \label{eqcsint} \ee
where the hat denotes the pull-back to the world-volume of the D4-brane
of a space-time field.

The expression \rref{eqcsint} is actually not exact (as we discuss
below), but it is
relevant because it allows us
to see that the Chern-Simons term indicates the presence on the world
volume of the D4-brane of a 2-form field strength, for which the
(string-like) boundaries of the D2-branes act as magnetic charges.
As we will discuss shortly, the gauge invariant combination is:
\be
{\cal F}_2=dV_1-\hat{B}_2.
\label{d2d4f2}
\ee

Considering now the world-volume action of the D2-brane,
we know that there is a minimal coupling to the RR 3-form potential:
\[ I_{D2}=\mu_2 \int_{W_3} \hat{A}_3 +\dots \]
When the D2-brane is open, the gauge transformation $\delta A_3=d\Lambda_2$
becomes anomalous:
\[ \delta I_{D2}=\mu_2 \int_{(\partial W)_2} \hat{\Lambda}_2. \]
The standard way to cancel this anomaly is by constraining the
boundary $(\partial W)_2$ to lie on the D4-brane world-volume
where a 2-form gauge potential $V_2$, transforming as $\delta V_2=
\hat{\Lambda}_2$, couples to it. The boundary of the D2-brane is now
an electric source for the 3-form field strength built out from
this potential. Again, the gauge invariant combination is given by:
\be
{\cal G}_3=dV_2-\hat{A}_3.
\label{d2d4g3}
\ee

The analysis of the Goldstone modes of broken supersymmetry and of
broken translation invariance, and the requirement that these bosonic
and fermionic modes still fit into a representation of the unbroken
supersymmetries, forces us to identify the two field strengths
${\cal F}_2$ and ${\cal G}_3$ by an electric-magnetic duality on the
D4-brane world-volume:
\be {\cal F}_2=\star {\cal G}_3. \label{d4wvduality} \ee

Moreover, we could have analyzed instead the (more familiar) configuration
of a fundamental string ending on the D4-brane. Preservation of the
gauge invariance would have led to the expression \rref{d2d4f2}.
We would have found
that the end point of the string behaves like an electric charge for the 2-form
field strength and like a magnetic charge for the 3-form field
strength. Thus we conclude that the boundaries of the
string and of the membrane are electric-magnetic dual objects on the
world-volume of the D4-brane.

We can now get back to the equation \rref{eqcsint}. The point is
that if we formulate the world-volume effective theory of the D4-brane
in terms of the
2-form field strength ${\cal F}_2$, then preservation of gauge invariance
of the D2-brane action, leading to \rref{d2d4g3},
and the electric-magnetic duality \rref{d4wvduality}
together imply the presence of a supplementary term:
\be
I_{D4}=\dots+\mu_4 \int_{W_5}{\cal F}_2 \wedge \hat{A}_3.
\label{d4wzterm}
\ee
This term is often called a Wess-Zumino term in the D4-brane action,
and its presence can also be traced back to more general considerations
on the gauge invariance of the D4-brane action in general backgrounds
(see for instance \cite{douglaswithin,greenhulltown}).

The presence of this term is crucial since it implies that there is an
additional term in the
r.h.s. of the equation \rref{eqcssource}, of the form $\mu_4 \delta_5 
\wedge {\cal F}_2$. This in turn means that the second term 
in the r.h.s. of the equation \rref{eqcssource2} now should read:
\[ -\mu_4 \delta_5 \wedge (B_2 +{\cal F}_2)\equiv -\mu_4 \delta_5\wedge
dV_1, \]
where we have used \rref{d2d4f2}.

The expression relating the tensions of the various objects can then 
be written as:
\be
\mu_2=\mu_4 \hat{\mu}_1, \qquad \qquad \hat{\mu}_1=\int_{S^2}dV_1.
\label{eqcsintcorr}
\ee
The difference between the equation above and \rref{eqcsint} 
is that now the singular part in ${\cal F}_2$ due to the string-like
source is carried by the purely world-volume field, and not by
the pull-back of the space-time field $B_2$. 

The equations of motion and Bianchi identities for the world-volume
field strength ${\cal F}_2$, completed with electric and magnetic sources,
are then:
\begin{eqnarray*}
d\star {\cal F}_2&=& \hat{\mu}_0 \hat{\delta}_4 -\hat{F}_4,  \\
d {\cal F}_2&=& \hat{\mu}_1 \hat{\delta}_3 -\hat{H}_3.
\end{eqnarray*}

\subsection{Charge conservation using the Chern-Simons terms}

We now go back to the general case, and we focus on the Chern-Simons
terms, as in \cite{townsurgery}, since we have seen in the example
above that they allow to precisely guess how the charge conservation
occurs.

We distinguish three different types of configurations ($p_a
\mapsto p_b$) of a $p_a$-brane ending on a $p_b$-brane: (electric $\mapsto$
magnetic), (electric $\mapsto$ electric) and (magnetic $\mapsto$ magnetic),
where by electric brane we mean a brane which couples minimally to a
field of supergravity (all the field strengths are taken by convention
to be of rank $n\leq D/2$).  Note that the (magnetic $\mapsto$
electric) never appears since it corresponds, in the
supergravities we consider here, to $p_a\geq p_b+2$.

Let us first consider the (electric $\mapsto$ magnetic) case in detail.
The equations of motion for the field strength corresponding to the
$p_a$-brane take the generic form
\begin{equation}
d*F_{p_a+2} = F_{D-p_b-2} \wedge F_{p_b-p_a+1} + \mu^e_a
\delta_{D-p_a-1}. \label{emcase}
\end{equation}
Here and in what follows wedge products are defined up to signs and
numerical factors irrelevant to our discussion.

In the r.h.s. of \rref{emcase} the first term comes from the
variation of the Chern-Simons term
\begin{equation}\label{cs}
\int A_{p_a+1}\wedge F_{D-p_b-2}\wedge F_{p_b-p_a+1},
\end{equation}
where $F_{D-p_b-2}$
is the field strength with magnetic coupling to the $p_b$-brane and
$F_{p_b-p_a+1}$ is a $(p_b-p_a+1)$-form field strength present in the
theory. Let us for the moment just postulate the presence of such a
Chern-Simons term. The second term in \rref{emcase}
is the $p_a$-brane charge density, where
$\delta_{D-p_a-1}$ is the Dirac delta function in the directions transverse
to the $p_a$-brane. It should also take into account the fact that
the brane is open.

The equation \rref{emcase} requires a couple of comments. We want to study
the deformation of intersecting closed brane configurations with zero
binding energy when one of the branes is sliced open along the intersection.
Firstly, notice that we have introduced an explicit source term for the
electric brane since, to study its opening, we want to extend the validity
of the usual closed brane solution on the branes.  This term is required
because the supergravity equations of motion from which the intersecting
brane solutions are derived do not contain any source term and are
therefore valid only outside the sources. Secondly we have to verify that,
in the limit where the open brane closes, the contribution from the
Chern-Simons term vanishes since the closed brane configuration depends
only on $F_{p_a+2}$ and $F_{D-p_b-2}$ and not on the third field
$F_{p_b-p_a+1}$.  The equation for the latter is
\[
d*F_{p_b-p_a+1}=F_{D-p_b-2}\wedge F_{p_a+2}.
\]
It is compatible with a solution where $F_{p_b-p_a+1}=0$ since, when the
two closed branes are orthogonal, and the fields are smeared over the
directions longitudinal to the branes (as in the classical
solutions), the electric $F_{p_a+2}$ and the
magnetic $F_{D-p_b-2}$ have necessarily a common index.

Taking into account that in the configuration considered,
\[
F_{p_b-p_a+1}=dA_{p_b-p_a},\] and \[
dF_{D-p_b-2}=\mu^m_b\delta_{D-p_b-1},
\]
one can rearrange \rref{emcase} in the following way:
\begin{equation}
d(*F_{p_a+2}- F_{D-p_b-2}\wedge A_{p_b-p_a})=\mu^e_a \delta_{D-p_a-1}
- \mu^m_b \delta_{D-p_b-1}\wedge A_{p_b-p_a}.
\label{emcase2}
\end{equation}
The integration of \rref{emcase2} over a $S^{D-p_a-1}$ sphere, which
intersects the $p_a$-brane only at a point and the $p_b$-brane on a
$S^{p_b-p_a}$ sphere which surrounds the intersection, gives
\[
0=\mu^e_a- \mu^m_b \int_{S^{p_b-p_a}}A_{p_b-p_a}.
\]
This equation can be rewritten as:
\begin{equation}
\mu^e_a=\mu^m_b \cdot \hat{\mu}_I,
\qquad \mbox{with} \qquad
\hat{\mu}_I\equiv\int_{S^{p_b-p_a}}A_{p_b-p_a}.
\label{effcharge}
\end{equation}
We see that the pull-back $\hat A^{(p_b+1)}_{p_b-p_a}$ of the potential
$A_{p_b-p_a}$ on the closed $p_b$-brane behaves like a $(p_b-p_a)$-form
field strength, magnetically coupled to the boundary.  We will see in the
next subsection that to preserve gauge invariance one has to add to this field
a $(p_b-p_a)$-form $dW_{p_b-p_a-1}$ defined on the world volume of the
$p_b$-brane. This procedure will actually change the definition of the 
charge $\hat{\mu}_I$, without nevertheless altering the present discussion. 
We thus define the field
strength
\begin{equation}\label{calGdef}
{\cal G}_{p_b-p_a}=dW_{p_b-p_a-1}-\hat A^{(p_b+1)}_{p_b-p_a}.
\end{equation}
The charge $\hat{\mu}_I$ is then simply a magnetic charge for $\cal G$.
On the $p_b$-brane, and outside sources like $\hat{\mu}_I$, the
field strength ${\cal G}_{p_b-p_a}$
satisfies the  Bianchi identity:
\[
d{\cal G}_{p_b-p_a}=-d\hat A^{(p_b+1)}_{p_b-p_a}=-\hat F^{(p_b+1)}_{p_b-p_a+1}.
\]

For the (electric $\mapsto$ electric) case the reasoning goes along the same
lines starting from the equation of motion:
\be
d*F_{p_a+2} = *F_{p_b+2} \wedge F_{p_b-p_a+1} + \mu^e_a \delta_{D-p_a-1}.
\label{eecase}
\ee
This time $d*F_{p_b+2} = \mu^e_b \delta_{D-p_b-1}$.

For the (magnetic $\mapsto$ magnetic) case we have to consider the Bianchi
identities instead of the equations of motion:
\be
dF_{D-p_a-2}=F_{D-p_b-2} \wedge F_{p_b-p_a+1} + \mu^m_a \delta_{D-p_a-1}.
\label{mmcase}
\ee
Here also, we have $dF_{D-p_b-2}=\mu^m_b \delta_{D-p_b-1}$.
In both of these two last cases, the Chern-Simons like term comes
from the additional terms that are present in the definitions of the
gauge invariant field strengths in type II 
supergravities.

We should now emphasize that for all the intersections
\rref{mopen}--\rref{dnsopen}, a Chern-Simons term of the type described
in \rref{emcase}, \rref{eecase} or \rref{mmcase} is always present.
This is really a non-trivial result since, as we had already stated, the
Chern-Simons terms played absolutely no r\^ole in deriving the classical
intersection rules. One
sees that requiring the possibility of open brane configurations
actually would lead to the introduction of these Chern-Simons terms which
from the supergravity point of view are only required by supersymmetry.

We will review in Section \ref{OPENREVsec} all
the relevant cases in 10 and 11 dimensional maximal supergravities.

\subsection{Electric-magnetic duality on the brane}

We saw in the preceding subsection how charge conservation leads to the
existence of a field $\cal G$ whose magnetic source is the charge of the
boundary. However there is an alternative description of this charge. It
arises from the consideration of gauge invariance for the $A_{p_a+1}$
potential and will lead   to the identification of the world-volume
degrees of freedom.

Consider the minimal coupling of the $A_{p_a+1}$ potential to the open
$p_a$-brane.  The boundary of the $p_a$-brane breaks gauge invariance under
$A_{p_a+1} \rightarrow A_{p_a+1}+d\Lambda_{p_a}$ as
\[
\delta \int_{W_{p_a+1}}A_{p_a+1}= \int_{(\partial W)_{p_a}}
\Lambda_{p_a}.
\]
The space-time volume $(\partial W)_{p_a}$ swept by the boundary of the
$p_a$-brane is the world-volume of a $(p_a-1)$-brane living on the host
$p_b$-brane. In order to restore gauge invariance one has to introduce
\cite{wittenbound} a field $V_{p_a}$ living on the $p_b$-brane and
transforming like $V_{p_a}\rightarrow V_{p_a} + \Lambda_{p_a}$.  The
resulting gauge invariant field strength on the $p_b$-brane is
\[
{\cal F}_{p_a+1}=dV_{p_a}-\hat A^{(p_b+1)}_{p_a+1}.
\]
In string theory, the existence of vector field $V_1$ degrees of freedom
stems from the zero mass excitations of open strings. In the context
of supergravity $V_{p_a}$ emerges from broken supersymmetry.
Indeed, the introduction of a $p_b$-brane breaks half of the
space-time supersymmetries. The broken supersymmetries give rise to eight
massless on-shell 
fermionic degrees of freedom of the $p_b$-brane and their bosonic
partners.  $D-p_b-1$ of them are the Goldstone translation modes of the
$p_b$-brane and are world-volume scalars. 
Because of the remaining (unbroken) supersymmetry,
we must have a total of eight on-shell bosonic degrees of freedom.
The field strength ${\cal
F}_{p_a+1}$ exactly accounts for the remaining Goldstone degrees of freedom.

The charge of the boundary of the $p_a$-brane is measured by
the integral:
\begin{equation}
\hat{\mu}_I=\int_{S^{p_b-p_a}}\star{\cal F}_{p_a+1}, \label{opelcharge}
\end{equation}
where $\star$ indicates the Hodge dual on the $p_b$-brane and
the $S^{p_b-p_a}$ encircles the $(p_a-1)$ boundary.
The $\hat{\mu}_I$ in \rref{opelcharge} is naturally 
identified with the one appearing
in \rref{effcharge}.  This identification is in fact required by
supersymmetry, as all massless degrees of freedom have already been
accounted for. Using
\rref{calGdef}, the two expressions for the charge $\hat{\mu}_I$,
\rref{opelcharge} and \rref{effcharge}, imply
\[
{\cal F}_{p_a+1}=\star {\cal G}_{p_b-p_a}. 
\]
It remains to justify why we introduced a potential  $W_{p_b-p_a-1}$ in
the definition of $\cal G$ given in \rref{calGdef}. To this end notice that
the Chern-Simons term \rref{cs} could have been written as
\[
\int A_{p_b-p_a}\wedge F_{D-p_b-2}\wedge F_{p_a+2}.
\]
In this form the Chern-Simons term is suitable to study the opening of a
$(p_b-p_a-1)$-brane on the $p_b$-brane. This study goes
along the same lines as before, with the interchange of
the r\^ole of $\cal F$ and $\cal G$. The introduction of $W$ is now
required to preserve gauge invariance under a transformation of
$A_{p_b-p_a}$, gauge invariance which otherwise would be broken by the
presence of the boundary of the $(p_b-p_a-1)$-brane.
For any `host' $p_b$-brane, 
open brane configurations always occur in dual pairs.
Their boundary charges are respectively electric
and magnetic sources for
$\cal F$, and vice-versa for $\cal G$.

To summarize, and taking into account the world-volume sources produced
by the boundaries of the open $p_a$- and $(p_b-p_a-1)$-brane, 
the  equations for the gauge field living on the host
$p_b$-brane are:
\begin{eqnarray*}
d\star {\cal F}_{p_a+1}&=&\hat{\mu}_{p_a-1}\hat{\delta}_{p_b-p_a+1}-
\hat F^{(p_b+1)}_{p_b-p_a+1},\\
d {\cal F}_{p_a+1}&=&\hat{\mu}_{p_b-p_a-2}\hat{\delta}_{p_a+2}-
\hat F^{(p_b+1)}_{p_a+2}.
\end{eqnarray*}

\markright{\protect \ref{OPENREVsec}. OPEN BRANES AND WORLD-VOLUME FIELDS}
\section{Open branes and world-volume fields for the host brane}
\label{OPENREVsec}
\markright{\protect \ref{OPENREVsec}. OPEN BRANES AND WORLD-VOLUME FIELDS}

In this section we review all the cases leading to open branes,
\rref{mopen}--\rref{dnsopen}, identifying for each one the
Chern-Simons term which participates to the charge conservation,
and the world-volume field on the host brane under which the 
boundary of the open brane is charged.

We take the actions of type II and 11 dimensional supergravities
as in \rref{iiaeinstein}, \rref{iibeinstein} and \rref{11action}
respectively. The precise definitions of the field strengths are also
given in Section \ref{WHYsec}.

\begin{table}[ht]
\[
\begin{array}{ccc} \hline \hline & & \\
F1^e\mapsto D2^e
&\qquad d*H_3=*F^{\prime}_4\wedge F_2
&\qquad d\star {\cal F}_2 =-\hat F^{(3)}_2\\
D0^e\mapsto D2^e
&\qquad d*F_2=*F^{\prime}_4\wedge H_3
&\qquad d{\cal F}_2 =-\hat H^{(3)}_3\\
& & \\
F1^e\mapsto D4^m
&\qquad  d*H_3= F^{\prime}_4\wedge F^{\prime}_4
&\qquad d\star {\cal F}_2 =-\hat F^{\prime (5)}_4\\
D2^e\mapsto D4^m
&\qquad d*F^{\prime}_4= F^{\prime}_4\wedge H_3
&\qquad d{\cal F}_2 =-\hat H^{(5)}_3\\
& & \\
F1^e\mapsto D6^m
&\qquad d*H_3= F_2 \wedge *F^{\prime}_4
&\qquad d\star {\cal F}_2 =-*\hat F^{\prime (7)}_4\\
D4^m\mapsto D6^m
&\qquad dF^{\prime}_4=  F_2 \wedge H_3
&\qquad d{\cal F}_2 =-\hat H^{(7)}_3\\
& & \\
\hline
& & \\
D0^e\mapsto NS5^m
&\qquad d*F_2= H_3 \wedge  *F^{\prime}_4
&\qquad d\star {\cal G}_1 =-*\hat F^{\prime (6)}_4\\
D4^m\mapsto NS5^m
&\qquad dF^{\prime}_4= H_3 \wedge F_2
&\qquad d{\cal G}_1 =-\hat F^{(6)}_2\\
& & \\
D2^e\mapsto NS5^m
&\qquad d*F^{\prime}_4 =H_3 \wedge F^{\prime}_4
&\qquad d\star {\cal G}_3=d{\cal G}_3=-\hat F^{\prime (6)}_4\\ 
& & \\ \hline\hline
\end{array}
\]
\caption{Open branes in type IIA}
\label{tableiiaopen}
\end{table}

\begin{table}[ht]
\[
\begin{array}{ccc} \hline \hline & & \\
F1^e\mapsto D3^*
&\qquad d*H_3=F'_5 \wedge F^{\prime}_3
&\qquad d\star {\cal F}_2 =-\hat F^{\prime (4)}_3\\
D1^e\mapsto D3^*
&\qquad d*F^{\prime}_3=F'_5 \wedge H_3
&\qquad d{\cal F}_2 =-\hat H^{(4)}_3\\
& & \\
F1^e\mapsto D5^m
&\qquad d*H_3= F^{\prime}_3 \wedge F'_5
&\qquad d\star {\cal F}_2 =-\hat F^{\prime (6)}_5\\
D3^*\mapsto D5^m
&\qquad dF'_5= F^{\prime}_3 \wedge H_3
&\qquad d{\cal F}_2 =-\hat H^{(6)}_3\\
& & \\
F1^e\mapsto D7^m
&\qquad d*H_3= F_1 \wedge *F^{\prime}_3
&\qquad d\star {\cal F}_2 =-*\hat F^{\prime (8)}_3\\
D5^m\mapsto D7^m
&\qquad dF^{\prime}_3= F_1 \wedge H_3
&\qquad d{\cal F}_2= -\hat H^{(8)}_3\\
& & \\
\hline
& & \\
D1^e\mapsto NS5^m
&\qquad  d*F^{\prime}_3= H_3 \wedge F'_5
&\qquad d\star {\cal G}_2 =-\hat F^{\prime (6)}_5\\
D3^*\mapsto NS5^m
&\qquad dF'_5= H_3 \wedge F^{\prime}_3
&\qquad d{\cal G}_2 =-\hat F^{\prime (6)}_3\\
& & \\
\hline
& & \\
F1^e\mapsto D1^e
&\qquad d*H_3= *F^{\prime}_3 \wedge F_1
&\qquad d\star {\cal F}_2  =-\hat F^{(2)}_1\\
& & \\
D5^m\mapsto NS5^m
&\qquad dF^{\prime}_3 = H_3 \wedge F_1
&\qquad d\star {\cal G}_6 =-\hat F^{(6)}_1\\
& & \\ \hline\hline
\end{array}
\]
\caption{Open branes in type IIB}
\label{tableiibopen}
\end{table}

\begin{table}[ht]
\[
\begin{array}{ccc} \hline \hline & & \\
M2^e \mapsto M5^m
&\qquad d*F_4=F_4 \wedge F_4
&\qquad d\star {\cal G}_3=d{\cal G}_3=-\hat F^{(6)}_4 \\
& & \\ \hline\hline
\end{array}
\]
\caption{Open branes in $D=11$ supergravity}
\label{tablemopen}
\end{table}

The open brane configurations of type IIA, type IIB and $D=11$ supergravities
are listed respectively in Tables \ref{tableiiaopen}, \ref{tableiibopen}
and \ref{tablemopen}. In each table we list in the first column 
the configurations, in the second column the relevant equations of motion
or Bianchi identities, and in the third column the equations for the
world-volume field strength (we neglect the sources in all the equations).
The configurations are arranged by dual electric-magnetic pairs with 
respect to the host branes.
In the tables, a superscript recalls the electric or magnetic nature 
of each brane (the D3-brane being self-dual), and in the 
type IIB case we have defined the RR 1-form field strength $F_1=d\chi$.

The configurations of type II theories fall into
two classes. The first is associated with the configuration with the
fundamental string ending on D$p$-branes, 
its dual being always a D$(p-2)$-brane
\cite{stromopen}.  The second class consists of D-branes ending on
the solitonic NS5-brane. These are the only branes which can end on
the NS5-brane.

Notice in Table \ref{tableiiaopen} the ``opening'' of a D0-brane on the
D2-brane and on the NS5-brane. It corresponds to an intersection of
dimension $-1$ which does not make obvious sense.  However we have seen
in the previous chapter
that the relation \rref{intformula} stills holds for the D0-brane
world-line intersections in the Euclidean provided the electric fields are
given imaginary values. 

In type IIB theory, we have seen in Section \ref{DUALMsec} that there
exists an $SL(2,R)$ duality symmetry at the level of the supergravity 
which is broken to $SL(2,Z)$ in the full string theory.
The last two cases of Table \ref{tableiibopen} are built up from objects
which are $SL(2,R)$-dual to each other. It follows that
the configurations $D1^e\mapsto F1^e$ and $NS5^m\mapsto D5^m$
exist by S-duality. In fact there is a continuum of dyonic configurations
\cite{schwarz}
obtained by acting with the $SL(2,R)$ group\footnote{
Actually, due to the existence of $(p,q)$ 1- and 5-branes, it is more
convenient in this cases to think about string networks \cite{stringnet}
and webs of $(p,q)$ 5-branes \cite{5braneweb} instead of
strings ending on strings and 5-branes ending on 5-branes respectively.}. These
configurations do not have duals. Indeed, the dual configurations
would involve intersections with a $(-1)$-brane, which do not make   sense
even in the Euclidean.

Let us now review the outcome of the analysis above from the point of view
of the world-volume effective theory of the host brane. The approach based on
the Chern-Simons terms actually allows us to guess the field content
of this
world-volume theory, and the guess is always confirmed when the theory
is known by other means.

All the D$p$-branes have a world-volume effective theory which can
be formulated in terms of a 2-form field
strength ${\cal F}_2$. The electric charges are the end points of the
fundamental strings, while the magnetic charges are the boundaries
of the D$(p-2)$-branes (as in \cite{stromopen}). Note the interesting case
of the D3-brane on the world-volume of which the S-duality between
fundamental strings and D-strings becomes electric-magnetic duality
between their end points. In the case of the D-branes, the presence of 
the 2-form field strength
is supported by the quantum stringy computation
which gives super-Yang-Mills as the low energy effective action of
the D-branes.

On the world-volume of the IIA NS5-brane, we can have the boundaries
of the D2- and D4-brane. The boundary of the D2-brane is self-dual
and thus couples to a self dual 3-form field-strength ${\cal G}^+_3$, while
the boundary of the D4-brane couples magnetically to a
1-form ${\cal G}_1$ deriving from
a scalar potential.
This scalar potential is nothing else than the scalar associated with
the 11th direction which
remains after `vertical' reduction of the M5-brane action.
Indeed, Table \ref{tablemopen} shows that
there is a self-dual 3-form living on the world-volume of the M5-brane.

For the IIB NS5-brane, again all the IIB D-branes can have boundaries
on it. The D1- and the D3-brane boundaries are respectively the electric
and magnetic charge related to a 2-form field strength
${\cal G}_2=d\tilde{V}_1-\hat{A}_2$
which can be considered the S-dual of the ${\cal F}_2$ field on the
D5-brane. The boundary of the D5-brane couples electrically to a
(non-propagating) 6-form field strength ${\cal G}_6$. This 6-form
should be related to the mass term in the IIB NS5-brane action as discussed
in \cite{town_sl2z}, and could play a r\^ole in the definition of
an $SL(2,Z)$ invariant IIB 5-brane action. 

This world-volume field content of the 
NS5-branes was derived in \cite{callharvstro} 
by directly identifying the Goldstone modes of the supergravity solution.

\section{Two more cases of open branes}
\label{TWOMOREsec}

In this section we consider the two limiting cases of \rref{fundopen}
and \rref{dnsopen}, which derive from consistent closed classical
solutions, but seem at first sight inconsistent with charge conservation
if there really is an open brane.

These two cases are F1$\mapsto$D0
and D6$\mapsto$NS5. It is straightforward to note that the arguments
given in Section \ref{CHCONSsec} do not apply to these cases, and this
can be seen in two ways: first of all, there is no longer a topological
obstruction for sliding off the $S^7$ (respectively the $S^2$) sphere,
since the host brane now coincides with the boundary of the open brane;
the charge seems thus no longer conserved. Secondly, gauge invariance
is necessarily broken because in both cases one would need a world-volume
gauge potential with rank equal to the dimension of the world-volume
itself, and this has clearly no dynamics.

The resolution of this problem is to go to the framework of 
massive type IIA supergravity \cite{romans}. In this theory, the 
1-form (RR) potential is no longer dynamical, but it is rather `eaten up'
by the 2-form (NSNS) potential that becomes massive. One can still write
the theory in terms of a 1-form potential, but the gauge invariant
2-form field strength is modified to be:
\be
F_2'=dA_1 +m B_2.
\label{massivef2}
\ee
This supergravity has a cosmological constant which is proportional to
$m^2$, and many more non-trivial terms with respect to the `massless'
IIA supergravity, but we will not need them for the present discussion.

Since the kinetic term for $A_1$ will be proportional to ${F_2'}^2$, the
structure of \rref{massivef2} is enough to ensure charge conservation.
However, before moving to the explicit proof, we should justify the use
of massive type IIA supergravity. This can be seen as follows.
One can go to a formulation in which $m$ is no longer an arbitrary constant, 
but it is dynamically fixed to be a constant. This can be done as
in \cite{seveneight} by the introduction of a (non propagating)
9-form potential with 10-form field strength. The 0-form dual field strength
then satisfies $dF_0=0$, fixing $m$ to be $F_0=m$. One can now see,
in this formulation, that 8 dimensional objects can in fact be sources
for the `cosmological' mass. The equation for the 0-form field can
indeed be modified by the addition of a source term:
\[
dF_0=\mu_8 \delta_1.
\]
The source term will produce a discontinuous jump of magnitude $\mu_8$ 
in the mass
$m$ when the 8-brane, which separates space-time in two disconnected
parts, is crossed. In the context of type IIA theory, this domain
wall is nothing else than a D8-brane. We thus see that massive IIA
supergravity naturally appears as the low-energy effective
action of type IIA string theory in a background containing D8-branes\footnote{
On one side of the D8-brane the cosmological mass $m$ can actually
vanish, giving back the usual massless type IIA theory in that region.}.

It is now a simple matter to prove charge conservation
(see for instance \cite{polcstro,bergablif,hanazaff}), actually
simpler than for the cases treated in Section \ref{CHCONSsec}.

Consider first the fundamental string ending on a D0-brane, F1$\mapsto$D0.
because of the presence of $B_2$ in the expression \rref{massivef2},
the equation of motion of $B_2$ is modified by an additional term
in the r.h.s.:
\be
d*H_3=\mu_1 \delta_8 - m *\! F_2'.
\label{massivef1d0}
\ee
We can take $m=\mu_8$ (we are always putting all numerical and
sign factors to one, and neglecting the dilaton), signaling the
presence of a D8-brane. The F1 can actually end on the D8-brane by
the usual mechanism. The presence of the D0-brane at the end of the
F1 is translated into the fact that:
\be
d*F_2'=\mu_0 \delta_9.
\label{massived0}
\ee
Integrating \rref{massivef1d0} over an $S^8$ sphere which intersects
the string only once, we obtain that \cite{polcstro}:
\[
0=\mu_1 - \mu_8 \int_{S^8}*F_2'.
\]
The integral in the second term gives nothing else than the charge of
the D0-brane, $\mu_0$. We thus have $\mu_1=\mu_8 \mu_0$, and
the charge of the fundamental string is conserved. This can actually
be seen the other way round, stating that in a non-trivial D8-brane
background, a D0-brane can exist only if a string ends on it.
The reverse is of course true when $m=0$.

The case of the D6-brane ending on the NS5-brane is similar 
\cite{hanazaff}, except
that now it is based on the Bianchi identities for the 2-form $F_2'$.
These read:
\be
dF_2'=\mu_6 \delta_3 + m H_3.
\label{massived6ns5}
\ee
Identifying $m$ with $\mu_8$ and integrating over an $S^3$ cutting
the D6 once, we obtain:
\[
0=\mu_6 +\mu_8 \int_{S^3} H_3,
\]
with the integral being the charge of the NS5-brane $\mu_5$.

We can also check that gauge invariance of the open brane is still
preserved in these two cases. Here the compensating gauge variation
will come from the bulk action instead than coming from the world
volume of the host brane. 

Under the gauge variation $\delta B_2 =d\Lambda_1$, the open string
produces an anomalous term along the world-line of the D0-brane:
\[
\delta \left(\mu_1 \int_{W_2}B_2\right) =\mu_1 \int_{W_1} \hat{\Lambda}_1.
\]
The variation of the bulk kinetic term for $F_2'$ exactly compensates it:
\[
\delta \left({1\over 2}\int_{M_{10}} *F_2' \wedge F_2'\right)=\int_{M_{10}} 
*F_2'
\wedge m \delta B_2 =\mu_8 \int_{M_{10}} d*F_2' \wedge \Lambda_1
=\mu_8 \mu_0 \int_{W_1} \hat{\Lambda}_1, 
\]
where we have used \rref{massived0}.
In the case of the D6-brane, although we have to reformulate
the kinetic term in terms of the dual field strength $F_8$, the conclusion
is again similar.

The phenomenon leading to the charge conservation in the cases above
is actually related to brane creation processes introduced
in \cite{hananywitten} and further discussed in \cite{bergablif,branecreation}.
This can be seen easily as follows. Suppose there is a D8-brane
which is such that we have $m=0$ on its left and $m=\mu_8$ on its right.
If we start with a configuration with an isolated D0-brane on the left, 
and we try to pass it through the D8-brane, we observe that a fundamental
string stretching from the D8 to the D0-brane has to be created for
the configuration with the D0-brane on the right to be stable.
This actually happens because the D0-brane induces a charge
on the world-volume of the D8-brane, and the induced charge changes sign
when the D0-brane goes from one side to the other. The relevant
equation on the world-volume of the D8-brane reads:
\[ d\star {\cal F}_2=\hat{\mu}_0 \hat{\delta}_8 -*\hat{F}^{\prime (9)}_2, \]
and we see that the change in sign of the second term in the r.h.s.
can be compensated by the introduction of a point like source,
i.e. the end of a fundamental string.

More generally, the r.h.s. of the world-volume equations
collected in the Tables \ref{tableiiaopen}--\ref{tablemopen}
gives rise to an induced charge on the `host'
brane only if the brane producing the charge has a well-defined
orientation in space-time. The configuration of these two
branes (like the D0 and the D8-brane in the example above) can always
be reconducted by dualities to a configuration of D-branes
with $\nu=8$ Neumann-Dirichlet directions. The common characteristic
of all these configurations (which coincide with the ones discussed
at the classical level in \cite{edelstein}) is that the two branes
cannot avoid each other in space-time if one tries to move one
past the other. A linking number of one brane with respect to the
other can thus be defined \cite{hananywitten}.

\section{Conclusive speculations}
\label{OPENCONCLUsec}

We have shown in this chapter that all the configurations with a closed
$p_a$-brane intersecting a closed $p_b$-brane on a $p_a-1$ dimensional
intersection with vanishing binding energy can be opened.  Charge
conservation and gauge invariance are simultaneously ensured by Goldstone field
strengths appearing on the world-volume of the host $p_b$-brane.  Each such
intersection is related to another one involving the same host $p_b$-brane,
and such that the respective boundaries of the open branes carry
dual charges. Note that the field strengths in the branes
have been generated from supergravity without any
appeal to the apparatus of string theories.

We have seen that all D-branes carry fields ${\cal F}_2$ related
to each other through dimensional reduction along the world-volume of
the brane. The reason behind this
universality is that all D-branes can receive charges from string
boundaries.  But this is not the case for the solitonic five-branes, who
can host only `dual' charges stemming from D-brane boundaries. 

A particular case is that where a fundamental string opens on a
D-brane. One can then consider the reverse process where open strings
collapse to form closed strings with zero binding energy to the D-brane.
Energy conservation permits the closed strings to separate from the
brane. This process can be viewed as the classical limit of the well known
quantum process of closed string emission by D-branes 
\cite{callanmaldacena,emission}.
Our analysis shows, at the classical level, that this string process can be
extended to the emission of closed branes. Any host brane supports
traveling open branes attached to it.  These configurations of open branes
can collapse through merging of the two boundaries
into configurations containing closed branes with zero binding
energy. This leads to the emission of supersymmetric closed branes.

The generality of this phenomenon points towards the existence of an
underlying quantum theory where the emission of closed branes and that of
closed strings occur as similar phenomena. This should in fact be
M-theory.
However, to describe in a more detailed way
the emission of higher branes, a quantum theory of the latter is 
nevertheless still lacking 
(see however \cite{openmatrix} for a proposal on
open membranes in Matrix theory).

\chapter{Little theories on the world-volume of branes}
\label{LITTLEchap}
\markboth{CHAPTER \protect \ref{LITTLEchap}. LITTLE THEORIES}{}

The branes of M-theory are interesting objects not only for the
r\^ole they play in 11 or 10 dimensional physics, but also 
because they are objects endowed with non-trivial dynamics on their
world-volume. As we have seen in the preceding chapter, the branes
have well-defined interactions with other branes, and these interactions
result in the definition of world-volume charged objects and of
world-volume effective theories. These world-volume theories, which
can be ordinary field theories or, as we will see in this chapter,
more complicated theories of extended objects, are especially
interesting if they can be studied on their own. For this to be 
possible one generically tries to define a limit in M or in type II
theories in which (most of) the bulk physics decouples, i.e. there
are no interactions between the world-volume of the brane and the
space-time in which it is embedded.
Note that an important feature common to all the world-volume theories is that 
they do not contain gravity.

In this chapter we focus on theories in 6 and 7 dimensions, and which are
moreover supersymmetric (with 16 or 8 supercharges). These are in
some sense the highest dimensional non-trivial theories, and all the
lower dimensional ones can in principle be recovered as particular limits.

The chapter is a slightly extended version of the paper \cite{little}.

\section{Introduction}

With the so-called Second Superstring Revolution, branes came to play an
important r\^ole, as reviewed in Chapter \ref{MINTROchap}. 
Accordingly, the attention on their world-volume dynamics grew up.
It soon became apparent that, in addition to the conventional 
`embedding' Nambu-Goto-like action, most of them
have a far richer dynamics, differing
in this respect from the fundamental type II superstrings and the
11 dimensional supermembrane. Namely, using a Goldstone mode analysis
(see also Chapter \ref{OPENchap}) it appears that the solitonic 5-branes
of type II theories, or NS5-branes, have world-volume low-energy
effective actions given by two different 6 dimensional supersymmetric field
theories, one chiral and containing a self-dual tensor field\footnote{
Strictly speaking, the theory should be defined by its equations of motion.
Because of the self-duality condition, it is difficult
to write a covariant action for this tensor field. See the recent
progress on this issue in \cite{refondualaction}. In the following,
the use of the word `action' should be considered merely as a shorthand.}, the
other non-chiral and with a vector field, in addition to the usual
embedding scalars \cite{callharvstro}. We will discuss extensively
these theories in this chapter.

In the following years, after the identification of the $p$-branes
carrying Ramond-Ramond charge with the D-branes \cite{polcdbranes},
it became clear that the world-volume low energy theory of these objects was
simply given by the dimensional reduction to $p+1$ dimensions 
of 10 dimensional super Yang-Mills (SYM) theory \cite{wittenbound}. 
The possibility
to use in this case string perturbation theory techniques allowed
actually to identify the effective theory of $N$ coinciding branes,
giving a SYM based on the group $U(N)$, while the Goldstone analysis
could at most give the $U(1)$ `center of mass' part.

By dualities, the effective 
theory of $N$ coinciding M5-branes (and NS5-branes of
type IIA), which are individually theories of self-dual tensors with
no non-abelian generalization, was postulated by Strominger \cite{stromopen}
to involve tensionless strings, arising actually from open M2-branes
stretched between the M5-branes. The theory so obtained was closely
related to the theory discussed in \cite{wten}. It has now to be stressed
that in order to consider this and the other world-volume theories on their own,
one had generically to decouple the bulk effects by taking some limits
in the 10 or 11 dimensional theory. However taking too much of a strong
limit can result in the omission of some higher energy effects
which are nevertheless still confined to the world-volume of the brane.

A pioneering work on non-critical string theories living in 6 dimensions
appeared in \cite{dijk}. These theories will be called `little string theories'
because they have presumably much less degrees of freedom than the
10 dimensional, critical ones. In \cite{sei2}, Seiberg defined two sorts
of little string theories as the theories which are left
on the world-volume of the NS5-branes of type IIA and IIB theories
when the string coupling is taken to vanish but the string scale
(the string length $l_s$ for instance) is kept fixed. In this limit
it was argued that the bulk physics decouples, but nevertheless
at energies of order $m_s$ there is non-trivial `stringy' dynamics
in 6 dimensions, coming from the
fundamental strings `trapped' to the world-volume of the branes. 
The two little string theories
were shown to be related by an analog of T-duality. They differ by their
low-energy field theory limit which is defined by the one of the respective
NS5-brane.

In the paper of Losev, Moore and Shatashvili \cite{mms}, it was pointed out
that adding transverse compact directions to the 5-branes, one could
define in some limits
little string theories with an additional dimensionless parameter, 
taken to be the coupling of the little string theory. Moreover, these
theories could now be mapped by dualities to a 7 dimensional theory,
with finite-energy membrane excitations, called $m$-theory. This
7 dimensional little theory was also discussed, in a more Matrix 
theory-oriented framework, in \cite{brun,hana}.

In this chapter, we study in more details the little string theories
and $m$-theory. We first revisit the theories in 6 and 7 dimensions with
16 supercharges, which is the amount of supersymmetry preserved by a
BPS brane in a maximally supersymmetric theory like M-theory or type II string
theories. These theories lead to the definition of $iia$, $iib$ little
string theories and $m$-theory. The names beared by these theories
will be justified by analogies with the 10 and 11 dimensional theories
with capital letters.
The different ways to obtain these little theories are analyzed. We start
from 5 and 6
dimensional extended objects defined in M or type II
theories and we take limits in which bulk modes decouple.
This leads nevertheless to a non-trivial theory without gravity
defined on the world-volume of the extended objects. We show the web of
dualities
between these little theories which exactly reproduces the scheme of
the ``big"
theories in 10 and 11 dimensions. 
The spectrum of the BPS extended
objects of these little theories is investigated and it is shown that it
agrees
with the U-duality group of M-theory compactified on $T^6$.
This will turn out to be of key importance in the application to the Matrix
description of M-theory compactifications on higher dimensional tori.
Also, for this application to be possible, the little theories as introduced
above must be defined generically taking $N$ parallel `host' branes,
although this feature will not enter into the precise definition
of the little theories.

We will then turn to theories in 6 dimensions with 8 supercharges. 
These theories have
(1,0) supersymmetry (i.e. the minimum in 6 dimensions), 
do not contain gravity and may have an additional integral parameter
leading possibly to a gauge
symmetry. Our strategy is to obtain them from the theories with 16
supercharges.
We mimic the 10 dimensional procedure in which type I theory
is obtained from IIB theory introducing an $\Omega 9$ orientifold and 16
D9-branes
\cite{orientifold,polcdbranes,tasi} (see also \cite{hull}). 
The two heterotic string theories are then found by chains of dualities. 
This procedure is reviewed in Chapter \ref{MINTROchap}.
Applying the same procedure to the 6 dimensional
theories, we find one theory with open strings and two with closed strings,
which
we call respectively type $i$, $h_b$ and $h_a$ theories. These are in fact
classes
of theories. Unlike the 10 dimensional case, the group structure is  not
totally constrained, but also more subtle to define, especially for
the $h_a$ ``little heterotic" theories.
As a consistency check of the picture, the $h_a$ theory can also be related to
a particular compactification of $m$-theory.

The chapter is organized as follows. In Sections \ref{16THsec} and
\ref{8THsec} we study respectively the 
theories with 16 and with 8 supercharges. We emphasize the web of dualities
which relates all these theories, and its similarity with the dualities
constituting M-theory. We also comment on the low-energy
effective theories and on the decoupling of the little theories from the
bulk. In Section \ref{APPLICsec} we apply the little theories to the
Matrix theory description of M-theory on $T^6$. We conclude in Section
\ref{DISCUsec} with
a discussion, mainly on the attempt to formulate the little theories
using another Matrix model.

\section{Theories with 16 supercharges}
\label{16THsec}

Supersymmetric theories with 16 supercharges naturally appear in type II 
string
theories and M-theory as the effective theory on the world-volume of BPS 
branes. 
In order to have well-defined theories on world-volumes one has to take a 
limit
in which the bulk modes decouple. This is achieved by sending the Planck mass, 
defined with respect to the non-compact space, to infinity. 

We will consider here theories defined on the world-volume of 5 branes
and 6 branes, and such that at least three of the transverse directions are
non-compact (in order to keep the space asymptotically flat and thus the
number $N$ of parallel branes arbitrary). 
This allows for extra transverse compact directions, which will
actually play a key r\^ole in defining the parameters of the little theories.
Note that we could also consider lower dimensional branes, but they will
in general have more transverse compact directions, eventually leading
to a complicated theory which is nothing else than a toroidal compactification
of one of the theories discussed below. The SYM theories are recovered
as low energy limits of these more complicated theories.

In M-theory, we have the following two objects:
\begin{itemize}
\item M5-brane, with up to 2 transverse compact directions parametrized by 
$R_1$ and $R_2$.
\item KK6-brane, which has naturally a transverse compact direction, the
so-called NUT direction (see \cite{senkk} for a recent review on KK monopoles,
and also Subsection \ref{MBRANESssec}).
\end{itemize}
In type IIA theory we have the following three objects:
\begin{itemize}
\item NS5(A)-brane, with 1 transverse compact direction parametrized by 
its radius $R_A$.
\item KK5(A)-brane, with its transverse NUT compact direction.
\item D6-brane, with no compact transverse directions.
\end{itemize}
The objects we have in type IIB theory are:
\begin{itemize}
\item NS5(B)-brane, with 1 transverse compact direction.
\item KK5(B)-brane, with its NUT compact direction.
\item D5-brane, with 1 compact transverse direction.
\end{itemize}
Note that we will sometimes call the above branes the `host' branes, since
they will host the dynamics of the little theories.

All these branes are related by the usual dualities relating type II and 
M-theory.
We will however distinguish between dualities which leave the world-volume
of the branes unaffected, as transverse T-dualities for NS branes 
(NS5 and KK5), 
IIB S-duality and transverse
compactifications, and dualities which on the other hand act on the 
world-volume,
as T-dualities and compactifications along a world-volume direction of 
NS branes or KK monopoles and T-dualities for D-branes (see Appendix
\ref{ZOOapp}).

Considering the dualities leaving the world-volume unaffected leads to three
different families of branes each one defining one theory:
\begin{itemize}
\item{$iia$:} KK5(A) $\leftrightarrow$ NS5(B) $\leftrightarrow$ D5
\item{$iib$:} KK5(B) $\leftrightarrow$ NS5(A) $\leftrightarrow$ M5
\item{$m$:} KK6 $\leftrightarrow$ D6
\end{itemize}
These three theories are related by dualities which affect the world-volume 
of the branes. A T-duality along the world-volume of a NS5 or a KK5
changes from IIA to IIB and thus also from $iia$ to $iib$. Compactification of
the KK6 on one of its world-volume directions yields the KK5(A), thus relating
$iia$ and $m$ theories via compactification. The same duality between little
theories is obtained acting with a T-duality on the world-volume of the D6,
which gives the D5. Note also that the D4-brane, which defines a theory in
5 dimensions, can be obtained either by a T-duality from the D5, or by
compactification from the M5. This shows that once compactified, there is no
longer difference between $iia$ and $iib$ theories in 5 dimensions.

Although the relations discussed above are rather formal at this stage, they
exactly reproduce the same pattern of dualities of the 10 and 11 dimensional
theories. We will show hereafter that in the proper limits in which
the above little theories make sense (i.e. when they decouple from the bulk), 
this structure still holds and acquires even more evidence.

We now turn to the description of the different little theories.

\subsection{$iia$ theory}

As explained above, there are three ways to define type $iia$ theory 
\cite{mms}. The six dimensional supersymmetry is $(1,1)$. 
This is most easily found for the D5 brane
from dimensional reduction of the $N=1$ supersymmetry in $D=10$ \cite{tasi}
(see also Section \ref{DBRANEsec}).
For the NS5(B) and the KK5(A) it has been discussed respectively in 
\cite{callharvstro}
and \cite{senkk}. The type $iia$ theory is thus non-chiral.

The first approach is based on the D5 with a 
transverse compact direction of radius $R_B$. 

We look for all the objects which from the D5 world-volume point of view have
a finite tension, i.e. we rule out branes extending in transverse non-compact
directions. The relevant configurations of branes intersecting with the D5, 
and breaking further $1/2$ of
the supersymmetry are: D1$\subset $D5, F1$\mapsto$D5, D3$\mapsto$D5, 
NS5(B)$\mapsto$D5 and KK5(B)$\parallel$D5. The F1, D3 and NS5 can have a 
boundary on
the D5, as discussed in Chapter \ref{OPENchap}, 
and their only dimension 
transverse to it wraps around the transverse compact direction.
 
Generically, supergravity solutions preserving $1/4$ of the supersymmetries
and representing two intersecting branes can
be computed, following for instance the computations performed 
in Section \ref{INTERsec} (see e.g. 
\cite{tseytlinharm,gaunkasttras,bdejv2,rules}). Their existence can
be deduced by the compatibility of the two supersymmetry projections which 
characterize the configuration. 
The supersymmetry projections characterizing the branes are discussed in 
Appendix \ref{ZOOapp}, where the notations are also recalled.

Before taking the limit in which the bulk decouples, we have to fix the 
tension and the coupling of the little string theory on the world-volume
of the D5-brane. Since we have three parameters at hand, namely the string
length $l_s$, the string coupling of IIB theory $g_B$ and the radius $R_B$, it
will be possible to send the 9 dimensional Planck mass\footnote{
We have to consider the Planck mass in 9 dimensions because one of the 
transverse directions is compact. Furthermore its radius will
be sent to zero in the limit discussed above. Note also that this limit
does not depend on the size of the 
directions longitudinal to the D5-brane. For simplicity, we take them 
to be infinite.
} to infinity while
keeping a non-trivial little theory on the brane characterized by two
parameters, one of which dimensionless.

The only string-like object which lives on the D5-brane is the D1 string 
trapped to its world-volume \cite{douglaswithin}.
We take it to define the fundamental little string of $iia$ theory.
Accordingly, its tension is defined by (using \rref{tdbrane} and neglecting
here and in the following all numerical factors):
\be t_a\equiv T_{D1}={1 \over g_B l_s^2}. \label{ltfa} \ee
The boundaries of the F1, D3 and NS5, which are respectively 0-, 2- and 
4-dimensional closed objects, act as little ``d-branes" for the $f1$
little string. Their tension is postulated to be inversely proportional to the
coupling of the little string theory $g_a$ \cite{mms}. We have:
\be
\begin{array}{rcccccl}
t_{d0}&\equiv& T_{F1}R_B & = & {R_B \over l_s^2} & \equiv & 
{t_a^{1 / 2} \over g_a} \\
t_{d2}&\equiv& T_{D3}R_B & = & {R_B \over g_B l_s^4} & \equiv & 
{t_a^{3 / 2} \over g_a} \\
t_{d4}&\equiv& T_{NS5}R_B & = & {R_B \over g_B^2 l_s^6} & \equiv & 
{t_a^{5 / 2} \over g_a} 
\end{array}
\label{ltda}
\ee
The above definitions are consistent and, taking into account \rref{ltfa} 
we have:
\be g_a={l_s \over g_B^{1 / 2} R_B}. \label{lga} \ee

The last object to consider is the KK5, which actually fills the world-volume
of the D5. We can nevertheless define its tension using \rref{tkk5}:
\be t_{s5}\equiv T_{KK5}={R_B^2 \over g_B^2 l_s^8}={t_a^3 \over g_a^2 }. 
\label{tsa}
\ee
The $d$4 and $s$5 branes were overlooked in the analysis of \cite{mms},
they are however defined by perfectly well-behaved 10 dimensional
configurations. They are important in the identification of this little
theory as a model for a toroidal compactification of Matrix theory as we
discuss at the end of this chapter.

It is worth pointing out that we were able to define the tensions 
of five little branes using only two combinations of the parameters
$g_B$, $l_s$ and $R_B$. This is really the crucial point which allows us
to formulate the little theories.

We have now defined the string tension $t_a$ and the  string coupling $g_a$ of 
the little theory. In order for this $iia$ theory to make sense, we have to
take a limit in which the bulk modes decouple i.e. a limit in which 
the nine dimensional Planck Mass $M_p$ is going to infinity at fixed  $t_a$ 
and $g_a$. The Planck Mass is given by:
\be
M_p^7 = {R_B \over g_B^2 l_s^8}=
{g_B t_a^{7 / 2} \over g_a}.
\label{mplanck}
\ee
The limit defining type $iia$ is thus characterized by:
\be 
g_B \rightarrow \infty , \qquad l_s \rightarrow 0 , \qquad R_B \rightarrow 0
\label{liia}
\ee

We can also find the $iia$ theory starting with the NS5(B)-brane with one
transverse compact direction of radius $\tilde{R}_B$. 
We call, in this case, the string
coupling of type IIB ${\tilde g}_B$ and the string length ${\tilde l}_s$. 
The 10 dimensional configurations breaking 1/4 supersymmetry which define 
the BPS objects living in 6 dimensions
are simply obtained by S-duality from the ones discussed in the preceding
approach. They are the following: F1$\subset $NS5(B), D1$\mapsto$NS5(B), 
D3$\mapsto$NS5(B), D5$\mapsto$NS5(B) and KK5(B)$\parallel$NS5(B). 
The little $iia$
string is identified to the fundamental string of type IIB theory. Its
tension is simply given by: 
$t_a = \tilde{l}_s^{-2}$. The little string coupling is:
\be g_a= {{\tilde g}_B \tilde{l}_s \over \tilde{R}_B}. \label{gns5b} \ee 
It can be obtained for instance computing $t_{d0}=T_{D1} \tilde{R}_B$.
The limit in which the Planck mass goes to infinity is defined by 
${\tilde g}_B 
\rightarrow 0$, $\tilde{R}_B \rightarrow 0$ and $\tilde{l}_s$ constant. 
This result is 
consistent with the S-duality transformations:
$g_B \rightarrow {\tilde g}_B= 1/ g_B,
 l_s^2 \rightarrow \tilde{l}_s^2= g_B l_s^2$ and $\tilde{R}_B=R_B$ left 
unchanged.

This picture is maybe the heuristically more appealing, since the little
strings are indentified to fundamental type IIB strings trapped inside
a NS5-brane, at vanishing
type IIB coupling. Note also that when we decompactify the transverse
direction, i.e. we take $\tilde{R}_B \rightarrow \infty$ instead
of $\tilde{R}_B \rightarrow 0$, we obtain the zero coupling limit
of the little type $iia$ string theory, which thus coincides with the
definition of
one of the two little string theories of Seiberg \cite{sei2}. Indeed, all
the BPS states but the $f$1 acquire an infinite tension and 
decouple.

The third object with a 6 dimensional (1,1) supersymmetric world-volume theory
which can be used to define $iia$ theory is the KK5 monopole of type
IIA string theory, obtained by a T-duality on the transverse compact
direction from the NS5(B)-brane. This direction becomes the NUT direction
of the Euclidean Taub-NUT space transverse to the KK5 world-volume \cite{senkk}.
It appears that in this picture all the relevant configurations which
preserve $1/4$ supersymmetries are made up from branes of type IIA
inside the world-volume of the KK5(A) \cite{mms}: F1$\subset$KK5(A),
D0$\subset$KK5(A), D2$\subset$KK5(A), D4$\subset$KK5(A) and
NS5(A)$\subset$KK5(A). This makes the identification of $t_a$ and $g_a$
straightforward. The fundamental $iia$ string coincides now with 
type IIA's F1, and thus $t_a={\tilde l}_s^{-2}$. 
Since here also the ``little"
$d$-branes coincide with the D-branes of type IIA (with $p\leq 4$),
also the little string coupling is given by the IIA one: $g_a=g_A$.
It is easy to find by T-duality from the NS5(B) picture the limit
in which the KK5 decouples from the bulk. Since under T-duality 
$g_B \rightarrow g_A={g_B {\tilde l}_s \over R_B}$, $R_B \rightarrow R_A=
{{\tilde l}_s^2 \over R_B}$ and ${\tilde l}_s$ is unchanged, 
in the KK5(A) picture
we have $g_A$ constant and $R_A \equiv R_{NUT} \rightarrow \infty$.
The Riemann tensor of the Taub-NUT geometry vanishes in this limit,
an indication that the KK monopole decouples from bulk physics
(we comment on the decoupling issue at the end of the section).

We recapitulate the BPS spectrum of type $iia$ theory in Table \ref{iiabps}.
We  list the different little branes and their mass considering now a compact
world-volume characterized by radii $\Sigma_i$ with $i=1 \dots 5$ and volume 
$\tilde{V}_5=\Sigma_1 \dots \Sigma_5$.
We include for later convenience the KK momenta $w$.
\begin{table}[tb]
\[
\begin{array}{|c|c|} \hline 
\mbox{Brane} & \mbox{Mass} \\ \hline
w & {1 \over \Sigma_i} \\
& \\
f1 & \Sigma_i t_a \\
& \\
d0 & {t_a^{1 / 2} \over g_a} \\
& \\
d2 & {\Sigma_i \Sigma_j t_a^{3 / 2} \over g_a} \\
& \\
d4 & {\tilde{V}_5 t_a^{5 / 2} \over \Sigma_i g_a} \\
& \\
s5 & {\tilde{V}_5 t_a^3 \over g_a^2} \\
& \\ \hline
\end{array} 
\]
\caption{Mass of the BPS objects in $iia$ theory.}
\label{iiabps}
\end{table}

We also summarize in Table \ref{iialimits} the different ways 
to obtain $iia$ theory and the relation between the parameters.
\begin{table}[tp]
\[
\begin{array}{|c|c|c|c|} \hline & & & \\
& D5 & NS5(B) & KK5(A) \\  
& {1 \over g_B}, R_B, l_s \rightarrow 0 & {\tilde g}_B, \tilde{R}_B 
\rightarrow 0 
& R_{NUT} \rightarrow \infty \\ & & & \\ \hline & & & \\
t_a & {1 \over g_B l_s^2} & {1 \over \tilde{l}_s^2} & 
{1 \over {\tilde l}_s^2} \\ & & & \\
g_a & {l_s \over g_B^{1 / 2} R_B} & {{\tilde g}_B \tilde{l}_s \over 
\tilde{R}_B} & g_A \\ & & & \\
\hline \end{array} 
\]
\caption{Definitions of $iia$ parameters.}
\label{iialimits}
\end{table}

To summarize, we have defined type $iia$ little string theory as a 6
dimensional string theory characterized by a tension $t_a$ and
a coupling $g_a$, and by a $(1,1)$ non-chiral supersymmetry. The low
energy effective action of the little string theory is identified with
the low-energy world-volume action of the brane on which it is defined,
in this case 5+1 dimensional SYM. As anticipated, it does not
contain gravity. If there are $N$ host
branes, the theory is based on the group $U(N)$. We collect at the end of
this section the remarks on the low-energy limits of the little theories.

\subsection{$iib$ theory}

We recall that there are three approaches to this 6 dimensional theory,
using respectively the M5-brane with two transverse compact directions,
the NS5-brane of type IIA with one compact transverse direction and
the KK5 monopole of type IIB \cite{mms}. These three different branes all have 
a world-volume theory with (2,0) chiral supersymmetry 
\cite{callharvstro,wten,stromopen,senkk}.

The procedure by which we analyze the structure of $iib$ little string
theory is similar to the one described in the preceding subsection.
We will however meet here an interesting structure of $iib$ which is
its $s$-duality. We begin with the M5 approach, where this duality
is geometric.

The M5-brane set up is characterized by the 11 dimensional Planck length
$L_p$ and the two radii $R_1$ and $R_2$ of the two transverse compact
directions. The configurations breaking 1/4 supersymmetry in M-theory 
leading to finite tension
objects on the world-volume of the M5 are the following:
M2$\mapsto$M5 with the M2 direction orthogonal to the M5 wrapping either
$R_1$ or $R_2$; M5$\cap$M5=3; KK6$\supset$M5 with the NUT direction of
the KK6 identified either with $R_1$ or $R_2$. 

The boundaries of the M2-branes are strings on the M5, but we cannot
immediately identify the fundamental $iib$ little string because 
we have two different kinds of them. We simply choose one of the two
(say, the boundary of the
M2 wrapped on $R_1$) to be the fundamental and thus to have
tension $t_b$, and the other to be the little $d$1 brane with tension
${t_b \over g_b}$. This defines $g_b$. $s$-duality of $iib$ is then
simply the interchange in M-theory of $R_1$ and $R_2$ (this can actually
be extended to a full $SL(2,Z)$ duality group considering M2-branes
wrapped on $(p,q)$ cycles of the torus). We have thus (cfr. \rref{tmbranes}):
\be
\begin{array}{rcl}
t_{f1} =& T_{M2} R_1 = {R_1 \over L_p^3} &\equiv t_b, \\
t_{d1} =& T_{M2} R_2 = {R_2 \over L_p^3} &\equiv {t_b \over g_b}.
\end{array} \label{tfb}
\ee
The little string coupling is then given by:
\be g_b={R_1 \over R_2}. \label{lgb} \ee
We can now identify the other world-volume objects by their tension:
\be
\begin{array}{rcl}
T_{M5} R_1 R_2 = &{R_1 R_2 \over L_p^6}= {t_b^2 \over g_b} &\equiv t_{d3} \\
T_{KK6} R_2 =& {R_1^2 R_2 \over L_p^9} = {t_b^3 \over g_b} &
\equiv t_{d5} \\
T_{KK6} R_1 =& {R_1 R_2^2 \over L_p^9} = {t_b^3 \over g_b^2} &
\equiv t_{s5} 
\end{array} \label{tdb} \ee
Note that under $s$-duality the $d$3 is inert and the $d$5 and $s$5 are
exchanged. 

We still have to find the limit in which the bulk physics decouples.
Keeping $t_b$ and $g_b$ finite, the Planck mass in 9 dimensions
is given by:
\[ M_p^7={R_1 R_2 \over L_p^9}=\left({t_b^2\over g_b}\right) {1 \over 
L_p^3}, \]
and goes to infinity when $L_p \rightarrow 0$. To keep the parameters
of $iib$ finite, we also have to take $R_1, R_2 \rightarrow 0$.

We now consider the NS5(A) approach. The parameters are the string length
$\tilde{l}_s$, the string coupling $\tilde{g}_A$ of type 
IIA theory and the radius $\tilde{R}_A$ of
the compact direction. The configurations, breaking 1/4 supersymmetry, 
leading to finite tension objects
in the world-volume of the NS5(A) are: F1$\cap$NS5(A), D2$\mapsto$NS5(A), 
D4$\mapsto$NS(A), D6$\mapsto$NS(5) and KK5(A)$\parallel$NS5(A).
In this framework the little string tension $t_b$ is defined by the fundamental
string F1, namely $t_b=\tilde{l}_s^{-2}$. 
The little string coupling $g_b$ is found
by identifying the tension of the $d$1-brane from the configuration with 
the D2. We have: 
\be
t_{d1}=T_{D2}\tilde{R}_A={\tilde{R}_A \over \tilde{g}_A \tilde{l}_s^3} 
\equiv {t_b \over g_b}, \qquad \qquad
g_b={\tilde{g}_A \tilde{l}_s \over \tilde{R}_A}
\label{gnsb}
\ee
We obtain this picture from the previous one by dimensional reduction
on $R_1$, $R_1=\tilde{g}_A \tilde{l}_s$. 
The $\tilde{R}_A$ here is the previous $R_2$.
In this case the limit is taken performing $\tilde{g}_A \rightarrow 0$ and
$\tilde{R}_A \rightarrow 0$ at fixed $t_b$ and $g_b$. 
Note that the $s$-duality
in this picture is less straightforward to obtain from 10 dimensional
string dualities (one has to operate a TST duality chain).

As for the type $iia$ strings, the limit $\tilde{R}_A \rightarrow \infty$
of the picture above produces the decoupling of all the little `solitons',
leaving the second of the theories discussed by Seiberg \cite{sei2}.
Since in this case $g_b=0$, $s$-duality ceases to be a symmetry of the theory
(in the same way as there is no S-duality in perturbative 
type IIB string theory).

Turning now to the KK5(B) picture, we find that, as in the type $iia$
case, the little string theory is the reduction to the world-volume
of the KK5 of the physics of the objects that fit inside it. Thus
we simply identify $t_b$ with ${\tilde l}_s^{-2}$, $g_b$ with $g_B$, 
$s$-duality with S-duality, $f$1 with F1
and so on. As in the previous KK5 case, the limit in which the bulk
decouples involves taking the radius of the NUT direction to infinity.

We recapitulate the BPS spectrum of type $iib$ theory in Table \ref{iibbps}.
As for the $iia$ case, we list the different little branes and their mass 
considering now a compact
world-volume characterized by radii $\Sigma_i$ with $i=1 \dots 5$ and volume 
$\tilde{V}_5=\Sigma_1 \dots \Sigma_5$.
\begin{table}[tb]
\[
\begin{array}{|c|c|} \hline 
\mbox{Brane} & \mbox{Mass} \\ \hline
w & {1 \over \Sigma_i} \\
& \\
f1 & \Sigma_i t_b \\
& \\
d1 & { \Sigma_i t_b \over g_b} \\
& \\
d3 & {\tilde{V}_5 t_b^2 \over \Sigma_i \Sigma_j g_b} \\
& \\
d5 & {\tilde{V}_5 t_b^3 \over  g_b} \\
& \\
s5 & {\tilde{V}_5 t_b^3 \over g_b^2} \\
& \\ \hline
\end{array} 
\]
\caption{Mass of the BPS objects in $iib$ theory.}
\label{iibbps}
\end{table}

We also summarize in Table \ref{iiblimits} the different ways to 
obtain $iib$ theory and the relation between the parameters.
\begin{table}[tb]
\[
\begin{array}{|c|c|c|c|} \hline & & & \\
& M5 & NS5(A) & KK5(B) \\  
& L_p, R_1, R_2 \rightarrow 0 & \tilde{g}_A, \tilde{R}_A \rightarrow 0 
& R_{NUT} \rightarrow \infty \\ & & & \\ \hline & & & \\
t_b & {R_1 \over L_p^3} & {1 \over \tilde{l}_s^2} & {1 \over {\tilde l}_s^2} 
\\ & & & \\
g_b & {R_1 \over R_2} & {\tilde{g}_A \tilde{l}_s \over \tilde{R}_A} & g_B 
\\ & & & \\
\hline \end{array} 
\]
\caption{Definitions of $iib$ parameters.}
\label{iiblimits}
\end{table}

To recollect, type $iib$ little string theory is also a non-critical
6 dimensional string theory characterized by a tension $t_b$ and
a coupling $g_b$, and with chiral $(2,0)$ supersymmetry. It also displays
a symmetry, $s$-duality, which acts on its BPS spectrum and which
exchanges $g_b$ with $1/g_b$. The low-energy effective theory
is defined by the self-dual tensor multiplet. When there are $N$
host branes defining the theory, we have $N$ such multiplets, but
the non-abelian generalization is still lacking, contrary to the
former $iia$ case. We will discuss this further shortly.

There is a $t$-duality relating $iia$ and $iib$ little string theories,
as most easily seen in the pictures using the NS5 or the KK5 branes.
It is simply the 10 dimensional T-duality between IIA and IIB, applied
on a direction longitudinal to the world-volume of the above-mentioned
branes. To be more specific, application of such a longitudinal T-duality
maps, say, the NS5(A) picture of $iib$ theory to the NS5(B) picture
of $iia$ theory, and similarly for the KK5 pictures.
The behaviour of the BPS objects is the same as in type II string theories:
KK momenta are exchanged with wound $f$1 strings (as in \cite{sei2}), 
the $s$5 brane of one
theory is mapped to the one of the other theory, and
$dp$-branes become $d(p+1)$- or $d(p-1)$-branes for transverse or 
longitudinal $t$-dualities respectively. $iia$ and $iib$ theories
are thus equivalent when reduced to 5 space-time dimensions or less.

\subsection{$m$-theory}

As stated at the beginning of this section, there are two objects with
7 dimensional world-volume in M/type II theories: the D6-brane in type
IIA and the KK6 monopole in M-theory. The supersymmetry algebra is unique
and obviously non-chiral.

We first consider the D6 approach. Note that for the transverse space
to be asymptotically flat, we cannot have any compact transverse
dimension. The free parameters are thus the string length $l_s$ and
type IIA string coupling $g_A$. Already at this stage we know that
the theory on the world-volume will be characterized by only one
parameter (one is lost taking the appropriate limit which decouples the
bulk).

In this case, we have to consider configurations preserving $1/4$
supersymmetries with a brane within the D6-brane. The only branes of type
IIA for which this works are the D2- and the NS5-brane.
We identify them with the $m$2 and $m$5 branes. As it is necessary for
the definition of $m$-theory, only one parameter suffices to define 
both their tensions. Indeed we have:
\be
\begin{array}{c}
t_{m2}\equiv T_{D2} = {1\over g_A l_s^3}\equiv {1\over l_m^3} \\
\\
t_{m5}\equiv T_{NS5} = {1\over g_A^2 l_s^6} = {1\over l_m^6}
\end{array} \label{tm} \ee
$l_m$ is thus the characteristic length of $m$-theory, the analog of the 
Planck length in M-theory.

In order to decouple gravity, we send the 10 dimensional Planck mass
to infinity. Keeping $l_m$ finite, we have:
\[ M_p^8={1\over g_A^2 l_s^8} = {1 \over (l_m^6) l_s^2}, \]
and thus we have to take $l_s \rightarrow 0$ and $g_A \rightarrow \infty$.

In the KK6 approach, there are two configurations preserving $1/4$ of 
supersymmetry: M2$\subset$KK6 and M5$\subset$KK6. M2 and M5 are thus
respectively identified to $m2$ and $m5$, and $l_m=\tilde{L}_p$ where 
$\tilde{L}_p$ is
the eleven dimensional Planck length.
The KK6 monopole can be seen as the M-theoretic origin (and thus the
strong coupling limit) of the D6-brane. The radius of the NUT direction
is thus given by $R_{NUT}=g_A l_s=g_A^{2/3} \tilde{L}_p$. 
Therefore, the limit above
$g_A \rightarrow \infty$ becomes $R_{NUT} \rightarrow \infty$. Again, in this
limit the geometry becomes that of flat space.

It is interesting to note that here as in the former cases of $iia$ and
$iib$ theories, the KK monopole description is the more ``economic" one,
in the sense that one has to take only one limit. However, the other 
descriptions will be useful, for instance, to make contact with Matrix theory
compactifications.

In Table \ref{litmbps} the masses of the different BPS objects of $m$-theory
are listed. Again we consider a compact volume $\tilde{V}_6=\Sigma_1 \dots 
\Sigma_6$. 
\begin{table}[tb]
\[
\begin{array}{|c|c|} \hline 
\mbox{Brane} & \mbox{Mass} \\ \hline
w & {1 \over \Sigma_i} \\
& \\
m2 &{ \Sigma_i\Sigma_j \over l_m^3} \\
& \\
m5 & {\tilde{V}_6 \over \Sigma_i l_m^6} \\
& \\
\hline
\end{array} 
\]
\caption{Mass of the BPS objects in $m$-theory.}
\label{litmbps}
\end{table}

The different ways to obtain $m$-theory are shown in Table \ref{litmlimits}, 
along with the relation between the parameters.
\begin{table}[tb]
\[
\begin{array}{|c|c|c|} \hline & &  \\
& D6 & KK6  \\  
& {1\over g_A}, l_s \rightarrow 0 
& R_{NUT} \rightarrow \infty \\ & &  \\ \hline & & \\
l_m & g_A^{{1\over3}}l_s & \tilde{L}_p \\ & &  \\
\hline \end{array} 
\]
\caption{Definitions of $m$-theory parameters.}
\label{litmlimits}
\end{table}

$m$-theory is thus a 7 dimensional theory in which the lowest dimensional
object is a membrane\footnote{As observed in \cite{brun,hana}, 7 is one
of the dimensions in which a supermembrane can be defined \cite{7dmembr}.}.
Its low-energy effective field theory limit is taken to be defined by the
world-volume low-energy field theory of the D6-brane, that is 6+1 
dimensional SYM, with $U(N)$ group if there are $N$ D6-branes.

The duality between $m$-theory and $iia$ theory can now be made more 
precise. The relations between the parameters of $m$-theory 
compactified on the ``7th" direction of radius $R_c$ and $iia$ theory
are easily found comparing the tensions of the wrapped and unwrapped
$m$2 brane on one side, and of the $f1$ and $d2$ branes on the other
side. One finds no surprises:
\[ t_a={R_c \over l_m^3}, \qquad \qquad g_a=\left({R_c\over l_m}
\right)^{3/2}.\]
In the KK5(A) and KK6 picture, this is a direct consequence of the similar
relations between M and IIA theories. It is more amusing to see
that they indeed correspond to T-duality relations between IIA and IIB 
when one goes to the D5/D6 picture.

We have thus completed the pattern of dualities existing between the
little theories with 16 supercharges: $m$-theory is related to $iia$
theory by compactification, $iia$ and $iib$ theories are related by
$t$-duality, and finally $iib$ theory has an $s$-duality. Accordingly,
these three theories can be considered as three different `phases',
or corners in the moduli space, of the same theory. This is very
remininscent of the dualities defining M-theory, as they are reviewed
in Chapter \ref{MINTROchap}. The analogy will be further developped
by the introduction of the little theories with 8 supercharges.

We can actually guess the maximal duality group of $m$-theory
(or of any one of the type $ii$ theories) when compactified down to,
say, $p+1$ dimensions, with now $p<5$. Since the branes of $m$-theory
are actually the same (in number and in extension) as the ones
of M-theory which can be accomodated on a 6-dimensional torus $T^6$, 
both fit into the same duality multiplets when a compactification
on $T^{6-p}$ is considered. It is tempting to speculate that like
M-theory on $T^6$, the fully compactified $m$-theory has a 
$u$-duality based on the group $E_6(Z)$, as discussed in \cite{mms}
(see also \cite{eliz}).
The minimal evidence of this duality, i.e. that the BPS states fit into
representations of $E_6(Z)$, will be important in providing the link
between these little theories and Matrix theory compactified on $T^6$.

\subsection{Some comments on the low-energy effective actions of the 
little theories}

Unlike 10 dimensional critical string theories, in the case of the little
string theories it is impossible at the present state of the knowledge
to directly compute their massless spectrum and the interactions, and then
formulate the low-energy effective action that governs them. We have
however another way to precisely guess these low-energy effective actions,
even if this way is somewhat indirect and relies on the fact that 
the definition of the little theories uses some host brane.
As anticipated when reviewing the little theories above, we assume
that the low-energy effective action of the little theories coincides
with the world-volume low-energy effective action of the host branes.
It is actually a matter of consistency that the two low-energy limits
coincide, the guess being that the low-energy limits of the
little theories exist.

Let us begin reviewing the low-energy effective action of type $iib$ little
string theory.
If we take the M5 picture to define $iib$ strings, then we know
that the low-energy theory is given by the $(2,0)$ multiplet containing
a self-dual tensor $H^+$ and 5 scalars $\phi_i$
(and the fermions which we do not consider here and in the following).
Two of the scalars, say $\phi_1$ and $\phi_2$, have a finite range
since they correspond to the compact directions, while the three others
have an infinite range.
Due to the self-duality of the tensor, this theory is believed not
to have a coupling at all \cite{sei1} (it is fixed to $g=1$). Accordingly
all the fields have mass dimension 2 and the range of the scalars
can be given by $0\leq\phi_{1,2}\leq R_{1,2}/L_p^3$, as in \cite{sethi}.
In type $iib$ variables, the ranges can be rewritten as:
\be
0\leq \phi_1\leq t_b, \qquad\qquad 0\leq \phi_2 \leq {t_b \over g_b}.
\label{rangephim5}
\ee
We already see at this stage that unlike `conventional' string theories,
here the little string coupling appears in the low-energy action in a
non-trivial way. Note that when considering a NS5(A)-brane instead of a M5,
the asymmetry\footnote{
Note that in \cite{sethi} the fact that the range of $\phi_2$ is proportional
to $R_A/g_A$
was taken as an indication that the direction effectively
decompactifies when $g_A\rightarrow 0$. Here however the little
string limit is such that the range is precisely kept finite.}
between $\phi_1$ and $\phi_2$ can be ascribed to the
fact that the former comes from the RR sector while the second from
the NSNS one \cite{sethi}.

We can now ask ourselves the question of how we recover in the low-energy
theory the BPS objects as solitons, in the same way as each brane of
M-theory corresponds to a supergravity solution. Naively one could think
that the 3-form $H^+$ naturally couples to a (self-dual) string, but
then we run into problems because this would give rise to only one string
soliton instead of the required two, according to Table \ref{iibbps}.
The resolution of this problem can be found in \cite{hlw1} (see also
\cite{death}). For a solution to be BPS, it cannot carry only one charge,
but there should be a balance of two forces that cancel in presence
of two parallel BPS objects. The second field which carries
a charge in this case is either $\phi_1$ or $\phi_2$ \cite{hlw1}.
The construction of \cite{hlw1} is however not enough for our purpose,
since in our case the same M2-brane `starts' and ends on the same M5.
One might be afraid of having an unstable situation with opposite charges,
but this does not occur \cite{death}: when both the field $H^+$ and
the scalar have the opposite sign, the state is still BPS and the 
energies add, with no dependence on the distance between the two objects.
Moreover the world-volume energy is proportional to the length
of the intersecting brane \cite{death} and thus in this case it should
presumably be proportional to the 
range of the scalar. This is also why we restricted the solitons to 
have as non-trivial scalar fields only the ones with finite range.
Taking into account \rref{rangephim5}, it is straightforward
to identify the $f$1 solution with the soliton charged under $\phi_1$
and the $d$1 to be the one charged under $\phi_2$. Note that
in the NS5(A) picture, $\phi_1$ is the scalar which does not correspond
to an embedding coordinate.
Let us stress that although we did not carry out any calculation 
to support all the above discussion, we find it likely to be correct.

The $s$-duality of $iib$ theory then reflects in the low-energy action
in the permutation of $\phi_1$ and $\phi_2$, and it is actually
a subgroup of the R-symmetry (which is $SO(5)$ \cite{sei1}).
The $d$3-brane soliton can be found in a similar way \cite{hlw2}.
This solution is now charged under the two scalars $\phi_1$ and $\phi_2$
at the same time, while the 3-form $H^+$ is taken to vanish. The
classical solution is thus also clearly inert under (classical) $s$-duality.

Concerning the little 5-branes, the $d$5 and the $s$5, we do not expect
them to correspond to classical solitons since they would have to couple
to non-dynamical 6-form potentials. They have a similar status as the
D9-brane in type IIB string theory.

The last remark about the $(2,0)$ low-energy effective theory is that
when there are $N$ coinciding M5-branes, the theory is conjectured
to possess tensionless strings \cite{wten,stromopen} which, upon
dimensional reduction to 4+1 dimensions, yield the massless gauge
bosons of a non-abelian $U(N)$ SYM theory. These tensionless strings 
arise from M2-branes stretched between two adjacent M5-branes, and
should not be confused with the little string solitons, which have
a finite tension since they originate from M2-branes wrapping on a transverse 
compact direction.

Going now to the definition of the low-energy effective action of $iia$
theory, we have already observed that it should coincide with
the 5+1 SYM effective theory of the D5-brane. We can go a little further:
in the case of the D5-brane, the SYM coupling is given by $g^2_{YM}=
g_B l_s^2$. Translating to $iia$ variables as given in Table \ref{iialimits},
this gives for the coupling of the $iia$ low-energy action $g^2_{YM}=1/t_a$.
One of the 4 scalars of the SYM multiplet, call it $\phi_B$, represents
a compact direction and its range is given by:
\be
0\leq \phi_B \leq {R_B\over l^2_s}\equiv {t_a^{1/2}\over g_a } .
\label{rangephid5}
\ee
Going to the NS5(B) picture, the $1/\tilde{g}_B$ behaviour in the
range of $\phi_B$ derives from its NSNS origin \cite{sethi}.
As in the $iib$ case, here also $g_a$ does not appear in front of
the action but rather in the definition of one of the fields.

The mapping between the $iia$ BPS objects (see Table \ref{iiabps}) and
classical solitons is as follows. The $f$1 corresponds to a SYM instanton
and accordingly has a tension given by $1/g^2_{YM}=t_a$. The tension
is independent of $g_a$ since no scalar participates to the instanton
solution. The $d$0 and the $d$2 are the electric and the magnetic charge
with respect to the SYM vector field, and if they are to be BPS the
scalar $\phi_B$ must balance their charge, as in \cite{death}. If we 
have $N$ D5-branes, the above little branes
simply correspond to the massive gauge boson
of broken gauge symmetry and to the 't Hooft-Polyakov monopole respectively,
the mass of both of which depends on the expectation value (and thus here
on the range) of a scalar
field. Here the original $SO(4)$ R-symmetry is broken completely at the
level of the solutions since the other scalars, being of infinite range,
give rise to infinitely massive objects.
The $d$4 brane should couple to the still conjectural 6-form field strength,
which was also considered in the previous chapter to occur in the
world-volume of the NS5(B)-brane. 

The low-energy effective action of $m$-theory is defined by the world-volume
low-energy effective theory of the D6-brane, which is 6+1 dimensional
SYM. The SYM coupling is now given by $g^2_{YM}=g_A l_s^3\equiv l_m^3$
(see Table \ref{litmlimits}), while all of the three scalars belonging
to the SYM multiplet have an infinite range. Accordingly, the usual classical
electric and magnetic charged objects have here an infinite mass,
and this is why there are no 0- and 3-branes in $m$-theory, contrary to
what one could have expected from the low-energy point of view. Rather, 
the $m$2 brane is given by the instanton, with tension $1/g^2_{YM}=1/l_m^3$.
It is less clear what is the SYM origin of the $m$5 brane, which
has a tension going like $1/g_{YM}^4$. It could couple to a non-propagating
7-form field strength. Indeed, in \cite{town_sl2z} it was conjectured
that a $(p+1)$-form field strength existed on the world-volume of every
D$p$-brane.

Let us end this subsection observing that there is yet another way
to find the BPS content of the little theories. It is simply based
on the knowledge of the supersymmetry algebras with 16 supercharges
in 6 and 7 dimesions, including the $p$-form `central' charges \cite{gomis}.
This procedure is very similar to the one which allows to find
all the BPS objects in M-theory \cite{democracy,hull}. In the
present case however one has to take into account that most of the
automorphism group of the supersymmetry algebras
is broken due to the presence of compact transverse directions
which single out some of the (little) branes.

\subsection{Some comments on the decoupling from the bulk}

A potential problem in the definition of the little theories was pointed
out in \cite{senmatrix,seibergmatrix} in a Matrix theory context. The
argument, applied to the little theories, is basically the following:
in the D6-brane definition of $m$-theory as summarized in Table 
\ref{litmlimits}, the mass of the D0-brane present in this type IIA theory
vanishes. Indeed, in the $m$-theory limit, it is:
\[ M_{D0}={1\over g_A l_s}={l_s^2\over l_m^3}\rightarrow 0. \]
This could be taken as a signal that there are (infinitely many) additional
massless degrees of freedom, which moreover are free to propagate in the
bulk. The  $m$-theory limit should eventually represent a 6+1 dimensional
theory coupled to the bulk through D0-branes.

The same problem can be reformulated for every little theory in any one
of its formulations, simply acting with dualities on the D0--D6-brane
system. In the pictures using the KK monopoles for instance, the
problem is seen to arise since, as can be seen in Tables \ref{iialimits},
\ref{iiblimits} and \ref{litmlimits}, they are always embedded
in theories where all the couplings ($g_A$, $g_B$ or $\tilde{L}_p$)
are finite.

We find however that the problem is subtler that it may seem at a first
sight. Indeed there are several hints that all point towards the
conclusion that the decoupling actually takes place. Going back to
the D0--D6-brane system, it is well known that they do not form
a supersymmetric bound state. Rather, the only supergravity solution
carrying at the same time D0 and D6-brane charge is not a stable state
(see Subsection \ref{D0D6ssec}), i.e. it has an excess of energy with
respect to the masses of the separated D0 and D6-branes. This means that
a D0-brane in the bulk will interact (at least in this way) with the 
D6-brane only if it is supplied some energy to do so. The same conclusion
comes from the stringy computation of the interaction between 
D0- and D6-branes \cite{lifshytz}, where it is found that there is
a repulsive force.

In the KK monopole picture of each little theory, the problem is 
reformulated as follows. In the limit $R_{NUT}\rightarrow \infty$, the geometry
of a KK monopole approximates the one of an ALE space 
(see \cite{refontaubnut2,senkk}) with a singularity
at the core if there are $N$ coinciding KK monopoles. Moreover
the ALE space appears effectively 4 dimensional since the NUT direction
decompactifies. Since the string 
coupling, or the 11 dimensional Planck length, are finite, one has
to check whether the bulk gravitons interact with the singularity
(where the little theory should be located) or not. Heuristically,
one can argue, following \cite{mms}, that the gravitons propagating
along the NUT direction feel, even if $R_{NUT}$ is very large, a vanishingly
small radius when they approach the core (see the metric of a KK monopole
\rref{kkmmetric}). They are thus effectively very massive there and
presumably do not penetrate the singularity, i.e. they decouple. 
Concerning the truly massless gravitons in the three remaining 
transverse directions, they see an effective 9 dimensional Planck mass 
which is proportional to $R_{NUT}^{1/7}$, like for instance in
\rref{mplanck}. It is thus very big and these gravitons also should 
decouple.
These are certainly not definitive statements, and one should search
for more quantitative evidence, but the above facts seem to point out
that something subtle can happen which ensures the decoupling.

Related to the decoupling issue discussed above, there is also
the problem of making sure that the little objects do not leave
the host brane. In fact, all the configurations defining the little
BPS objects are marginal bound states of the host brane with some
other brane. There is thus no energy barrier which prevents
the two branes to depart from each other.
One way to address this problem is to consider, for instance in the
case of D1$\subset$D5 or D2$\subset$D6 defining respectively the
little $iia$ string and the $m$2 of $m$-theory, the low-energy
effective theory of a D$p$-brane inside a D$(p+4)$-brane \cite{douglaswithin}.
This was also considered in \cite{mms}. It appears that in this low-energy
theory, the scalars belonging to the hypermultiplets (the matter multiplets,
arising from the open strings going from the D$p$-brane to the D$(p+4)$-brane)
describe the motion of the smaller brane inside the bigger, while
the scalars belonging to the vector multiplet describe the motion
of the smaller brane away from the bigger, in the bulk. In this 
terminology, the Coulomb branch of the D$p$-brane theory describes
motion in the bulk, while the Higgs branch describes motion trapped
to the world-volume of the D$(p+4)$-brane. A trapped brane thus
cannot escape to the bulk if the Higgs and Coulomb branches are decoupled,
and this is what seems to happen in the little theory limit \cite{mms}.
This is a tentative explanation (which also applies to the little string 
theories of Seiberg \cite{sei2}), and we will also mention in the
concluding section of this chapter a similar mechanism which seems
to ensure the decoupling of the little theories from the bulk.

A different solution to this problem could be to take the NS5-brane picture
to define the little string theories, and then to abandon the idea
of having a transverse compact direction, the radius of which eventually
goes to zero. Intead, one could take two (sets of) NS5-branes in flat space
at a distance $L$, and then suspend the D-branes between them
(as in a Hanany-Witten configuration \cite{hananywitten})
in order to give rise to the little branes. The little type $ii$ strings
would still be given by type II strings trapped inside the NS5-branes.
The limit defining the little string theory would then be the same,
provided $L$ replaces the r\^ole of $R$, except in the
Planck mass that is now the 10 dimensional one. This approach has however some
serious flaws: on one hand, there seems to be no direct way to
recover the $s$5 branes in the little string theories (but this might
not be a problem since we could conceivably discard them); 
on the other hand, and more seriously, there is no
directly equivalent definition of $m$-theory. This is why we preferred in
this chapter
to define the little theories using a transverse compact direction,
signaling however above the potential problems of this definition.

We should also mention the issue raised in \cite{throat} and
which applies also to the little string theories of \cite{sei2}. 
Stated very briefly, a NS5-brane has a non-vanishing Hawking temperature
even in the extremal limit (see equation \rref{blackhawking} and the
discussion which follows it). Moreover, the dilaton, and thus the
string coupling, explode near the location of the NS5-brane
(see the equations \rref{ns5metric} and \rref{ns5smetric}).
Both of these facts would be an indication
that the theory does not decouple from the bulk. However the background
of an extremal NS5-brane (in the string frame) has also the 
peculiarity of having an infinite `throat' which effectively disconnects
the location of the brane from asymptotic infinity. 
The radiation which emanates
from the brane seems thus not to reach the bulk of the flat 
spacetime surrounding the brane \cite{throat}.

The issues discussed in this subsection should be an indication
that the exact mechanism which allows us to consider the little
theories as decoupled from the bulk clearly deserves a more detailed
analysis. However, in our opinion it is not hopeless to think that the little
theories treated in this chapter will eventually make sense.

\section{Theories with 8 supercharges}
\label{8THsec}

We propose in this section to define the little string theories with
(1,0) supersymmetry in 6 dimensions. Note that this is the highest
dimension in which a theory with 8 supercharges can live. 
We construct the (1,0) theories by analogy with the 10 dimensional
relation between $N=1$ and $N=2$ string theories.

In 10 dimensions, type I open string theory can be obtained from type
IIB string theory \cite{orientifold,polcdbranes}. 
One adds to the IIB theory an $\Omega 9$
orientifold yielding $SO$ open strings \cite{tasi}, 
and then adds 16 D9-branes to
have a vanishing net flux of D9 RR charge. This leads to an $N=1$
supersymmetric theory
with open strings carrying $SO(32)$ Chan-Paton factors. 
The two heterotic string theories are then obtained by dualities.
The $SO(32)$ heterotic theory is found by S-duality from the type I
(identifying the D1-brane in the latter to the fundamental heterotic
string of the former \cite{polwitt}). The $E_8 \times E_8$ heterotic
theory is obtained by T-duality from the $SO(32)$ one. 
The $E_8 \times E_8$ theory can also be derived from
M-theory compactified on $S^1/Z_2$ \cite{horavawitten}.

Our strategy is the following: we define the theories with 8 supercharges
using the 5-branes of the type $ii$ little string theories, and we then
show that the same pattern of dualities as in 10 dimensions arises.

\subsection{Type $i$, $h_b$ and $h_a$ theories}

Let us start with the $iib$ little string theory, where we can define 
a procedure very close to that of \cite{orientifold,polcdbranes}. 
In this theory we have
$d$5-branes (cfr. Table \ref{iibbps}), 
which are Dirichlet branes for the little
$iib$ fundamental strings, filling the 6-dimensional
space-time. They are thus the analog of the D9-branes of type IIB
theory. We now go to one of the precise pictures defining $iib$ strings to
analyze the structure of the theory defined by $iib$ in presence 
of a certain number $n$ of $d$5-branes.

If we take the KK5(B) picture (see Table \ref{iiblimits}), 
the $d$5-brane arises
from the $D=10$ D5-brane with its world-volume inside the KK5.
It is now straightforward to identify which BPS states of the $iib$
theory survive the ``projection" due to the presence of the $d$5-branes.
From the 10-dimensional supersymmetry relations listed in Appendix 
\ref{ZOOapp}, we can see that only D1-branes can live at the same time within
the KK5 and the D5-branes. The closed $f$1, coinciding with the F1, is no
longer a BPS state, and the same occurs to the $d$3 and the $s$5. We
are thus left with a theory of open little strings (the open
IIB strings within the D5-brane), with a $d$1-brane BPS state. We 
propose to call this theory type $i$.

Note that along with the $n$ D5-branes, 
one can also add an $\Omega5$ orientifold plane\footnote{
Much in the same way as it was introduced in \cite{evan} in the 
context of brane configurations describing field theory dualities involving
$SO$ and $Sp$ groups.} 
without breaking further supersymmetry. Since there are still 3
non-compact transverse directions, the $SO$ or $Sp$ nature of the
orientifold and the number of D5-branes is not fixed by simple charge
flux arguments. Therefore, unlike the 10 dimensional case, here we can
have a priori arbitrary $U(n)$, $SO(2n)$ or $Sp(2n)$ gauge groups on 
the D5-branes.
The $\Omega 5$ defines an $\omega 5$ little orientifold plane for
the $iib$ theory. 

If there is only one KK5 brane, the gauge group discussed above
corresponds to the gauge group of the little type $i$ string theory.
On the other hand, if there are $N$ coinciding KK5 branes
this issue is more subtle.
Note also that we could have carried out all the above discussion starting
from another configuration, for instance the one with $N$ NS5(A)-branes
within $n$ D6-branes. The latter configuration is closer to a 
Hanany-Witten set up \cite{hananywitten}, and could thus be useful
when considering the low-energy effective action of the type $i$
strings. We return on this at the end of the section.

In order to define a (1,0) closed string theory, we can simply apply
the $s$-duality of $iib$ strings to the type $i$ theory. This
duality maps the $d$1 branes to the $f$1 little strings, and most
notably the $d$5-branes to the $s$5-branes. The only BPS states
of this theory are thus the $f$1. We call this theory $h_b$. We could
have directly found this $h_b$ theory from the $iib$ one
by piling up $n$ $s$5-branes. If we are allowed to define the $s$-dual
of the $\omega5$ orientifold, then this procedure is reminiscent
of the one used by Hull \cite{hull} to obtain the heterotic $SO(32)$ theory
from type IIB. The possible groups characterizing the $h_b$ theory should be
the same as the ones for type $i$ theory.

For future convenience, we recall that the above theory can be defined
by a configuration with generically $N$ NS5(A)-branes (the host branes)
parallel to $n$ KK5(A) monopoles (giving rise to the $s$5-branes).

There is still a 5-dimensional object in the little string theories
that could be used to define a new (1,0) theory, namely the $s$5-brane
of the type $iia$ theory.
Taking the NS5(B) as the host brane, 
we obtain this theory piling up $n$
KK5(B)-monopoles parallel to its world-volume.  
In this case the low-energy effective action, and its gauge symmetries,
is more subtle to define, as we will see at the end of the section.
We call this little string theory $h_a$. It is $t$-dual to the
$h_b$ one. 

Using the duality between $iia$ and $m$-theory, we can see that $iia$
in presence of $s$5-branes is dual to $m$-theory with $m$5-branes,
which are domain walls, or boundaries of the 7-dimensional `little' space-time.
Thus the $h_a$ theory can be seen as an $m$-theory
compactification in presence of $m$5-branes.
This description is very rough and schematic, 
but could be related
to a 7-dimensional analog of the Horava-Witten mechanism 
\cite{horavawitten}
to obtain the $E_8 \times E_8$ heterotic string theory
(although in \cite{horavawitten} the 9-dimensional objects are 
really boundaries rather than branes). 
Note also that the $m$-theory configuration giving $h_a$ theory can
also be seen as NS5-branes within D6-branes, but this time we
have generically $n$ NS5-branes and $N$ (host) D6-branes, that is the 
opposite than for the configuration leading to type $i$ theory.

We thus see that the pattern of dualities that arises between the
theories with 8 and 16 supercharges is very similar to the one
between $N=1$ and $N=2$ string theories in 10 dimensions. We list in 
Table \ref{ihahbdef} the main characteristics of the (1,0) 
little string theories.
\begin{table}[tb]
\[
\begin{array}{|c|c|c|} \hline 
\mbox{Theory} & \mbox{Defined by:} & \mbox{BPS objects}   \\ \hline
i  & iib + d5 & d1 \\
h_a  & iia + s5 & f1 \\
h_b  & iib + s5 & f1 \\
\hline
\end{array} 
\]
\caption{Main characteristics of the theories with 8 supercharges}
\label{ihahbdef}
\end{table}

\subsection{The low-energy effective actions and further comments}

We now turn to the discussion of some speculative points related
to the theories discussed above.

We would like to find the candidates for the low-energy effective
theories of the little theories with $(1,0)$ supersymmetry. This task
is however more subtle than for the theories with 16 supercharges.
Let us begin with $h_a$ theory. Its low energy theory is a theory that
has actually been extensively studied in the literature 
\cite{blum,intri,brukar}. It can be alternatively seen as arising from one 
of the following two configurations, represented schematically as
$(N)$ NS5(B)$\parallel (n)$KK5(B) and $(N)$D6$\mapsto (n)$NS5(A).
Since the background of $n$ KK5 monopoles can be related to an ALE space
with a $A_{n-1}$ singularity \cite{senkk}, the first configuration
makes the link with the theories discussed in \cite{blum,intri}.
Actually, in \cite{intri} two little string theories are defined
as the theories on the NS5-branes at a $A_{n-1}$ singularity. They are
very likely to coincide with our $h_b$ and $h_a$ theories.

The second configuration with the D6-branes suspended between the
NS5(A)-branes is closer to a Hanany-Witten configuration, and has indeed
been considered in \cite{brukar}, with results coinciding with
\cite{blum,intri}. As usual in this kind of configurations, the theory
on the D6-brane is reduced to a 5+1 dimensional theory by the presence
of the NS5-branes. In this case however there are two differences
with respect to the cases considered in \cite{hananywitten}.
First of all, D6-brane charge conservation imposes that on every NS5-brane
there are as many D6-branes on the right than on the left. This is clearly
the case in our configuration since the $N$ D6-branes actually wrap a
compact direction. Secondly, since the world-volume of
the NS5-brane is also 5+1 dimensional, it also contributes to the
resulting effective field theory.

The low-energy field theory, for a generic set up in which the
NS5-branes are at finite distance on the D6-branes,
thus contains the $(1,0)$ vector multiplets yielding a gauge group $U(N)^n$
(this is true if the direction
of the D6-branes perpendicular to the NS5-branes is compact,
as considered for lower dimensional D-branes in e.g. \cite{hori,wittenm}),
some matter hypermultiplets coming from the massless modes of the
open strings stretching between different D6-branes, and $n$ self-dual tensor
multiplets coming from the NS5-branes. This is precisely the
right matter content to make the field theory above anomaly free
(the account of the precise mechanism can be found in \cite{blum}).
The naive argument that there was no constraint on $n$ because the
net flux could be non-vanishing was thus enough to ensure anomaly freedom
at the level of the low-energy effective action.

In the limit in which all the NS5-branes coincide, the scalars in the 
tensor multiplets have vanishing expectation value (except one) and all but
one of the SYM couplings become infinite. It is not clear what kind
of theory is left, but there should be a restoration of a (global)
$U(n)$ symmetry and tensionless strings (coming from D2-branes stretching 
from one NS5-brane to another) should also come into play.

Note also that in the case where there is only one host brane ($N=1$),
the effective theory should contain a single and decoupled $U(1)$ vector
field and $n$ tensor fields coming from the NS5-branes, but an
explicit check of the anomaly cancellation should be performed.

Let us now consider the low-energy effective action of $h_b$ theory.
This should actually coincide with the low-energy action of
the type $i$ theory, since $s$-duality does not affect the world-volume
theories. Depending on whether we see it as representing type $i$ or
$h_b$ theory, the relevant configuration is respectively
$(N)$NS5(A)$\subset (n)$D6 or $(N)$NS5(A)$\parallel (n)$KK5(A).
If we perform a T-duality along the NUT direction of the
second configuration, we obtain $(N)$KK5(B)$\parallel (n)$NS5(B).
We thus obtain that the low-energy effective action of $h_b/i$ theories
is exactly the same as the one for $h_a$ theory, but with $N$ and $n$
interchanged (see also \cite{intri,ganorsethi}). 
The theory has a $U(n)^N$ gauge group. It is now more compelling to understand
what happens when all the host branes coincide. The difficulty
is related to the existence of a non-abelian generalization of the
theory of $N$ self-dual tensors.
Note that in this latter case, the configuration with only one host
brane is actually the simplest precisely because the problem above
is eluded. We recover a $U(n)$ gauge theory with a (decoupled) tensor
multiplet.

The specularity of these two low-energy effective actions is puzzling
and could well have a physical meaning. Indeed, because of the
$t$-duality relating the $h_a$ and $h_b$ theories, also the
two low-energy theories should coincide upon compactification
on a circle. To help in clarifying this issue, 
one should certainly have a better
understanding of the above mentioned limit in which all but one of the
couplings take an infinite value.

Let us end this section with two further comments.
Seiberg \cite{sei2} defines (1,0) little string theories from the
world-volume of the 5-branes in the two heterotic string
theories. These little theories have however a global $SO(32)$ or
$E_8 \times E_8$ symmetry, which is unlikely to arise in our
cases. The (1,0) theories of \cite{sei2} seem thus different from those
discussed in this section (in the sense that it should not be possible
to derive them from a pure type $ii$ little string framework).

As a side remark, it is worth noting that
the 5-branes of the little theories play a crucial r\^ole in the
interplay between theories with 16 and 8 supercharges.
By analogy, 9-branes in M-theory and in type
II theories might be interesting to study.
The existence of an M9-brane and NS-like 9-branes of type IIA and IIB theory
was indeed discussed in \cite{hull}.

\section{Application to Matrix theory compactification}
\label{APPLICsec}

The little theories discussed above are relevant to the description 
of Matrix theory compactified on higher dimensional tori. We refer
the reader to Section \ref{MATRIXsec} for an introduction to Matrix theory.

In the original conjecture \cite{bfss}, M-theory in the infinite momentum
frame (IMF) is described
by the Matrix theory of a system of $N$ D0-branes, in the large $N$ limit.
If some of the remaining 9 space directions are compactified (on $T^d$ say),
one has to correctly include in the Matrix description
the additional BPS states that will fit into representations
of the U-duality group of compactified M-theory. A way to achieve this
is to take the system of D0-branes on $T^d$ and transform it into a
system of $N$  D$d$-branes completely wrapped on the dual
torus. Then one could
hope that all the physics of M-theory on $T^d$ would be captured
by the SYM theory in $d+1$ dimensions which is the low-energy effective
action of this system of D$d$-branes. 
This has been called the SYM prescription for Matrix theory compactifications
\cite{bfss,tay1,snd,tay2}.

We will follow in this section Seiberg's formulation of Matrix theory 
\cite{seibergmatrix} (see Section \ref{MATRIXsec}). We recall
here that the two basic relations are \rref{rmp2} and \rref{rimp}:
\[
{R_s \over\tilde{l}_p^2}= {R \over l_p^2}, \qquad\qquad
{\tilde{L}_i \over \tilde{l}_p}= {L_i \over l_p},
\]
where $R$, $l_p$ and $L_i$ are the (finite) parameter of the original M-theory,
and the Matrix theory limit is $R_s \rightarrow 0$, 
together with the large $N$ limit.
For completeness we list here
the relations between quantities in the string theory in which the 
D$d$-branes live, and Matrix theory variables (see 
\cite{senmatrix,seibergmatrix} and equations \rref{msbehav}--\rref{gymbehav}):
\bea
l_s^2&=& {\tilde{l}_p^3\over R_s}=R_s^{1/2}{l_p^3\over R^{3/2}}, \nonumber \\
\Sigma_i&=&{l_s^2\over \tilde{L}_i}={l_p^3\over RL_i}, \nonumber \\
g_s&=&R_s^{3-d \over 4}{R^{3-d \over 4} l_p^{{3\over 2}(d-1)} \over V_d}, 
\nonumber \\
g_{YM}^2&=&g_s l_s^{d-3}={R^{3-d} l_p^{3(d-2)}\over V_d}. \label{genform}
\eea
$l_s$ and $g_s$ are respectively the string length and coupling;
$L_i$ and $\Sigma_i$ are the sizes of the torus in M-theory and in 
the auxiliary (T-dualized) 
string theory picture respectively, and $V_d=L_1\dots L_d$; 
$g_{YM}^2$ is the SYM coupling in $d+1$ dimensions, 
and it is dimensionful in $d\neq 3$.
We have singled out in all the quantities above the dependence in $R_s$.
In the Matrix theory limit, we always have that $\Sigma_i$ and $g_{YM}^2$ are
finite and $l_s \rightarrow 0$, but the string coupling blows up
for $d\geq 4$.

Now for $d\geq 4$ the SYM is ill-defined because non-renormalizable 
(see e.g. \cite{sei1}),
and thus the SYM prescription for Matrix compactification seems
to break down. However, what we should consider as a model for the 
description of M-theory on a torus is really the ``theory on the 
D-brane" and not only its low-energy field theory limit. 
Furthermore, to be able
to consider a system of $N$ D$d$-branes on its own, one has to take a 
limit in which the bulk physics in the auxiliary string theory decouples.
This limit has to be compatible with the Matrix theory limit.

For Matrix theory on $T^4$, it turns out \cite{roza1}
that the theory of D4-branes
at strong string coupling coincides with the 6 dimensional (2,0)
supersymmetric field theory (see \cite{sei1}), 
which is the theory of $N$ M5-branes
in flat space (i.e. at $L_p\rightarrow 0$). For Matrix theory on $T^5$, 
the theory of D5-branes
at strong coupling is mapped \cite{sei2} by a IIB S-duality to the theory of
$N$ NS5-branes at weak coupling but finite string tension, 
a theory which has string-like 
excitations. Finally, Matrix theory on $T^6$ is a theory of D6-branes
which, at strong coupling, becomes a theory of KK6-monopoles
\cite{brun,hana}. This 7-dimensional theory has membranes and, as we showed
in Section \ref{16THsec}, 
has a well-defined structure which has been called $m$-theory.

We will show in the remainder of this section how all the ``phases" of
$m$-theory (i.e. its 7- and 6-dimensional versions) describe M-theory
on $T^6$, and how some particular limits of them yield back the
compactifications on $T^5$ and $T^4$. In other words, we find the
theories mentioned above \cite{roza1,sei2}
as limits of the $iia$ and $iib$ little string theories.

Specializing now to $d=6$, we consider first $m$-theory in the D6-brane
picture. We have for the string coupling:
\be g_A=R_s^{-3/4}{l_p^{15/2}\over R^{3/4} V_6}. \label{gA} \ee
For the $m$-theory to be well-defined, its length scale $l_m$ has to be a fixed
parameter. Picking its value from Table \ref{litmlimits}, 
it takes the following expression in Matrix theory variables:
\be l_m^3=g_A l_s^3={l_p^{12}\over R^3 V_6}. \label{lm} \ee
We thus see that Matrix theory on $T^6$ is described by the theory 
on the D6-branes at $g_A \rightarrow \infty$, $l_s \rightarrow 0$
and keeping $l_m$ finite. This is exactly the limit used to define
$m$-theory. We conclude that $m$-theory is the candidate for the
Matrix theory description of M-theory on $T^6$.

Knowing \rref{lm} and the relations between $\Sigma$'s and $L$'s, we
can now translate the masses of the BPS states in $m$-theory into
masses of M-theory objects. We know in advance to which kind of objects
they will map to: since the BPS states break half of the supersymmetries
of the little theories, they correspond to objects of M-theory in the IMF
which break 1/4 of the supersymmetries. These are branes with travelling
waves in the 11th direction, i.e. longitudinal branes. The
remaining dimensions of these branes are wrapped on the $T^6$. One could 
also have deduced this from the fact that the IMF energies of these states
will be proportional to $n$ the number of BPS little branes, and
independent of $N$. Since these objects are string-like in the 5 dimensional
supergravity to which M-theory is reduced, they should carry the 27
magnetic charges of this theory (i.e. they should fit into
the ${\bf \overline {27}}$ of the U-duality group $E_6(Z)$ 
\cite{hulltownsend}). 
We indeed find the following identifications (see Table \ref{litmbps}, 
and Appendix \ref{ZOOapp} for the tensions of the M-branes): 
\[
M_{m2}={\Sigma_i \Sigma_j\over l_m^3}={R V_6\over L_i L_j l_p^6},
\]
the 15 $m$2 wrapped membranes
are mapped to longitudinal M5-branes; 
\[ M_w={1\over \Sigma_i}={RL_i\over l_p^3}, \]
the 6 momenta $w$ are mapped to longitudinal M2-branes; 
\[ M_{m5}={\tilde{V}_6 \over \Sigma_i l_m^6}={RV_6 L_i\over l_p^9} \]
the 6 $m$5 states are longitudinally wrapped
KK6 monopoles (the NUT direction being always on the $T^6$). 
All these 27 states can be found also
in the $iia$ and $iib$ pictures to be discussed below, although the
identification is less straightforward. This clearly convinces that 
the little string theories are 6-dimensional phases of a description
of M-theory on $T^6$.

We would also like to obtain the spectrum of the 27 electric charges in 5
dimensional supergravity (fitting into the {\bf 27} of $E_6(Z)$). These
correspond to completely wrapped branes in M-theory, or transverse branes
in the Matrix theory language (they can be represented as boosted branes). 
These objects preserve 16 supercharges in the Matrix model, 
and thus are totally supersymmetric states of the little theory. 
In the low-energy
SYM picture of the little theories, some of these transverse branes can
be associated to the electric and magnetic fluxes of the SYM 
\cite{tay2,gura,eliz}. However
the transverse M5-branes are missing from this description, which is
thus incomplete (note also that in the previous case there are no 
BPS states in the SYM which
would represent the longitudinal KK6, or $m$-theory's $m$5). Going
back to the D6-brane picture, one can find all these half-supersymmetric
states by embedding in the D6-branes other branes of type IIA theory
in a way that they make a non-threshold bound state
(the archetype of these states is the supergravity solution of 
\cite{izquierdo}).
These states can be found by chains of dualities from \cite{izquierdo}
and are: F1$\subset$D6, D4$\subset$D6 and KK5$\subset$D6.
The energy of the F1, for instance, is given by:
\[
E_{F1}=\sqrt{M_{D6}^2 +M_{F1}^2}-M_{D6}={M_{F1}^2\over 2 M_{D6}},
\]
since in the Matrix theory limit $M_{D6}\gg M_{F1}$. We obtain:
\[
E_{F1}= {\Sigma_i^2 l_m^3 \over 2 \tilde{V}_6}={R\over 2 L_i^2},
\]
where in the first equality we have already substituted for the $m$-theory
variables. This energy is to be compared with the energy of an object of
mass $M$ in the IMF, which is $E={1\over 2}R M^2$. The F1$\subset$D6
thus corresponds to transverse KK momenta on the $T^6$.
Much in the same way we can identify the D4$\subset$D6 and the
KK5$\subset$D6 as corresponding respectively to the transverse M2 and M5.
The energy of these states can be found e.g. in \cite{eliz}. When there
are $N$ D6-branes and $n$ other branes inside them, the IMF energy goes
like $n^2/N$. 

We now discuss the other pictures and the other little theories, along
with the relations between their parameters and the Matrix theory 
variables. It is clear that the parameters of the little theories,
once expressed in Matrix variables, will no longer depend on the
picture by which the little theory was defined. 

The $iia$ theory is most easily obtained going from the D6 to the D5
picture by T-duality. The reason to do this could be that one of the
radii of the original $T^6$ is much bigger than the others, and we might want
to decompactify it eventually. Then the parameters characterizing
the IIB auxiliary theory in which the D5 lives are given by:
\be g_B=R_s^{-1/2}{l_p^6\over R^{1/2} V_5}, \qquad l_s^2=R_s^{1/2}
{l_p^3\over R^{3/2}}, \qquad
R_B=R_s^{1/2} {L_6\over R^{1/2}}(=\tilde{L}_6), \label{d5param} \ee
where $V_5=L_1\dots L_5$. The parameters of the little $iia$ string theory
can be easily extracted using Table \ref{iialimits}:
\be g_a={V_5^{1/2}\over l_p^{3/2} L_6}, \qquad t_a={R^2 V_5\over l_p^9}.
\label{iiaparam} \ee
In the Matrix theory limit $R_s \rightarrow 0$ the $iia$ parameters
$g_a$ and $t_a$ are thus finite. Note however
that if $L_6\rightarrow \infty$ instead, then $t_a$ remains fixed
while $g_a$ inevitably goes to zero. In this limit all the branes of $iia$
except the $f$1 aquire an infinite tension and thus decouple. We are
left with a little string theory at zero coupling, which has exactly
the right number of states to describe Matrix theory on $T^5$. It has
indeed 5 winding plus 5 momentum BPS states, which together make up
the 10 longitudinal states of Matrix theory on $T^5$ \cite{sei2}.

To show more directly that the $ii$ strings tend exactly to the description of
Matrix theory on $T^5$ given by Seiberg \cite{sei2}, we go to
the NS5(B) picture of $iia$ strings. This is performed by an S-duality,
and we obtain for the IIB parameters:
\be \tilde{g}_B={1\over g_B}=R_s^{1/2}{R^{1/2} V_5\over l_p^6}, \quad
\tilde{l}_s^2=g_B l_s^2={l_p^9\over R^2 V_5}, \quad
\tilde{R}_B=R_B=R_s^{1/2} {L_6\over R^{1/2}}. \label{ns5bpar}\ee
Now $\tilde{g}_B\rightarrow 0$ and $\tilde{l}_s$ is finite in the
Matrix theory limit.
This was the original setting for the definition of the little string 
theories proposed by Seiberg to describe Matrix theory on $T^5$.
Consistently with the picture of Section \ref{16THsec},
they are thus the zero coupling limit of the more
complete type $ii$ little string theories that describe Matrix theory on $T^6$.

In order to go to the $iib$ theory, we perform a T-duality along, say,
the $\hat{5}$ direction. We obtain a NS5-brane in a IIA theory
characterized by:
\be
\tilde{g}_A={\tilde{g}_B \tilde{l}_s \over \Sigma_5}=
R_s^{1/2}{R^{1/2} V_4^{1/2}L_5^{3/2}
\over l_p^{9/2}}, \quad  \tilde{l}_s^2={l_p^9\over R^2 V_4 L_5}, \quad
\tilde{R}_A=R_s^{1/2} {L_6\over R^{1/2}}, \label{ns5apar} \ee
with $V_4=L_1\dots L_4$. It is worth noting that from the $iib$ point
of view, the 5th direction has a radius:
\be \Sigma_5'={\tilde{l}_s^2\over \Sigma_5}={l_p^6\over R V_4}.
\label{5prime} \ee
This expression is finite and has forgotten all dependence on $L_5$.
Thus, we should no longer think of the fifth direction of the NS5(A) brane
as related to the fifth direction of the original $T^6$. Moreover, from
\rref{genform}
we can identify $\Sigma_5'=g_{YM(4+1)}^2$, as in \cite{roza1}.
The parameters of the $iib$ theory are given by:
\be
g_b={L_5\over L_6}, \qquad t_b={R^2 V_4 L_5\over l_p^9}. \label{iibparam}\ee
Of course, we could have computed these parameters without leaving
the little string theories, by $t$-duality from the $iia$-theory.
For $L_6 \rightarrow \infty$ and $L_5$ finite, $t_b$ can be fixed
but $g_b \rightarrow 0$ and we recover again the second string theory
with 16 supercharges proposed by Seiberg \cite{sei2}. 
When $L_5, L_6 \rightarrow \infty$,
i.e. when we wish to consider a compactification on $T^4$,
the tension of the little strings becomes very large, only the massless
modes contribute,
and we are left with a field theory of a special kind, which is however
still 6 dimensional. 
To properly identify it, we can go to the M5 picture, since when
$L_5$ is non compact we are at strong IIA coupling.

The M5 picture is easily obtained by decompactification of a new
direction in the auxiliary M-theory, the radius of which we denote as $R_1$.
The parameters are thus:
\be R_1=\tilde{g}_A \tilde{l}_s=\tilde{L}_5, \quad R_2=\tilde{L}_6, \quad
L_p^3=R_s^{1/2}{l_p^9 \over R^{5/2} V_4}. \label{mpar}\ee
We clearly have $L_p\rightarrow 0$, which means that
Matrix theory on $T^4$ is described by the theory on M5-branes in flat
space.  We have thus reproduced the results of \cite{roza1}.

As a last remark on this issue, note that we could have gone to the M5
picture from the D5 one through a T-duality on $\hat{5}$ which
would have transformed the D5 into a D4, and then elevating the latter
to an M5-brane. Though the labelling of the directions in the auxiliary
theory is clearly different in this M5 from the one of the previous
paragraph, when expressed in Matrix variables the quantities are exactly
the same. This is related to the fact shown in \rref{5prime}
that in the $iib$ picture
the ``base space" does not refer any more to the original $L_5$.

\section{Discussion}
\label{DISCUsec}

We have given in this chapter a description of little theories in 6 and
7 dimensions. Our analysis is entirely
based on the spectrum of BPS states present in each one of these
theories. The focus on BPS states was partly motivated by the
application of these little theories to the Matrix theory description
of M-theory compactifications, and to the necessity to recover
the right U-duality group. 
It would be however very interesting to pursue the study
of these non-critical string theories and $m$-theory beyond the BPS analysis.
A full quantum and possibly non-perturbative formulation of these theories will
elucidate the direct relation between the little string theories or 
$m$-theory and
their low-energy effective actions, which were argued not to contain gravity.
In other words, this formulation should reproduce the low-energy effective
action of the host branes used to define the little theories. It may also help
in understanding the full structure of the (2,0) field theory in 6 dimensions.

A way to approach these theories is to give them a Matrix-like formulation.
This means that one would hope to study them non-perturbatively
by formulating them in the infinite momentum 
frame or in the discrete light cone quantization. This is particularly
appealing for the $(2,0)$ theory since the low-energy field theory
has no perturbative limit. This approach has been initiated in
\cite{aha1,withiggs,hana} and was then applied to several other
related theories.
We will here present it only very schematically.

The basic idea is to build the Matrix model for $m$-theory along
the same lines as the Matrix model for M-theory, i.e. Matrix theory.
In this line of thought, the $d$0-branes of type $iia$ little strings
could be the partons of $m$-theory in the IMF (or in the DLCQ).
We can actually guess the hamiltonian governing their dynamics,
following \cite{hana}. If we take for instance the NS5(B) picture
for the $iia$ theory, then the $d$0-branes are given by the
D1-branes winding around the transverse compact direction. Generically,
we have $N$ NS5-branes with $n$ D1-branes suspended between each pair 
of NS5-branes. This is again a configuration similar to a Hanany-Witten
set up \cite{hananywitten}, and the theory of the $d$0-branes
can then be formulated along the same lines. 
The theory will have 8 supercharges, and scalars both in the adjoint of the
gauge groups, describing the separation of the D1-branes between themselves
(parametrizing the Coulomb branch), and in the (bi)fundamental, representing the
motion away from the NS5-branes (parametrizing the Higgs branch).
The Matrix model of $m$-theory is then the quantum mechanics on the
Coulomb branch of the theory above. For the model to be well defined,
the Coulomb branch has to decouple from the Higgs branch (which is 
equivalent in this picture to asking that the D1-branes do not leak
into the bulk), and this limit has to coincide with the region
of validity of $m$-theory.

Let us only mention that the approach of \cite{aha1,withiggs} was
somewhat different, also because they attempted to describe
the $(2,0)$ field theory and the related little string theory of \cite{sei2}
respectively. There the partons of the IMF formulation of the
$(2,0)$ field theory were taken to be the instantons in 4+1 dimensional
SYM, i.e. the theory reduces to the dynamics of D0-branes within D4-branes. 
This corresponds
to the quantum mechanics on the Higgs branch of a definite theory.
It is possible to relate the two approaches by acting with some
dualities on the former one and then taking some limits, as discussed
in \cite{hana}.

The little theories described in this chapter are clearly not yet 
firmly established, nevertheless they seem not only interesting
on their own, but also important in the more vast context of 
M-theory. The knowledge of the little theories could be crucial
in understanding some particular problem in M-theory, or alternatively
they could be used as a tool for studying some other issue. 
Their last, but in our opinion not the least, attractive feature
is that, as they seem to closely reproduce the web of dualities
of M-theory, they could be used as a fairly accurate toy model for it.

\chapter{The four dimensional Schwarzschild black hole as an extremal brane 
configuration}
\label{SCHWARZchap}
\markboth{CHAPTER \protect \ref{SCHWARZchap}. THE SCHWARZSCHILD BLACK HOLE}{}

In Chapter \ref{PBRANEchap} we have shown that using intersecting brane
configurations, one can build five and four dimensional black holes 
which, despite
being extremal and supersymmetric, have a non-vanishing horizon area.
The latter in turn  results in a non-vanishing Bekenstein-Hawking entropy.
These black holes can be considered from the standpoint of an equivalent
configuration (i.e. a configuration defined by the same conserved charges)
in the framework of perturbative string theory with D-branes. Precisely
the supersymmetric properties of these black holes allow a computation
of their statistical entropy in the weakly coupled D-brane picture.
The two entropies thus computed match exactly.

This being an enormous advance in the understanding of black hole entropy,
it is however still disappointing to have such an explanation only
for extremal (or near-extremal) black holes. After all, one would like
to understand black hole entropy in the simplest case in which it
arises, namely for the Schwarzschild black hole.

In this chapter we show that it is possible, performing several
boosts and dualities on a neutral black brane of M-theory to which
the Schwarzschild black hole is related by trivial compactification,
to map it onto a configuration of intersecting
branes with four charges. The infinite boost limit is well-defined and
corresponds to extremality for the intersecting brane configuration.
The Bekenstein-Hawking entropy of the four dimensional Schwarzschild
black hole is thus exactly reproduced by the statistical entropy
of the D-brane configuration.

The chapter closely follows the paper \cite{schwarzbh}.

\section{Black hole entropy in M-theory}

The formulation of some long-standing problems of black hole quantum physics
in the new framework of M-theory has proven successful.
Using D-brane techniques a statistical explanation of the
entropy for some black holes has been discovered. The entropy has
been computed in terms of the degeneracy of
D-brane configurations describing at weak coupling, charged 
black holes in the extremal and near-extremal limit 
\cite{stromingervafa,callanmaldacena,horstrom,maldastrom,johnkhuri,%
horolow,balalars}.
Unfortunately, this systematic approach cannot be applied directly to
neutral Schwarzschild black holes. Microscopic considerations
based on Matrix theory
have however been discussed recently \cite{bfksa,klesu,li,homa,bfksb,ohtazhou}.
Other considerations \cite{skend} involve a connection between the
Schwarzschild black hole and the 2+1 dimensional BTZ black hole \cite{btz},
which has been given a microscopic description in \cite{carlip,strom,birm,%
maldastrombtz,martinecbtz,behrndt}.

In \cite{das}, it is proposed to view a
Schwarzschild black hole as the compactification of a black brane
in 11 dimensional supergravity and to relate it to a charged black
hole with the same thermodynamic entropy. The charged black hole is
obtained by subjecting the black brane to a boost \cite{tseyt} in an internal
uncompactified direction followed by Kaluza-Klein reduction
on a different radius \cite{das} (see also \cite{homa}).
Following this idea,
a quantitative analysis of Schwarzschild black holes applying some M-theory
concepts has been suggested in \cite{fran}.
There, a near extremal limit is
defined, in which the Schwarzschild radius remains arbitrarily large
at infinite boost. It is proposed to use this limit to obtain the entropy of
Schwarzschild black holes from the microscopic entropy of the charged ones,
viewed as systems of D-branes. 
This was applied in \cite{fran} to the seven dimensional black hole, 
mapped onto a near extremal system of D3-branes.

In this chapter we apply this proposal to four dimensional Schwarzschild
black holes.
In order to relate a four dimensional Schwarzschild black hole to a
``countable" D-brane configuration, the procedure is more involved
because we will have to perform several boosts and dualities \cite{das}. More
precisely, each boost creates a (Ramond-Ramond) charge, and we will end
up with a configuration of intersecting branes with four charges. 
In this case, the infinite boost limit leads exactly to extremality
where the configuration is marginal and a standard microscopic
counting of the entropy can be safely performed (because it is protected
by BPS arguments). In this way
the Bekenstein-Hawking entropy of the four dimensional Schwarzschild 
black hole is exactly recovered.

\markright{\protect \ref{MAPPsec}. MAPPING A NEUTRAL BLACK BRANE...}
\section{Mapping a neutral black brane on a configuration with four charges}
\label{MAPPsec}
\markright{\protect \ref{MAPPsec}. MAPPING A NEUTRAL BLACK BRANE...}

The metric of a four dimensional Schwarzschild black hole is:
\be
ds^2=-f dt^2+f^{-1} dr^2+r^2 d\Omega_2^2, \qquad f=1-{r_0\over r}.
\label{schwarz} \ee
This metric can be trivially embedded in an 11 dimensional space-time
simply by taking its product with a flat 7-dimensional (compact) space.
It thus corresponds to a neutral black seven-brane\footnote{
Note that we will always take $r_0\gg R, L_i$. In this regime the black
brane is entropically favoured with respect to a higher dimensional
black hole (see for a recent discussion \cite{homa,das}).
} compactified on
a $T^6 \times S^1$ characterized by sizes $L_i$ ($i=1\dots 6$) and $2\pi R$:
\be
ds^2=-f dt^2+dz^2+\sum_{i=1}^6 dx_i^2 +f^{-1} dr^2+r^2 d\Omega_2^2.
\label{schwarzbrane}
\ee
The four and eleven dimensional Newton constants, respectively
$G_4$ and $G_{11}\equiv l_p^9$, are related by:
\be 
G_4={l_p^9 \over 2\pi L_1\dots L_6 R} \label{newton}.
\ee
The Bekenstein-Hawking entropy of the black hole described by \rref{schwarz},
or equivalently by \rref{schwarzbrane},
is given by:
\be 
S_{BH}={A_h\over 4 G_4}\equiv {\pi r_0^2\over G_4}=
2\pi^2 {L_1\dots L_6 R \over l_p^9} r_0^2.
\label{sbh} \ee

We now consider this neutral black seven-brane in the framework of M-theory.
Most generally, we recall the precise relation between the parameters
of M-theory compactified on a circle $S^1$ (i.e. the eleven dimensional
Planck length $l_p$ and the radius $R$) and the parameters of type IIA
string theory, i.e. the string coupling $g_s$ and the string length
$l_s\equiv \sqrt{\alpha'}$ (see Section \ref{DUALMsec}):
\begin{eqnarray}
R&=&g_s l_s, \label{r11} \\
l_p^3&=& 2^{4/3} \pi^{7/3} g_s l_s^3. \label{lpl11}
\end{eqnarray}
This gives the 10 dimensional Newton coupling constant:
\be 
G_{10}={l_p^9 \over 2\pi R}=8\pi^6 g_s^2 l_s^8. \label{G10}
\ee

Using boosts (in a sense to be defined below) and dualities, we will
map the above black brane onto a configuration of intersecting branes
carrying 4 Ramond-Ramond charges. 
We will then show that there exists a limit in which the latter
configuration approaches extremality in such a way that the 
statistical evaluation of its entropy is well-defined.
 
We now proceed to the careful description of all the steps leading to the
final configuration corresponding to the intersection
D4$\cap$D4$\cap$D4$\cap$D0, which is a marginal bound state in the
extremal limit.

Suppose that the coordinate $z$ is (momentarily) non-compact. 
For instance, we can
simply consider that we have gone to the covering space of the $S^1$ factor
of the compact space over which the neutral black brane is wrapped.
We can now perform a boost of rapidity $\alpha$ in that direction:
\be
\begin{array}{rcl}
t&=& \cosh\alpha\ t' +\sinh\alpha\ z', \\
z&=& \sinh\alpha\ t' +\cosh\alpha\ z'.
\end{array}
\label{schboost}
\ee
The metric of the 
Schwarzschild black hole embedded in 11 dimensions is of course
not invariant under the boost \rref{schboost}. 
It acquires a non-vanishing off-diagonal $g_{t'z'}$ component:
\[
ds^2=-H_\alpha^{-1} f {dt'}^2 +H_\alpha\left(dz'+H_\alpha^{-1} 
{r_0\over r}\sinh\alpha \cosh\alpha
dt'\right)^2+ \sum_{i=1}^6 dx_i^2 +f^{-1} dr^2+r^2 d\Omega_2^2, \]
\be 
H_\alpha=1+{r_0\over r} \sinh^2 \alpha.
\label{boostedsch}
\ee
The resulting configuration, as viewed in 10 dimensions, is a non-extremal
set of D0$_1$-branes (smeared on the $T^6$); the subscript indicates
that the D0-branes have been created by the first boost. The 10 dimensional
fields are the following:
\[
ds^2=-H_\alpha^{-7/8}f dt^2 +H_\alpha^{1/8}\left(\sum_{i=1}^6 dx_i^2 +
f^{-1} dr^2+r^2 d\Omega_2^2\right),
\]
\be
e^\phi =H_\alpha^{3/4}, \qquad\qquad A_t=H_\alpha^{-1}
{r_0\over r}\sinh\alpha \cosh\alpha, \label{blackd0}
\ee
where the metric is in the Einstein frame and we have dropped the primes.
The fields above are the ones of a black brane carrying electric charge
under the 1-form RR potential.

We have now to pay attention to the physical meaning of the boost,
remembering that the coordinates $z$ and $z'$ are eventually compactified.
The boost \rref{schboost} is a coordinate transformation if $z$ parametrizes
the covering space of the $S^1$, but identifications at constant
$t$ or $t'$ clearly give two different compactified 10 dimensional theories.

If we compactify at constant $t'$, i.e.
in the boosted frame, the length $R$ is rescaled to the value:
\be R'={R\over \cosh \alpha}. \label{rprime} \ee
We thus define a new compactification identifying the boosted coordinate $z'$
on intervals of length $2\pi R'$. This is not a constant time compactification
as viewed from the unboosted $z$ frame. The 10 dimensional theories defined 
respectively by compactification on $R$ and $R'$ are thus different (for
instance, the 10 dimensional Newton constants will differ).
They nevertheless describe the same physics
at the horizon where all time intervals
are blueshifted to zero. This can be seen also from the fact that
if we compute the proper length of the circle parametrized by $z'$
in \rref{boostedsch}, we find $2\pi R$ precisely if the calculation
is performed at the horizon.

Transformations like \rref{schboost} are thus not coordinate transformations
for the 10 dimensional theory, neither ordinary dualities since 
from the 10 dimensional point of view a charge is created, i.e. a
new parameter comes into the solution. 
This can also be seen by the fact that in 11 dimensions they
mix a spatial coordinate with time.
Note that these transformations can 
nevertheless be considered as dualities if one compactifies also
the time coordinate \cite{tseyt,timedual}.

This `new' IIA string theory is
characterized by parameters $g_s'$ and $l_s'$ which have a well-defined
dependence on the boost parameter $\alpha$ given
by \rref{r11} and \rref{lpl11} where $R$ is replaced by $R'$.

In the following step we perform T-dualities in the directions, say,
$\hat{1} \hat{2} \hat{3} \hat{4}$ of $T^6$ to obtain a non-extremal
configuration of D4$_1$-branes. We use the standard T-duality relations:
\be
L_i \rightarrow {4\pi^2 l_s^2 \over L_i}, \qquad
g_s \rightarrow g_s {2\pi l_s \over L_i}. \label{schtduality}
\ee

We now uplift this IIA configuration to 11 dimensions. Note that this
is a ``new" M-theory in the sense that the Planck length is now a function
of $\alpha$ and the dependence on $\alpha$ of the radius
of compactification has changed. We have now a non-extremal M5-brane.

To create a second charge, we perform a boost of parameter
$\beta$ on the eleventh direction, which is longitudinal to the M5-brane. 
The M5-brane is however not invariant under the boost due to its 
non-extremality.  Following the same procedure
as for the first boost, the radius of compactification of the new
M-theory is rescaled
by $1/\cosh\beta$. After compactification, the resulting IIA configuration
corresponds to the non-extremal version of the marginal bound state 
D4$_1\cap$D0$_2$.

We then T-dualize on the $\hat{1} \hat{2} \hat{5} \hat{6}$ directions,
leading to a D4$_1\cap$D4$_2$ configuration. The first set of D4-branes
lies now in the $\hat{3} \hat{4} \hat{5} \hat{6}$ directions.
Note that the T-dualities are always performed in order to obtain a
configuration such that, when going to M-theory, the 11th direction
is common to all the branes in the 11 dimensional configuration.

Uplifting to eleven dimension  for the second time, the parameters
now depend on the two boosts $\alpha$ and $\beta$. We are now ready
to create a third charge, performing a third boost of parameter
$\gamma$. 

This results, after compactification and T-dualities over 
$\hat{1} \hat{2} \hat{3} \hat{4}$, to a non-extremal configuration
D4$_1\cap$D4$_2\cap$D4$_3$, lying respectively in the
$\hat{1} \hat{2} \hat{5} \hat{6}$, $\hat{3} \hat{4} \hat{5} \hat{6}$
and $\hat{1} \hat{2} \hat{3} \hat{4}$ directions.

A last uplift--boost--compactification procedure characterized
by a boost parameter $\delta$ leads to our final configuration
D4$_1\cap$D4$_2\cap$D4$_3\cap$D0$_4$. The corresponding metric in the
Einstein frame is (see e.g. \cite{cvetictseytlin,ivanov,ohta}):
\begin{eqnarray}
ds^2&=&-H_{\alpha}^{-{3\over 8}}H_{\beta}^{-{3\over 8}}
H_{\gamma}^{-{3\over 8}}H_{\delta}^{-{7\over 8}}f dt^2+
H_{\alpha}^{-{3\over 8}}H_{\beta}^{5\over 8}
H_{\gamma}^{-{3\over 8}}H_{\delta}^{1\over 8}(dy_1^2+dy_2^2) \nonumber \\
&&+H_{\alpha}^{5\over 8}H_{\beta}^{-{3\over 8}}
H_{\gamma}^{-{3\over 8}}H_{\delta}^{1\over 8}(dy_3^2+dy_4^2) +
H_{\alpha}^{-{3\over 8}}H_{\beta}^{-{3\over 8}}
H_{\gamma}^{5\over 8}H_{\delta}^{1\over 8}(dy_5^2+dy_6^2) \nonumber \\
&&+H_{\alpha}^{5\over 8}H_{\beta}^{5\over 8}
H_{\gamma}^{5\over 8}H_{\delta}^{1\over 8}(f^{-1}dr^2+r^2d\Omega_2^2),
\label{conf}
\end{eqnarray}
where
\be
f=1-{r_0\over r}, \qquad H_\alpha=1+{r_0\over r} \sinh^2 \alpha,
\label{harmfns} \ee
and similarly for $H_{\beta}$, $H_{\gamma}$ and $H_{\delta}$. The non-trivial
components of the RR field strengths are:
\begin{eqnarray}
\tilde{F}_{ty_1y_2y_5y_6r}&=&-\partial_r \left(H_\alpha^{-1} {r_0\over r}
\cosh\alpha \sinh\alpha\right), \nonumber \\
\tilde{F}_{ty_3y_4y_5y_6r}&=&-\partial_r \left(H_\beta^{-1} {r_0\over r}
\cosh\beta \sinh\beta\right),  \\
\tilde{F}_{ty_1y_2y_3y_4r}&=&-\partial_r \left(H_\gamma^{-1} {r_0\over r}
\cosh\gamma \sinh\gamma\right), \nonumber \\
F_{tr}&=&-\partial_r\left(H_\delta^{-1} {r_0\over r}
\cosh\delta \sinh\delta\right), \nonumber 
\label{fields}
\end{eqnarray}
where $\tilde{F}_6$ is the 10 dimensional Hodge dual of the 4-form
RR field strength.

The string coupling and string length of the type $\widehat{
\mbox{IIA}}$ theory in which
this configuration is embedded are, in terms of the original quantities
appearing in \rref{newton} and \rref{sbh}:
\begin{eqnarray}
\hat{g}_s&=&4\pi^{1\over2} {l_p^{9\over2}\over L_1L_2L_5L_6 R^{1\over2}}
\left({\cosh\alpha \cosh\beta \cosh\gamma\over \cosh^3\delta}\right)^{1\over2},
\label{ghat}\\
\hat{l}_s^2&=&{1\over 2^{4\over3}\pi^{7\over3}} {l_p^3\over R}
\cosh\alpha \cosh\beta \cosh\gamma \cosh\delta \label{lshat}
\quad (=l_s^2 \cosh\alpha \dots \cosh\delta) .
\end{eqnarray}
The lengths of the final 6-torus over which the above configuration
is wrapped are:
\begin{eqnarray}
\hat{L}_{1,2}&=&{2^{2\over3}\over\pi^{1\over3}}{l_p^3\over R L_{1,2}}
\cosh\alpha \cosh\gamma, \nonumber \\ 
\hat{L}_{3,4}&=&L_{3,4}\cosh\beta \cosh\gamma, \label{Lhat}\\
\hat{L}_{5,6}&=&{2^{2\over3}\over\pi^{1\over3}}{l_p^3\over R L_{5,6}}
\cosh\alpha \cosh\beta. \nonumber 
\end{eqnarray}

We now compute the charge densities of the D-branes in this configuration:
\begin{eqnarray}
Q_{D4_1}&=& {1 \over 16 \pi \hat{G}_{10}} \int_{\hat{T}^2_{(3,4)} \times
S^2} F_4 = {\pi^{7\over3} \over 2^{11 \over 3}} 
{L_1^2 L_2^2 L_3 L_4 L_5^2 L_6^2 R^5 r_0 \tanh \alpha\over l_p^{21} 
\cosh^3 \alpha \cosh^3 \beta \cosh^3 \gamma \cosh \delta}, \nonumber \\
Q_{D4_2}&=& {1 \over 16 \pi \hat{G}_{10}} \int_{\hat{T}^2_{(1,2)} \times
S^2} F_4 = {\pi^{5\over3} \over 2^{7 \over 3}} 
{L_1 L_2  L_5^2 L_6^2 R^3 r_0 \tanh \beta \over l_p^{15}
\cosh^3 \alpha \cosh^3 \beta \cosh^3 \gamma \cosh \delta}, \nonumber \\
Q_{D4_3}&=& {1 \over 16 \pi \hat{G}_{10}} \int_{\hat{T}^2_{(5,6)} \times
S^2} F_4 = {\pi^{5\over3} \over 2^{7 \over 3}} 
{L_1^2 L_2^2  L_5 L_6 R^3 r_0 \tanh \gamma \over l_p^{15} 
\cosh^3 \alpha \cosh^3 \beta \cosh^3 \gamma \cosh \delta}, 
\label{charges} \\
Q_{D0_4}&=& {1 \over 16 \pi \hat{G}_{10}} \int_{\hat{T}^6 \times
S^2}\star F_2 = {\pi \over 2}
{L_1 \dots L_6 R r_0 \sinh \delta \over l_p^{9}
\cosh \alpha \cosh \beta \cosh \gamma }. \nonumber
\end{eqnarray}
The charge densities above are normalized in such a way that
the elementary D-branes have a charge density equal to their 
tension \cite{tasi}.

The tensions of elementary D0 and D4-branes are given by (see Appendix 
\ref{ZOOapp}):
\begin{eqnarray}
T_{D0} &=& {1 \over \hat{g}_s \hat{l}_s} = {\pi^{2\over3} \over 2^{4 \over 3}}
{L_1 L_2  L_5 L_6 R \over l_p^{6}} {\cosh \delta
\over \cosh \alpha \cosh \beta \cosh \gamma }, \label{tensions}\\
T_{D4} &=& {1 \over 2^4 \pi^4  \hat{g}_s \hat{l}_s^5} = 
{\pi^{4\over3} \over 2^{8 \over 3}}{L_1 L_2  L_5 L_6 R^3 \over l_p^{12}}
{1 \over \cosh^3 \alpha \cosh^3 \beta \cosh^3 \gamma \cosh \delta}. \nonumber
\end{eqnarray}

Using \rref{charges} and \rref{tensions}, we can now compute the different
numbers of constituent D-branes of each type:
\begin{eqnarray}
N_1&=&Q_{D4_1} T_{D4}^{-1} = {\pi \over 2}{L_1 \dots L_6 R^2 \over l_p^{9}}
r_0 \tanh \alpha, \nonumber \\
N_2&=&Q_{D4_2} T_{D4}^{-1} =( 2\pi)^{1 \over 3}
{L_5  L_6 \over l_p^{3}} 
r_0 \tanh \beta, \nonumber \\
N_3&=&Q_{D4_3} T_{D4}^{-1} =( 2\pi)^{1 \over 3}
{L_1  L_2 \over l_p^{3}} 
r_0 \tanh \gamma, \label{numbers} \\
N_4&=&Q_{D0_4} T_{D0}^{-1} =( 2\pi)^{1 \over 3}
{L_3  L_4 \over l_p^{3}} 
r_0 \tanh \delta. \nonumber 
\end{eqnarray}

Strictly speaking,
these numbers represent the number of branes only in the extremal limit,
but can be interpreted more generally as the difference between the number
of branes and anti-branes.
Note that it is also
possible to compute the numbers \rref{numbers} by evaluating after every boost 
the number of D0 branes created.
Indeed, the numbers are strictly invariant under all the subsequent
dualities and further boosts.

\markright{\protect \ref{BOOsec}. STATISTICAL ENTROPY}
\section{Infinite boost limit, extremality, and statistical entropy}
\label{BOOsec}
\markright{\protect \ref{BOOsec}. STATISTICAL ENTROPY}

We will now show that taking all the boost parameters to infinity 
with $r_0$ kept fixed is
equivalent to taking the extremal limit on the intersecting D-brane
configuration. In order to do this, we compute the ADM mass
applying \rref{admmassformula} to \rref{conf}:
\be 
M={\pi\over4}{L_1\dots L_6 R\over l_p^9} r_0
{\cosh 2\alpha+\cosh 2\beta +\cosh 2\gamma +\cosh 2\delta\over
\cosh \alpha \cosh \beta \cosh \gamma \cosh \delta}.\\
\label{mass}
\ee
This formula is to be compared with the extremal value derived from the
charge densities \rref{charges}:
\bea
M_{ext}&=&Q_{D4_1}\hat{L}_1\hat{L}_2\hat{L}_5\hat{L}_6+
Q_{D4_2}\hat{L}_3\hat{L}_4\hat{L}_5\hat{L}_6+
Q_{D4_3}\hat{L}_1\hat{L}_2\hat{L}_3\hat{L}_4+
Q_{D0_4} {{}\over {}}\nonumber \\
&=&{\pi\over4}{L_1\dots L_6 R\over l_p^9} r_0
{\sinh 2\alpha+\sinh 2\beta +\sinh 2\gamma +\sinh 2\delta\over
\cosh \alpha \cosh \beta \cosh \gamma \cosh \delta}.
\label{mext}
\eea
It is then easy to show that when all the boost parameters are equal and
are taken to infinity, the departure from
extremality rapidly goes to zero as:
\be
{M-M_{ext}\over M_{ext}}\sim e^{-4\alpha}. \label{extremality}
\ee
Note that in the same limit both $M$ and $M_{ext}$ go to zero
as $e^{-2\alpha}$. However we have to take into account the formulas
\rref{ghat} and \rref{lshat} which tell us that $\hat{g}_s$ remains
finite and $\hat{m}_s\equiv \hat{l}_s^{-1}$ goes to zero as
$e^{-2\alpha}$. Thus the masses above are finite in string units.
Note also that for all the internal directions the ratio
$\hat{L}_i/\hat{l}_s$ is finite\footnote{This result and the
finiteness of $\hat{g}_s$ imply that the string units and the
Planck units in 4 dimensions are of the same order in the boost
parameters $\alpha$.}.

Despite the fact that $\hat{m}_s$ is vanishingly small in the limit
discussed, we can neglect the massive string modes because the
Hawking temperature goes to zero even faster. Indeed, 
applying \rref{hawkingtemp} to \rref{conf}, we have:
\be
T_H={1 \over 4\pi r_0 \cosh \alpha \cosh \beta \cosh \gamma \cosh \delta},
\label{hawking}
\ee
which gives a $T_H/\hat{m}_s$ of order $e^{-2\alpha}$ in the extremal limit.

For the supergravity description to be valid and not corrected
by higher order curvature terms, we have to check that the typical
length scale derived from the Riemann tensor is much bigger than
the string length. This can indeed be computed, and it holds as soon
as $r_0\gg l_p$. 

In this infinite boost limit, the numbers of constituent D-branes 
\rref{numbers} tend to a finite value, which moreover is large if
we take $r_0\gg l_p$ as above and we assume that $R$ and all $L_i$
are of order $l_p$ or bigger.
The fact that the $N$'s are large but tend to a finite value
is a crucial element for the validity of our mapping procedure.

We are now in position to compute the microscopic degeneracy of this
extremal D4$\cap$D4$\cap$D4$\cap$D0 configuration. This configuration
can be related to the one considered in \cite{maldastrom} 
(see also \cite{maldathesis}) by 
a series of T- and S-dualities (see Appendix \ref{ZOOapp}). 
Indeed, performing a T-duality in 
the $\hat{5}$ direction, followed by a type IIB S-duality, and then by
three more T-dualities in the directions labelled by $\hat{5}$, $\hat{1}$
and $\hat{2}$, one ends up with a configuration
consisting of $N_2$ D6-branes wrapped on the whole $T^6$, $N_3$ 
NS5-branes lying in the $\hat{1} \hat{2} \hat{3} \hat{4} \hat{5}$ directions,
$N_1$ D2-branes in the $\hat{5} \hat{6}$ directions (which become actually
$N_1 N_3$ after breaking on the NS5-branes) and finally $N_4$ quanta of
momentum in the $\hat{5}$ direction. Going to flat space, the degeneracy
of these momentum excitations can be computed as in \cite{maldastrom}.

Let us review very briefly how this counting goes (a complete and 
introductory account 
can be found in \cite{maldathesis}; it is also useful to keep in mind
the D-brane physics, as reviewed in Section \ref{DBRANEsec}).
Consider first the system in which $N_{D2}$ D2-branes lie parallel
to $N_{D6}$ D6-branes. There will be all sorts of open strings
connecting the various D-branes. The low-energy theory associated
with this system will be based on the gauge group $U(N_{D2})\times
U(N_{D6})$, and will have 8 supercharges (thus corresponding to a
$\cal N$=2 theory in $D=4$). We now search for the massless modes that
can contribute to the $n$ units of momentum, which flow along one of the
directions of the D2-branes. The scalars in the vector multiplets
will parametrize motion in which the D2-branes are away from the D6-branes,
while the scalars in the hypermultiplets will parametrize motion of
the D2-branes within the D6-branes. It turns out (see \cite{maldathesis})
that one can take into account only the hypermultiplets in the bifundamental,
which correspond to strings going from D2-branes to D6-branes and vice-versa.
A careful study of the quantization of open strings with $\nu=4$
Neumann-Dirichlet directions tells us that there is a total of 
$N_B=4N_{D2}N_{D6}$
such massless bosonic degrees of freedom, accompanied by an equal number $N_F$ 
of fermionic degrees of freedom.

Now the $N_{NS5}$ NS5-branes come into play. The configuration is such that
the D2-branes can open and end on them (see Chapter \ref{OPENchap}).
We thus end up having a total amount of $N_{NS5}N_{D2}$ D2-branes, each
one suspended between two NS5-branes. The momentum flows along the 
intersection. We now have $N_B=N_F=4N_{NS5}N_{D2}N_{D6}$ massless modes
which contribute to it. The logarithm of the degeneracy of this 
2 dimensional gas of massless modes is then given, for $n\gg N_B$, by 
(see e.g. \cite{gsw}):
\be
\ln d(n)=2\pi \sqrt{{1\over 6}\left(N_B+{1\over 2}N_F\right)n}\equiv S_{micro}. 
\label{degener}
\ee

Reinserting the numbers of each set of component branes as given before, 
we have that $N_B=N_F=4N_1 N_2 N_3$ and $n=N_4$.
The statistical entropy is then given for large $N$'s by\footnote{
The condition $N_4\gg N_1 N_2 N_3$ can be released following the usual
trick of considering multiply wrapped branes \cite{maldasuss,maldathesis}.}:
\be
S_{micro}=2\pi \sqrt{N_1 N_2 N_3 N_4}. \label{smicro}
\ee
Note that all the $N$'s, which by \rref{numbers} are proportional to 
$r_0$, can be taken arbitrarily large because our limit has been 
defined keeping $r_0$ fixed and, in fact, arbitrarily large.
Using the values \rref{numbers} in the infinite boost limit,
we find:
\be
S_{micro}=2\pi^2 {L_1\dots L_6 R \over l_p^9} r_0^2= {\pi r_0^2\over G_4}
={A_h\over 4 G_4},
\ee
in perfect agreement with the Bekenstein-Hawking entropy of the
original four dimensional Schwarzschild black hole given
in \rref{sbh}.

We end this chapter with some comments on the above result.

At first sight, counting states of a Schwarzschild black hole through
a mapping onto an extreme BPS black hole seems a very indirect procedure.
However the mapping crucially rests on the relation between 
compactification radii (see e.g. Eq. \rref{rprime}) 
ensuring equality of semi-classical thermodynamic
entropies. As we have seen, this mapping is equivalent,
on the horizon only, to a coordinate transformation 
in eleven dimensions.
The physics outside the horizon is different and in spacetime we have
two distinct physical systems. Nevertheless, the very fact that entropy
was obtained by a counting of quantum states strongly suggests that the
two different systems can be related by a reshuffling of degrees of
freedom defined on the horizon.

Another crucial element of our computation is the fast convergence
in the infinite boost limit to a configuration of extremal BPS D-branes.
This is in sharp contradistinction to the case examined in \cite{fran}
where all the entropy comes from a departure from extremality. This 
arises because the slow vanishing of the excess mass $\Delta M\equiv M-M_{ext}$
in that case
is exactly compensated by the growth of the internal volume to give 
a finite value to the entropy of the non-BPS excitations of the D3-branes.
In our case the situation is different. The product $\Delta M L$ goes
to zero in the limit, leaving a pure BPS state which can then be safely
extrapolated to flat space.

Let us conclude by mentioning that in \cite{mathur} the mapping procedure
described in this chapter
between a Schwarzschild black hole and a black hole with four charges
was used to relate at low energies neutral Hawking radiation in the former to
D0-brane emission in the latter. The low-energy absorption cross-sections
were shown to agree, however the grey body factors were argued to be
out of reach essentially because of the strong near-extremality 
condition.

\appendix

\chapter{Kaluza-Klein reduction and the resulting action}
\label{KKapp}
\markboth{APPENDIX \protect \ref{KKapp}. KALUZA-KLEIN REDUCTION}{}

In this appendix we start from a theory including gravity in $D+p$
dimensions and we reduce it on a general $p$-dimensional torus with
non-trivial metric.
The main assumption in doing this is that all the fields do not 
depend on the $p$ `internal' coordinates, i.e. we will only 
consider the zero modes of these reduced fields.
We call this procedure `Kaluza-Klein reduction', and take this as a
synonym of dimensional reduction.

The aim is to derive the general action in $D$ dimensions which
includes the new fields arising from the reduction procedure, namely
the field components with part of their indices in the compactified space.

Let us start with a theory in $D+p$ dimensions containing only gravity.
If we single out the $p$ internal directions, we can recast the metric
in the following form:
\be
ds^2\equiv G_{MN}dz^M dz^N= g_{\mu \nu}dx^\mu dx^\nu + h_{ij}
(dy^i+A_\mu^i dx^\mu)(dy^j+A_\nu^j dx^\nu). \label{kkmetric}
\ee
Here $z^M=(x^\mu, y^i)$, $\mu=0\dots D-1$ and $i=1\dots p$;
all the functions appearing in \rref{kkmetric} depend only on $x^\mu$. 
The components of the $(D+p)$-dimensional metric $G_{MN}$ can thus be 
expressed as:
\be
G_{\mu\nu}=g_{\mu\nu}+h_{ij}A_\mu^i A_\nu^j, \qquad
G_{\mu i}=h_{ij} A_\mu^j, \qquad G_{ij}=h_{ij}. \label{kkG}
\ee
The inverse metric components $G^{MN}$, which are such that 
$G^{MP}G_{PN}=\delta^M_N$, can be easily computed to be:
\be
G^{\mu\nu}=g^{\mu\nu},\qquad G^{\mu i}=-A^{i\mu}\equiv -g^{\mu\nu} A^i_\nu,
\qquad G^{ij}=h^{ij}+A^{i\lambda}A^j_\lambda. \label{kkGinv}
\ee
The inverse metric components $g^{\mu\nu}$ and $h^{ij}$ are such that,
respectively, $g^{\mu \lambda}g_{\lambda \nu}=\delta^\mu_\nu$ and
$h^{ik}h_{kj}=\delta^i_j$.

The starting point is the following action for pure gravity in $D+p$
dimensions:
\be
I =\int d^{D+p}z \sqrt{-G} R[G]. \label{kkaction}
\ee
$R[G]$ denotes the curvature scalar constructed from the metric $G_{MN}$.
The relation between the determinants is straightforward, namely:
\be
G=g h, \qquad \leftrightarrow \qquad \sqrt{-G}=\sqrt{-g} \sqrt{h}.
\label{kkdet} \ee
The curvature scalar can also be recast in $D$-dimensional quantities.
Using the equations \rref{kkG} and \rref{kkGinv}, one obtains after
quite a lengthy computation:
\be
R[G]=R[g]-{3\over 4}\partial_\mu h^{ij} \partial^\mu h_{ij}
-{1\over 4}h^{ij}\partial_\mu h_{ij} \ h^{kl} \partial^\mu h_{kl}
-h^{ij} \Box h_{ij} -{1\over 4}h_{ij}F^i_{\mu\nu}F^{j\mu\nu},
\label{kkmaster}
\ee
where $R[g]$ and $\Box$ are the $D$-dimensional curvature and Dalembertian
operator respectively, built from the metric $g_{\mu\nu}$, the field
strength is given by $F^i_{\mu\nu}=\partial_\mu A^i_\nu-\partial_\nu
A^i_\mu$ and all the greek indices are raised with $g^{\mu\nu}$.

It is in general better to get rid of the term containing the Dalembertian,
transforming it into a total divergence. In order to achieve this, we note
that in \rref{kkaction} the above expression is multiplied by $\sqrt{h}$,
which satisfies:
\be
{1\over \sqrt{h}}\Box \sqrt{h}={1\over2}h^{ij} \Box h_{ij}+
{1\over2}\partial_\mu h^{ij} \partial^\mu h_{ij}+
{1\over 4}h^{ij}\partial_\mu h_{ij} \ h^{kl} \partial^\mu h_{kl}.
\label{deth}
\ee
The action \rref{kkaction} is thus equivalent, modulo a total divergence,
to:
\be
I=\int d^D x \sqrt{-g}\sqrt{h}\left\{ R[g]+{1\over 4}
\partial_\mu h^{ij} \partial^\mu h_{ij}
+{1\over 4}h^{ij}\partial_\mu h_{ij} \ h^{kl} \partial^\mu h_{kl}
-{1\over 4}h_{ij}F^i_{\mu\nu}F^{j\mu\nu}\right\} .
\label{kkredaction}
\ee

In order to put this action into its canonical form (i.e. to go to the
`Einstein frame'), we now want to eliminate the factor of $\sqrt{h}$
in front of $R[g]$. We thus have to operate a Weyl rescaling of the
metric. Most generally, if one transforms the metric by:
\be
g_{\mu\nu}=e^{2\varphi}\tilde{g}_{\mu\nu}, \label{weyl}
\ee
the scalar curvature in $D$ dimensions transforms accordingly and gives:
\be
R[g]=e^{-2\varphi} \left\{ R[\tilde{g}]-2(D-1)\Box \varphi
-(D-1)(D-2) \partial_\mu \varphi \partial^\mu \varphi \right\},
\label{weylcurv}
\ee
where the metric entering in the quantities on the r.h.s. is always 
$\tilde{g}_{\mu\nu}$.

The Weyl rescaling we have to do in \rref{kkredaction} in order to go to
the canonical action is:
\be
g_{\mu\nu}=e^{-{1\over D-2} \log h}\tilde{g}_{\mu\nu}.
\label{kkweyl}
\ee
We can again drop a total divergence arising from the term proportional
to $\Box \log h$, and if we use $\partial_\mu \log h=h^{ij}\partial_\mu h_{ij}$,
we finally obtain the Kaluza-Klein reduced action in $D$ dimension,
in its canonical form (for readability we drop the tildes):
\be
I=\int d^D x \sqrt{-g}\left\{ R
+{1\over 4} \partial_\mu h^{ij} \partial^\mu h_{ij}
-{1\over 4(D-2)} h^{ij}\partial_\mu h_{ij} \ h^{kl} \partial^\mu h_{kl}
-{1\over 4} h^{1\over D-2} h_{ij}F^i_{\mu\nu}F^{j\mu\nu}\right\}.
\label{kkweylaction}
\ee

We can now specialize to the simplest case in which we reduce on a single
direction $y$. This case will also turn out to be the most useful.
Here we can write $h_{yy}\equiv h =e^{2\sigma}$, and the action
\rref{kkweylaction} for this case becomes:
\be
I=\int d^D x \sqrt{-g}\left\{ R-{D-1 \over D-2} \partial_\mu \sigma
\partial^\mu \sigma -{1\over 4} e^{2{D-1 \over D-2} \sigma}
F_{\mu\nu}F^{\mu\nu}\right\}.
\label{kksingle}
\ee
The kinetic term of the scalar can be put to its canonical form simply
defining a new scalar $\phi=a \sigma$ such that its kinetic term is
multiplied by ${1\over 2}$. This requirement gives for $a$ the value:
\be
a=\sqrt{2 (D-1) \over D-2},
\label{kkcoupling}
\ee
and the resulting action is simply:
\be
I=\int d^D x \sqrt{-g}\left\{ R -{1\over 2} \partial_\mu \phi
\partial^\mu \phi -{1\over 4} e^{a\phi} F_{\mu\nu}F^{\mu\nu}\right\}.
\label{kksinglecan}
\ee

It is worthwhile at this stage to recall two physically relevant cases for the
expression \rref{kkcoupling}. The first one is when 4 dimensional
physics is seen as a reduction from 5 dimensions. In this case $D=4$
and thus $a=\sqrt{3}$. It is this class of 4 dimensional dilatonic
black holes that will be relevant to the study of Kaluza-Klein monopoles.

The second case is the reduction of 11 dimensional supergravity to 10 
dimensions. Here we have $D=10$ and thus $a={3\over 2}$. In this case
however we have a physical interpretation of the scalars
involved: $\langle e^\phi \rangle=g_s$ the string coupling constant,
and $\langle e^\sigma \rangle \sim R_{11}/l_p$ the size of the 11th direction
in 11 dimensional Planck units.
The relation $\phi={3\over2}\sigma$ thus entails $R_{11}\sim g_s^{2/3} l_p$.

For the sake of completeness, we also have to find how the matter fields,
and in particular the antisymmetric tensor fields, reduce under
Kaluza-Klein compactification. For simplicity, we proceed in the simplest 
case of the reduction on a single direction.

Suppose in $D+1$ dimensions we have a $n$-form field strength deriving
from a potential:
\be
H_{M_1 \dots M_n}=n \partial_{[M_1}C_{M_2\dots M_n]}=
\partial_{M_1}C_{M_2\dots M_n} + cyclic\  permutations.
\label{kknform}
\ee
Upon reduction on $y$, the $(n-1)$-form potential $C$ gives rise to
two potentials $A^{(n-1)}$ and $A^{(n-2)}$, respectively an $(n-1)$-
and an $(n-2)$-form. The relation is simply:
\be
A^{(n-1)}_{\mu_1\dots \mu_{n-1}}=C_{\mu_1\dots \mu_{n-1}}, \qquad
A^{(n-2)}_{\mu_1\dots \mu_{n-2}}=C_{\mu_1\dots \mu_{n-2}y}.
\label{kkredpot}
\ee
The corresponding field strengths are also easily found to be:
\be
F^{(n)}_{\mu_1\dots \mu_{n}}=n \partial_{[\mu_1}
A^{(n-1)}_{\mu_2\dots \mu_{n}]}=n \partial_{[\mu_1}
C_{\mu_2\dots \mu_{n}]}=H_{\mu_1\dots \mu_{n}}, 
\label{kkredfs1}
\ee
\bea 
\lefteqn{F^{(n-1)}_{\mu_1\dots \mu_{n-1}}= (n-1)\partial_{[\mu_1}
A^{(n-2)}_{\mu_2\dots \mu_{n-1}]}=} \nonumber \\
&=& (n-1)\partial_{[\mu_1} C_{\mu_2\dots \mu_{n-1}]y}=   
n\partial_{[\mu_1} C_{\mu_2\dots \mu_{n-1}y]}=
H_{\mu_1\dots \mu_{n-1}y}.
\label{kkredfs2}
\eea

The starting Lagrangian in $D+1$ dimensions is:
\be
L=-{1 \over 2 n!} H_{M_1\dots M_n}H^{M_1\dots M_n}.
\label{kklag}
\ee
Using again the expressions \rref{kkGinv} for $p=1$, $h_{yy}=e^{2\sigma}$,
and the above formulas \rref{kkredfs1} and \rref{kkredfs2},
we find:
\be
L=-{1\over 2 (n-1)!}e^{-2\sigma}F^{(n-1)}_{\mu_1\dots\mu_{n-1}}
F^{(n-1)\mu_1\dots\mu_{n-1}}-{1\over 2 n!} F'^{(n)}_{\mu_1\dots\mu_n}
F'^{(n)\mu_1\dots\mu_n}, 
\label{kkredlag}
\ee
where we have defined the following modified $n$-form field strength:
\be
F'^{(n)}_{\mu_1\dots\mu_n}=F^{(n)}_{\mu_1\dots\mu_n} -n
F^{(n-1)}_{[\mu_1\dots\mu_{n-1}}A_{\mu_n]}. 
\label{kkcsterm}
\ee
The Kaluza-Klein procedure thus introduces this Chern-Simons-like coupling.

The action for the $n$-form can then be reexpressed in the Einstein
frame like in \rref{kksinglecan}:
\bea
I_{forms}&=&\int d^{D+1}z \sqrt{-G}\left\{-{1 \over 2 n!} H^2\right\}
\nonumber \\
&=& \int d^D x \sqrt{-g} \left\{ -{1\over 2 (n-1)!} e^{a_{n-1}\phi} F_{(n-1)}^2
-{1 \over 2 n!} e^{a_n \phi} F_{(n)}^{'2}\right\},
\label{kkformaction}
\eea
where the couplings to the scalar are:
\be
a_{n-1}=-(D-n) \sqrt{2\over(D-1)(D-2)},\qquad
a_n=(n-1) \sqrt{2\over(D-1)(D-2)}.
\label{kkformcouplings}
\ee
Note that the metric appearing in \rref{kkformaction} is the Weyl
rescaled one, namely the $\tilde{g}_{\mu\nu}$ in \rref{kkweyl}.

Again, taking the example of the reduction of the 4-form appearing
in 11 dimensional supergravity, we obtain for the 3- and 4-form
of the 10-dimensional type IIA supergravity respectively
$a_3=-1$ and $a_4={1\over2}$.

\chapter{Summary of the properties of the branes in 10 
and 11 dimensions}
\label{ZOOapp}
\markboth{APPENDIX \protect \ref{ZOOapp}. PROPERTIES OF THE BRANES}{}

In this appendix, we give a list of the projections 
imposed on the supersymmetric 
parameters of the theory when there is a brane in the background. We also
give the tensions of the branes, and their transformation laws under
the various dualities. 

In 11 dimensions, the Majorana condition can be imposed on the 
spinors\footnote{We refer the reader to e.g. \cite{strathdee} 
for a general discussion on the
spinor representations and the extended supersymmetry algebras in 
arbitrary $D$ space-time dimensions.}.
Moreover, a representation of the Clifford algebra exists in which all
the $\Gamma_M$ matrices are real. The charge conjugation matrix $C$,
which is such that $C\Gamma_M^T=-\Gamma_M C$ and $C^T=-C$, can be taken
to be $C=\Gamma_0$. The
$\Gamma_M$ matrices are such that the one corresponding to the 11th direction
satisfies $\Gamma_{11}=\Gamma_0 \dots \Gamma_9$.

The supersymmetry algebra of 11 dimensional supergravity (or of M-theory), 
including all possible $p$-form charges, is given by (see for instance
\cite{democracy,hull}):
\be
\{Q, Q^T\}=\Gamma^M C P_M+\Gamma^{MN}C Z_{MN}+\Gamma^{MNPQR}CZ_{MNPQR},
\label{11susyalgebra}
\ee
where $Q$ is the supersymmetry generator, which is Majorana, and
$\Gamma^{M_1\dots M_p}=\Gamma^{[M_1}\dots \Gamma^{M_p]}$. Note that
we could still add $p$-form charges for $p=6,9,10$ but these
can be accounted for by a Hodge duality followed by a redefinition
of the charges with $p=5,2,1$ respectively.

We call BPS states those for which the matrix $\{Q, Q^T\}$ has some zero
eigenvalues. This means that they are supersymmetric since
some combinations of the supercharges are represented trivially by zero
on them. 
Taking the determinant of the r.h.s. of 
\rref{11susyalgebra} and asking that it vanishes, 
one can see (see e.g. \cite{towncargese}) that the supersymmetry
condition will impose at the same time an equality between the mass
and one component (or more) of a $p$-form charge, and a supersymmetry
projection on the supersymmetric parameter of the theory.

In 11 dimensional supergravity (or in M-theory),
we have one Majorana supersymmetric parameter $\epsilon$. 
The outcome of the analysis discussed above leads to the following 
relations (the numbers between brackets 
indicate the directions longitudinal to
the brane, and W[1] stands for a travelling wave or KK momentum in the 
direction $\hat 1$):
\begin{eqnarray}
\mbox{W[1]:} & \qquad & \epsilon=\Gamma_0 \Gamma_1 \epsilon \nonumber\\
\mbox{M2[1,2]:} & \qquad & \epsilon=\Gamma_0 \Gamma_1 \Gamma_2 \epsilon 
\nonumber\\
\mbox{M5[1..5]:} & \qquad & \epsilon=\Gamma_0 \dots \Gamma_5 \epsilon 
\label{msusy}\\
\mbox{KK6[1..6]:} & \qquad & \epsilon=\Gamma_0 \dots \Gamma_6 \epsilon 
\nonumber\\
\mbox{M9[1..9]:} & \qquad & \epsilon=\Gamma_0 \dots \Gamma_9 \epsilon 
\nonumber
\end{eqnarray}
Note that there are no other combinations of the $\Gamma_M$ matrices which
square to the identity. Also, the signs in the relations above have
been arbitrarily chosen and fix the convention which makes the
distinction between a brane and an anti-brane.

The tensions of these objects are given as follows.
The quantum of mass of a KK momentum on a compact direction of radius $R$ is:
\be M_{W}={1 \over R}. \label{mwave} \ee
If $L_p$ is the 11 dimensional Planck length (related to the 11 dimensional
Newton constant by $G_{11}=L_p^9$), the tensions of the M2 and M5
branes are:
\be T_{M2}= {\pi^{1/3} \over 2^{2/3} L_p^3}, \qquad T_{M5}={1 \over 
2^{7/3}\pi^{1/3} L_p^6}. \label{tmbranes}
\ee
The precise numerical factors are derived by dualities from type II
string theories. 
The tension of a KK6 monopole with a transverse NUT direction of radius 
$R_N$ is:
\be T_{KK6}={\pi R_N^2 \over 4 L_p^9}. \label{tkk6} \ee
This can be easily obtained from the tension of a D6-brane, which
is given below. We will not discuss here the tension of the M9.

In type II theories, there are 2 Majorana-Weyl spinors $\epsilon_L$ and
$\epsilon_R$ (with reference to the string origin of the relative 
supersymmetry generators). 
They satisfy the chirality conditions:
\[ \epsilon_L=\Gamma_{11} \epsilon_L, \qquad \epsilon_R=\eta \Gamma_{11} 
\epsilon_R, \]
with $\eta=+1$ for IIB theory and $\eta=-1$ for IIA theory
(the overall sign is again fixed by convention). The supersymmetry
projections are the following (we denote by F1 the fundamental strings of each
theory):
\begin{eqnarray}
\mbox{F1[1]:} & \qquad & \left\{ \begin{array}{rcl} \epsilon_L&=&\Gamma_0
\Gamma_1 \epsilon_L \\ \epsilon_R&=&-\Gamma_0 \Gamma_1 \epsilon_R 
\end{array} \right. \nonumber \\
\mbox{W[1]:} & \qquad & \left\{ \begin{array}{rcl} \epsilon_L&=&\Gamma_0
\Gamma_1 \epsilon_L \\ \epsilon_R&=&\Gamma_0 \Gamma_1 \epsilon_R
\end{array} \right. \nonumber \\
\mbox{NS5[1..5]:} & \qquad & \left\{ \begin{array}{rcl} \epsilon_L&=&\Gamma_0 
\dots
\Gamma_5 \epsilon_L \\ \epsilon_R&=&-\eta \Gamma_0 \dots \Gamma_5 \epsilon_R
\end{array} \right. \label{iisusy} \\
\mbox{KK5[1..5]:} & \qquad & \left\{ \begin{array}{rcl} \epsilon_L&=&\Gamma_0 
\dots
\Gamma_5 \epsilon_L \\ \epsilon_R&=&\eta \Gamma_0 \dots \Gamma_5 \epsilon_R
\end{array} \right. \nonumber \\
\mbox{D$p$[1..$p$]:} & \qquad & \epsilon_L=\Gamma_0 \dots \Gamma_p \epsilon_R
\nonumber
\end{eqnarray}
Note that the relations for IIA theory are obtained from those of M-theory
compactifying on the 11th direction. $\Gamma_{11}$ plays thus the r\^ole of
the chiral projector in 10 dimensions, and the supersymmetry parameters are
related by $\epsilon_{L(R)}={1 \over 2}(1\pm\Gamma_{11})\epsilon$.
Also the relations of IIA and IIB theories are related by T-duality,
namely under a T-duality over the $\hat{\imath}$ direction the supersymmetry
parameters transform (see e.g. \cite{tasi} and Section \ref{DBRANEsec})
as $\epsilon_L \rightarrow \epsilon_L$ and $\epsilon_R \rightarrow
\Gamma_i \epsilon_R$ (up to a sign). We recapitulate at the end of this 
appendix how all the branes are related by compactification and dualities.

The mass of a KK mode W is as in \rref{mwave}. Type II string theories are
both characterized by the string length $l_s=\sqrt{\alpha'}$ and by the string
coupling constant $g$. The tension of the fundamental string is:
\be T_{F1}={1 \over 2\pi l_s^2}. \label{tfund} \ee
The tension of the solitonic NS5 branes is given by:
\be T_{NS5}={1 \over (2\pi)^5 g^2 l_s^6}. \label{tns} \ee
The KK5 monopole has a tension of:
\be T_{KK5}={R_N^2 \over (2\pi)^5 g^2 l_s^8}, \label{tkk5} \ee
where $R_N$ is the radius of the NUT direction. Finally the tensions of the
D$p$-branes are given by:
\be T_{Dp}={1 \over (2\pi)^p g l_s^{p+1}}. \label{tdbrane} \ee

Let us comment on how the precise numerical factors in the tensions
are found. The expression \rref{tfund} is taken as a definition.
Then the dualities are precisely defined as follows. The type IIB S-duality
acts as:
\[ g\rightarrow {1\over g}, \qquad \qquad l^2_s \rightarrow g l_s^2, \]
while T-duality on a circle of radius $R$ (and thus of length $L=2\pi R$)
acts as:
\[ R \rightarrow {l_s^2 \over R} \qquad \qquad g \rightarrow g{l_s \over R}. \]
Using S- and T-dualities, all the tensions \rref{tns}, \rref{tkk5}
and \rref{tdbrane} are found. Note that if one considers the NUT direction
of the KK5 monopole as one of the directions of its world-volume, then
one has to define a new tension dividing \rref{tkk5} by $2\pi R_N$.

The tensions above completely fix the 10 dimensional Newton constant.
Indeed, the tensions of all electric-magnetic dual objects must satisfy:
\be
T_p T_{D-p-4}={2\pi\over 16\pi G_D}.
\label{quantcond}
\ee
In 10 dimensions we thus have $G_{10}=8\pi^6 g^2 l_s^8$.

Going to 11 dimensions, one has the two conditions $G_{10}={G_{11}\over 
2\pi R}$ and $M_{D0}=M_W^{(11)}$ (which gives $R=gl_s$), where $R$ is now
the radius of the 11th direction. The expressions \rref{tmbranes} and
\rref{tkk6} are thus recovered, and can be shown to satisfy \rref{quantcond}
with $G_{11}=L_p^9$.

We end with a summary of how the various branes are related by dualities
and compactifications from 11 to 10 dimensions.

Under a T-duality along a direction longitudinal to the world-volume
of the brane the winding modes and the KK momentum of a fundamental string
are exchanged, i.e. the F1 and the W are mapped to each other; a D$p$-brane
becomes a D$(p-1)$-brane; the NS5-brane and the KK5-monopole are inert.

Under a T-duality along a direction transverse to the world-volume 
of the brane the F1 and the W are inert; a D$p$-brane is mapped to a
D$(p+1)$-brane; the NS5-brane is mapped to a KK5-monopole and vice-versa.
Note that when an object in IIA (IIB) theory is said to be inert 
under T-duality, it exactly means that it is mapped to its analog in
IIB (IIA) theory.

Under S-duality in IIB theory the F1 and the D1-brane are mapped to each
other; the D5- and the NS5-branes are mapped to each other; the D3-brane,
the wave W and the KK5-monopole are inert. We will not discuss the action
of S-duality on the higher branes, namely the D7 and the D9.

Compactification along one of the world-volume directions or transverse
to them gives, for the 11 dimensional wave W, the D0-brane and the 10
dimensional wave W respectively; for the M2-brane, the F1 and the D2;
for the M5, the D4 and the NS5; for the KK6, the KK5 and the D6 (in this
last case the compactification is performed along the NUT direction).
We will not consider the compactification of the hypothetical M9.

\chapter{Computation of curvature tensors}
\label{RICCIapp}

In this appendix we work out the explicit form of the components 
of the Ricci tensor for a general metric which is compatible with
the expected form of the sought for $p$-brane solutions.  

We will consider general  diagonal metrics with
components depending on only $d$ coordinates, the other $p$ and
the time being translational invariant. To begin with we
consider a set of coordinates for which the metric is isotropic
in the $d$-dimensional space, but we do not impose any symmetry
in that space. Then we consider a special case in which we have indeed
spherical symmetry. In this latter case it is possible, by a reparametrization
of the radial coordinate, to generalize the Ricci tensor to 
a non-isotropic set of coordinates.

We will use the vielbein formalism because it has several 
interesting by-products, such as giving without extra effort the
full Riemann tensor in orthonormal coordinates, and it also
uses the spin connection which will be needed 
when considering supersymmetry.

\subsubsection*{The vielbein formalism}

The first step in the vielbein formalism is to rewrite the metric
in terms of a $D\times D$ matrix ${e^{\hat{\mu}}}_\nu$, where the `hatted'
indices are indices of the $D$ dimensional Lorentz group, while the
`unhatted' ones are covariant indices for $D$ dimensional diffeomorphisms.
The relations between the metric and the vielbein are the following:
\bea
g_{\mu\nu}&=&\eta_{\hat{\rho}\hat{\sigma}}{e^{\hat{\rho}}}_\mu 
{e^{\hat{\sigma}}}_\nu,
\label{vieldirect} \\
\eta^{\hat{\rho}\hat{\sigma}}&=&g^{\mu\nu}
{e^{\hat{\rho}}}_\mu{e^{\hat{\sigma}}}_\nu,
\label{vielinverse}
\eea
where $\eta_{\hat{\rho}\hat{\sigma}}$ is the metric of flat Minkowski
space-time.

If we define the 1-forms:
\be
e^{\hat{\mu}}={e^{\hat{\mu}}}_\nu dx^\nu,
\label{1forms}
\ee
the metric can be rewritten in the orthonormal frame:
\be
ds^2=g_{\mu\nu}dx^\mu dx^\nu =\eta_{\hat{\mu}\hat{\nu}} e^{\hat{\mu}}
e^{\hat{\nu}}.
\label{orthonormal}
\ee

We now introduce the spin connection ${\omega^{\hat{\mu}\hat{\nu}}}_\rho$,
which is the `Yang-Mills' connection for the local Lorentz
transformations. This is because the Lorentz transformations acting
on the vielbein have to be gauged away in order to keep the same
number of physical degrees of freedom. Since it belongs to the 
adjoint representation, the spin connection is antisymmetric
in the Lorentz indices.

We impose as a constraint on the vielbein that it is covariantly constant:
\be
\hat{D}_\lambda {e^{\hat{\mu}}}_\nu=\partial_\lambda {e^{\hat{\mu}}}_\nu
+{\omega^{\hat{\mu}}}_{\hat{\rho}\lambda} {e^{\hat{\rho}}}_\nu -
\Gamma^\sigma_{\lambda\nu} {e^{\hat{\mu}}}_\sigma =0.
\label{covariantd}
\ee
This equation implies that the metric is  covariantly constant, as it 
should be. 

Introducing also the 1-form version of the spin connection:
\be
{\omega^{\hat{\mu}}}_{\hat{\nu}}={\omega^{\hat{\mu}}}_{\hat{\nu}\lambda}
dx^\lambda, \label{1formspin}
\ee
the relation \rref{covariantd} implies the following equation:
\be
de^{\hat{\mu}}+{\omega^{\hat{\mu}}}_{\hat{\nu}}\wedge e^{\hat{\nu}}=0,
\label{vielspin}
\ee
which completely determines the spin connection.

The solution to the equation above can be easily computed using that:
\be
de^{\hat{\mu}}={h^{\hat{\mu}}}_{\hat{\nu}\hat{\rho}} e^{\hat{\nu}}\wedge
e^{\hat{\rho}}, \qquad  h_{\hat{\mu}\hat{\nu}\hat{\rho}}=
-h_{\hat{\mu}\hat{\rho}\hat{\nu}}.
\label{detspin1}
\ee
Then writing ${\omega^{\hat{\mu}}}_{\hat{\nu}}={f^{\hat{\mu}}}_{\hat{\nu}
\hat{\rho}}e^{\hat{\rho}}$ the equation \rref{vielspin} becomes:
\be
h_{\hat{\mu}\hat{\nu}\hat{\rho}}={1\over2}(f_{\hat{\mu}\hat{\nu}\hat{\rho}}
-f_{\hat{\mu}\hat{\rho}\hat{\nu}})={1\over2}(f_{\hat{\mu}\hat{\nu}\hat{\rho}}
+f_{\hat{\rho}\hat{\mu}\hat{\nu}}),
\label{detspin2}
\ee
which gives:
\be
f_{\hat{\mu}\hat{\nu}\hat{\rho}}=h_{\hat{\mu}\hat{\nu}\hat{\rho}}
-h_{\hat{\nu}\hat{\mu}\hat{\rho}}-h_{\hat{\rho}\hat{\mu}\hat{\nu}}.
\label{detspin3}
\ee
The final expression for the spin connection is thus:
\be
{\omega^{\hat{\mu}}}_{\hat{\nu}}={1\over2}\left\{ {e_{\hat{\nu}}}^\sigma
{e_{\hat{\rho}}}^\lambda(\partial_\sigma {e^{\hat{\mu}}}_\lambda
-\partial_\lambda {e^{\hat{\mu}}}_\sigma)-
e^{\hat{\mu}\sigma}{e_{\hat{\rho}}}^\lambda(\partial_\sigma
e_{\hat{\nu}\lambda}-\partial_\lambda e_{\hat{\nu}\sigma})
-e^{\hat{\mu}\sigma}{e_{\hat{\nu}}}^\lambda(\partial_\sigma 
e_{\hat{\rho}\lambda}-\partial_\lambda e_{\hat{\rho}\sigma})\right\}
e^{\hat{\rho}}.
\label{detspin4}
\ee
However, it is often easier to determine the spin connection simply
inspecting the expression for $de^{\hat{\mu}}$ and guessing 
which ${\omega^{\hat{\mu}}}_{\hat{\nu}}$ would satisfy eq. \rref{vielspin}.

Having the spin connection, it is straightforward to determine the
curvature, which is defined by:
\be
{\R^{\hat{\mu}}}_{\hat{\nu}}=d{\omega^{\hat{\mu}}}_{\hat{\nu}}
+{\omega^{\hat{\mu}}}_{\hat{\rho}}\wedge {\omega^{\hat{\rho}}}_{\hat{\nu}},
\qquad \R_{\hat{\mu}\hat{\nu}}=-\R_{\hat{\nu}\hat{\mu}}.
\label{riemannform}
\ee
This curvature tensor conveys exactly the same information as the
Riemann tensor. It is indeed proportional to it. Writing
${\R^{\hat{\mu}}}_{\hat{\nu}}={1\over2}{\R^{\hat{\mu}}}_{\hat{\nu}
\sigma\rho}dx^\sigma \wedge dx^\rho$, one can show, using
only \rref{covariantd} and \rref{vieldirect}--\rref{vielinverse}, 
that the following equality holds:
\be
{e_{\hat{\mu}}}^{\alpha} {e^{\hat{\nu}}}_\beta {\R^{\hat{\mu}}}_{\hat{\nu}
\sigma\rho}=\partial_\sigma \Gamma^\alpha_{\beta\rho} -
\partial_\rho \Gamma^\alpha_{\beta\sigma }+
\Gamma^\alpha_{\gamma\sigma} \Gamma^\gamma_{\beta\rho}-
\Gamma^\alpha_{\gamma\rho}\Gamma^\gamma_{\beta\sigma}\equiv
{\R^\alpha}_{\beta\sigma\rho}.
\label{riemanntrue}
\ee

Note that we can also express the components of the curvature tensor
in the orthonormal frame, 
${\R^{\hat{\mu}}}_{\hat{\nu}}={1\over2}{\R^{\hat{\mu}}}_{\hat{\nu}
\hat{\sigma}\hat{\rho}} e^{\hat{\sigma}} \wedge e^{\hat{\rho}}$. The
Ricci tensor, which enters the Einstein equations, is then defined
by:
\be
R_{\hat{\mu}\hat{\nu}}= {\R^{\hat{\lambda}}}_{\hat{\mu}\hat{\lambda}
\hat{\nu}},\qquad R_{\hat{\mu}\hat{\nu}}=R_{\hat{\nu}\hat{\mu}}.
\label{riccidef}
\ee

\subsubsection*{The curvature in a special case of interest}

We now specialize to the computation of the curvature for a
particular metric relevant to the study of $p$-branes. The aim of the 
rest of this appendix is to provide an expression for the l.h.s.
of the Einstein equations \rref{eomeinstein}.

The general metric we consider is the following:
\be
ds^2=-B^2 dt^2 +\sum_i C_i^2 dy_i^2 +G^2 \delta_{ab} dx^a dx^b.
\label{isotropicmetric}
\ee
All the functions $B$, $C_i$ ($i=1\dots p$) and $G$ depend only
on the $x^a$ ($a=1\dots d$) coordinates, though no additional
symmetry is postulated in the $d$ dimensional space spanned by
the $x$'s. We refer to Chapter \ref{PBRANEchap} for the physical motivation 
leading to a metric of the form above. We call these $x^a$ coordinates
isotropic since the metric of the $d$ dimensional subspace is
the flat metric multiplied by an overall function.

Since the metric \rref{isotropicmetric} is diagonal, the vielbeins
are very easily determined:
\be
e^{\hat{t}}=Bdt, \qquad e^{\hat{\imath}}=C_i dy^i, \qquad e^{\hat{a}}=Gdx^a.
\label{isoviel}  
\ee
The non-vanishing spin connection components are then the following: 
\bea
{\omega^{\hat{t}}}_{\hat{a}}&=& {1\over G} \partial_a \ln B 
\ e^{\hat{t}}, \label{isospinconta} \\
{\omega^{\hat{\imath}}}_{\hat{a}}&=& {1\over G} \partial_a \ln C_i
\ e^{\hat{\imath}}, \label{isospinconia} \\
{\omega^{\hat{a}}}_{\hat{b}}&=& {1\over G}\left( \partial_b \ln G
\ e^{\hat{a}} -\partial_a \ln G \ e^{\hat{b}}\right). \label{isospinconab}
\eea
The Riemann tensor components in the orthonormal frame are:
\bea
{\R^{\hat{t}\hat{\imath} }}_{\hat{t} \hat{\jmath}}&=& -{1\over G^2}
\partial_a \ln B\ \partial_a \ln C_i \ \delta_{ij}, \label{isoriemtitj} \\
{\R^{\hat{t}\hat{a}}}_{\hat{t} \hat{b}} &=&{1\over G^2} \left\{
-\partial_a \partial_b \ln B -\partial_a \ln B\ \partial_b \ln B
+\partial_a \ln B\ \partial_b \ln G +\partial_b \ln B\ \partial_a \ln G 
\right. \nonumber \\ & &  \qquad \qquad\left.
- \delta^a_b \partial_c \ln B\ \partial_c \ln G \right\}, \label{isoriemtatb}\\
{\R^{\hat{\imath}\hat{\jmath}}}_{\hat{k} \hat{l}} &=& -{1\over G^2}
\partial_a \ln C_i\ \partial_a \ln C_j \ (\delta^i_k \delta^j_l -
\delta^i_l \delta^j_k), \label{isoriemijkl} \\
{\R^{\hat{\imath}\hat{a}}}_{\hat{\jmath}\hat{b}} &=&{1\over G^2} \left\{
-\partial_a \partial_b \ln C_i -\partial_a \ln C_i\ \partial_b \ln C_i
+\partial_a \ln C_i\ \partial_b \ln G +\partial_b \ln C_i\ \partial_a \ln G
\right. \nonumber \\ & &  \qquad \qquad\left.
- \delta^a_b \partial_c \ln C_i \partial_c\ \ln G \right\} 
\delta^i_j, \label{isoriemiajb} \\
{\R^{\hat{a}\hat{b}}}_{\hat{c}\hat{d}}&=&{1\over G^2} \left\{
-\delta^a_c \partial_b \partial_d \ln G +\delta^a_d \partial_b \partial_c
\ln G +\delta^b_c \partial_a \partial_d \ln G -\delta^b_d \partial_a
\partial_c \ln G 
\right. \nonumber \\ & &
+ \delta^a_c \partial_b \ln G\ \partial_d \ln G -\delta^a_d \partial_b \ln G\
\partial_c \ln G -\delta^b_c \partial_a \ln G\ \partial_d \ln G +\delta^b_d 
\partial_a \ln G\ \partial_c \ln G
\nonumber \\ & & \left.
-(\delta^a_c \delta^b_d - \delta^a_d \delta^b_c) \partial_e \ln G\ 
\partial_e \ln G \right\}. \label{isoriemabcd}
\eea
Repeated indices relative to the $x^a$ coordinates are summed. Note
also that for a geometry with a metric like \rref{isotropicmetric},
one can build from the Riemann tensor and the metric a scalar
showing singular behaviour if and only if at least one of the components above
also has a singular behaviour. The above expressions are thus
useful in searching for the singularites of a determined geometry.

We now proceed to the computation of the Ricci tensor \rref{riccidef}.
For our simplified problem, only three sets of components will be 
non-vanishing. They are the following (remember that $\delta^a_a=d$):
\bea
R^{\hat{t}}_{\hat{t}}&=&{1\over G^2} \left\{-\partial_a\partial_a \ln B
-\partial_a \ln B \ \partial_a 
\left[ \ln B +\sum_i \ln C_i + (d-2) \ln G \right] \right\}, 
\label{isoriccitt} \\
R^{\hat{\imath}}_{\hat{\jmath}}&=&{1\over G^2} \left\{-\partial_a\partial_a
\ln C_i -\partial_a \ln C_i \ \partial_a 
\left[ \ln B +\sum_i \ln C_i + (d-2) \ln G \right] \right\} 
\delta^{\hat{\imath}}_{\hat{\jmath}}, \label{isoricciij} \\
R^{\hat{a}}_{\hat{b}}&=&{1\over G^2} \left\{-\partial_a\partial_b
\left[ \ln B +\sum_i \ln C_i + (d-2) \ln G \right] 
\right. \nonumber \\ & & + \partial_a \ln G \
\partial_b \left[ \ln B +\sum_i \ln C_i + (d-2) \ln G \right] 
\nonumber \\ & &+ \partial_b
\ln G \ \partial_a \left[ \ln B +\sum_i \ln C_i + (d-2) \ln G \right]
\nonumber \\ & & -\partial_a \ln B \ \partial_b \ln B - \sum_i
\partial_a \ln C_i \ \partial_b \ln C_i -(d-2)
\partial_a \ln G \ \partial_b \ln G 
\nonumber \\ & & \left.- \delta^a_b \partial_c \partial_c
\ln G - \delta^a_b \partial_c \ln G\ \partial_c
\left[ \ln B +\sum_i \ln C_i + (d-2) \ln G \right] \right\}. \label{isoricciab}
\eea
The quantity $\varphi \equiv \ln B +\sum_i \ln C_i + (d-2) \ln G$ appears
ubiquitously in the expressions above. We will indeed see that it plays
an important r\^ole in the simplification of the Einstein equations,
in some particular cases (i.e. for extremal branes).
Note also that the only restriction on $d$ is that it should be 1 or
bigger, otherwise the metric is constant.

Since the metric is of the special form \rref{isotropicmetric}, the Ricci
tensor with Lorentz (hatted) indices actually coincides with the
one carrying covariant (unhatted) indices.

The Ricci tensor components displayed in \rref{isoriccitt}--\rref{isoricciab}
are the ones which will be used when searching for multi-$p$-brane solutions,
including the intersecting configurations. 

These equations are thus most suitable for finding solutions which are 
static despite the fact that there might be several `objects' ($p$-branes
of different kinds which are pointlike from the point of view
of the $d$-dimensional space of the $x$'s) at a finite distance from
each other. This kind of configuration is likely to reflect a 
balance of forces typical of a BPS saturated state, i.e. of extremal
$p$-branes.

On the other hand, a non-extremal configuration most likely consists
of a single collapsed object in the whole of $d$-space. It might thus
turn out to be useful to reformulate the problem requiring the
solution to be spherically symmetric. 

\subsubsection*{The Ricci tensor for a metric with spherical symmetry}

The metric \rref{isotropicmetric} actually incorporates also the 
spherically symmetric case, and the non-isotropic one by a simple
change of coordinates. However in General Relativity a good choice
of coordinates can often substantially simplify the equations of motion
and the search for their solutions. It is thus not a worthless task
to rewrite the Ricci tensor components for a metric belonging to the 
same class of geometries, but in a slighly different coordinate 
system.

The metric we will consider is the following:
\be
ds^2=-B^2 dt^2 +\sum_i C_i^2 dy_i^2 + F^2 dr^2 +G^2 r^2 d\Omega^2_{d-1}.
\label{sphmetric}
\ee
It has spherical symmetry in the $d$ space, and whenever $F\neq G$
it is not isotropic in the sense that the metric \rref{isotropicmetric} was.
All the functions of course depend only on $r$, and $d\Omega^2_{d-1}$
is the metric on the $(d-1)$-sphere $S^{d-1}$.

We could compute the Ricci tensor using the same straightforward 
strategy as before. We will instead use the results 
\rref{isoriccitt}--\rref{isoricciab} and perform the change of coordinates,
in agreement with the fact that physically this new case is a particular
case and not a generalization of the previous one.
We will operate in two steps: first specialize to spherical symmetry, and
secondly perform a reparametrization of the radial coordinate to go
to a metric like \rref{sphmetric}.

The first step is to take the isotropic metric \rref{isotropicmetric}
and to go to spherical coordinates. The transformation between
the set of $x^a$ ($a=1\dots d$) coordinates and the set $\{ r, \theta_\alpha
(\alpha=1\dots d-1)\}$ can be written in condensed form:
\bea
x_a&=&r \sin \theta_1 \dots \sin \theta_{a-1} \cos \theta_a, \qquad a=1\dots
d-1\nonumber \\
x_d&=&r\sin \theta_1 \dots \sin \theta_{d-1}, \label{sphtox}
\eea
or inversely:
\bea
r&=& \sqrt{x_1^2 +\dots +x_d^2}\equiv \sqrt{x_a x_a}, \label{xtor} \\
\cot \theta_\alpha &=& {x_\alpha \over\sqrt{x_{\alpha+1}^2+\dots+ x_d^2}},
\qquad \alpha=1\dots d-1. \label{xtotheta}
\eea
The range of the angular variables is $0\leq \theta_1,\dots,
\theta_{d-2}\leq \pi$ and $0\leq \theta_{d-1}\leq 2\pi$. Actually
to cover the full $d$ dimensional space $R^d$ and not half of it,
one should add to \rref{xtotheta} the supplementary equation
$\sin \theta_{d-1}={x_d\over \sqrt{x_{d-1}^2+x_d^2}}$, in order to take
into account the sign of $x_d$. However it will be not necessary here
since we focus on a local change of coordinates.

If $f$ is a function of $r$, then the following relations hold for its
derivatives, where ${}'\equiv {d\over dr}$:
\bea
\partial_a f&=& {x_a\over r}f', \nonumber \\
\partial_a \partial_b f&=& \delta^a_b {1\over r}f' +{x_a x_b\over r^2}
\left(f''-{1\over r}f'\right), \label{radderiv} \\
\partial_a \partial_a f&=& f'' + {d-1\over r}f'. \nonumber 
\eea

The Ricci tensor components \rref{isoriccitt}--\rref{isoricciab} can
thus be specialized to the spherically symmetric case:
\bea
R^{\hat{t}}_{\hat{t}}&=&{1\over G^2} \left\{-(\ln B)''-{d-1\over r}
(\ln B)'-(\ln B)'\varphi'\right\}, \label{sphisortt} \\
R^{\hat{\imath}}_{\hat{\jmath}}&=&{1\over G^2} \left\{-(\ln C_i)''-{d-1\over r}
(\ln C_i)'-(\ln C_i)'\varphi'\right\}\delta^i_j, \label{sphisorij} \\
R^{\hat{a}}_{\hat{b}}&=&\delta^a_b {1\over G^2} \left\{-(\ln G)''
-{d-1\over r} (\ln G)' -(\ln G)' \varphi' -{1\over r} \varphi' \right\}
\label{sphisorab} \\ & &
+{x_a x_b\over r^2} {1\over G^2} \left\{ -\varphi''+{1\over r}\varphi'
+2 (\ln G)' \varphi' - {(\ln B)'}^2 -\sum_i {(\ln C_i)'}^2 
-(d-2) {(\ln G)'}^2 \right\}, \nonumber
\eea
where $\varphi$ was defined just after eq. \rref{isoricciab}.

We now want to write the components of the Ricci tensor in terms of the
indices relative to the spherical
coordinates. This only involves the $R^{\hat{a}}_{\hat{b}}$ components.
Noting that the latter are of the form $R^{\hat{a}}_{\hat{b}}=
R_{(1)}\delta^a_b + R_{(2)} {x_a x_b\over r^2}$, it is easy to show,
using the relations \rref{sphtox}, \rref{xtor} and \rref{xtotheta}, that
the components in spherical coordinates are:
\[
R^{\hat{r}}_{\hat{r}}=R_{(1)}+R_{(2)}, \qquad R^{\hat{\alpha}}_{
\hat{\beta}}=R_{(1)}\delta^\alpha_\beta \quad (\alpha, \beta=1\dots d-1),
\qquad R^{\hat{r}}_{\hat{\alpha}}=0,
\]
thus giving:
\bea
R^{\hat{r}}_{\hat{r}}&=&{1\over G^2} \left\{ -\varphi''
+(\ln G)' \varphi' -(\ln G)'' -{d-1\over r} (\ln G)' 
\right. \nonumber \\ & & \left.
- {(\ln B)'}^2 -\sum_i {(\ln C_i)'}^2 -(d-2) {(\ln G)'}^2 \right\},
\label{sphisorrr} \\
R^{\hat{\alpha}}_{\hat{\beta}}&=&{1\over G^2} \left\{ -(\ln G)''
-{d-1\over r} (\ln G)' -(\ln G)' \varphi' -{1\over r} \varphi' \right\}
\delta^\alpha_\beta. \label{sphisortheta}
\eea

\newcommand{\g}{\bar{G}}
\newcommand{\r}{\bar{r}}

We are now ready to proceed to the second step, which is a reparametrization
of the radial coordinate $r$. We have to go from the (isotropic)
metric:
\[
ds^2=\dots +G^2\left(dr^2+r^2 d\Omega^2_{d-1}\right)
\]
to the more general metric:
\[
ds^2=\dots +F^2 d\r^2 +\g^2 \r^2 d\Omega^2_{d-1}.
\]
We will drop the bars only at the end of the computation, when it will be
no longer ambiguous. We also define for the time being $\dot{f}\equiv
{d\over d\r}f$.

The relation above between $r$ and $\r$ is thus such that:
\be
Gr=\g\r, \qquad F=G{dr\over d\r}. \label{reparamr}
\ee
This implies the following relations for some of the quantities 
appearing in the Ricci tensor components ($f$ is an arbitrary function 
independent from $G$; note that $\varphi$ is not a function of this kind):
\begin{eqnarray*}
f'&=&{G\over F} \dot{f} \\
(\ln G)'+{1\over r}&=&{G\over F} \left[ (\ln \g)\dot{}+{1\over \r}\right] \\
f''+{1\over r}f'&=&{G^2\over F^2} \left[\ddot{f}+\dot{f} (\ln \g)\dot{}
+{1\over \r}\dot{f}-(\ln F)\dot{} \dot{f}\right] \\
(\ln G)''+{1\over r}(\ln G)'&=&{G^2\over F^2} \left[(\ln \g)\ddot{}
+{2\over\r}(\ln \g)\dot{}+{(\ln \g)\dot{}}^2-(\ln F)\dot{}(\ln \g)\dot{}
-{1\over\r}(\ln F)\dot{}\right]. 
\end{eqnarray*}
These relations can now be plugged into the Ricci tensor components
\rref{sphisortt}, \rref{sphisorij}, \rref{sphisorrr} and
\rref{sphisortheta}. 

We write the result dropping the bars (thus conforming to the notation
of \rref{sphmetric}):
\bea
R^{\hat{t}}_{\hat{t}}&=&{1\over F^2}\left\{-(\ln B)''-(\ln B)'\left[
\ln B+\sum_i \ln C_i -\ln F +(d-1)\left(\ln G +\ln r\right)\right]'
\right\}, \nonumber \\
R^{\hat{\imath}}_{\hat{\jmath}}&=&\delta^i_j {1\over F^2}\left\{-(\ln C_i)''
-(\ln C_i)'\left[
\ln B+\sum_i \ln C_i -\ln F +(d-1)\left(\ln G +\ln r
\right)\right]'
\right\}, \nonumber \\
R^{\hat{r}}_{\hat{r}}&=&{1\over F^2}\left\{-(\ln B)''-\sum_i(\ln C_i)''
- {(\ln B)'}^2 -\sum_i {(\ln C_i)'}^2 +(\ln B)'(\ln F)' 
\right. \nonumber \\ & & \left.
+\sum_i(\ln C_i)'(\ln F)'
-(d-1)\left[ (\ln G)'' +{(\ln G)'}^2
+{2\over r} (\ln G)'
\right.\right. \nonumber \\ & & \left.  \left. 
-(\ln G)'(\ln F)' -{1\over r}(\ln F)'\right] \right\}, \nonumber\\
R^{\hat{\alpha}}_{\hat{\beta}}&=&\delta^\alpha_\beta  {1\over F^2}
\left\{-\left[(\ln G)'+{1\over r}\right] \left[
\ln B+\sum_i \ln C_i -\ln F +(d-1)\left(\ln G +\ln r\right)\right]'
\right. \nonumber \\ & & \left.
-(\ln G)''+{1\over r^2} +(d-2){F^2\over r^2 G^2}\right\}.
\label{sphricciunik}
\eea

These are the Ricci tensor components which will be used when searching
for general black $p$-brane solutions. Here also the ans\"atze made
to solve the equations will strongly depend on the structure of the
expressions above.

\end{document}